\documentclass[11pt]{article}
\pdfoutput=1

\usepackage{jheppub}
\usepackage[T1]{fontenc}
\usepackage[normalem]{ulem}
\usepackage{jheppub}
\usepackage{url,comment}
\usepackage{times}
\usepackage{latexsym}
\usepackage{graphicx, graphics, hyperref, amsmath, amssymb, slashed, color, bbm,amsthm, array}
\usepackage[usenames,dvipsnames]{xcolor}
 \usepackage{subfigure}
\usepackage{mdwlist}
\usepackage{multirow}
\usepackage[e]{esvect}
\usepackage{slashed}

\usepackage{cancel}
\usepackage{pstricks}
\usepackage[toc,page]{appendix}
\usepackage{enumerate}

\usepackage{comment}

\allowdisplaybreaks

\subheader{\hfill \rm OU-HET-1158}

\title{
Effective field theory in light of relative entropy
}
\author{
  Qing-Hong Cao$^{1,2,3}$,
  Naoto Kan$^{4}$
  ,
   and 
  Daiki Ueda$^{3}$
}
\affiliation{\vspace{2mm} $^1$
Department of Physics and State Key Laboratory of Nuclear Physics and Technology, Peking University, Beijing 100871, China \\
$^2$Collaborative Innovation Center of Quantum Matter, Beijing 100871, China
\\
$^3$Center for High Energy Physics, Peking University, Beijing 100871, China \\
$^4$ Department of Physics, Osaka University, Toyonaka, Osaka 560-0043, Japan}
\abstract{
We study constraints on the effective field theory (EFT) from the relative entropy between two theories: we refer to these as target and reference theories.
The consequence of the non-negativity of the relative entropy is investigated by choosing some reference theories for a given target theory involving field theories, quantum mechanical models, etc.
It is found that the constraints on EFTs, e.g., the single massless scalar field with the dimension-eight operator, and SMEFT dimension-eight $SU(N)$ gauge bosonic operators, are consistent with the positivity bounds from the unitarity and causality when the higher-derivative operators are generated by the interaction between heavy and light fields. 
The constraints on Einstein-Maxwell theory with higher-derivative operators from the non-negativity of relative entropy are also investigated.
The constraints on such EFTs from the relative entropy hold under an assumption that perturbative corrections from the interaction involving higher-derivative operators of light fields are not dominant in the EFTs.
The consequence of this study on the weak gravity conjecture and the second law of thermodynamics is also discussed. 
}
\emailAdd{qinghongcao@pku.edu.cn}
\emailAdd{naotokan000@gmail.com}
\emailAdd{ueda@pku.edu.cn}

\begin{document}
\maketitle

\section{Introduction}
\label{sec:introduction}
Effective field theory (EFT) is a fundamental framework for describing low-energy phenomena. 
Information about the high energy regime is transferred to the EFT by integrating out heavy degrees of freedom and can be extracted by determination of the parameters of the EFTs.  
Extracting the nature of the information about the high-energy regime would be a significant scientific goal, and the EFT approach is actively studied from both experimental and theoretical points of view.

From an experimental point of view, there is growing attention to the EFT approach to describe physics beyond the Standard Model (SM). 
The CERN large hadron collider (LHC) has discovered the Higgs boson~\cite{ATLAS:2012yve,CMS:2012qbp} and strengthened the foundation of the SM.
Overwhelming evidence and hints require physics beyond the SM. 
Still, the intensive searches for new particles at the weak scale or heavier have yet to find convincing evidence of such new particles.
In these circumstances, information about the new particles is transferred to the EFT involving the SM fields by integrating out the new particles.
The EFTs such as the Standard Model Effective Field Theory (SMEFT)~\cite{Grzadkowski:2010es,Henning:2015alf,Jenkins:2013zja,Jenkins:2013wua,Alonso:2013hga,Brivio:2017vri,Li:2020gnx,Murphy:2020rsh} are actively studied in this situation.   
Various observables provide constraints on the SMEFT Wilson coefficients~\cite{Han:2004az,Pomarol:2013zra,Corbett:2012ja,Ellis:2014jta,Dumont:2014lca,Corbett:2013pja,Chang:2013cia,Elias-Miro:2013mua,Boos:2013mqa,Ellis:2014dva,Falkowski:2014tna,Berthier:2016tkq,Banerjee:2019twi,Biekotter:2020flu,Efrati:2015eaa,Silvestrini:2018dos,Descotes-Genon:2018foz,Aebischer:2018iyb,Hurth:2019ula,Aebischer:2020dsw,Aoude:2020dwv,Faroughy:2020ina,Falkowski:2015krw,Falkowski:2017pss,Falkowski:2020pma}, which could point us to the UV completion of the SMEFT in the future.

From a theoretical point of view, to exclude particular EFTs, the Weak Gravity Conjecture (WGC)~\cite{Arkani-Hamed:2006emk} (see also \cite{Harlow:2022gzl} for a review) is actively studied.
The string theory yields a vast landscape of four-dimensional EFTs \cite{Taylor:2015xtz}.
In contrast to the landscape, the set of EFTs which cannot be generated from quantum gravity is called the swampland~\cite{Vafa:2005ui}\footnote{Criteria of the swampland are studied as the swampland conjectures~\cite{Banks:2010zn,Arkani-Hamed:2006emk,Ooguri:2006in,Grimm:2018ohb,Ooguri:2016pdq,Freivogel:2016qwc,Obied:2018sgi,Ooguri:2018wrx,Garg:2018reu}. The WGC is one of the swampland conjectures.}.
Predicting quantitative probability distribution about what kind of EFTs belong to the landscape is a challenge.
A simpler version of this challenge is suggested as the WGC, i.e., a rule to distinguish the landscape from the swampland. 
A mild version of the WGC states that the $U(1)$ charge-to-mass ratio of extremally charged black holes is larger than unity in any gravitational EFT that admits a consistent UV completion~\cite{Arkani-Hamed:2006emk,Goon:2019faz}.
Some attempted derivations for this statement have been made using black holes and entropy consideration~\cite{Cheung:2018cwt, Cheung:2019cwi,Goon:2019faz} or positivity bounds~\cite{Adams:2006sv} from unitarity and causality~\cite{Bellazzini:2019xts,Hamada:2018dde}.
In particular, Refs.~\cite{Cheung:2018cwt, Cheung:2019cwi,Goon:2019faz} are based on a positivity of entropy difference between 
Einstein-Maxwell theories with and without perturbative corrections from the higher-dimensional operators.
These works imply a close connection of the positive entropy difference with the positivity bounds from unitarity and causality.
Although the WGC is suggested in the context of quantum gravity, the methodology to exclude particular EFTs that cannot be UV completed is useful in various EFTs with and without gravity, and such a close connection naturally leads us to consider a new approach to constraints on the EFTs. 
Recently, inspired by the connection between the entropy and positivity bounds, a new approach~\cite{Cao:2022iqh} has been proposed to constrain the EFTs by a property of the relative entropy~\cite{10.1214/aoms/1177729694,10.2996/kmj/1138844604,RevModPhys.50.221}.
The relative entropy defined by two probability distribution functions is a non-negative quantity, which is often used as a distance-like concept between the two probability distribution functions.
In Ref.~\cite{Cao:2022iqh}, consequences of the non-negativity of the relative entropy have been studied by defining probability distribution function for various theories.
They mainly considered the distance between theories with and without the interaction between heavy and light degrees of freedom. They showed that the relative entropy yields constraints on some EFTs such as the single massless scalar field with the dimension-eight operator, dimension-eight $SU(N)$ gauge bosonic operators in the SMEFT, and Einstein-Maxwell theory with higher-derivative operators when the higher-derivative operators are generated by the interaction between heavy and light fields.
These arguments for the constraints on the Wilson coefficients hold under an assumption that perturbative corrections from the interaction involving higher-derivative operators of light fields are not dominant in the EFTs.
The connection of the non-negativity of the relative entropy with the WGC and the second law of thermodynamics is also discussed. 
The key role of the distance-like concept in the constraints on the EFTs would lead to an interest in considering the relative entropy between a given theory and various theories.
For example, in Euclidean space, the distances between a given point and various points yield information about the coordinate of the given point.
Similarly, we can evaluate the relative entropy between the given theory and several theories and study their distances.
We refer to the given theory that one wants to extract its information as a {\it target theory}.
Also, we refer to the other theories as {\it reference theories}.
The relative entropy between the target and reference theories would provide various information about the target theory depending on the reference theory.
The appropriate reference theory should be selected depending on the information one wants to extract.
Then, the reference theory generally describes quite different physics from the target theory.
In Ref.~\cite{Cao:2022iqh}, the theory with the interaction between heavy and light degrees of freedom denotes the target theory, and the theory without the interaction is a reference theory.
In this paper, we refer to the reference theory of Ref.~\cite{Cao:2022iqh} as the {\it non-interacting reference theory}.

In this paper, we provide the details of Ref.~\cite{Cao:2022iqh} and update the results in Ref.~\cite{Cao:2022iqh} by considering more target theories and new reference theories.
We provide some new reference theories such as {\it massive free field reference theory}, which also yields the constraints on perturbative corrections from the heavy degrees of freedom to the Euclidean effective action.
Each reference theory would have different advantages depending on the target theory.
For each reference theory, we provide calculation methods of the relative entropy between the target theory and the reference theory.
The relative entropy is calculated by the Euclidean path integral method, and therefore our following discussions are based on the validity of the Euclidean path integral method.
We adopt the top-down approach for consistency checks and evaluate the relative entropy for various target theories containing heavy degrees of freedom.
Also, we adopt the bottom-up approach and investigate the consequence of the non-negativity of relative entropy in EFTs such as the single massless scalar field with the dimension-eight operator, SMEFT dimension-eight $SU(N)$ gauge bosonic operators, and Einstein-Maxwell theory with higher-derivative operators.
In addition, we will discuss connections of this study with some inequality such as causality, the second law of thermodynamics, the WGC, etc.

This paper is organized as follows.
In Sec.~\ref{sec:entr}, we review the details of the main idea of the entropy constraint on EFT of Ref.~\cite{Cao:2022iqh} and provide the procedures to calculate the relative entropy by introducing some new reference theories.
In Sec.~\ref{sec:top} we follow the top-down approach and consider various target theories to perform consistency checks of the entropy constraint.
In Sec.~\ref{sec:bot} we follow the bottom-up approach and provide the bounds on some EFTs from the relative entropy.
In Secs.~\ref{sec:WGC} and \ref{sec:imp}, we discuss connections between the entropy constraint and some inequalities in physics.
We finish with the summary of the paper in Sec.~\ref{sec:sum}.

\section{Entropy constraint on Euclidean effective action}
\label{sec:entr}
In this section, for the sake of being self-contained, we start with a review of the entropy constraint~\cite{Cao:2022iqh} and then update the discussion of Ref.~\cite{Cao:2022iqh}.
Inequalities satisfied by the Euclidean effective actions of the two different theories or systems are provided from the non-negativity of the relative entropy.
In Sec.~\ref{sec:relative}, we explain the main idea of the entropy constraint in two ways: the field theoretical approach and the quantum mechanical approach.
In Sec.~\ref{sec:defTH}, some reference theories addressed in this paper are listed.
In Sec.~\ref{sec:shif}, we focus on the field theory and provide some inequalities satisfied by the Euclidean effective action. 
In Sec.~\ref{sec:sum_en}, we summarize some properties of the entropy constraint.

\subsection{Main idea}
\label{sec:relative}
The relative entropy is defined by two probability distribution functions $\rho_{\rm R}$ and $\rho_{\rm T}$ as follows:
\begin{align}
    S(\rho_{\rm R}||\rho_{\rm T})\equiv {\rm Tr}\left[\rho_{\rm R}\ln \rho_{\rm R}-\rho_{\rm R}\ln \rho_{\rm T} \right],
\end{align}
where $\rho_{\rm R}$ and $\rho_{\rm T}$ satisfy $\rho_{\rm R,T}=\rho_{\rm  R,T}^{\dagger}$, and ${\rm Tr}[\rho_{\rm  R,T}]=1$ because they are probability distribution functions.
One of the important properties of the
relative entropy is non-negativity.
For convenience, we provide brief proof of the non-negativity of the relative entropy.
Consider a convex function $f(x)$, which satisfies $f(x_{\rm R})-f(x_{\rm T})\leq (x_{\rm R}-x_{\rm T})\cdot df(x_{\rm T})/dx$.
For $f(x)\to x \ln x$, $x\to \rho_{\rm R}$, and ${x}_{\rm T}\to {\rho}_{\rm T}$, the definition of convex function yields
\begin{align}
    S(\rho_{\rm R} ||\rho_{\rm T})={\rm Tr}\left[\rho_{\rm R}\ln \rho_{\rm R}-\rho_{\rm R}\ln \rho_{\rm T} \right]\geq 0.\label{eq:non-neg}
\end{align}
Note here that, in Eq.~\eqref{eq:non-neg}, the equality holds if and only if $\rho_{\rm R}=\rho_{\rm T}$ by the definition of convex function.
Therefore, the relative entropy characterizes differences between two probability distribution functions $\rho_{\rm R}$ and $\rho_{\rm T}$ and is often used as a distance between $\rho_{\rm R}$ and $\rho_{\rm T}$ even though it is not a symmetric function of the two sets of probabilities $S(\rho_{\rm R}||\rho_{\rm T})\neq S(\rho_{\rm T}||\rho_{\rm R})$.

\begin{figure*}[t]
\centering
\includegraphics[width=0.8\textwidth]{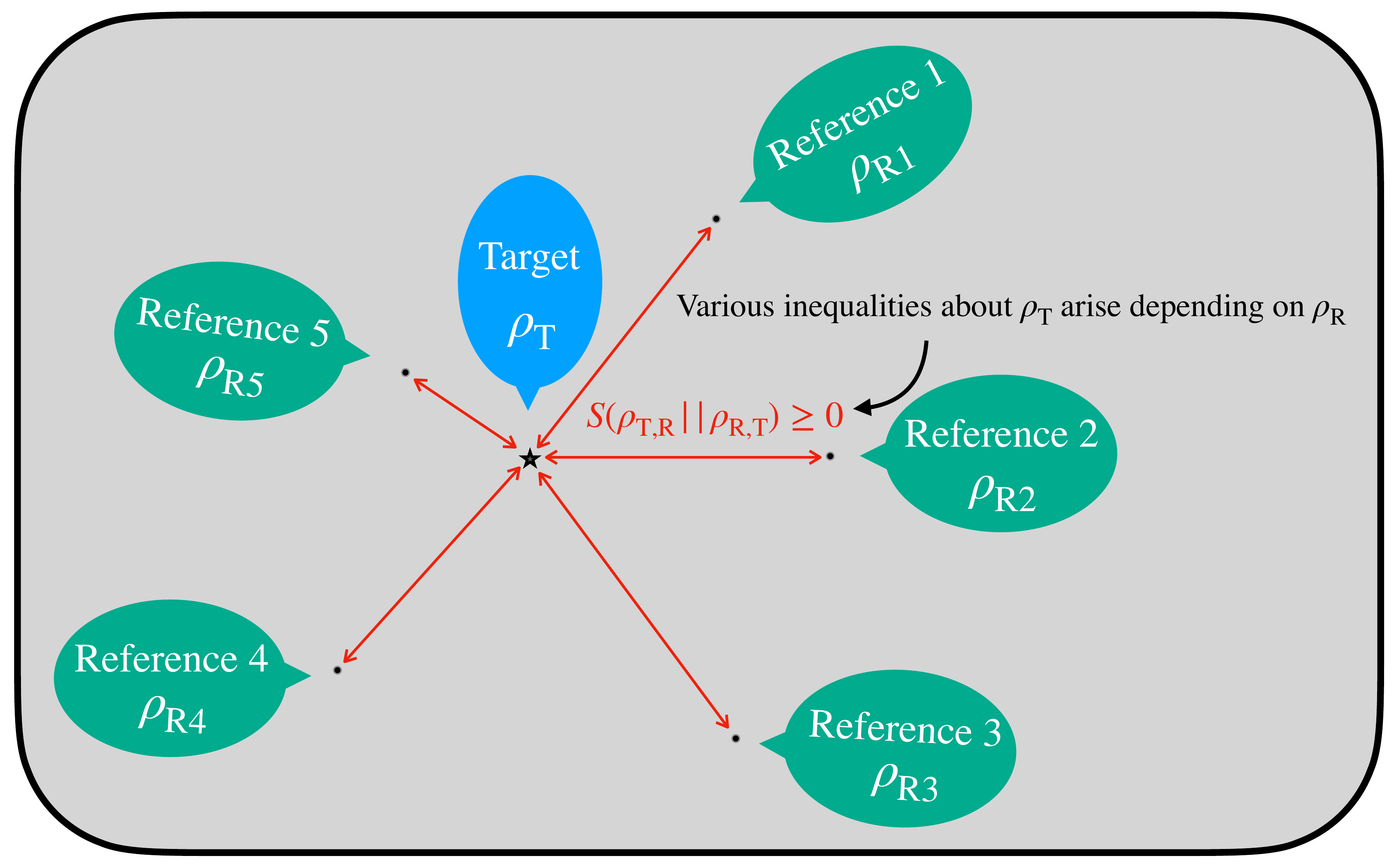}
\caption{
A schematic picture of the main idea.
Relative entropy is a distance-like quantity between two probability distribution functions. 
For a given probability distribution function of the target theory $\rho_{\rm T}$, the relative entropy between $\rho_{\rm T}$ and reference probability distribution functions $\rho_{\rm R}$ yields inequalities about $\rho_{\rm T}$.
Various information about $\rho_{\rm T}$ can be extracted by changing $\rho_{\rm R}$. 
The appropriate reference probability distribution function is selected, depending on the information one wants to extract.
For example, for a reference theory $\rho_{\rm R1}$, the relative entropy represents the constraints on the EFTs generated from the target theory; see Secs.~\ref{sec:top} and \ref{sec:bot}, and for $\rho_{\rm R2}$, the relative entropy denotes the second law of thermodynamics in the system described by the target theory; see Sec.~\ref{sec:second}, and so on. 
\label{fig:idea}
}
\end{figure*}

In the entropy constraint~\cite{Cao:2022iqh}, we define the probability distribution function $\rho_{\rm T}$ for the theory or system $\rm T$ which is the target from which one wants to extract its information.
In this work, we mainly focus on perturbation corrections generated by heavy degrees of freedom of the theory $\rm T$ and attempt to extract the information about their properties. 
We refer to the theory $\rm T$ as the {\it target theory}.
On the other hand, we define $\rho_{\rm R}$ for a reference system $\rm R$, which is an auxiliary system to extract the information about the theory $\rm T$ by comparing $\rm T$ with $\rm R$.
We refer to the theory $\rm R$ as the {\it reference theory}.
The main idea of the entropy constraint is to {\it evaluate the relative entropy between the target theory and suitable reference theory and extract the information about the target theory}.
In Fig.~\ref{fig:idea}, we schematically describes the main idea of the entropy constraint.
In Ref.~\cite{Cao:2022iqh}, a non-interacting theory are chosen as the reference theory, and a connection between the non-negativity of the relative entropy and the positivity bounds on EFTs has been studied.
It should be noted that the reference theory is on the same Hilbert space as the target theory but is generally not relevant to the target theory and describes different physics from the target theory.
The point is that the extracted information from the non-negativity of the relative entropy changes depending on the reference theory, even if the target theory does not change.
In other words, we need to choose the suitable reference theory depending on the information one wants to extract from the target theory.

First, consider the system described by the field theory and evaluate the relative entropy between two theories, $\rm R$ and $\rm T$ by defining the probability distribution function of the two theories.
Although its definition is not unique, in this paper, we mainly consider probability distribution functions defined as follows:
\begin{align}
    P_{\rm R}[\Phi]\equiv\frac{e^{-I_{\rm R}[\Phi]}}{Z_{\rm R}},~~~P_{\rm T}[\Phi]\equiv\frac{e^{-I_{\rm T}[\Phi]}}{Z_{\rm T}},\label{eq:FPR}
\end{align}
where $I_{\rm R}$ and $I_{\rm T}$ are Euclidean actions of the theories, $\rm R$ and $\rm T$, respectively, $\Phi$ denotes degrees of freedom of the field theoretical dynamics, and the partition functions are defined as
\begin{align}
    Z_{\rm R}\equiv\int d[\Phi]e^{-I_{\rm R}[\Phi]},~~~Z_{\rm T}\equiv\int d[\Phi]e^{-I_{\rm T}[\Phi]}.
\end{align}
The relative entropy between theories $\rm R$ and $\rm T$ is calculated as follows:
\begin{align}
    S(P_{\rm R}||P_{\rm T})&\equiv\int d[\Phi]\left[P_{\rm R}[\Phi]\ln P_{\rm R}[\Phi]-P_{\rm R}[\Phi]\ln P_{\rm T}[\Phi] \right]
    \\
    &=\int d[\Phi]\left[P_{\rm R}[\Phi]\left(-\ln Z_{\rm R}- I_{\rm R}[\Phi]\right)-P_{\rm R}[\Phi]\left(-\ln Z_{\rm T}-I_{\rm T}[\Phi]\right) \right]
    \\
    &=-\ln Z_{\rm R}+\ln Z_{\rm T} +\int d[\Phi] P_{\rm R}[\Phi]\left(I_{\rm T}[\Phi]-I_{\rm R}[\Phi]\right)\label{eq:sabmid}
    \\
    &=W_{\rm R} -W_{\rm T}+{\langle I_{\rm T}-I_{\rm R}\rangle}_{\rm R}\geq 0,\label{eq:sABF}
\end{align}
where $W_{\rm R}\equiv -\ln Z_{\rm R}$, $W_{\rm T}\equiv -\ln Z_{\rm T}$, and ${\langle I_{\rm T}-I_{\rm R}\rangle}_{\rm R}\equiv \int d[\Phi] P_{\rm R}[\Phi]\left(I_{\rm T}[\Phi]-I_{\rm R}[\Phi]\right)$.
In the second line, $\ln P_{\rm R,T}=-\ln Z_{\rm R,T}-I_{\rm R,T}[\Phi]$ is used, and $\int d[\Phi]P_{\rm R,T}=1$ yields the third line.
The last line arises from the non-negativity of the relative entropy.
Therefore, it follows from Eq.~\eqref{eq:sABF} that the upper bound on $ W_{\rm T}-W_{\rm R}$ is expressed as
\begin{align}
     {\langle I_{\rm T}-I_{\rm R}\rangle}_{\rm R}\geq W_{\rm T}-W_{\rm R}.\label{eq:uppFT}
\end{align}
Similar to the above procedures, another choice of the relative entropy is calculated as follows:
\begin{align}
    S(P_{\rm T}||P_{\rm R})&\equiv\int d[\Phi]\left[P_{\rm T}[\Phi]\ln P_{\rm T}[\Phi]-P_{\rm T}[\Phi]\ln P_{\rm R}[\Phi] \right]\notag
    \\
    &=\int d[\Phi]\left[
    P_{\rm T}[\Phi]\left(-\ln Z_{\rm T} -I_{\rm T}[\Phi]\right)
    -P_{\rm T}[\Phi]\left(-\ln Z_{\rm R} -I_{\rm R}[\Phi] \right)
    \right]\notag
    \\
    &=-\ln Z_{\rm T} +\ln Z_{\rm R} -\int d[\Phi] P_{\rm T}[\Phi]\left(I_{\rm T}[\Phi]-I_{\rm R}[\Phi]\right)\notag
    \\
    &=W_{\rm T} -W_{\rm R} -{\langle I_{\rm T}-I_{\rm R} \rangle}_{\rm T}\geq 0,\label{eq:sBAF}
\end{align}
where ${\langle I_{\rm T}-I_{\rm R} \rangle}_{\rm T}\equiv \int d[\Phi] P_{\rm T}[\Phi]\left(I_{\rm T}[\Phi]-I_{\rm R}[\Phi]\right)$.
Then, Eq.~\eqref{eq:sBAF} yields the lower bound on $W_{\rm T}-W_{\rm R}$ as
\begin{align}
    W_{\rm T} -W_{\rm R} \geq {\langle I_{\rm T}-I_{\rm R} \rangle}_{\rm T}.\label{eq:lowFT}
\end{align}
The Euclidean actions $I_{\rm R}$ and $I_{\rm T}$ are determined by the Wick-rotated Lagrangian and boundary conditions, and the above explanations are valid even for finite temperature systems.
Our strategy is to extract the information about $I_{\rm T}$ from the inequalities of \eqref{eq:uppFT} and \eqref{eq:lowFT} by choosing the suitable reference theory $I_{\rm R}$.

Next, consider the system described by quantum mechanical dynamics and evaluate the relative entropy between the two systems $\rm R$ and $\rm T$.
The above discussions do not rely on the Lorentz symmetry, so similar inequalities are derived for the quantum mechanical dynamics. 
Equation~\eqref{eq:FPR} corresponds to the following density operators.
\begin{align}
    \rho_{\rm R} \equiv\frac{e^{-\beta H_{\rm R}}}{Z_{\rm R}},~~~\rho_{\rm T} \equiv\frac{e^{-\beta H_{\rm T}}}{Z_{\rm T}},\label{eq:quP}
\end{align}
where $H_{\rm R}$ and $H_{\rm T}$ are Hamiltonians of $\rm R$ and $\rm T$, respectively, $\beta$ is an inverse temperature of the systems, and the partition functions are defined as
\begin{align}
    Z_{\rm R}\equiv{\rm Tr}[e^{-\beta H_{\rm R}}],~~~Z_{\rm T}\equiv{\rm Tr}[e^{-\beta H_{\rm T}}].\label{eq:quPar}
\end{align}
The above definition of the probability distribution functions of Eq.~\eqref{eq:quP} is one example, and other choices are also possible.
We will show different choices in a later section.
Equation~\eqref{eq:non-neg}, \eqref{eq:quP}, and \eqref{eq:quPar} yield the relative entropy between theories $\rm R$ and $\rm T$ as follows:
\begin{align}
    S(\rho_{\rm R}||\rho_{\rm T})&\equiv{\rm Tr}\left[\rho_{\rm R}\ln \rho_{\rm R}-\rho_{\rm R}\ln \rho_{\rm T}\right]\notag
    \\
    &={\rm Tr}\left[\rho_{\rm R}\left(-\ln Z_{\rm R}-\beta H_{\rm R}\right)-\rho_{\rm R}\left(-\ln Z_{\rm T}-\beta H_{\rm T}\right)\right]\notag
    \\
    &=-\ln Z_{\rm R} +\ln Z_{\rm T} +\beta\cdot{\rm Tr}\left[\rho_{\rm R} (H_{\rm T}-H_{\rm R})\right]\notag
    \\
    &=W_{\rm R}-W_{\rm T}+\beta\cdot{\langle H_{\rm T}-H_{\rm R}\rangle}_{\rm R}\geq 0,\label{eq:sABQ}
\end{align}
where $W_{\rm R}= -\ln Z_{\rm R}$, $W_{\rm T}= -\ln Z_{\rm T}$, and ${\langle H_{\rm T}-H_{\rm R}\rangle}_{\rm R}\equiv {\rm Tr}\left[\rho_{\rm R} (H_{\rm T}-H_{\rm R})\right]$.
In the second line, $\ln \rho_{\rm R,T}=-\ln Z_{\rm R,T}-\beta H_{\rm R,T}$ is used, and ${\rm Tr} [\rho_{\rm R,T}]=1$ yields the third line.
The last line arises from the non-negativity of the relative entropy.
Similar to the field theoretical approach, Eq.~\eqref{eq:sABQ} yields the upper bound on $W_{\rm T}-W_{\rm R}$ as
\begin{align}
    \beta\cdot{\langle H_{\rm T}-H_{\rm R} \rangle}_{\rm R} \geq W_{\rm T}-W_{\rm R}.
\end{align}
Another choice of the relative entropy is also calculated as follows:
\begin{align}
    S(\rho_{\rm T}||\rho_{\rm R})&\equiv{\rm Tr}\left[\rho_{\rm T}\ln \rho_{\rm T}-\rho_{\rm T}\ln \rho_{\rm R} \right]\notag
    \\
    &={\rm Tr}\left[\rho_{\rm T} \left(-\ln Z_{\rm T}-\beta H_{\rm T}\right)-\rho_{\rm T} \left(-\ln Z_{\rm R} -\beta H_{\rm R}\right) \right]\notag
    \\
    &=-\ln Z_{\rm T} +\ln Z_{\rm R} -\beta\cdot {\rm Tr}\left[\rho_{\rm T} \left(H_{\rm T}-H_{\rm R}\right)\right]\notag
    \\
    &=W_{\rm T}-W_{\rm R} -\beta\cdot {\langle H_{\rm T}-H_{\rm R}\rangle}_{\rm T}\geq 0,\label{eq:sBAQ}
\end{align}
where ${\langle H_{\rm T}-H_{\rm R}\rangle}_{\rm T}\equiv {\rm Tr}\left[\rho_{\rm T} \left(H_{\rm T}-H_{\rm R}\right)\right]$.
Eq.~\eqref{eq:sBAQ} yields the lower bound on $W_{\rm T}-W_{\rm R}$ as
\begin{align}
    W_{\rm T}-W_{\rm R}\geq \beta\cdot {\langle H_{\rm T}-H_{\rm R}\rangle}_{\rm T}.
\end{align}
Consequently, we obtain the lower and upper bounds on $W_{\rm T}-W_{\rm R}$ from the non-negativity of the relative entropy by both the field theoretical and quantum mechanical approaches.
In the next section, we provide some examples of the reference theory to derive constraints on perturbative corrections from heavy degrees of freedom to the Euclidean effective action of the target theory.

\subsection{Examples of reference theories}
\label{sec:defTH}
For the target theories consisting of heavy and light fields, we consider EFTs generated by integrating out the heavy fields.
Throughout this section, $\Phi$'s and $\phi$'s denote the heavy and light fields in the field theoretical dynamics, respectively. 
The Euclidean action of the target theory is expressed as follows:
\begin{align}
    I_{\rm T}[\phi,\Phi]=I_0 [\phi,\Phi]+I_{\rm I} [\phi,\Phi],\label{eq:tarTH}
\end{align}
where $I_0$ does not include the interaction between $\Phi$'s and $\phi$'s, and $I_{\rm I}$ is the interacting term.
From Eq.~\eqref{eq:FPR}, for the background fields, $\phi$'s, the probability distribution functions of the theories $\rm R$ and $\rm T$ are defined as follows:
\begin{align}
    P_{\rm R}[\Phi]\equiv\frac{e^{-I_{\rm R}[\phi,\Phi]}}{Z_{\rm R}[\phi]},~~~P_{\rm T}[\Phi]\equiv\frac{e^{-I_{\rm T}[\phi,\Phi]}}{Z_{\rm T}[\phi]},
\end{align}
with the partition functions,
\begin{align}
    Z_{\rm R}[\phi]\equiv\int d[\Phi]e^{-I_{\rm R}[\phi,\Phi]},~~~Z_{\rm T}[\phi]\equiv\int d[\Phi]e^{-I_{\rm T}[\phi,\Phi]}.\label{eq:parBG}
\end{align}
The relative entropy between $P_{\rm R}$ and $P_{\rm T}$ is calculated as in Eq.~\eqref{eq:sABF} and \eqref{eq:sBAF}.
The path integral is performed only over the dynamical heavy field because $\phi$ is the background field.

Even for dynamical light fields $\phi$'s\footnote{We have to be careful with the validity of the Euclidean path integral over $\phi$'s. When we treat $\phi$'s as the dynamical fields, the path integral needs to be performed around a local minimum. If not, the saddle point approximation breaks down. We need not require such validity if $\phi$'s are the background fields.
}, the probability distribution functions of the theories $\rm R$ and $\rm T$ can be defined as follows:
\begin{align}
    P_{\rm R}[\phi,\Phi]\equiv\frac{e^{-I_{\rm R}[\phi,\Phi]}}{Z_{\rm R}[\widetilde{\phi}_{\rm R},\widetilde{\Phi}_{\rm R}]},~~~P_{\rm T}[\phi,\Phi]\equiv\frac{e^{-I_{\rm T}[\phi,\Phi]}}{Z_{\rm T}[\widetilde{\phi}_{\rm T},\widetilde{\Phi}_{\rm T}]},\label{eq:proABdyn}
\end{align}
where $(\widetilde{\phi}_{\rm R},\widetilde{\Phi}_{\rm R})$ and $(\widetilde{\phi}_{\rm T},\widetilde{\Phi}_{\rm T})$ are sets of classical solutions of $I_{\rm R}$ and $I_{\rm T}$, respectively.
The partition functions are given by
\begin{align}
    Z_{\rm R}[\widetilde{\phi}_{\rm R},\widetilde{\Phi}_{\rm R}]\equiv\int d[\phi]d[\Phi]e^{-I_{\rm R}[\phi,\Phi]},~~~Z_{\rm T}[\widetilde{\phi}_{\rm T},\widetilde{\Phi}_{\rm T}]\equiv\int d[\phi]d[\Phi]e^{-I_{\rm T}[\phi,\Phi]}.\label{eq:parABdyn}
\end{align}
The relative entropy is calculated as in Eq.~\eqref{eq:sABF} by replacing $d[\Phi]$ with $d[\phi]d[\Phi]$ as follows:
\begin{align}
    S(P_{\rm R}||P_{\rm T})&\equiv\int d[\phi]d[\Phi] \left[
    P_{\rm R}[\phi,\Phi] \ln P_{\rm R}[\phi,\Phi]-P_{\rm R}[\phi,\Phi] \ln P_{\rm T}[\phi,\Phi] 
    \right]\notag
    \\
    &=\int d[\phi]d[\Phi] \left[
    P_{\rm R}[\phi,\Phi] \left(-\ln Z_{\rm R} -I_{\rm R}[\phi,\Phi]\right)-P_{\rm R}[\phi,\Phi] \left(-\ln Z_{\rm T}-I_{\rm T}[\phi,\Phi]\right)
    \right]\notag
    \\
    &=-\ln Z_{\rm R} +\ln Z_{\rm T}+\int d[\phi]d[\Phi] P_{\rm R}[\phi,\Phi]\left(I_{\rm T}[\phi,\Phi]-I_{\rm R}[\phi,\Phi]\right)\notag
    \\
    &=W_{\rm R}-W_{\rm T} +{\langle I_{\rm T}-I_{\rm R}\rangle}_{\rm R}\geq 0,\label{eq:dyrel1}
\end{align}
where $Z_{\rm R}\equiv Z_{\rm R}[\widetilde{\phi}_{\rm R},\widetilde{\Phi}_{\rm R}]$, $Z_{\rm T}\equiv Z_{\rm T}[\widetilde{\phi}_{\rm T},\widetilde{\Phi}_{\rm T}]$, $W_{\rm R}\equiv -\ln Z_{\rm R}[\widetilde{\phi}_{\rm R},\widetilde{\Phi}_{\rm R}] $, $W_{\rm T}\equiv -\ln Z_{\rm T}[\widetilde{\phi}_{\rm T},\widetilde{\Phi}_{\rm T}] $, and  
\begin{align}
    {\langle I_{\rm T}-I_{\rm R}\rangle}_{\rm R}\equiv \int d[\phi]d[\Phi] P_{\rm R}[\phi,\Phi]\left(I_{\rm T}[\phi,\Phi]-I_{\rm R}[\phi,\Phi]\right).
\end{align}
For the dynamical light fields, the relative entropy of \eqref{eq:sBAF} is given by
\begin{align}
    S(P_{\rm T}||P_{\rm R})&\equiv\int d[\phi]d[\Phi] \left[
    P_{\rm T}[\phi,\Phi]\ln P_{\rm T}[\phi,\Phi]-P_{\rm T}[\phi,\Phi] \ln P_{\rm R}[\phi,\Phi]
    \right]\notag
    \\
    &=\int d[\phi]d[\Phi] \left[
    P_{\rm T}[\phi,\Phi]\left(-\ln Z_{\rm T}-I_{\rm T}[\phi,\Phi] \right)
    -P_{\rm T}[\phi,\Phi] \left(-\ln Z_{\rm R} -I_{\rm R}[\phi,\Phi]\right)
    \right]\notag
    \\
    &=-\ln Z_{\rm T} +\ln Z_{\rm R} -\int d[\phi]d[\Phi] P_{\rm T}[\phi,\Phi]\left(I_{\rm T}[\phi,\Phi]-I_{\rm R}[\phi,\Phi] \right)\notag
    \\
    &=W_{\rm T}-W_{\rm R}-{\langle I_{\rm T}-I_{\rm R}\rangle}_{\rm T}\geq 0,\label{eq:dyrel2}
\end{align}
where
\begin{align}
    {\langle I_{\rm T}-I_{\rm R}\rangle}_{\rm T}\equiv \int d[\phi]d[\Phi] P_{\rm T}[\phi,\Phi]\left(I_{\rm T}[\phi,\Phi]-I_{\rm R}[\phi,\Phi] \right).
\end{align}
Equations~\eqref{eq:dyrel1} and \eqref{eq:dyrel2} are the same as the form of Eq.~\eqref{eq:sABF} and \eqref{eq:sBAF}, respectively, which do not depend on whether the light fields are dynamical or not.
To clarify procedures of the wave function renormalization of the light fields, we assume the dynamical light fields in Sec.~\ref{sec:Massle},~\ref{sec:EH}, and~\ref{sec:bot} but the light background fields in the other sections.

In the following, we list some reference theories to derive information about the target theory.
The first three examples are relevant to the constraints on the corrections to $W_{\rm T}$ from $\Phi$'s.
In particular, the first reference theory plays an important role in deriving the constraints on EFTs in the bottom-up approach, which is discussed in Sec.~\ref{sec:bot}.
The last one is connected with the second law of thermodynamics; see Sec.~\ref{sec:second}.

\begin{itemize}

    \item {Non-interacting reference theory ---}
In Ref.~\cite{Cao:2022iqh}, the Euclidean action of the reference theory is defined as a non-interacting theory as follows:
\begin{align}
    I_{\rm NI}[\phi,\Phi]\equiv I_0[\phi,\Phi],
\end{align}
where $I_0$ is the same as the first term of Eq.~\eqref{eq:tarTH}.
We refer to this reference theory as the non-interacting reference theory (NIRT).
For the background light fields $\phi$'s, the probability distribution function of the NIRT is defined as
\begin{align}
    P_{\rm NI}[\Phi]\equiv\frac{e^{-I_{\rm NI}[\phi,\Phi]}}{Z_{\rm NI}[\phi]},
\end{align}
with the partition function,
\begin{align}
    Z_{\rm NI}[\phi]\equiv\int d[\Phi]e^{-I_{\rm NI}[\phi,\Phi]}.
\end{align}
For the dynamical light fields $\phi$'s, we defne the probability distribution function of the NIRT as
\begin{align}
    P_{\rm NI}[\phi,\Phi]\equiv\frac{e^{-I_{\rm NI}[\phi,\Phi]}}{Z_{\rm NI}[\widetilde{\phi}_{\rm NI},\widetilde{\Phi}_{\rm NI}]},
\end{align}
where the partition function is given by
\begin{align}
    Z_{\rm NI}[\widetilde{\phi}_{\rm NI},\widetilde{\Phi}_{\rm NI}]\equiv\int d[\phi]d[\Phi]e^{-I_{\rm NI}[\phi,\Phi]},
\end{align}
where $\widetilde{\phi}_{\rm NI}$ and $\widetilde{\Phi}_{\rm NI}$ are classical solutions of $I_{\rm NI}$.

    \item{Massive free field reference theory ---}
We propose a reference theory defined by an Euclidean action,
\begin{align}
    I_{\rm MF}[\phi,\Phi]\equiv I_{\phi}[\phi]+I_{\Phi}[\Phi],
\end{align}
where $I_{\phi}[\phi]\equiv I_{0}[\phi,0]$.
$I_{\Phi}$ denotes only the kinetic and mass terms of $\Phi$, and its mass term is the same as that of $I_0$.
In contrast to the NIRT, the self-interacting terms of $\Phi$ do not include in $I_{\rm MF}$.
We refer this reference theory as the massive free field reference theory (MFFRT).
When $\phi$'s are assumed to be background fields, we perform the path integral only over $\Phi$'s.
Then, the probability distribution functions of the MFFRT is defined as follows:
\begin{align}
    P_{\rm MF}[\Phi]&\equiv\frac{e^{-I_{\rm MF}[\phi,\Phi]}}{Z_{\rm MF}[\phi]},
\end{align}
with the partition functions,
\begin{align}
    Z_{\rm MF}[\phi]&\equiv\int d[\Phi]e^{-I_{\rm MF}[\phi,\Phi]},
\end{align}
where solutions of $I_{\rm MF}$ for the heavy fields take zero values.
For the dynamical light fields $\phi$'s, the probability distribution functions of the MFFRT is defined as follows:
\begin{align}
    P_{\rm MF}[\phi,\Phi]&\equiv\frac{e^{-I_{\rm MF}[\phi,\Phi]}}{Z_{\rm MF}[\widetilde{\phi}_{\rm MF},\widetilde{\Phi}_{\rm MF}]},
\end{align}
where the partition function is given by
\begin{align}
    Z_{\rm MF}[\widetilde{\phi}_{\rm MF},\widetilde{\Phi}_{\rm MF}]\equiv\int d[\phi]d[\Phi]e^{-I_{\rm MF}[\phi,\Phi]},
\end{align}
where $\widetilde{\phi}_{\rm MF}$ and $\widetilde{\Phi}_{\rm MF}$ are classical solutions of $I_{\rm MF}$.
$\widetilde{\Phi}_{\rm MF}$ can take zero values by absorbing the plane wave solutions into quantum fluctuations.

    \item {Infinite heavy mass reference theory ---}
    As the reference theory, we consider a theory with the same form of the action as the target theory with the infinite mass of $\Phi$,
    \begin{align}
        I_{\rm IH}[\phi,\Phi]\equiv\lim_{m_{\Phi}\to \infty} I_{\rm T}[\phi,\Phi].
    \end{align}
We refer to this reference theory as the infinite heavy mass reference theory (IHMRT).
Throughout this work, for the IHMRT, we focus on the background light fields and perform the path integral only over $\Phi$'s.
The probability distribution of the IHMRT is defined as,
\begin{align}
    P_{\rm IH}[\Phi]\equiv\frac{e^{-I_{\rm IH}[\phi,\Phi]}}{Z_{\rm IH}[\phi]},
\end{align}
with the partition function,
\begin{align}
    Z_{\rm IH}[\phi]\equiv\int d[\Phi]e^{-I_{\rm IH}[\phi,\Phi]}.
\end{align}
In Sec.~\ref{sec:top}, we study the IHMRT only in the tree level calculations. 

\item Thermal reference theory --- Consider a system consisting of a thermodynamic system $\rm S$ and a heat bath system $\rm B$.
We assume both heavy and light degrees of freedom are included in the thermodynamic system $\rm S$.
The Hamiltonian of the whole system is expressed as $ H_{\rm T}= H_{\rm S} +H_{\rm B} +H_{\rm SB}$, where $H_{\rm S}$ is the Hamiltonian of the thermodynamic system $\rm S$, and $H_{\rm B}$ is that of the heat bath system $\rm B$.
The interacting term $H_{\rm SB}$ denotes the interaction between $\rm S$ and $\rm B$ and can generally depends on time.
At the initial time, assume the quantum state of the whole system is expressed as
\begin{align}
   \rho_{\rm T}\equiv \rho_{\rm ini}=\rho_{\rm ini,S} \otimes e^{-\beta H_{\rm B}}/Z_{\rm B}(\beta),
\end{align}
where $\rho_{\rm ini,S}$ is the initial state of $\rm S$, and $e^{-\beta H_{\rm B}}/Z_{\rm B}(\beta)$ is that of $\rm B$, $\beta$ is an inverse temperature of the heat bath system at the initial time, and $Z_{\rm B}(\beta)\equiv {\rm Tr}_{\rm B}[e^{-\beta H_{\rm B}}]$ is defined by tracing over the heat bath degrees of freedom.
Note here that the specific form of $\rho_{\rm ini,S}$ is irrelevant to this discussion.
We assume the probability distribution of the target theory is defined by the initial state of the whole system.
After the time evolution described by a unitary operator $U$, the final state of the whole system is expressed as
\begin{align}
     \rho_{\rm fin}\equiv U \rho_{\rm ini} U^{\dagger}.
\end{align}
By tracing out the heat bath degrees of freedom, the final state of S is calculated as
\begin{align}
    \rho_{\rm fin,S}\equiv {\rm Tr}_{\rm B}[\rho_{\rm fin}].
\end{align}
Then, define the reference probability distribution function as follows:
\begin{align}
    \rho_{\rm R}\equiv U^{\dagger}\rho_{\rm fin,S}\otimes e^{-\beta H_{\rm B}}/Z_{\rm B}(\beta) U.
\end{align}
We refer to this reference theory as the thermal reference theory in this work.
The thermal reference theory is useful to see a connection between the non-negativity of relative entropy and the second law of thermodynamics~\cite{2000cond.mat..9244T,2012}.
In the other reference theories discussed before, it is supposed that the target and reference theory do not include the heat bath degrees of freedom.
However, in Sec.~\ref{sec:second}, we will demonstrate that the relative entropy between the target and reference theories does not change even if the heat bath degrees of freedom are added to both the theories.

\end{itemize}

One of our main interests is the constraints on EFT generated by the target theory, and the above first three reference theories are mainly considered in the following sections.
Here, we would like to emphasize that the above definitions of the probability distributions of the target theory and reference theory are not unique, and are part of examples.

\subsection{Inequalities satisfied by Euclidean effective action in field theory}
\label{sec:shif}

We have discussed the general properties of the relative entropy between two probability distribution functions so far.
In this section, we focus on the systems described by field theoretical dynamics and provide inequalities satisfied by heavy field corrections to the Euclidean effective action by using the non-negativity of the relative entropy.
For each reference theory, we provide inequalities satisfied by the Euclidean effective action of the target theory in the following.

\subsubsection{Non-interacting reference theory}
\label{sec:noninref}
\begin{figure*}[t]
\centering
\includegraphics[width=0.65\textwidth]{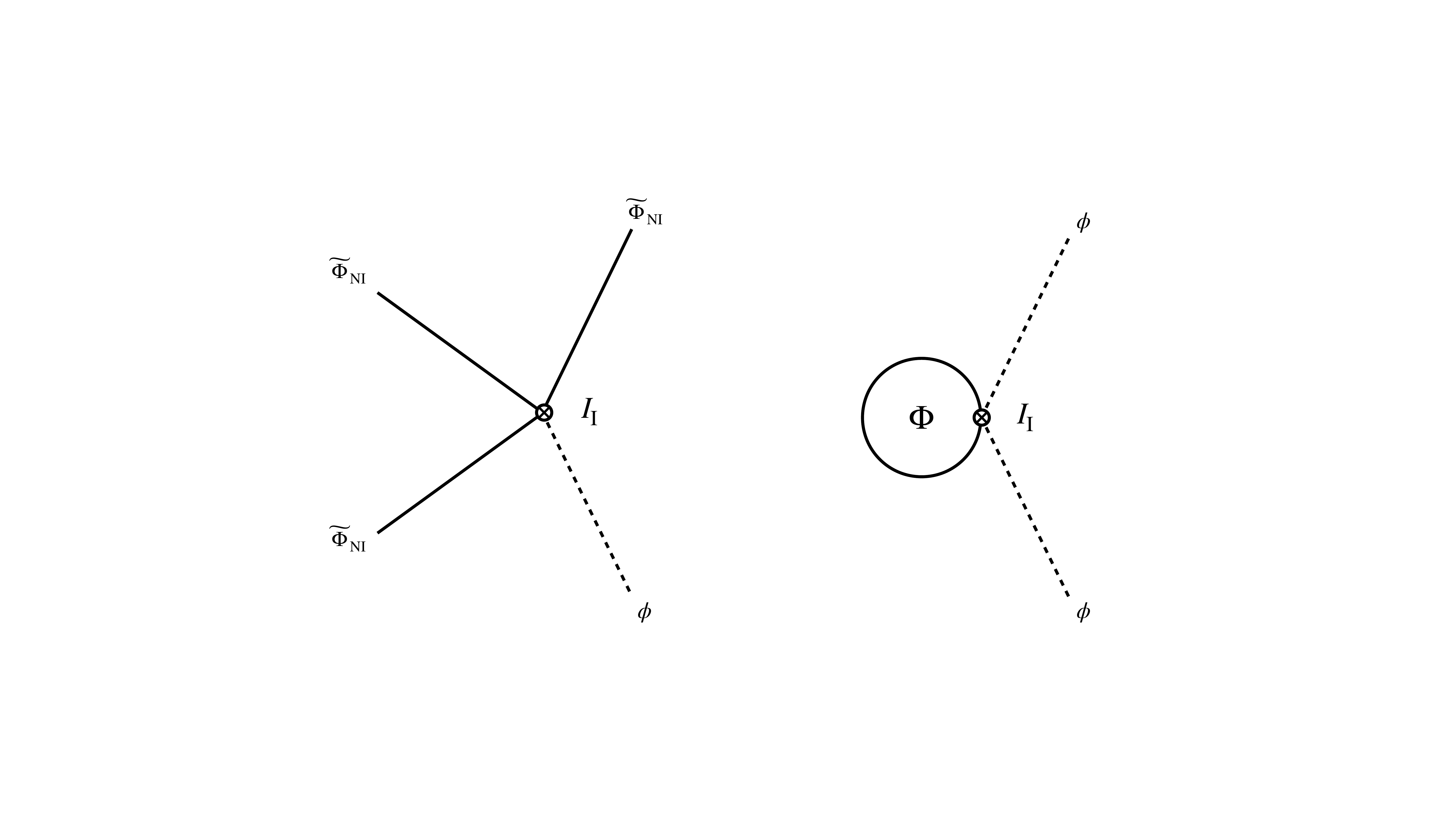}
\caption{
Feynman diagrams for ${\langle I_{\rm I}\rangle}_{\rm NI}$ at the tree and one-loop level.
The vertices denote the interacting term $I_{\rm I}$. 
The solid and dashed lines are the heavy and the background light fields, respectively. 
The left and right diagrams denote tree and one-loop level corrections, respectively.
In the non-interacting reference theory, the classical solution of the heavy field may take a non-zero value because of the linear term of $\Phi$, and the left diagram generally does not vanish but can be eliminated by redefinitions of $\Phi$; see Sec.~\ref{sec:lin}.
\label{fig:NIdiag1}
}
\end{figure*}

By introducing an auxiliary parameter $g$, we define,
\begin{align}
I_g[\phi,\Phi]\equiv I_0[\phi,\Phi]+g\cdot I_{\rm I}[\phi,\Phi].\label{eq:Ignon}
\end{align}
By changing the parameter $g$, the target and reference theories are given as,
\begin{align}
    &I_{\rm T}[\phi,\Phi]=\lim_{g\to 1} I_g[\phi,\Phi],~~~I_{\rm NI}[\phi,\Phi]=\lim_{g\to 0} I_g[\phi,\Phi].
\end{align}
For the background light fields, the partition function and effective action of $I_g$ are respectively defined as follows:
\begin{align}
    &Z_{g}[{\phi}]\equiv\int d[\Phi] e^{-I_g[\phi,\Phi]},
    \\
    &W_{g}[{\phi}]\equiv-\ln Z_{g}[{\phi}],
\end{align}
From Eqs.~\eqref{eq:sABF} and \eqref{eq:sBAF}, by defining a probability distribution function,
\begin{align}
    P_g[\Phi]\equiv \frac{e^{-I_g[\phi,\Phi]}}{Z_g[\phi]},
\end{align}
the relative entropy between $P_0$ and $P_g$ is calculated as follows:
\begin{align}
    S(P_0||P_g)&=W_0[{\phi}]-W_g[{\phi}]+ g\cdot {\langle I_{\rm I}\rangle}_{g=0}\geq 0\quad\Rightarrow\quad W_g[\phi]-W_0[\phi]\leq g\cdot {\langle I_{\rm I}\rangle}_{g=0},\label{eq:upg}
    \\
    S(P_g||P_0)&=W_g[{\phi}]-W_0[{\phi}]- g\cdot {\langle I_{\rm I}\rangle}_{g}\geq 0\quad\Rightarrow\quad g\cdot {\langle I_{\rm I}\rangle}_{g} \leq  W_g[{\phi}]-W_0[{\phi}],\label{eq:lowg}
\end{align}
with 
\begin{align}
    {\langle I_{\rm I}\rangle}_{g}\equiv \int d[\Phi] I_{\rm I}[\phi,\Phi]\cdot \frac{e^{-I_g[\phi,\Phi]}}{Z_{g}[{\phi}]}=\frac{\partial W_{g}[{\phi}]}{\partial g},
\end{align}
where the partial derivative means differentiating by $g$ while keeping ${\phi}$. 
Equations~\eqref{eq:upg} and \eqref{eq:lowg} yield
\begin{align}
    g\cdot {\langle I_{\rm I}\rangle}_{g=0}\geq W_g[\phi]-W_0[\phi]\geq g\cdot {\langle I_{\rm I}\rangle}_{g} \quad\Rightarrow\quad {\langle I_{\rm I}\rangle}_{\rm NI}\geq W_{\rm T}[\phi]-W_{\rm NI}[\phi]\geq {\langle I_{\rm I}\rangle}_{\rm T}~~{\rm for}~g=1.\label{eq:uplowBG}
\end{align}
Here, we used $W_{\rm T}=W_{g=1}$, $W_{\rm NI}=W_0$, and the following relations.
\begin{align}
    {\langle I_{\rm I}\rangle}_{g=0}&=\left(\frac{\partial W_{g}[\phi]}{\partial g}\right)_{g=0}\notag
    \\
    &=\int d[\Phi] I_{\rm I}[\phi,\Phi]\cdot \frac{e^{-I_{0}[\phi,\Phi]}}{Z_{0}[{\phi}]}\notag
    \\
    &=\int d[\Phi] I_{\rm I}[\phi,\Phi]\cdot \frac{e^{-I_{\rm NI}[\phi,\Phi]}}{Z_{\rm NI}[{\phi}]}\notag
    \\
    &={\langle I_{\rm I}\rangle}_{\rm NI},\label{eq:g0NI}
    \\
    {\langle I_{\rm I}\rangle}_{g=1}&=\left(\frac{\partial W_{g}[\phi]}{\partial g}\right)_{g=1}\notag
    \\
    &=\int d[\Phi] I_{\rm I}[\phi,\Phi]\cdot \frac{e^{-I_{g=1}[\phi,\Phi]}}{Z_{g=1}[\phi]}\notag
    \\
    &=\int d[\Phi] I_{\rm I}[\phi,\Phi]\cdot \frac{e^{-I_{\rm T}[\phi,\Phi]}}{Z_{\rm T}[\phi]}\notag
    \\
    &={\langle I_{\rm I}\rangle}_{\rm T},\label{eq:g1NI}
\end{align}
where, 
in particular, Eq.~\eqref{eq:g0NI} denotes the Feynman diagrams of Fig.~\ref{fig:NIdiag1}.
Note here that the Euclidean effective action $W_{\rm NI}[\phi]$ generally includes the corrections from the self-interacting terms of $\Phi$.
For ease of understanding, let us schematically express the Euclidean effective actions as follows:
\begin{align}
    W_{\rm NI}[{\phi}]&=I_0[{\phi},0]+({\rm vacuum~energy})',
    \\
    W_{\rm T}[{\phi}]&=I_0[{\phi},0]+({\rm vacuum~energy})'+({\rm renormalizable~terms~of~{\phi}})+({\rm non\text{-}renormalizable~terms~of~{\phi}}),\label{eq:schNI}
\end{align}
where $I_{\rm T}[{\phi},0]=I_0[{\phi},0]$ and $I_{\rm NI}[{\phi},0]=I_0[{\phi},0]$ are used.
Here, $({\rm vacuum~energy})'$ denotes the vacuum energy coming from the dynamical fields, which may include corrections from the self-interacting terms of $\Phi$.
Note that $({\rm vacuum~energy})'$ is independent of the background fields $\phi$'s.
Also, the third term of the right-hand side of Eq.~\eqref{eq:schNI} denotes the corrections from the interacting term $I_{\rm I}$ to the renormalizable terms of ${\phi}$, and their fourth term is the corrections from the interacting term $I_{\rm I}$ to the non-renormalizable terms of ${\phi}$.
Therefore, $W_{\rm T}[\phi]-W_{\rm NI}[\phi]$ represents the perturbative corrections from the interaction between heavy and light degrees of freedom to the Euclidean effective action of the target theory other than the vacuum energy as follows:
\begin{align}
    W_{\rm T}[\phi]-W_{\rm NI}[\phi]=({\rm renormalizable~terms~of~\phi})+({\rm non\text{-}renormalizable~terms~of~\phi}).\label{eq:schWBA}
\end{align}
The point is that the right-hand side of Eq.~\eqref{eq:schWBA} does not include $\phi$ independent terms.
Equations~\eqref{eq:uplowBG} and \eqref{eq:schWBA} imply that {\it the expectation values of the interaction yield bounds on the perturbative corrections from the interacting term $I_{\rm I}$ to the Euclidean effective action of the target theory.}
For example, $W_{\rm T}[\phi]-W_{\rm NI}[\phi]$ is increased in the theory with ${\langle I_{\rm I}\rangle}_{\rm T}\geq 0$ but decreased in the theory with ${\langle I_{\rm I}\rangle}_{\rm NI}\leq 0$.
In Ref.~\cite{Cao:2022iqh}, the theory satisfying ${\langle I_{\rm I}\rangle}_{\rm NI}\leq 0$ is referred to as the {\it non-positive interacting} theory.

For convenience, we explain the meaning of the upper bound of Eq.~\eqref{eq:uplowBG}.
Expand $W_{g}$ with respect to $g$ as follows:
\begin{align}
    W_{g}[\phi]&=W_{0}[\phi]+g\cdot \left(\frac{\partial W_{g}[\phi] }{\partial g}\right)_{g=0}
    +\frac{g^2}{2}\cdot \left(\frac{\partial^2 W_{g}[\phi] }{\partial g^2}\right)_{g=0}+\mathcal{O}(g^3),\notag
    \\
    &=W_{{\rm NI}}[\phi]+g\cdot {\langle I_{\rm I}\rangle}_{\rm NI}+\frac{g^2}{2}\cdot \left(\frac{\partial^2 W_{g}[\phi] }{\partial g^2}\right)_{g=0}+\mathcal{O}(g^3),\notag
    \\
    &=W_{{\rm NI}}[\phi]+g\cdot {\langle I_{\rm I}\rangle}_{\rm NI}+ \Delta W_{g}^{(2)},
    \label{eq:niexp}
\end{align}
where $W_{0}[\phi]=W_{{\rm NI}}[\phi]$ is used, and we defined the corrections for the second or higher order for $g$ as
\begin{align}
    \Delta W_{g}^{(2)}\equiv \frac{g^2}{2}\cdot \left(\frac{\partial^2 W_{g}}{\partial g^2}\right)_{g=0}+\mathcal{O}(g^3).
\end{align}
Combining Eq.~\eqref{eq:niexp}, and the upper bound of Eq.~\eqref{eq:uplowBG}, we obtain
\begin{align}
    \Delta W_{g}^{(2)}\leq 0 \quad\Rightarrow\quad \Delta W_{g=1}^{(2)}\leq 0~~{\rm for}~g=1 .\label{eq:Wsec}
\end{align}
Note here that $g\cdot {\langle I_{\rm I}\rangle}_{\rm NI}$ cancels in the upper bound of Eq.~\eqref{eq:uplowBG}.
Consequently, the upper bound of Eq.~\eqref{eq:uplowBG} means that the Euclidean effective action decreases by the second or higher order corrections for the interaction $I_{\rm I}$.
Also, according to Eq.~\eqref{eq:g0NI}, the non-positive interacting theory is a class of theories in which the Euclidean effective action is unchanged, or reduced at the first order of the interaction.
For the non-positive interacting theory, the sign of the shift of the Euclidean effective action is the same as that of the second or higher order corrections for the interaction.
In other words, the non-positive interaction, i.e., ${\langle I_{\rm I}\rangle}_{\rm NI}\leq 0$, is a sufficient condition to reduce the Euclidean effective action by the interaction $I_{\rm I}$.

Focusing on the NIRT, we explain an important property of the relative entropy, i.e., the invariance of Eq.~\eqref{eq:uppFT} under the field redefinition to eliminate the linear term of $\Phi$.
    Consider a target theory with the linear term in the Euclidean space,
    \begin{align}
        I_{\rm T}[\phi,\Phi]=I^{\rm lin}_0[\phi,\Phi]+I_{\rm I}[\phi,\Phi],\label{eq:genI}
    \end{align}
    where $I^{\rm lin}_0$ is assumed to involve the linear term of $\Phi$, and $I_{\rm I}$ is the interacting term.
    Assume the classical solution of $I^{\rm lin}_0$ for $\Phi$ takes $v$, where indices of the classical solution, such as Lorentz indices, are omitted.
    Also, the classical solution of $I_{\rm T}$ for $\Phi$ takes $v+f(\phi)$, where $f(\phi)$ depends on the light field $\phi$ because of the interacting term $I_{\rm I}$. 
    Note here that $f$ vanishes in the limit of $I_{\rm I}\to 0$.
Define the action of NIRT as,
\begin{align}
I_{\rm NI}[\phi,\Phi]\equiv I_0^{\rm lin}[\phi,\Phi].
\end{align}
At the tree level, the Euclidean effective actions of the target and reference theories are respectively calculated as follows:
\begin{align}
    W_{\rm T}[\phi]&=I^{\rm lin}_0[\phi,v+f(\phi)]+ I_{\rm I}[\phi,v+f(\phi)],
    \\
    W_{\rm NI}[\phi]&=I^{\rm lin}_0[\phi,v].
\end{align}
The shift of the Euclidean effective action is calculated as
\begin{align}
 W_{\rm T}[\phi]-W_{\rm NI}[\phi]=I^{\rm lin}_0[\phi,v+f(\phi)]+ I_{\rm I}[\phi,v+f(\phi)]-I^{\rm lin}_0[\phi,v].\label{eq:delWGgen}
\end{align}
The expectation value of the interaction $I_{\rm I}$ in the Euclidean space is also calculated as
\begin{align}
    {\langle I_{\rm I}\rangle}_{\rm NI}=\int d[\Phi]P_{\rm NI}[\Phi] I_{\rm I}[\phi,\Phi]= I_{\rm I}[\phi,v],\label{eq:IIgn}
\end{align}
where $P_{\rm NI}[\Phi]\equiv e^{-I_{\rm NI}}/Z_{\rm NI}[\phi]$ with $Z_0[\phi]\equiv \int d[\Phi]e^{-I_0^{\rm lin}}$.
Equations~\eqref{eq:delWGgen}, \eqref{eq:IIgn}, and \eqref{eq:uplowBG} yield,
\begin{align}
    &W_{\rm T}[\phi]-W_{\rm NI}[\phi]\leq {\langle I_{\rm I}\rangle}_{\rm NI}\notag
    \\
    &\Rightarrow I^{\rm lin}_0[\phi,v+f(\phi)]+ I_{\rm I}[\phi,v+f(\phi)]-I_{\rm I}[\phi,v]-I^{\rm lin}_0[\phi,v]\leq 0.\label{eq:re1gn}
\end{align}
    Next, consider a field redefinition $\Phi\equiv \eta+v$.
Equation~\eqref{eq:genI} is expressed as
\begin{align}
    I'_{\rm T}[\phi,\eta]&\equiv I_{\rm T}[\phi,\eta+v]\notag
    \\
    &=I_0^{\rm lin}[\phi,\eta+v]+I_{\rm I}[\phi,\eta+v]\notag
    \\
    &=\left(I_0^{\rm lin}[\phi,\eta+v]+I_{\rm I}[\phi,v] \right)+\left(I_{\rm I}[\phi,\eta+v]-I_{\rm I}[\phi,v]\right).\label{eq:linrem1}
\end{align}
Here, we define
\begin{align}
    &I'_0[\phi,\eta]\equiv I_0^{\rm lin}[\phi,\eta+v]+I_{\rm I}[\phi,v],\label{eq:linrem2}
    \\
    &I'_{\rm I}[\phi,\eta]\equiv I_{\rm I}[\phi,\eta+v]-I_{\rm I}[\phi,v],\label{eq:linrem3}
\end{align}
where $I'_0$ does not include the linear term of $\eta$.
Also define the action of NIRT as
\begin{align}
    I'_{\rm NI}[\phi,\Phi]\equiv I'_0[\phi,\Phi].
\end{align}
At the tree level, the Euclidean effective actions of $I'_{\rm T}$ and $I'_{\rm NI}$ are respectively calculated as follows:
\begin{align}
    W'_{\rm T}[\phi]&=I'_0[\phi,f(\phi)]+ I'_{\rm I}[\phi,f(\phi)]=I_0^{\rm lin}[\phi,v+f(\phi)]+I_{\rm I}[\phi,v]+ I'_{\rm I}[\phi,f(\phi)],
    \\
    W'_{\rm NI}[\phi]&=I_0^{\rm lin}[\phi,v]+I_{\rm I}[\phi,v].
\end{align}
Note here that the classical solution of $I'_{\rm T}$ for $\eta$ takes $f(\phi)$ because that of $I_{\rm T}$ for $\Phi=v+\eta$ is $v+f(\phi)$.  
Similarly, the classical solution of $I'_{\rm NI}$ for $\eta$ takes a zero value because that of $I_{\rm NI}$ for $\Phi=v+\eta$ is $v$. 
Then, the shift of the Euclidean effective action is calculated as
\begin{align}
     W'_{\rm T}[\phi]-W'_{\rm NI}[\phi]&=I_0^{\rm lin}[\phi,v+f(\phi)]+I_{\rm I}[\phi,v]+ I'_I[\phi,f(\phi)]-I_0^{\rm lin}[\phi,v]-I_{\rm I}[\phi,v]\notag
    \\
    &=I_0^{\rm lin}[\phi,v+f(\phi)]+ I'_{\rm I}[\phi,f(\phi)]-I_0^{\rm lin}[\phi,v].\label{eq:DelWprige}
\end{align}
The expectation value of the interaction $I'_{\rm I}$ in the Euclidean space is calculated as
\begin{align}
     {\langle I'_{\rm I}\rangle}_{\rm NI}=\int d[\eta] P'_{\rm NI}[\eta] I'_{\rm I}[\phi,\eta]=0,\label{eq:Iprigen}
\end{align}
where $P'_{\rm NI}[\eta]\equiv e^{-I'_{\rm NI}[\phi,\eta]}/Z'_{\rm NI}[\phi]$ with $Z'_{\rm NI}[\phi]\equiv \int d[\eta]e^{-I'_{\rm NI}[\phi,\eta]}$.
Equations~\eqref{eq:DelWprige}, \eqref{eq:Iprigen}, and Eq.~\eqref{eq:uplowBG} yield,
\begin{align}
    &W'_{\rm T}[\phi]-W'_{\rm NI}[\phi]\leq {\langle I'_{\rm I}\rangle}_{\rm NI} \notag
    \\
    &\Rightarrow
    I_0^{\rm lin}[\phi,v+f(\phi)]+ I'_{\rm I}[\phi,f(\phi)]-I_0^{\rm lin}[\phi,v]\notag
    \\
    &=I_0^{\rm lin}[\phi,v+f(\phi)]+ I_{\rm I}[\phi,v+f(\phi)]-I_{\rm I}[\phi,v]-I_0^{\rm lin}[\phi,v]
    \leq 0.
\end{align}
This result is the same as Eq.~\eqref{eq:re1gn}.
Consequently, it is found that the upper bound of  Eq.~\eqref{eq:uplowBG} is invariant under the field redefinition to remove the linear term of $\Phi$.
This result means that, at the tree level, theories can be the non-positive interacting theory by the field redefinition.
In Sec.~\ref{sec:lin}, we will show an example of this result.
Since the calculations of the relative entropy becomes easier by the field redefinition to remove the linear term of $\Phi$, in Sec.~\ref{sec:bot}, we often use the procedures of Eq.~\eqref{eq:linrem1}, \eqref{eq:linrem2}, and \eqref{eq:linrem3}. 

We comment on some properties of the NIRT.
One of the main features of the NIRT is that the $\phi$-independent terms cancel in $W_{\rm T}[{\phi}]-W_{\rm NI}[{\phi}]$ as in Eq.~\eqref{eq:schWBA}.
In the context of the positivity bounds on EFTs, a class of EFTs that corrections to the renormalizable terms can be removed by the redefinition of the light fields, has been actively studied.
For such a class of theories, the cancellation of the $\phi$-independent terms is convenient to derive the constraints on the correction to higher-derivative terms by using Eq.~\eqref{eq:uplowBG}
\footnote{Even in finite temperature systems, field independent terms cancel in $W_{\rm T}[\phi]-W_{\rm NI}[\phi]$.}.
Indeed, in Sec. \ref{sec:top} and \ref{sec:bot}, we will provide constraints on the higher-dimensional operators of such theories.
We will see that the results are consistent with the positivity bounds obtained by the analyticity, causality and unitarity.

\subsubsection{Massive free field reference theory}
\begin{figure*}[t]
\centering
\includegraphics[width=0.65\textwidth]{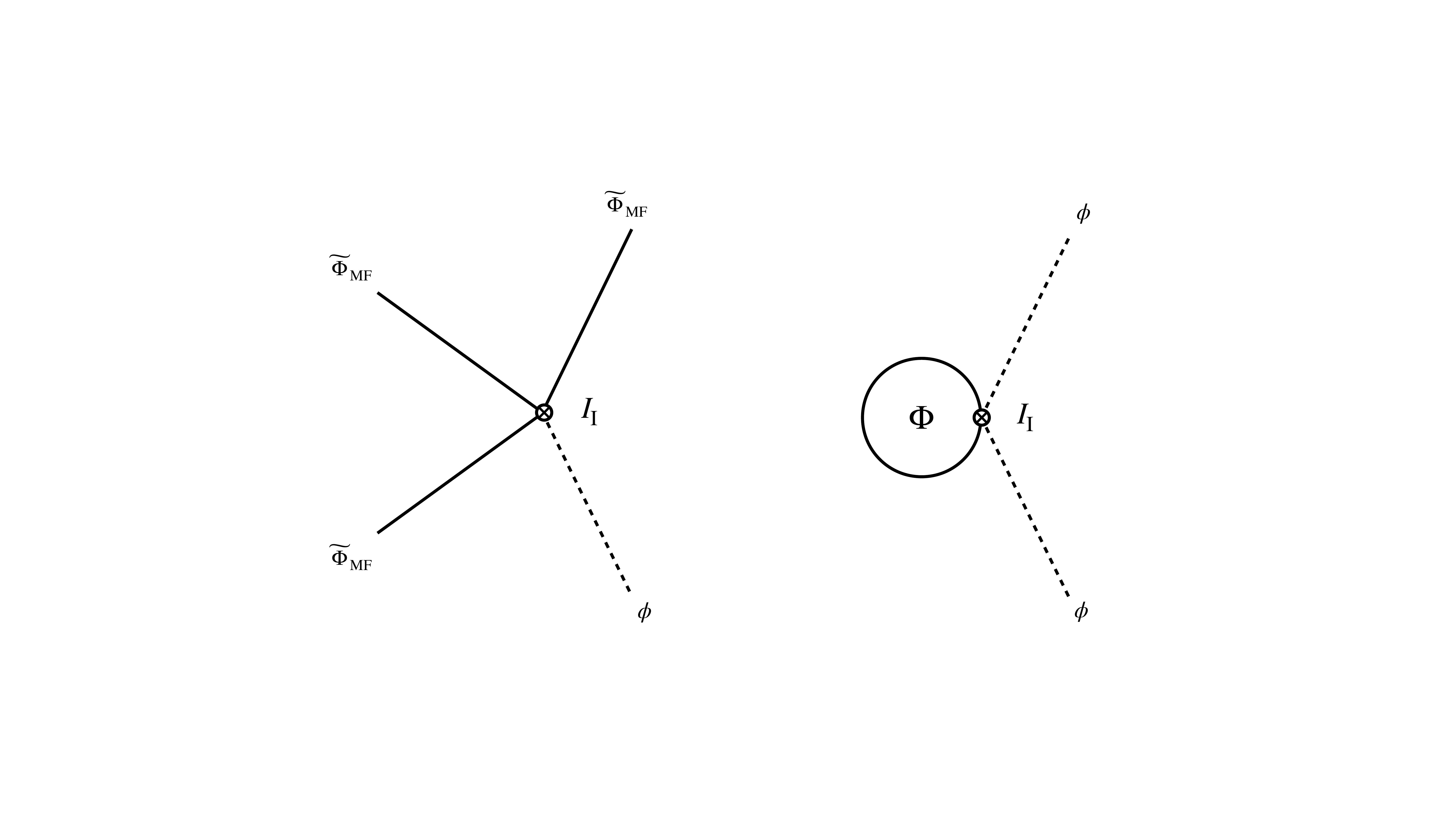}
\caption{
Feynman diagrams for ${\langle I_{\rm I}\rangle}_{\rm MF}$ at the tree and one-loop level.
The vertices denote the interacting term $I_{\rm I}$. 
The solid and dashed lines are the heavy and the background light fields, respectively. 
In contrast to the non-interacting reference theory, the classical solution of the heavy field can take zero value since the reference theory is the massive free field theory.
The left diagram is a tree-level correction, which vanishes because of zero values of the classical solution in the reference theory.
The right diagram denotes a one-loop level correction, which generally does not vanish.
\label{fig:diag1}
}
\end{figure*}

\begin{figure*}[t]
\centering
\includegraphics[width=0.65\textwidth]{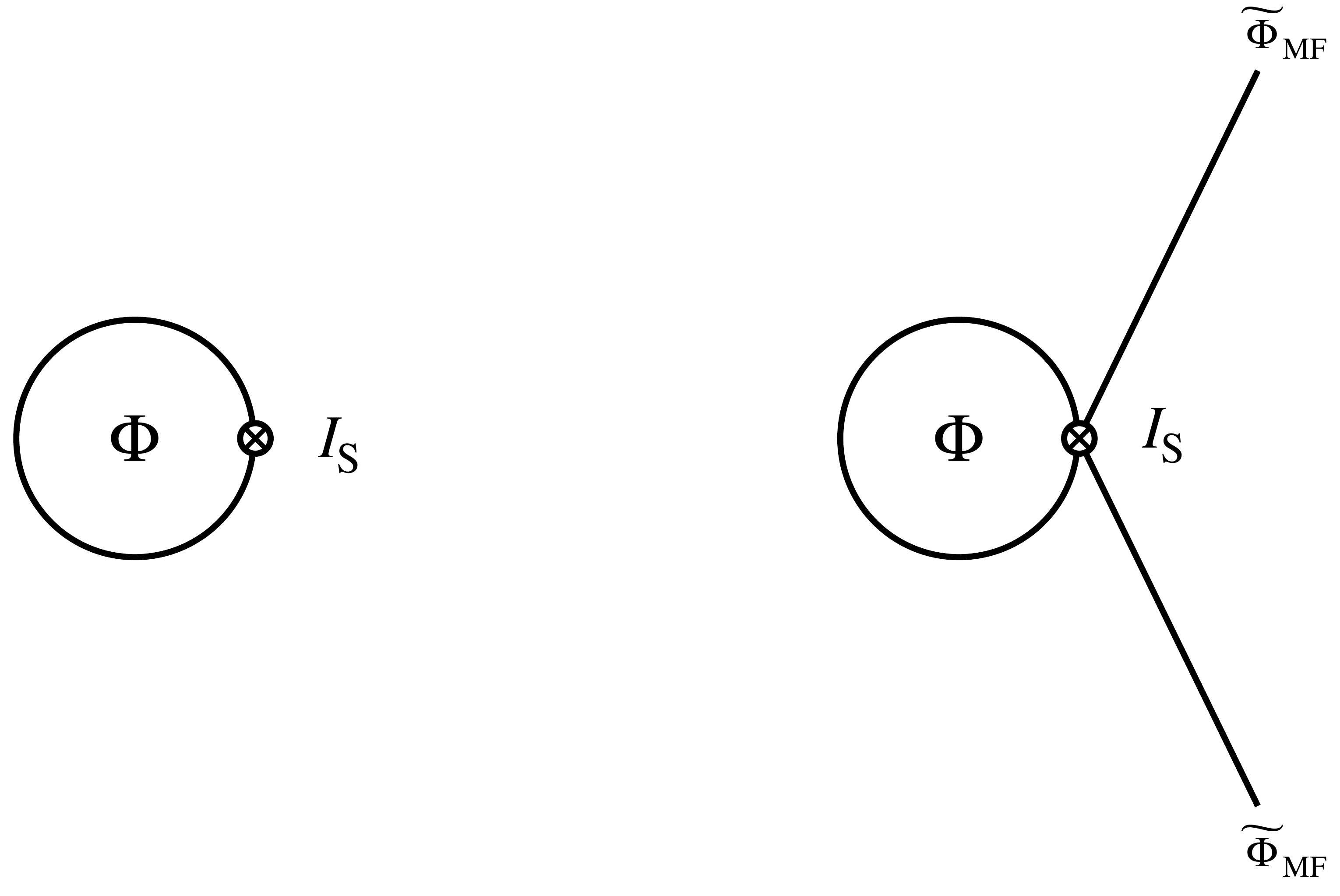}
\caption{
Feynman diagrams for ${\langle I_{\rm S}\rangle}_{\rm MF}$ at the one-loop level.
The vertices denote the self-interacting term $I_{\rm S}$.
The solid line denotes the heavy field. 
The right diagram vanishes because of the classical solution of the reference theory, but the left diagram may remain.
\label{fig:diag2}
}
\end{figure*}

We rewrite the action $I_0$ as follows:
\begin{align}
    I_0[\phi,\Phi]&=I_0[\phi,0]+I_{\Phi}[\Phi]+I_{\rm S}[\Phi],
\end{align}
where $I_{\Phi}$ is the action of free field $\Phi$, and $I_{\rm S}$ is the self-interacting term of $\Phi$.
By introducing an auxiliary parameters $g$, we define an action as follows:
\begin{align}
    J_{g}[\phi,\Phi]&\equiv I_{0}[\phi,0]+I_{\Phi}[\Phi]+g\cdot\left(  I_{\rm S}[\Phi] + I_{\rm I}[\phi,\Phi]\right).\label{eq:Ig}
\end{align}
The action of Eq.~\eqref{eq:Ig} satisfies,
\begin{align}
    &\lim_{g\to  0} J_g[\phi,\Phi]=I_{\rm MF}[\phi,\Phi],~~~\lim_{g\to 1} J_g[\phi,\Phi]=I_{\rm T}[\phi,\Phi].
\end{align}
When the light fields are the background fields, the partition function and effective actions of $J_g$ are defined as follows:
\begin{align}
    &z_{g}[\phi]\equiv \int d[\Phi]e^{-J_{g}[\phi,\Phi]},
    \\
    &w_{g}[\phi]\equiv -\ln z_{g}[\phi].
\end{align}
From Eqs.~\eqref{eq:sABF} and \eqref{eq:sBAF}, by defining a probability distribution function,
\begin{align}
    p_g[\Phi]\equiv \frac{e^{-J_g[\phi,\Phi]}}{z_g[\phi]},
\end{align}
the relative entropy between $p_0$ and $p_g$ is calculated as follows:
\begin{align}
    S(p_0||p_g)&=w_0[\phi]-w_g[\phi] +g \cdot {\langle I_{\rm S}+I_{\rm I}\rangle}_{g=0}\geq 0\Rightarrow w_g[\phi]-w_0[\phi] \leq g\cdot {\langle I_{\rm S}+I_{\rm I}\rangle}_{g=0},\label{eq:mfg}
    \\
    S(p_g||p_0)&= w_g[\phi]-w_0[\phi] -g\cdot {\langle I_{\rm S}+I_{\rm I}\rangle}_g \geq 0
    \Rightarrow g\cdot {\langle I_{\rm S}+I_{\rm I}\rangle}_g\leq w_g[\phi]-w_0[\phi],\label{eq:mf0}
\end{align}
with
\begin{align}
    {\langle I_{\rm S}+I_{\rm I}\rangle}_g \equiv \int d[\Phi] \left(I_{\rm S}[\Phi]+I_{\rm I}[\phi,\Phi]\right)\cdot \frac{e^{-J_g[\phi,\Phi]}}{z_g[\phi]}=\frac{\partial w_g[\phi]}{\partial g},
\end{align}
where the partial derivative denotes differentiating by $g$ while keeping $\phi$.
Combining Eq.~\eqref{eq:mfg} and \eqref{eq:mf0}, we obtain
\begin{align}
    &g\cdot {\langle I_{\rm S}+I_{\rm I}\rangle}_{g=0}\geq w_g[\phi]-w_0[\phi]\geq g\cdot {\langle I_{\rm S}+I_{\rm I}\rangle}_g \notag
    \\
    \Rightarrow 
    &{\langle I_{\rm S}+I_{\rm I}\rangle}_{\rm MF}\geq W_{\rm T}[\phi]-W_{\rm MF}[\phi]\geq  {\langle I_{\rm S}+I_{\rm I}\rangle}_{\rm T}~~{\rm for}~g=1.\label{eq:wt}
\end{align}
Here, we used $W_{\rm T}=w_{g=1}$, $W_{\rm MF}=w_{0}$, and the following relations,
\begin{align}
    {\langle I_{\rm S}+I_{\rm I}\rangle}_{g=0}&=\left(\frac{\partial w_g[\phi]}{\partial g}\right)_{g=0}\notag
    \\
    &=\int d[\Phi] \left(I_{\rm S}[\Phi]+I_{\rm I}[\phi,\Phi]\right)\cdot \frac{e^{-J_0[\phi,\Phi]}}{z_0[\phi]}\notag
    \\
    &=\int d[\Phi] \left(I_{\rm S}[\Phi]+I_{\rm I}[\phi,\Phi]\right)\cdot \frac{e^{-I_{\rm MF}[\phi,\Phi]}}{Z_{\rm MF}[\phi]}\notag
    \\
    &={\langle I_{\rm S}+I_{\rm I}\rangle}_{\rm MF},
    \\
     {\langle I_{\rm S}+I_{\rm I}\rangle}_{g=1}&=\left(\frac{\partial w_g[\phi]}{\partial g}\right)_{g=1}\notag
    \\
    &=\int d[\Phi] \left(I_{\rm S}[\Phi]+I_{\rm I}[\phi,\Phi]\right)\cdot \frac{e^{-J_{g=1}[\phi,\Phi]}}{z_{g=1}[\phi]}\notag
    \\
    &=\int d[\Phi] \left(I_{\rm S}[\Phi]+I_{\rm I}[\phi,\Phi]\right)\cdot \frac{e^{-I_{\rm T}[\phi,\Phi]}}{Z_{\rm T}[\phi]}\notag
    \\
    &={\langle I_{\rm S}+I_{\rm I}\rangle}_{\rm T},   
\end{align}
where $z_0=Z_{\rm MF}$ and $z_{g=1}=Z_{\rm T}$ are used. 
In particular, the Feynman diagrams for ${\langle I_{\rm S}\rangle}_{\rm MF}$ and ${\langle I_{\rm I}\rangle}_{\rm MF}$ are shown in Fig.~\ref{fig:diag1} and \ref{fig:diag2}, respectively.
The Euclidean effective actions are schematically expressed as follows:
\begin{align}
    W_{\rm MF}[{\phi}]&=I_0[{\phi},0]+({\rm vacuum~energy}),\label{eq:vacA}
    \\
    W_{\rm T}[\phi]&=I_0[\phi,0]+({\rm vacuum~energy})+({\rm corrections~from~\Phi}).\label{eq:vacB}
\end{align}
where $({\rm vacuum~energy})$ denotes the vacuum energy coming from $\Phi$ loop effects, and the third term of the right-hand side of Eq.~\eqref{eq:vacB} denotes the perturbative corrections from $\Phi$ other than the vacuum energy.
The shift of the Euclidean effective action is given by
\begin{align}
    W_{\rm T}[\phi]-W_{\rm MF}[\phi]=({\rm corrections~from~\Phi}).
\end{align}
Note here that the $({\rm corrections~from~\Phi})$ may include $\phi$ independent terms
because the self-interacting term of $\Phi$ in $I_{\rm T}$ appears. 
Therefore, $W_{\rm T}-W_{\rm MF}$ also includes the correction from the self-interacting terms of $\Phi$ in contrast to $W_{\rm T}-W_{\rm NI}$, and the right-hand side of Eq.~\eqref{eq:wt} may include the field-independent terms.
The point is that the inequality of \eqref{eq:wt} provides different information about the target theory from the NIRT.
Some applications are provided in the later section. 

{Here, we consider the meaning the upper bound of \eqref{eq:wt}.
Similar to the case of the NIRT, expand $w_{g}$ for $g$ as follows:
\begin{align}
    w_{g}[\phi]&=w_{0}[\phi]+g\cdot \left(\frac{\partial  w_{g}[\phi]}{\partial g}\right)_{g=0}+\frac{g^2}{2}\cdot \left(\frac{\partial^2 w_{g}[\phi] }{\partial g^2}\right)_{g=0}+\mathcal{O}(g^3),\notag
    \\
    &=W_{{\rm MF}}[\phi]+g\cdot {\langle I_{\rm S}+ I_{\rm I}\rangle}_{\rm MF}+\frac{g^2}{2}\cdot \left(\frac{\partial^2 w_{g}[\phi] }{\partial g^2}\right)_{g=0}+\mathcal{O}(g^3),\notag
    \\
    &=W_{{\rm MF}}[\phi]+g\cdot {\langle I_{\rm S}+ I_{\rm I}\rangle}_{\rm MF}+
    \Delta w^{(2)}_{g},\label{eq:wMF}
\end{align}
where $w_{0}[\phi]=W_{{\rm MF}}[\phi]$ is used, and we defined
\begin{align}
    \Delta w^{(2)}_{g}\equiv \frac{g^2}{2}\cdot \left(\frac{\partial^2 w_{g}[\phi] }{\partial g^2}\right)_{g=0}+\mathcal{O}(g^3).
\end{align}
From Eq.~\eqref{eq:wMF} and \eqref{eq:wt}, we get
\begin{align}
    \Delta w^{(2)}_{g}\leq 0.\label{eq:wmf2}
\end{align}
Therefore, the inequality of \eqref{eq:wt} means that the Euclidean effective action decrease by the second or higher order corrections for $I_{\rm S}$ and $I_{\rm I}$.
}

\subsubsection{Infinite heavy mass reference theory}
The upper bound of Eq.~\eqref{eq:uppFT} is calculated as,
\begin{align}
    {\langle I_{\rm T}-\lim_{m_{\Phi}\to\infty}I_{\rm T}\rangle}_{\rm IH}\geq W_{\rm T}[\phi]-W_{\rm IH}[\phi],\label{eq:IHuplow}
\end{align}
where $W_{\rm IH}[\phi]=-\ln Z_{\rm IH}[\phi]$.
The left-hand side of Eq.~\eqref{eq:IHuplow} is defined as follows:
\begin{align}
    &{\langle I_{\rm T}-\lim_{m_{\Phi}\to\infty}I_{\rm T}\rangle}_{\rm IH}=\int d[\Phi] P_{\rm IH}[\Phi]\left(I_{\rm T}[\phi,\Phi]-\lim_{m_{\Phi}\to\infty}I_{\rm T}[\phi,\Phi]\right).
\end{align}
For simplicity, we focus on the tree level UV completions.
Using the saddle point approximations, we obtain
\begin{align}
    &Z_{\rm IH}[\phi]=\int d[\Phi]e^{-I_{\rm IH}[\phi,\Phi]}=e^{-I_{\rm IH}[\phi,\widetilde{\Phi}_{\rm IH}]},
    \\
    &Z_{\rm T}[\phi]=\int d[\Phi]e^{-I_{\rm T}[\phi,\Phi]}=e^{-I_{\rm T}[\phi,\widetilde{\Phi}_{\rm T}]},
    \\
    &\int d[\Phi]P_{\rm IH}[\Phi] \left(I_{\rm T}[\phi,\Phi]-I_{\rm NI}[\phi,\Phi]\right)=I_{\rm T}[\phi,\widetilde{\Phi}_{\rm IH}]-\lim_{m_{\Phi}\to\infty}I_{\rm T}[\phi,\widetilde{\Phi}_{\rm IH}].
\end{align}
Similar to the NIRT, ${\langle I_{\rm T}-\lim_{m_{\Phi}\to\infty}I_{\rm T}\rangle}_{\rm IH}$ does not vanish when the linear term of $\Phi$ in $I_{\rm IH}$ remains.
However, ${\langle I_{\rm T}-\lim_{m_{\Phi}\to\infty}I_{\rm T}\rangle}_{\rm IH}=0$ holds after the field redefinition of $\Phi$ such that $\phi$ independent terms are removed in $\widetilde{\Phi}_{\rm IH}$.
For ${\langle I_{\rm T}-\lim_{m_{\Phi}\to\infty}I_{\rm T}\rangle}_{\rm IH}=0$, one obtains as follows:
\begin{align}
    0\geq W_{\rm T}[\phi]-W_{\rm IH}[\phi].
\end{align}
The above right-hand side denotes the correction from $\Phi$ to the Euclidean effective action of the target theory.

\subsection{Summary of entropy constraints}
\label{sec:sum_en}
In Secs.~\ref{sec:defTH} and \ref{sec:shif}, based on the Euclidean path integral method, we provide procedures to calculate the relative entropy in the field theory.
Given the target theory, the relative entropy can be calculated by the procedures explained in Sec.~\ref{sec:shif}.
Before summarizing the properties of the relative entropy, we list the assumptions used to derive the results in this section in the following way.
\begin{enumerate}[(I)]
    \item Hermiticity of probability distribution functions ---
    We assume the target and reference theory are represented by the Hermitian probability distribution functions.
    To derive the non-negativity of the relative entropy in Eq.~\eqref{eq:non-neg}, we used the Hermiticity of probability distribution functions, i.e., $\rho_{\rm R,T}=\rho_{\rm R,T}^{\dagger}$.
    The non-negativity of the relative entropy can be broken when this condition is not satisfied.

    \item Validity of Euclidean path integral method ---
    We assume the EFTs are generated from the solution of the local minimum.
    As shown in Sec.~\ref{sec:incon}, the non-negativity of the relative entropy can be broken when the Euclidean path integral method is not valid, i.e., the saddle point approximation does not work because of the solution not being the local minimum.

\end{enumerate}

Under these assumptions, for the NIRT and MFFRT, we obtained the properties that do not depend on the details of the target theories as follows:

\begin{itemize}
    \item {\bf Non-interacting reference theory} ---
    The upper and lower bounds on the perturbative corrections from the interaction between heavy and light degrees of freedom to the Euclidean effective action is denoted as Eq.~\eqref{eq:uplowBG},
    \begin{align}
        {\langle I_{\rm I} \rangle}_{\rm NI}\geq W_{\rm T}[\phi]-W_{\rm NI}[\phi]\geq {\langle I_{\rm I}\rangle}_{\rm T},\notag
    \end{align}
    where $W_{\rm T}-W_{\rm NI}$ is the perturbative correction to renormalizable and unrenormalizable terms for $\phi$, and the background independent terms are canceled.
    The upper bound of Eq.~\eqref{eq:uplowBG} is rewritten as Eq.~\eqref{eq:Wsec},
    \begin{align}
        \Delta W^{(2)}_{g}\leq 0.\notag
    \end{align}
    This inequality means that the Euclidean effective action decreases by the second or higher order corrections for the interaction between heavy and light degrees of freedom.
    In general, ${\langle I_{\rm I} \rangle}_{\rm NI}$ takes a non-zero value, but it can vanish at the tree level by the redefinition to eliminate the linear term of $\Phi$ in $I_{\rm T}$.
    The relative entropy is invariant under such a field redefinition.

    \item {\bf Massive free field reference theory} ---
    The upper and lower bounds on the perturbative corrections from the self-interaction of heavy fields and the interaction between heavy and light degrees of freedom to the Euclidean effective action is denoted as Eq.~\eqref{eq:wt},
    \begin{align}
        {\langle I_{\rm S}+I_{\rm I}\rangle}_{\rm MF}\geq W_{\rm T}[\phi]-W_{\rm MF}[\phi]\geq  {\langle I_{\rm S}+I_{\rm I}\rangle}_{\rm T}.\notag
    \end{align}
    For $W_{\rm T}-W_{\rm NI}$, the $\phi$ independent vacuum energy from the heavy field loop effects cancels, but the $\phi$ independent perturbative corrections from the self-interacting term of $\Phi$ may be included.
    The upper bound of Eq.~\eqref{eq:wt} is rewritten as Eq.~\eqref{eq:wmf2},
    \begin{align}
        \Delta w^{(2)}_{g}\leq 0.
    \end{align}
    This inequality means that the Euclidean effective action decreases by the second or higher order corrections for the self-interaction of heavy fields and the interaction between heavy and light degrees of freedom.
    On the left-hand side of the above inequality, $\phi$ independent perturbative corrections from the self-interacting term of $\Phi$ are generally included.
\end{itemize}

In the next section, we will calculate the relative entropies for various target theories and check the non-negativity of the relative entropy, i.e., the above properties.
Also, we often face a situation where the EFT is known while the target UV theory is unknown.
We refer to this situation as the bottom-up approach and consider such a situation in Sec.~\ref{sec:bot}.
In Sec.~\ref{sec:bot}, we focus on a class of EFTs that the corrections to non-higher derivative terms are removed by field redefinitions and apply the NIRT to derive the constraints on such EFTs.
The NIRT is more convenient than the MFFRT in the bottom-up approach because the background independent terms vanish in $W_{\rm T}-W_{\rm NI}$.
In Sec.~\ref{sec:bot}, we provide the constraints on some EFTs under an assumption that the corrections from the interactions involving higher-derivative operators of the light fields are not dominant in the EFTs.

\section{Top-down approach: relative entropy in various theories}
\label{sec:top}
In this section, for consistency checks of Sec.~\ref{sec:entr}, we evaluate Eqs.~\eqref{eq:uplowBG}, \eqref{eq:wt}, and \eqref{eq:IHuplow} in various theories.
In particular, we focus on the upper bound on $W_{\rm T}-W_{\rm R}$, which is relevant to the positivity bound on EFTs.
We adopt the top-down approach, i.e., UV theories including heavy degrees of freedom are assumed to be known, and evaluate the inequalities in the previous sections.
As a pedagogical example, we first consider probability distribution functions described by the Gaussian distribution function.
In Sec.~\ref{sec:spin}, the constraint on a quantum mechanical model consisting of spins is studied. 
In the other examples, we focus on weakly coupled field theoretical dynamics and consider up to four-derivative operators.
In Sec.~\ref{sec:incon}, we will also explain that the non-negativity of the relative entropy can be violated when the Euclidean path integral method does not work, i.e., the saddle point approximations are not valid.

\subsection{Gaussian distribution functions}
\label{sec:Gauss}
Firstly, we consider a target system described by the Gaussian distribution function defined as follows:
\begin{align}
    I_{\rm T}[x,X]\equiv m^2\cdot x^2 +M^2\cdot X^2+c\cdot x\cdot X,~~~~[{\rm \bf Target}]
\end{align}
where $x$ and $X$ are not fields, but variables corresponding to the light and heavy degrees of freedom, respectively, and $m, M$, and $c$ are real constant parameters.
Although this target theory is neither a field theoretical nor a quantum mechanical model, this is a pedagogical example to understand the procedure to calculate the relative entropy.
We assume $x$ is not an integral variable and behaves as a background field.
The free theory and the interaction between $x$ and $X$ are respectively denoted as
\begin{align}
    &I_0[x,X] \equiv m^2\cdot x^2 +M^2\cdot X^2,
    \\
    &I_{\rm I}[x,X]\equiv c\cdot x\cdot X.\label{eq:Gausint}
\end{align}
For this target theory, the NIRT is the same as the MFFRT.
    The action of the NIRT and MFFRT is defined as
    \begin{align}
        I_{\rm R}[x,X]\equiv I_0[x,X].~~~~[{\rm \bf Reference}]
    \end{align}
    Then, probability distribution functions are defined as follows:
    \begin{align}
    &P_{\rm R}[X]\equiv\frac{1}{Z_{\rm R}[x]}e^{-I_{\rm R}[x,X]},~~~P_{\rm T}[X]\equiv\frac{1}{Z_{\rm T}[x]}e^{-I_{\rm T}[x,X]},
    \end{align}
    with the partition functions
    \begin{align}
    &Z_{\rm R}[x]\equiv \int_{-\infty}^{\infty}dX e^{-I_{\rm R}[x,X]},~~~Z_{\rm T}[x]\equiv \int_{-\infty}^{\infty} dX e^{-I_{\rm T}[x,X]}.
    \end{align}
    From these partition functions, the effective actions are given by
    \begin{align}
        &W_{\rm R}[x]\equiv-\ln Z_{\rm R}[x],~~~W_{\rm T}[x]\equiv-\ln Z_{\rm T}[x].
    \end{align}
    By introducing an auxiliary parameter $g$, we define,
    \begin{align}
        I_g[x,X]\equiv I_0[x,X]+g\cdot I_{\rm I}[x,X].
    \end{align}
    By changing $g$, the target and reference theory are reproduced as follows:
    \begin{align}
        I_{\rm T}[x,X]=\lim_{g\to 1} I_g[x,X],~~~I_{\rm R}[x,X]=\lim_{g\to 0} I_g[x,X].
    \end{align}
    The partition function and effective action of $I_g$ are respectively calculated as follows:
    \begin{align}
        &Z_g[x]\equiv \int d[X] e^{-I_g[x,X]}=e^{-m^2 x^2}\sqrt{\frac{\pi}{M^2}}\cdot  e^{ g^2\cdot c^2 x^2/4 M^2},
        \\
        &W_g[x]\equiv -\ln Z_g[x]=-\frac{g^2\cdot c^2 x^2}{4 M^2}-\ln \left[e^{-m^2 x^2}\sqrt{\frac{\pi}{M^2}}\right].\label{eq:wggaus}
    \end{align}
    By defining a probability distribution function,
    \begin{align}
        P_g[X]\equiv \frac{e^{-I_g[x,X]}}{Z_g[x]},
    \end{align}
        the expectation value of the interaction is also calculated as
\begin{align}
    {\langle I_{\rm I}\rangle}_{g=0} =\int_{-\infty}^{\infty}dX P_{0}[X]\cdot I_{\rm I}[x,X]=0.\label{eq:IIgauss}
\end{align}
    From Eqs.~\eqref{eq:sABF}, \eqref{eq:wggaus}, and \eqref{eq:IIgauss}, the relative entropy between $P_0$ and $P_g$ is given by
    \begin{align}
     S(P_{0}||P_{g})&\equiv \int_{-\infty}^{\infty}dX \left(P_{0}\ln P_{0}- P_{0}\ln P_{g} \right),\notag
    \\
    &=-\ln Z_{0}[x] +\ln Z_{g}[x]+g\cdot {\langle I_{\rm I}\rangle}_{g=0},\notag
    \\
    &=-\ln Z_{0}[x] +\ln Z_{g}[x],\notag
    \\
    &=W_{0}[x]-W_{g}[x],\notag
    \\
    &=\frac{ g^2\cdot c^2 x^2}{4 M^2}\geq 0.
    \end{align}
    Note here that the relative entropy is invariant under the field redefinition of $X$.
The definition of the interaction of Eq.~\eqref{eq:Gausint} is not invariant under the redefinition of $X$, but the relative entropy, i.e., the integral of the Gaussian distributions, do not change under the field redefinition.
    By taking to be $g=1$, we obtain the relative entropy between $P_{\rm R}$ and $P_{\rm T}$ as follows:
\begin{align}
    S(P_{\rm R}||P_{\rm T})&=W_{\rm R}[x]-W_{\rm T}[x]=\frac{ c^2 x^2}{4 M^2}\geq 0,
\end{align}
where $P_{\rm R}=P_0$, $P_{\rm T}=P_{g=1}$, $W_{\rm R}=W_{g=0}$, and $W_{\rm T}=W_{g=1}$ are used.
The above inequality represents the upper bound of Eq.~\eqref{eq:uplowBG} and \eqref{eq:wt}.
The relative entropy takes a positive value, as we expected.
The above procedure of calculation of the relative entropy is the same as the field theoretical dynamics.

\subsection{A spin system in one dimension}
\label{sec:spin}
The entropy inequality is derived even in the quantum mechanical model.
Let us consider a spin system in one dimension defined by a Hamiltonian,
\begin{align}
    H_{\rm T}\equiv-J \sum_{i=1}^{N/2}\sigma_{2i-1}\sigma_{2i} -\mu M \sum_{i=1}^N \sigma_i,~~~~[{\rm \bf Target}]
\end{align}
where $\sigma_i=\pm 1$ denotes a spin on site $i$, $J$ is a coupling characterizing exchange interactions, $N$ is the number of sites, $\mu$ is a magnetic moment, and $M$ is an external magnetic field.
We assume $N$ is even so that the system consists of $N/2$ pairs of adjacent sites.
Then, $H_0$ and $H_{\rm I}$ are defined as follows:
\begin{align}
    &H_0\equiv- J \sum_{i=1}^{N/2} \sigma_{2i-1}\sigma_{2i},
    \\
    &H_{\rm I} \equiv-\mu M \sum_{i=1}^N \sigma_i.
\end{align}
The Hamiltonian of the NIRT is defined as
\begin{align}
    H_{\rm NI}\equiv H_0.~~~~[{\rm \bf Reference}]
\end{align}
The density operators of the target and reference systems are respectively given by
\begin{align}
    &\rho_{\rm NI} \equiv\frac{e^{-\beta H_{\rm NI}}}{Z_{\rm NI}(\beta)},~~~ \rho_{\rm T} \equiv\frac{e^{-\beta H_{\rm T}}}{Z_{\rm T} (\beta)},
\end{align}
with the inverse temperature $\beta$, and the partition functions
\begin{align}
    Z_{\rm NI}(\beta)\equiv{\rm Tr}[e^{-\beta H_{\rm NI}}],~~~Z_{\rm T}(\beta)\equiv{\rm Tr}[e^{-\beta H_{\rm T}}].
\end{align}
The effective actions are given by
\begin{align}
    &W_{\rm NI}(\beta)\equiv-\ln Z_{\rm NI}(\beta),~~~W_{\rm T}(\beta)\equiv-\ln Z_{\rm T}(\beta).
\end{align}
By introducing the parameter $g$, we define
\begin{align}
    H_g\equiv H_0 +g\cdot H_{\rm I}.
\end{align}
The target and reference theory are reproduced as follows:
\begin{align}
    H_{\rm T}=\lim_{g\to 1}H_g,~~~H_{\rm R}=\lim_{g\to 0}H_g.
\end{align}
The partition function and effective action of $H_g$ are respectively given as follows:
\begin{align}
    &Z_g(\beta)\equiv {\rm Tr}[e^{-\beta H_{g}}]=\left(2 \{e^{\beta J}\cosh (2\beta g \mu M)+e^{-\beta J} \} \right)^{N/2},
    \\
    &W_g(\beta)\equiv -\ln Z_g(\beta).
\end{align}
By defining a density operator
\begin{align}
    \rho_g(\beta)\equiv \frac{e^{-\beta H_g}}{Z_g(\beta)},
\end{align}
the expectation value of the interaction is calculated as
\begin{align}
{\rm Tr}[\rho_{0} H_{\rm I}]=0.
\end{align}
From Eq.~\eqref{eq:sABQ}, the relative entropy between $\rho_0$ and $\rho_g$ is given by
\begin{align}
    S(\rho_0||\rho_g)&\equiv {\rm Tr}[\rho_{0}\ln \rho_{0}-\rho_{0}\ln \rho_{g}],\notag
    \\
    &=-\ln Z_{0}(\beta)+\ln Z_{g}(\beta)+g\cdot {\rm Tr}[\rho_{0} H_{\rm I}],\notag
    \\
    &=-\ln Z_{0}(\beta)+\ln Z_{g}(\beta),\notag
    \\
    &=W_{0}(\beta)-W_{g}(\beta),\notag
    \\
    &=-\frac{N}{2}\ln \left[\frac{e^{\beta J}+e^{-\beta J}}{e^{\beta J}\cosh (2\beta g \mu M)+e^{-\beta J}}\right]\geq 0.
\end{align}
The first line denotes the definition of the relative entropy, and in the third line, ${\rm Tr}[\rho_{0} H_{\rm I}]=0$ is used.
By taking to be $g=1$, the relative entropy between $\rho_{\rm NI}$ and $\rho_{\rm T}$ is given by
\begin{align}
    S(\rho_{\rm NI}||\rho_{\rm T})&=W_{\rm NI}(\beta)-W_{\rm T}(\beta)=-\frac{N}{2}\ln \left[\frac{e^{\beta J}+e^{-\beta J}}{e^{\beta J}\cosh (2\beta \mu M)+e^{-\beta J}}\right]\geq 0,
\end{align}
where $\rho_{\rm NI}=\rho_0$, $\rho_{\rm T}=\rho_{g=1}$, $W_{\rm NI}=W_{g=0}$, and $W_{\rm T}=W_{g=1}$ are used.
This result represents the upper bound of Eq.~\eqref{eq:uplowBG}.
We see that the external magnetic field decreases the Euclidean effective action of the target system because of the non-negativity of the relative entropy.

\subsection{A tree level UV completion of single massless scalar field theory}
\label{sec:sfiftscal1}
Consider a theory in Minkowski space:
\begin{align}
    I_{\rm T}[\phi,\Phi]\equiv\int d^4 x \left(\frac{1}{2}(\partial_{\mu}\phi \partial^{\mu}\phi) +\frac{1}{2}(\partial_{\mu}\Phi \partial^{\mu}\Phi)-\frac{m^2}{2} \Phi^2 +\frac{\alpha}{\Lambda}\cdot \Phi (\partial_{\mu}\phi \partial^{\mu}\phi) \right),~[{\rm \bf Target}]\label{eq:scl_massless}
\end{align}
where $\phi$ denotes a massless scalar field, $\Phi$ is a heavy scalar field with mass $m$, and $\alpha/\Lambda$ is a dimensionful coupling constant.
The above theory involving a linear term of $\Phi$
will be studied later.
The action in the Euclidean space is expressed as
\begin{align}
        I_{\rm T}^{\rm (E)}[\phi,\Phi]&=\int (d^4 x)_{\rm E} \left(\frac{1}{2}(\partial_{I}\phi_{\rm E} \partial_{I}\phi_{\rm E}) +\frac{1}{2}(\partial_{I}\Phi_{\rm E} \partial_{I}\Phi_{\rm E})+\frac{m^2}{2} \Phi^2_{\rm E} +\frac{\alpha}{\Lambda}\cdot \Phi_{\rm E} (\partial_{I}\phi_{\rm E} \partial_{I}\phi_{\rm E}) \right),\notag
        \\
        &=\int (d^4 x)_{\rm E} \left(-\frac{1}{2}(\partial_{\mu}\phi \partial^{\mu}\phi) +\frac{1}{2}(\partial_{I}\Phi_{\rm E} \partial_{I}\Phi_{\rm E})+\frac{m^2}{2} \Phi^2_{\rm E} -\frac{\alpha}{\Lambda}\cdot \Phi_{\rm E} (\partial_{\mu}\phi \partial^{\mu}\phi) \right),\label{eq:tree1Tag}
\end{align}
where, in the second line, we assume the background field $\phi$ of the Euclidean space are defined from that of the Minkowski space; see Appendix~\ref{app:Wick_rotation}.
We define the actions $I_0$ and $I_{\rm I}$ in the Euclidean space as follows:
\begin{align}
I_0^{\rm (E)}[\phi,\Phi]&\equiv I_{\rm T}^{\rm (E)}[0,\Phi]+I_{\rm T}^{\rm (E)}[\phi,0]=\int (d^4 x)_{\rm E} \left(-\frac{1}{2}(\partial_{\mu}\phi \partial^{\mu}\phi) +\frac{1}{2}(\partial_{I}\Phi_{\rm E} \partial_{I}\Phi_{\rm E})+\frac{m^2}{2} \Phi^2_{\rm E} \right),
\\
I_{\rm I}^{\rm (E)}[\phi,\Phi]&\equiv I^{\rm (E)}_{\rm T}[\phi,\Phi]-I_0^{\rm (E)}[\phi,\Phi]=-\frac{\alpha}{\Lambda}\cdot\int (d^4 x)_{\rm E} \Phi_{\rm E} (\partial_{\mu}\phi \partial^{\mu}\phi).
\end{align}
For the background field $\phi$, by integrating out $\Phi_{\rm E}$ at the tree level, the partition function and effective action are defined as
\begin{align}
    &Z_{\rm T}[\phi]\equiv \int d[\Phi_{\rm E}] e^{-I_{\rm T}^{\rm (E)}[\phi,\Phi_{\rm E}]},
    \\
    &W_{\rm T}[\phi]\equiv -\ln Z_{\rm T}[\phi],
\end{align}
In this target theory, the action of the MFFRT is the same as that of the NIRT.
For each reference theory in Sec.~\ref{sec:defTH}, we consider the inequalities of \eqref{eq:uplowBG}, \eqref{eq:wt}, and \eqref{eq:IHuplow} in the following.
\begin{itemize}
    \item {\bf Non-interacting reference theory} ---
    The action of the NIRT in the Euclidean space is expressed as follows:
    \begin{align}
        I^{\rm (E)}_{\rm NI}[\phi,\Phi]\equiv \int (d^4x)_{\rm E} \left(-\frac{1}{2}(\partial_{\mu}\phi\partial^{\mu}\phi)+\frac{1}{2}(\partial_I \Phi_{\rm E}\partial_I \Phi_{\rm E})+\frac{m^2}{2}\Phi^2_{\rm E} \right).~~[{\rm \bf Reference}]
    \end{align}
The partition function and effective action of the NIRT are defined as
\begin{align}
    &Z_{\rm NI}[\phi]\equiv \int d[\Phi_{\rm E}] e^{-I^{\rm (E)}_{\rm NI}[\phi,\Phi]},
    \\
    &W_{\rm NI}[\phi]\equiv -\ln Z_{\rm NI}[\phi].
\end{align}

By using the parameter $g$, we define
\begin{align}
    I_g^{\rm (E)}[\phi,\Phi]\equiv I_0^{\rm (E)}[\phi,\Phi]+g\cdot I_{\rm I}^{\rm (E)}[\phi,\Phi].
\end{align}
The target and reference theories are respectively expressed as follows:
\begin{align}
    I_{\rm T}^{\rm (E)}[\phi,\Phi]=\lim_{g\to 1} I_g^{\rm (E)}[\phi,\Phi],~~~I_{\rm NI}^{\rm (E)}[\phi,\Phi]=\lim_{g\to 0} I_g^{\rm (E)}[\phi,\Phi].
\end{align}
The partition function and effective action of $I_g^{\rm (E)}$ are respectively calculated as follows:
\begin{align}
     &Z_{g}[\phi]\equiv \int d[\Phi_{\rm E}] e^{-I_{g}^{\rm (E)}[\phi,\Phi]}=e^{-I_{g}^{\rm (E)}[\phi,\widetilde{\Phi}_g]}
    \\
    &W_{g}[\phi]\equiv I_{g}^{\rm (E)}[\phi,\widetilde{\Phi}_g]=\int (d^4 x)_{\rm E} \left(-\frac{1}{2}(\partial_{\mu} \phi\partial^{\mu} \phi)-g^2\cdot\frac{\alpha^2}{2m^2 \Lambda^2}(\partial_{\mu} \phi\partial^{\mu} \phi)^2  \right),\label{eq:dim8TBG}
\end{align}
where $\widetilde{\Phi}_{g}$ is the classical solution of $I_{g}^{\rm (E)}$ and is calculated as
\begin{align}
    \widetilde{\Phi}_{g}=g\cdot\frac{\alpha}{m^2\Lambda}(\partial_{\mu}\phi \partial^{\mu}\phi) .
\end{align}
By defining the probability distribution
\begin{align}
    P_g[\Phi]\equiv \frac{e^{-I_g^{\rm (E)}[\phi,\Phi]}}{Z_g[\phi]},
\end{align}
the expectation value of the interacting term is calculated as
\begin{align}
    {\langle I_{\rm I}\rangle}_{g=0}&\equiv \int d[\Phi_{\rm E}]P_{0}[\Phi_{\rm E}] I_{\rm I}^{\rm (E)}[\phi,\Phi_{\rm E}]\notag
    \\
    &=I_{\rm I}^{\rm (E)}[\phi,\widetilde{\Phi}_{g=0}]\notag
    \\
    &=I_{\rm I}^{\rm (E)}[\phi,0]=0.
\end{align}
From Eqs.~\eqref{eq:sABF} and \eqref{eq:dim8TBG}, the relative entropy between $P_0$ and $P_g$ is calculated as follows:
\begin{align}
    S(P_0||P_g)&\equiv\int d[\Phi_{\rm E}] \left(
    P_0\ln P_0-P_0\ln P_g
    \right)\notag
    \\
    &=-\ln Z_0[\phi]+\ln Z_g[\phi] +g\cdot \int d[\Phi_{\rm E}] P_0[\Phi_{\rm E}] I^{\rm (E)}_{\rm I}[\phi,\Phi_{\rm E}]\notag
    \\
    &=-\ln Z_0[\phi]+\ln Z_g[\phi]\notag
    \\
    &=W_0[\phi]-W_g[\phi]\notag
    \\
    &=g^2\cdot \frac{\alpha^2}{2m^2\Lambda^2}\int (d^4x)_{\rm E} (\partial_{\mu}\phi\partial^{\mu}\phi)^2\geq 0.
\end{align}
The first line is the definition of the relative entropy, and in the third line, ${\langle I_{\rm I}\rangle}_{g=0}=0$ is used.
By taking to be $g=1$, the relative entropy between $P_{\rm NI}$ and $P_{\rm T}$ is given by
\begin{align}
    S(P_{\rm NI}||P_{\rm T})&=W_{\rm NI}[\phi]-W_{\rm T}[\phi]=\frac{\alpha^2}{2m^2\Lambda^2}\int (d^4x)_{\rm E} (\partial_{\mu}\phi\partial^{\mu}\phi)^2\geq 0\Rightarrow \frac{\alpha^2}{2m^2\Lambda^2}\geq 0,
\end{align}
where $P_{\rm NI}=P_{0}$, $P_{\rm T}=P_{g=1}$, $W_{\rm NI}=W_0$, and $W_{\rm T}=W_{g=1}$ are used.
We see that this result is consistent with the upper bound of Eqs.~\eqref{eq:uplowBG} and \eqref{eq:wt}. 
In this target theory, $W_{\rm T}-W_{\rm NI}$ denotes the dimension-eight operator, and the non-negativity of the relative entropy yields a constraint on its Wilson coefficient.
\item {\bf Infinite heavy mass reference theory} ---
The action of the IHMRT in the Euclidean space is expressed as follows:
\begin{align}
    I^{\rm (E)}_{\rm IH}[\phi,\Phi]\equiv \lim_{m\to \infty}I_{\rm T}[\phi,\Phi].~~~~[{\rm \bf Reference}]
\end{align}
The classical solution of $I_{\rm IH}$ satisfies $\widetilde{\Phi}_{\rm IH}=0$.
Using the saddle point approximations, at the tree level, we obtain
\begin{align}
    &Z_{\rm IH}[\phi]\equiv\int d[\Phi_{\rm E}]e^{-I^{\rm (E)}_{\rm IH}[\phi,\Phi]}=e^{-I^{\rm (E)}_{\rm IH}[\phi,0]},
    \\
    &W_{\rm IH}[\phi]\equiv-\ln Z_{\rm IH}[\phi]=I^{\rm (E)}_{\rm IH}[\phi,0],\label{eq:WIHsi}
    \\
    &{\langle I_{\rm T}-\lim_{m\to\infty}I_{\rm T} \rangle}_{\rm IH}=0.
\end{align}
From Eqs.~\eqref{eq:dim8TBG} and \eqref{eq:WIHsi}, the difference of the Euclidean effective action is given by
\begin{align}
    W_{\rm T}[\phi]-W_{\rm IH}[\phi]=-\frac{\alpha^2}{2m^2 \Lambda^2}\int (d^4x)_{\rm E} (\partial_{\mu}\phi\partial^{\mu}\phi)^2\leq 0 \Rightarrow \frac{\alpha^2}{2m^2 \Lambda^2}\geq 0,
\end{align}
where $W_{\rm T}=W_{g=1}$ and Eq.~\eqref{eq:WIHsi} are used.
The above inequality represents Eq.~\eqref{eq:IHuplow} and is consistent with the non-negativity of the relative entropy.
Therefore, even in the IHMRT, the non-negativity of the relative entropy yields the constraint on the Wilson coefficient.
\end{itemize}

\subsection{A tree level UV completion of single massless scalar field theory with linear term}
\label{sec:lin}
Consider the theory of Eq.~\eqref{eq:scl_massless} with a linear term of $\Phi$ in the Minkowski space:
\begin{align}
    I_{\rm T}[\phi,\Phi]\equiv \int d^4x \left(\frac{1}{2}(\partial_{\mu}\phi\partial^{\mu}\phi)+\frac{1}{2}(\partial_{\mu}\Phi\partial^{\mu}\Phi)-\frac{m^2}{2}\left(\Phi- v\right)^2 +\frac{\alpha}{\Lambda}\cdot \Phi (\partial_{\mu}\phi\partial^{\mu}\phi) \right),~[{\rm \bf Target}]\label{eq:lineUV}
\end{align}
where $v$ denotes a dimensionful parameter.
The action in the Euclidean space is expressed as
\begin{align}
    I_{\rm T}^{\rm (E)}[\phi,\Phi]&=\int (d^4x)_{\rm E} \left(\frac{1}{2}(\partial_{I}\phi_{\rm E}\partial_{I}\phi_{\rm E})+\frac{1}{2}(\partial_{I}\Phi_{\rm E}\partial_{I}\Phi_{\rm E})+\frac{m^2}{2}\left(\Phi_{\rm E}- v\right)^2 +\frac{\alpha}{\Lambda}\cdot \Phi_{\rm E} (\partial_{I}\phi_{\rm E}\partial_{I}\phi_{\rm E})  \right),\notag
    \\
    &=\int (d^4x)_{\rm E} \left(-\frac{1}{2}(\partial_{\mu}\phi\partial^{\mu}\phi)+\frac{1}{2}(\partial_{I}\Phi_{\rm E}\partial_{I}\Phi_{\rm E})+\frac{m^2}{2}\left(\Phi_{\rm E}- v\right)^2 -\frac{\alpha}{\Lambda}\cdot \Phi_{\rm E} (\partial_{\mu}\phi\partial^{\mu}\phi)  \right).\label{eq:IT34}
\end{align}
Similar to the previous example, $\phi$ is the background field.
As discussed in Appendix~\ref{app:Wick_rotation}, the background field of the Euclidean space is defined from that of the Minkowski space.
The actions $I_0$ and $I_{\rm I}$ in the Euclidean space are respectively defined as,
\begin{align}
    I^{\rm (E)}_0 [\phi,\Phi]&\equiv I^{\rm (E)}_{\rm T} [0,\Phi]+\left(I^{\rm (E)}_{\rm T} [\phi,0]-\int (d^4 x)_{\rm E} \frac{m^2v^2}{2}\right),\notag
    \\
    &=\int (d^4x)_{\rm E} \left(-\frac{1}{2}(\partial_{\mu}\phi\partial^{\mu}\phi)+\frac{1}{2}(\partial_{I}\Phi_{\rm E}\partial_{I}\Phi_{\rm E})+\frac{m^2}{2}\left(\Phi_{\rm E} - v\right)^2\right),
    \\
    I^{\rm (E)}_{\rm I} [\phi,\Phi]&\equiv I^{\rm (E)}_{\rm T} [\phi,\Phi]-I^{\rm (E)}_{0} [\phi,\Phi]= -\frac{\alpha}{\Lambda}\cdot \int (d^4 x)_{\rm E} \Phi_{\rm E} (\partial_{\mu}\phi\partial^{\mu}\phi).
\end{align}
The partition function and effective action are defined as
\begin{align}
    &Z_{\rm T}[\phi]\equiv \int d[\Phi_{\rm E}] e^{-I^{\rm (E)}_{\rm T}[\phi,\Phi_{\rm E}]},
    \\
    &W_{\rm T}[\phi]\equiv  -\ln Z_{\rm T}[\phi].
\end{align}
For each reference theory in Sec.~\ref{sec:defTH}, we consider the constraints on the Euclidean effective action in the following way.
\begin{itemize}
    \item {\bf Non-interacting reference theory} ---
    The action of NIRT in the Euclidean space is defined as follows:
    \begin{align}
        I^{\rm (E)}_{\rm NI}[\phi,\Phi]\equiv \int (d^4 x)_{\rm E} \left(-\frac{1}{2}(\partial_{\mu}\phi \partial^{\mu}\phi)+\frac{1}{2}(\partial_I\Phi_{\rm E} \partial_I \Phi_{\rm E})+\frac{m^2}{2}\left(\Phi^2_{\rm E} -v\right)^2 \right).~[{\rm \bf Reference}]
    \end{align}
    Note here that the linear term of $\Phi_{\rm E}$ arises from the third term of the right-hand side.
    The solution of the equation of motion of $I^{\rm (E)}_{\rm NI}$ for $\Phi_{\rm E}$ is calculated as $\widetilde{\Phi}_{\rm  NI}=v$.
    The partition function and effective action of the NIRT are respectively defined as follows:
    \begin{align}
        &Z_{\rm NI}[\phi]\equiv \int d[\Phi_{\rm E}] e^{-I^{\rm (E)}_{\rm NI}[\phi,\Phi_{\rm E}]},
        \\
        &W_{\rm NI}[\phi]\equiv -\ln Z_{\rm NI}[\phi].
    \end{align}
    By introducing the parameter $g$, we define
    \begin{align}
        I_g^{\rm (E)}[\phi,\Phi]\equiv I^{(\rm E)}_0[\phi,\Phi]+g\cdot I^{(\rm E)}_{\rm I}[\phi,\Phi]. 
    \end{align}
    The target and reference theories are given by,
    \begin{align}
        I^{(\rm E)}_{\rm T}[\phi,\Phi]=\lim_{g\to 1} I_g^{\rm (E)}[\phi,\Phi],~~~I^{(\rm E)}_{\rm NI}[\phi,\Phi]=\lim_{g\to 0} I_g^{\rm (E)}[\phi,\Phi].
    \end{align}
    The partition function and effective action of $I_g^{\rm (E)}$ are respectively calculated as follows:
    \begin{align}
        &Z_g[\phi]\equiv \int d[\Phi_{\rm E}] e^{-I_{g}^{\rm (E)}[\phi,\Phi_{\rm E}]} =e^{-I_{g}^{\rm (E)}[\phi,\widetilde{\Phi}_{g}]},
        \\
        &W_g[\phi]\equiv-\ln Z_g[\phi]= \int (d^4x)_{\rm E} \left(-\frac{1}{2}(\partial_{\mu}\phi\partial^{\mu}\phi)-g^2\cdot\frac{\alpha^2}{2m^2\Lambda^2}(\partial_{\mu}\phi\partial^{\mu}\phi)^2-g\cdot\frac{\alpha}{\Lambda}\cdot v (\partial_{\mu} \phi\partial^{\mu} \phi) \right),\label{eq:Wglin}
    \end{align}
    where $\widetilde{\Phi}_{g}$ is the classical solution of $I_{g}^{\rm (E)}$ and is calculated as follows:
\begin{align}
   \widetilde{\Phi}_{g}=v+g\cdot\frac{\alpha}{m^2\Lambda}(\partial_{\mu}\phi\partial^{\mu}\phi).
\end{align}
By defining the probability distribution function as
\begin{align}
    P_g[\Phi]\equiv \frac{e^{-I_g^{\rm (E)}[\phi,\Phi]}}{Z_g[\phi]},
\end{align}
the expectation value of the interacting term is calculated as
    \begin{align}
        {\langle I_{\rm I}\rangle}_{g=0}&=\int d[\Phi_{\rm E}] P_{0}[\Phi_{\rm E}] I^{\rm (E)}_{\rm I}[\phi,\Phi_{\rm E}]\notag
        \\
        &= I^{\rm (E)}_{\rm I}[\phi,\widetilde{\Phi}_0]\notag
        \\ &=-\frac{\alpha}{\Lambda}\cdot \int (d^4x)_{\rm E} v (\partial_{\mu}\phi\partial^{\mu}\phi).\label{eq:linIIg01}
    \end{align}
    From Eqs.~\eqref{eq:sABF}, \eqref{eq:Wglin}, and \eqref{eq:linIIg01}, the relative entropy between $P_0$ and $P_g$ is calculated as follows:
    \begin{align}
        S(P_0||P_g) &\equiv \int d[\Phi_{\rm E}] \left(
        P_0[\Phi_{\rm E}]\ln P_0[\Phi_{\rm E}]-P_0[\Phi_{\rm E}] \ln P_g[\Phi_{\rm E}]
        \right)\notag
        \\
        &=-\ln Z_0[\phi]+\ln Z_g[\phi] +g\cdot \int d[\Phi_{\rm E}] P_{g}[\Phi_{\rm E}] I^{\rm (E)}_{\rm I}[\phi,\Phi_{\rm E}]\notag
        \\
        &=-\ln Z_0[\phi]+\ln Z_g[\phi]+g\cdot {\langle I_{\rm I}\rangle}_{g=0}\notag
        \\ &=W_0[\phi]-W_g[\phi]-g\cdot \frac{\alpha}{\Lambda}\cdot \int (d^4x)_{\rm E} v (\partial_{\mu}\phi\partial^{\mu}\phi)\notag
        \\ &=g^2\cdot\frac{\alpha^2}{2m^2\Lambda^2}\int (d^4 x)_{\rm E}  (\partial_{\mu}\phi\partial^{\mu}\phi)^2\geq 0,
    \end{align}
    where the first line is the definition of the relative entropy, Eq.~\eqref{eq:linIIg01} was used in the third line, and $g\cdot {\langle I_{\rm I}\rangle}_{g=0}$ cancels in the last line.
    By taking to be $g=1$, the relative entropy between $P_{\rm NI}$ and $P_{\rm T}$ is given by
  \begin{align}
        S(P_{\rm NI}||P_{\rm T}) &=W_{\rm NI}[\phi]-W_{\rm T}[\phi]+ {\langle I_{\rm I}\rangle}_{\rm NI}=\frac{\alpha^2}{2m^2\Lambda^2}\int (d^4 x)_{\rm E}  (\partial_{\mu}\phi\partial^{\mu}\phi)^2\geq 0 \Rightarrow \frac{\alpha^2}{2m^2\Lambda^2}\geq 0,\label{eq:lires1}
    \end{align}
    where $P_{\rm NI}=P_0$, $P_{\rm T}=P_{g=1}$, $W_{\rm NI}=W_0$, $W_{\rm T}=W_{g=1}$, and ${\langle I_{\rm I}\rangle}_{\rm NI}={\langle I_{\rm I}\rangle}_{g=0}$ are used. 
    The above inequality yields a constraint on the Wilson coefficient of the dimension-eight operator, which is consistent with the non-negativity of the relative entropy. 

    We also show that the above result holds even if ${\langle I_{\rm I}\rangle}_{\rm NI}$ is eliminated by a redefinition of $\Phi$.
    We have already seen this fact in Sec. \ref{sec:noninref} generically.
    By defining a new field $\eta$ as $\Phi\equiv \eta+v$, the action of Eq.~\eqref{eq:IT34} is expressed as,
    \begin{align}
        I'_{\rm T}{}^{({\rm E})}[\phi,\eta]&\equiv I_{\rm T}^{(\rm E)}[\phi,\eta+v]\notag
        \\
        &=\int (d^4x)_{\rm E} \bigg(-\frac{1}{2}(\partial_{\mu}\phi\partial^{\mu}\phi)+\frac{1}{2}(\partial_{I}\eta_{\rm E}\partial_{I}\eta_{\rm E})+\frac{1}{2}m^2\eta_{\rm E}^2 \notag
        \\ &\quad\quad\quad\quad-\frac{\alpha}{\Lambda}\cdot v (\partial_{\mu}\phi\partial^{\mu}\phi)-\frac{\alpha}{\Lambda}\cdot \eta_{\rm E} (\partial_{\mu}\phi\partial^{\mu}\phi)\bigg).~~~~[{\rm \bf Target}]
    \end{align}
    Note here that the liner term of $\eta$ does not arise.
    Define the actions $I'_0$ and $I'_{\rm I}$ in the Euclidean space as follows:
    \begin{align}
        &{I'}_0^{\rm (E)}[\phi,\eta]\equiv \int (d^4x)_{\rm E} \left(
        -\frac{1}{2}(\partial_{\mu}\phi\partial^{\mu}\phi)+\frac{1}{2}(\partial_I\eta_{\rm E}\partial_I\eta_{\rm E})+\frac{m^2}{2}\eta_{\rm E}^2-\frac{\alpha}{\Lambda}\cdot v (\partial_{\mu}\phi\partial^{\mu}\phi)
        \right),
        \\
        &{I'}_{\rm I}^{\rm (E)}[\phi,\eta]\equiv -\frac{\alpha}{\Lambda}\cdot \int (d^4x)_{\rm E} \eta_{\rm E} (\partial_{\mu}\phi\partial^{\mu}\phi). 
    \end{align}
    The partition function and effective action of $I'_{\rm T}{}^{({\rm E})}$ are defined as
    \begin{align}
        &Z'_{\rm T}[\phi]\equiv \int d[\eta_{\rm E}] e^{-{I'}^{(\rm E)}_{\rm T}[\phi,\eta_{\rm E}]},
        \\
        &W'_{\rm T}[\phi]\equiv -\ln Z'_{\rm T}[\phi].
    \end{align}
    The action of NIRT for ${I'}^{(\rm E)}_{\rm T}$ is defined as follows:
    \begin{align}
        {I'}^{(\rm E)}_{\rm NI}[\phi,\eta]&\equiv {I'}^{\rm (E)}_0[\phi,\eta]\notag
        \\
        &=\int (d^4x)_{\rm E} \left(-\frac{1}{2}(\partial_{\mu}\phi\partial^{\mu}\phi)+\frac{1}{2}(\partial_I\eta_{\rm E}\partial_I\eta_{\rm E})+\frac{1}{2}m^2 \eta^2_{\rm E}-\frac{\alpha}{\Lambda}\cdot \eta_{\rm E} (\partial_{\mu}\phi\partial^{\mu}\phi)
        \right),~~~~[{\rm \bf Reference}]
    \end{align}
    where the liner term of $\eta$ does not arise.
    The partition function and effective action of the NIRT are respectively defined as follows:
    \begin{align}
        &Z'_{\rm NI}[\phi]\equiv \int d[\eta_{\rm E}] e^{-{I'}^{\rm (E)}_{\rm NI}[\phi,\eta_{\rm E}]},
        \\
        &W'_{\rm NI}[\phi]\equiv -\ln Z'_{\rm NI}[\phi].
    \end{align}
    By introducing the parameter $g$, we define
    \begin{align}
        {I'}^{\rm (E)}_g[\phi,\eta]\equiv {I'}^{\rm (E)}_0[\phi,\eta]+g\cdot {I'}^{\rm (E)}_{\rm I}[\phi,\eta]. 
    \end{align}
    The target and reference theories are expressed as follows:
    \begin{align}
        {I'}^{\rm (E)}_{\rm T}[\phi,\eta]=\lim_{g\to 1} {I'}^{\rm (E)}_g[\phi,\eta],~~~{I'}^{\rm (E)}_{\rm NI}[\phi,\eta]=\lim_{g\to 0} {I'}^{\rm (E)}_g[\phi,\eta].
    \end{align}
    The partition function and effective action of ${I'}^{\rm (E)}_g$ are respectively calculated as follows:
    \begin{align}
     &Z'_{g}[\phi]\equiv \int d[\eta_{\rm E}] e^{-{I'}^{\rm (E)}_{g}[\phi,\eta_{\rm E}]}=e^{-{I'}^{\rm (E)}_{g}[\phi,\widetilde{\eta}_{g}]},
        \\
    &W'_{g}[\phi]\equiv {I'}^{\rm (E)}_{g}[\phi,\widetilde{\eta}_{g}]=\int (d^4x)_{\rm E} \left(
     -\frac{1}{2}\left(1+2\cdot \frac{\alpha}{\Lambda}\cdot v\right)(\partial_{\mu}\phi\partial^{\mu}\phi)-g^2\cdot\frac{\alpha^2}{2\Lambda^2 m^2}(\partial_{\mu}\phi\partial^{\mu}\phi)^2
        \right),\label{eq:Wgpri}
    \end{align}
    where $\widetilde{\eta}_{g}$ is a classical solution of ${I'}^{\rm (E)}_{g}$ and is given by
    \begin{align}
        \widetilde{\eta}_{g}=g\cdot\frac{\alpha}{m^2 \Lambda} (\partial_{\mu}\phi\partial^{\mu}\phi).
    \end{align}
    By defining the probability distribution function as 
    \begin{align}
        P'_g[\eta]\equiv \frac{e^{-{I'}^{\rm (E)}_g[\phi,\eta]}}{Z'_g[\phi]},
    \end{align}
    the expectation value of the interacting term is calculated as
    \begin{align}
        {\langle I_{\rm I}\rangle}_{g=0}&=\int d[\eta_{\rm E}] P'_{0}[\eta_{\rm E}] I^{\rm (E)}_{\rm I}[\phi,\eta_{\rm E}]\notag
        \\
        &=I^{\rm (E)}_{\rm I}[\phi,\widetilde{\eta}_{0}]\notag
        \\
        &=0.\label{eq:IIpr}
    \end{align}
    From Eqs.~\eqref{eq:sABF}, \eqref{eq:Wgpri}, and \eqref{eq:IIpr}, the relative entropy between $P_0$ and $P_g$ is calculated as follows:
    \begin{align}
        S(P'_0||P'_g) &\equiv \int d[\eta_{\rm E}] \left(
        P'_0[\eta_{\rm E}]\ln P'_0[\eta_{\rm E}]-P'_0[\eta_{\rm E}] \ln P'_g[\eta_{\rm E}]
        \right)\notag
        \\
        &=-\ln Z'_0[\phi]+\ln Z'_g[\phi] +g\cdot \int d[\eta_{\rm E}]P_0[\eta_{\rm E}] I^{\rm (E)}_{\rm I}[\phi,\eta_{\rm E}]
        \\
        &=-\ln Z'_0[\phi]+\ln Z'_g[\phi]
        \\
        &=W'_0[\phi]-W'_g[\phi]\notag
        \\
        &=g^2\cdot\frac{\alpha^2}{2\Lambda^2 m^2}\int (d^4 x)_{\rm E}(\partial_{\mu}\phi\partial^{\mu}\phi)^2\geq 0,
    \end{align}
    where the first line is the definition of the relative entropy, and ${\langle I_{\rm I}\rangle}_{g=0}=0$ is used in the third line.
    By taking to be $g=1$, the relative entropy between $P'_{\rm NI}\equiv P'_{0}$ and $P'_{\rm T}\equiv P'_{g=1}$ is given by
    \begin{align}
        S(P'_{\rm NI}||P'_{\rm T}) =W'_{\rm NI}[\phi]-W'_{\rm T}[\phi]=\frac{\alpha^2}{2\Lambda^2 m^2}\int (d^4 x)_{\rm E}(\partial_{\mu}\phi\partial^{\mu}\phi)^2\geq 0\Rightarrow  \frac{\alpha^2}{2\Lambda^2 m^2}\geq 0,
    \end{align}
    where $W'_{\rm NI}\equiv W'_{0}$, and $W'_{\rm T}\equiv W'_{g=1}$ are used.
    This result is the same as Eq.~\eqref{eq:lires1}, and we found that Eq.~\eqref{eq:uppFT} is invariant under the field redefinition to remove the linear term of $\Phi$.
    Therefore, it is found that the constraint on the EFT does not depend on the condition of vanishing the linear term.
    As explained in the details in Sec.~\ref{sec:noninref}, this is because the linear term proportional to $g$ cancels in the relative entropy as shown in Eq.~\eqref{eq:Wsec}.

    \item {\bf Massive free field reference theory} ---
    The action of the MFFRT in the Euclidean space is defined as follows:
    \begin{align}
        I^{\rm (E)}_{\rm MF}[\phi,\Phi]\equiv \int (d^4x)_{\rm E} \left(-\frac{1}{2}(\partial_{\mu} \phi\partial_{\mu} \phi) +\frac{1}{2} (\partial_I \Phi_{\rm E}\partial_I \Phi_{\rm E})+\frac{m^2}{2}\Phi^2_{\rm E}+\frac{m^2 v^2}{2} \right).~[{\rm \bf Reference}]
    \end{align}
    Note here that the MFFRT does not include self-interacting terms of $\Phi_{\rm E}$.
    The solution of the equation of motion of $I^{\rm (E)}_{\rm MF}$ for $\Phi_{\rm E}$ is calculated as $\widetilde{\Phi}_{\rm MF}=0$.
    The partition function and effective action of the MFFRT are respectively defined as follows:
    \begin{align}
        &Z_{\rm MF}[\phi]\equiv \int d[\Phi_{\rm E}] e^{-I^{(\rm E)}_{\rm MF}[\phi,\Phi_{\rm E}]},
        \\
        &W_{\rm MF}[\phi]\equiv -\ln Z_{\rm MF}[\phi].
    \end{align}
    By introducing the parameter $g$, we define
    \begin{align}
        J_g^{\rm (E)}[\phi,\Phi]\equiv I_0^{\rm (E)}[\phi,0]+I_{\Phi}^{\rm (E)}[\Phi] +g\cdot \left(I_{\rm S}^{\rm (E)}[\Phi]+I_{\rm I}^{\rm (E)}[\phi,\Phi]\right),
    \end{align}
    with
    \begin{align}
        &I_{\Phi}^{\rm (E)}[\Phi]\equiv \int (d^4x)_{\rm E}\left(
        \frac{1}{2}(\partial_I \Phi_{\rm E} \partial_I \Phi_{\rm E}) +\frac{m^2}{2}\Phi^2_{\rm E}
        \right),
        \\
        &I_{\rm S}^{\rm (E)}[\Phi]\equiv -v\cdot m^2 \int (d^4x)_{\rm E} \Phi_{\rm E}.
    \end{align}
    The target and reference theories are expressed as follows:
    \begin{align}
        I^{\rm (E)}_{\rm T}[\phi,\Phi]=\lim_{g\to 1}J_g^{\rm (E)}[\phi,\Phi],~~~I^{\rm (E)}_{\rm MF}[\phi,\Phi]=\lim_{g\to 0}J_g^{\rm (E)}[\phi,\Phi].
    \end{align}
    The partition function and effective action of $J_g^{\rm (E)}$ are respectively calculated as follows:
    \begin{align}
        &z_g[\phi]\equiv \int d[\Phi_{\rm E}] e^{-J_g^{\rm (E)}[\phi,\Phi_{\rm E}]} =e^{-J_g^{\rm (E)}[\phi,\widetilde{\Phi}_{g}]},
        \\
        &w_g[\phi]\equiv -\ln z_g[\phi]\notag
        \\
        &=\int (d^4x)_{\rm E}\left(-\frac{1}{2}(\partial_{\mu}\phi\partial^{\mu}\phi)-g^2\cdot \frac{\alpha^2}{2m^2 \Lambda^2}(\partial_{\mu}\phi\partial^{\mu}\phi)^2-g^2\cdot \frac{\alpha}{\Lambda}\cdot v (\partial_{\mu}\phi\partial^{\mu}\phi)
        +(1-g^2)\cdot \frac{m^2 v^2}{2}\right),\label{eq:wgJg}
    \end{align}
    where $\widetilde{\Phi}_g$ is the classical solution of $J_g^{\rm (E)}$ and is given as
    \begin{align}
       \widetilde{\Phi}_g=g\cdot v+g\cdot\frac{\alpha}{m^2\Lambda} (\partial_{\mu}\phi\partial^{\mu}\phi).\label{eq:phigJg}
    \end{align}
    By defining the probability distribution function as
    \begin{align}
        p_g[\Phi]\equiv \frac{e^{-J_g^{\rm (E)}[\phi,\Phi]}}{z_g[\phi]},
    \end{align}
    we obtain
    \begin{align}
        {\langle I_{\rm S}+I_{\rm I}\rangle}_{g=0}&\equiv \int d[\Phi_{\rm E}] p_0[\Phi_{\rm E}] \left(I_{\rm S}[\Phi_{\rm E}]+I_{\rm I}[\phi,\Phi_{\rm E}]\right)\notag
        \\
        &=I_{\rm S}[\widetilde{\Phi}_{0}]+I_{\rm I}[\phi,\widetilde{\Phi}_{0}]\notag
        \\
        &=0,\label{eq:ISII}
    \end{align}
    where $\widetilde{\Phi}_0=0$ holds from Eq.~\eqref{eq:phigJg}.
    From Eqs.~\eqref{eq:sABF}, \eqref{eq:wgJg}, and \eqref{eq:ISII}, the relative entropy between $p_0$ and $p_g$ is calculated as follows:
    \begin{align}
        S(p_0||p_g)&\equiv \int d[\Phi_{\rm E}]\left(p_0[\Phi_{\rm E}]\ln p_0[\Phi_{\rm E}]-p_0[\Phi_{\rm E}] \ln p_g[\Phi_{\rm E}] \right)\notag
        \\
        &=-\ln z_0[\phi]+\ln z_g[\phi] +g\cdot \int d[\Phi_{\rm E}] p_0[\Phi_{\rm E}] \left(I_{\rm S}[\Phi_{\rm E}]+I_{\rm I}[\phi,\Phi_{\rm E}]\right)\notag
        \\
        &=-\ln z_0[\phi]+\ln z_g[\phi]\notag
        \\
        &=w_0[\phi]-w_g[\phi]\notag
        \\
        &=g^2\cdot\frac{1}{2m^2}\int (d^4x)_{\rm E} \left(\frac{\alpha}{\Lambda} (\partial_{\mu}\phi\partial^{\mu}\phi)+m^2v^2\right)^2\geq 0.
    \end{align}
    By taking to be $g=1$, the relative entropy between $P_{\rm MF}$ and $P_{\rm T}$ is given by
    \begin{align}
        S(P_{\rm MF}||P_{\rm T})=W_{\rm MF}[\phi]-W_{\rm T}[\phi]=\frac{1}{2m^2}\int (d^4x)_{\rm E} \left(\frac{\alpha}{\Lambda} (\partial_{\mu}\phi\partial^{\mu}\phi)+m^2v^2\right)^2\geq 0,
    \end{align}
    where $P_{\rm MF}= p_0$, $P_{\rm T}= p_{g=1}$, $W_{\rm MF}=w_0$, and $W_{\rm T}=w_1$ are used.
    This inequality denotes Eq.~\eqref{eq:wt} and is consistent with the non-negativity of relative entropy.
    Because of the self-interacting term of $\Phi$, the shift of the Euclidean effective action includes the term independent of $\phi$.
\end{itemize}

\subsection{Neutral bosons interacting with photon}
\label{sec:neutral}
We consider heavy neutral bosons such as the dilaton and the axion.
For each model, we evaluate the relative entropy in the following way.
\subsubsection{Dilaton}
\label{sec:dilat}
    The action of the dilaton in the Minkowski space is expressed as
       \begin{align}
           I_{\rm T}[A,\phi]\equiv\int d^4 x \left[-\frac{1}{4}F_{\mu\nu}F^{\mu\nu}+\frac{1}{2}(\partial_{\mu}\phi)^2 -\frac{m_{\phi}^2}{2} \phi^2 +\frac{\phi}{f_{\phi}}F_{\mu\nu}F^{\mu\nu}\right],~~~~[{\rm \bf Target}]\label{eq:Iphi}
       \end{align}
       where $m_{\phi}$ and $f_{\phi}$ are the mass and the decay constant of the heavy neutral scalar boson, respectively, and $F_{\mu\nu}$ is the field strength of photon field defined by $F_{\mu\nu}=\partial_{\mu}A_{\nu}-\partial_{\nu}A_{\mu}$.
       Based on the procedure in Appendix \ref{app:Wick_rotation}, the action in the Euclidean space is obtained as
       \begin{align}
           I_{\rm T}^{\rm (E)}[A,\phi_{\rm E}]&=\int (d^4 x)_{\rm E} \left(\frac{1}{4}F_{{\rm E},IJ} F_{{\rm E},IJ}+\frac{1}{2}\partial_I \phi_{\rm E}\partial_I \phi_{\rm E} +\frac{m_{\phi}^2}{2}  (\phi_{\rm E})^2 -\frac{\phi_{\rm E}}{f_{\phi}}F_{{\rm E},IJ}F_{{\rm E},IJ}\right),\notag
           \\
           &=\int (d^4 x)_{\rm E} \left(\frac{1}{4}F_{\mu\nu}F^{\mu\nu}+\frac{1}{2}\partial_I \phi_{\rm E}\partial_I \phi_{\rm E} +\frac{m_{\phi}^2}{2}  (\phi_{\rm E})^2 -\frac{\phi_{\rm E}}{f_{\phi}}F_{\mu\nu}F^{\mu\nu}\right),
       \end{align}
       where $F_{\mu\nu}$ is a background field in the Minkowski pace.
       Then, $I_0$ and $I_{\rm I}$ are defined as follows:
       \begin{align}
           I_0^{\rm (E)}[A,\phi_{\rm E}]&\equiv I_{\rm T}^{\rm (E)}[A,0]+I_{\rm T}^{\rm (E)}[0,\phi_{\rm E}]=\int (d^4 x)_{\rm E} \left(\frac{1}{4}F_{\mu\nu}F^{\mu\nu}+\frac{1}{2}\partial_I \phi_{\rm E}\partial_I \phi_{\rm E} +\frac{m_{\phi}^2}{2}  (\phi_{\rm E})^2 \right),
           \\
           I_{\rm I}^{\rm (E)}[A,\phi_{\rm E}]&\equiv I_{\rm T}^{\rm (E)}[A,\phi_{\rm E}]-I_0[A,\phi_{\rm E}]=\int (d^4 x)_{\rm E} \left(-\frac{\phi_{\rm E}}{f_{\phi}}F_{\mu\nu}F^{\mu\nu}\right).
       \end{align}
       The partition function and effective action are defined as
       \begin{align}
           &Z_{\rm T}[A]\equiv \int d[\phi_{\rm E}] e^{-I_{\rm T}^{\rm (E)}[A,\phi_{\rm E}]},
           \\
           &W_{\rm T}[A]\equiv -\ln Z_{\rm T}[A],
       \end{align}
       In this target theory, the NIRT is the same as the MFFRT, and the action of the NIRT and MFFRT in the Euclidean space is defined as
    \begin{align}
        I_{\rm NI}^{\rm (E)}[A,\phi_{\rm E}]\equiv \int (d^4x)_{\rm E} \left(\frac{1}{4}F_{\mu\nu}F^{\mu\nu}+\frac{1}{2}\partial_I \phi_{\rm E}\partial_I \phi_{\rm E} +\frac{m^2_{\phi}}{2}(\phi_{\rm E})^2\right).~~~~[{\rm \bf Reference}]
    \end{align}
       The solution of the equation of motion of $I_{\rm NI}^{\rm (E)}$ is calculated as $\widetilde{\phi}_{\rm NI}=0$.
       The partition function and effective action of the reference theory are respectively defined as follows:
       \begin{align}
            &Z_{\rm NI}[A]\equiv \int d[\phi_{\rm E}] e^{- I_{\rm NI}^{\rm (E)}[A,\phi_{\rm E}]},
            \\
            &W_{\rm NI}[A]\equiv -\ln Z_{\rm NI}[A].
       \end{align}
       By introducing the parameter $g$, we define
       \begin{align}
           I_{g}^{\rm (E)}[A,\phi]\equiv I_0^{\rm (E)}[A,\phi]+g\cdot I_{\rm I}^{\rm (E)}[A,\phi].
       \end{align}
       The target theory and reference theories are expressed as follows:
       \begin{align}
           I^{\rm (E)}_{\rm T}[A,\phi]=\lim_{g\to 1}I_{g}^{\rm (E)}[A,\phi],~~~I^{\rm (E)}_{\rm NI}[A,\phi]=\lim_{g\to 0}I_{g}^{\rm (E)}[A,\phi].
       \end{align}
       The partition function and effective action of $I_g^{\rm (E)}$ are respectively calculated as follows:
       \begin{align}
           Z_g[A]&\equiv\int d[\phi_{\rm E}] e^{-I_g^{\rm (E)}[A,{\phi}_{\rm E}]}=e^{-I_g^{\rm (E)}[A,\widetilde{\phi}_g]},
           \\
           W_g[A]&\equiv -\ln Z_g[A]=\int (d^4x)_{\rm E} \left(\frac{1}{4}F_{\mu\nu}F^{\mu\nu}-g^2\cdot \frac{1}{2 f^2_{\phi}m_{\phi}^2}(F_{\mu\nu}F^{\mu\nu})^2  \right),\label{eq:Wgdila}
       \end{align}
       where the solution of the equation of motion of $I_{g}^{\rm (E)}$ for $\phi_{\rm E}$ with the heavy mass is calculated as
       \begin{align}
           \widetilde{\phi}_g=g\cdot \frac{1}{f_{\phi}m_{\phi}^2}F_{\mu\nu}F^{\mu\nu}.\label{eq:phidsila}
       \end{align}
       By defining the probability distribution function 
       \begin{align}
           P_g[\phi]\equiv \frac{e^{-I_g^{\rm (E)}[A,\phi]}}{Z_g[A]},
       \end{align}
       we obtain
       \begin{align}
           {\langle I_{\rm I}\rangle}_{g=0}&=\int d[\phi_{\rm E}]P_0[\phi_{\rm E}] I_{\rm I}[A,\phi_{\rm E}]\notag
           \\
           &=I_{\rm I}[A,\widetilde{\phi}_{0}]\notag
           \\
           &=0,\label{eq:IIdila}
       \end{align}
       where $\widetilde{\phi}_0=0$ holds from Eq.~\eqref{eq:phidsila}.
       From Eqs.~\eqref{eq:sABF}, \eqref{eq:Wgdila}, and \eqref{eq:phidsila}, the relative entropy between $P_0$ and $P_g$ is calculated as follows:
       \begin{align}
           S(P_0||P_g)&\equiv \int d[\phi_{\rm E}] \left(
           P_0[\phi_{\rm E}]\ln P_0[\phi_{\rm E}]-P_0[\phi_{\rm E}]\ln P_g[\phi_{\rm E}] 
           \right)\notag
           \\
           &=-\ln Z_0[A] +\ln Z_g[A] +g\cdot \int d[\phi_{\rm E}] P_0[\phi_{\rm E}] I_{\rm I}[A,\phi_{\rm E}]\notag
           \\
           &=-\ln Z_0[A] +\ln Z_g[A]\notag
           \\
           &=W_0[A]-W_g[A]\notag
           \\
           &=g^2\cdot\frac{1}{2 f^2_{\phi}m_{\phi}^2}\int (d^4 x)_{\rm E}(F_{\mu\nu}F^{\mu\nu})^2\geq 0,
       \end{align}
       where the first line is the definition of the relative entropy, Eq.~\eqref{eq:IIdila} is used in the third line, the fourth line is the definition of the effective action, and Eq.~\eqref{eq:Wgdila} and the non-negativity of the relative entropy are used in the last line.
       By taking to be $g=1$, the relative entropy between $P_{\rm NI}$ and $P_{\rm T}$ is given by
       \begin{align}
          S(P_{\rm NI}||P_{\rm T})&=W_{\rm NI}[A]-W_{\rm T}[A]=\frac{1}{2 f^2_{\phi}m_{\phi}^2}\int (d^4 x)_{\rm E}(F_{\mu\nu}F^{\mu\nu})^2\geq 0\Rightarrow \frac{1}{2 f^2_{\phi}m_{\phi}^2}\geq 0,
       \end{align}
where $P_{\rm NI}=P_0$, $P_{\rm T}=P_{g=1}$, $W_{\rm NI}=W_0$, and $W_{\rm T}=W_{g=1}$ are used.
        The relative entropy denotes the dimension-eight term and yields the constraints on the Wilson coefficients of the dimension-eight operator. 
\subsubsection{Axion}
\label{sec:axion}
The action of the axion in the Minkowski space is expressed as
    \begin{align}
        I_{\rm T}[A,a]\equiv \int d^4 x \left[-\frac{1}{4}F_{\mu\nu}F^{\mu\nu}+\frac{1}{2}(\partial_{\mu}a)^2 -\frac{m_{a}^2}{2} a^2 +\frac{a}{f_{a}}F_{\mu\nu}\widetilde{F}^{\mu\nu}\right],~~~~[{\rm\bf Target}]\label{eq:Ia}
    \end{align} 
       where $m_{a}$ and $f_{a}$ are the mass and the decay constant of the heavy neutral pseudo-scalar boson, respectively, and the dual field strength is defined as $\widetilde{F}^{\mu\nu}=\frac{1}{2}\epsilon^{\mu\nu\rho\sigma}F_{\rho\sigma}$.
       The action in the Euclidean space is given by
       \begin{align}
           I^{\rm (E)}_{\rm T}[A,a_{\rm E}]&=\int (d^4x)_{\rm E} \left(\frac{1}{4}F_{{\rm E},IJ}F_{{\rm E},IJ}+\frac{1}{2}\partial_I a_{\rm E}\partial_I a_{\rm E} +\frac{m_a^2}{2}(a_{\rm E})^2-i\frac{a_{\rm E}}{f_a}F_{{\rm E},IJ}\widetilde{F}_{{\rm E},IJ}\right),\notag
           \\
           &=\int (d^4x)_{\rm E} \left(\frac{1}{4}F_{\mu\nu}F^{\mu\nu}+\frac{1}{2}\partial_I a_{\rm E}\partial_I a_{\rm E} +\frac{m_a^2}{2}(a_{\rm E})^2-\frac{a_{\rm E}}{f_a}F_{\mu\nu}\widetilde{F}^{\mu\nu}\right),
       \end{align}
       where the background field of the Euclidean space  $F_{{\rm E},IJ}$ is defined from that of the Minkowski space; see Appendix~\ref{app:Wick_rotation}. 
       We define $I_0$ and $I_{\rm I}$ as follows:
       \begin{align}
           &I_0^{\rm (E)}[A,a_{\rm E}]\equiv I_{\rm T}^{\rm (E)}[A,0]+I_{\rm T}^{\rm (E)}[0,a_{\rm E}]=\int (d^4x)_{\rm E} \left(\frac{1}{4}F_{\mu\nu}F^{\mu\nu}+\frac{1}{2}\partial_I a_{\rm E}\partial_I a_{\rm E} +\frac{m_a^2}{2}(a_{\rm E})^2\right),
           \\
           &I_{\rm I}^{\rm (E)}[A,a_{\rm E}]\equiv I_{\rm T}^{\rm (E)}[A,a_{\rm E}]-I_0[A,a_{\rm E}]=\int (d^4x)_{\rm E} \left(-\frac{a_{\rm E}}{f_a}F_{\mu\nu}\widetilde{F}^{\mu\nu}\right).
       \end{align}
       The partition function and effective action are defined as
       \begin{align}
           &Z_{\rm T}[A]\equiv \int d[a_{\rm E}] e^{-I^{\rm (E)}_{\rm T}[A,a_{\rm E}]},
           \\
           &W_{\rm T}[A]\equiv -\ln Z_{\rm T}[A].
       \end{align}
       The action of the NIRT and MFFRT in the Euclidean space is defined as
       \begin{align}
           I^{\rm (E)}_{\rm NI}[A,a_{\rm E}]\equiv \int (d^4x)_{\rm E} \left(\frac{1}{4}F_{\mu\nu}F^{\mu\nu}+\frac{1}{2}\partial_I a_{\rm E}\partial_I a_{\rm E} +\frac{m_a^2}{2}(a_{\rm E})^2\right).~~~~[{\rm\bf Reference}]
       \end{align}
       The solution of the equation of motion of $I^{\rm (E)}_{\rm NI}$ is calculated as $\widetilde{a}_{\rm NI}=0$.
       The partition function and effective action of the reference theory are defined as
       \begin{align}
           &Z_{\rm NI}[A]\equiv \int d[a_{\rm E}] e^{-I^{\rm (E)}_{\rm NI}[A,a_{\rm E}]},
           \\
           &W_{\rm NI}[A]\equiv -\ln Z_{\rm NI}[A].
       \end{align}
       By introducing the parameter $g$, we define
       \begin{align}
           I_g^{\rm (E)}[A,a]\equiv I^{\rm (E)}_{0}[A,a]+g\cdot I^{\rm (E)}_{\rm I}[A,a]. 
       \end{align}
       The target and reference theories are expressed as follows:
       \begin{align}
           I^{\rm (E)}_{\rm T}[A,a]=\lim_{g\to 1}I_g^{\rm (E)}[A,a],~~~I^{\rm (E)}_{\rm NI}[A,a]=\lim_{g\to 0}I_g^{\rm (E)}[A,a].
       \end{align}
       The partition function and effective action of $I_g^{\rm (E)}$ are respectively calculated as follows:
       \begin{align}
           &Z_g[A]\equiv \int d[a_{\rm E}] e^{-I_g^{\rm (E)}[A,a_{\rm E}]}=e^{-I_g^{\rm (E)}[A,\widetilde{a}_{g}]},
           \\
           &W_g[A]\equiv -\ln Z_g[A] =\int (d^4x)_{\rm E} \left(\frac{1}{4}F_{\mu\nu}F^{\mu\nu}-g^2\cdot\frac{1}{2 f_a^2 m^2_a}(F_{\mu\nu}\widetilde{F}^{\mu\nu})^2 \right),\label{eq:Wgaxon}
       \end{align}
       where the solution of the equation of motion of $I_{g}^{\rm (E)}$ for $a_{\rm E}$ is calculated as
       \begin{align}
           \widetilde{a}_g=g\cdot\frac{1}{f_a m^2_a}F_{\mu\nu}\widetilde{F}^{\mu\nu}.\label{eq:agaxion}
       \end{align}
       By defining the probability distribution function as
       \begin{align}
       P_g[a]\equiv \frac{e^{-I_g^{\rm (E)}[A,a]}}{Z_g[A]},
       \end{align}
       the expectation value of the interaction is calculated as follows:
       \begin{align}
           {\langle I_{\rm I}\rangle}_{g=0}&\equiv \int d[a_{\rm E}] P_0[a_{\rm E}] I_{\rm I}^{\rm (E)}[A,a_{\rm E}]\notag
           \\
           &=I_{\rm I}^{\rm (E)}[A,\widetilde{a}_{0}]\notag
           \\
           &=0,\label{eq:IIaxion}
       \end{align}
       where $\widetilde{a}_0=0$ holds from Eq.~\eqref{eq:agaxion}.
       From Eqs.~\eqref{eq:sABF}, \eqref{eq:Wgaxon}, and \eqref{eq:IIaxion}, the relative entropy between $P_0$ and $P_g$ is calculated as follows:
       \begin{align}
           S(P_0||P_g)&\equiv \int d[a_{\rm E}] \left(
           P_0[a_{\rm E}] \ln P_0[a_{\rm E}]-P_0[a_{\rm E}]\ln P_g[a_{\rm E}] 
           \right)\notag
           \\
           &=-\ln Z_0[A]+\ln Z_g[A]+g\cdot \int d[a_{\rm E}] P_0[a_{\rm E}] I_{\rm I}^{\rm (E)}[A,a_{\rm E}]\notag
           \\
           &=-\ln Z_0[A]+\ln Z_g[A]\notag
           \\
           &=W_0[A]-W_g[A]\notag
           \\
           &=g^2\cdot\frac{1}{2 f_a^2 m^2_a}\int (d^4x)_{\rm E}(F_{\mu\nu}\widetilde{F}^{\mu\nu})^2\geq 0,
       \end{align}
       where the first line is the definition of the relative entropy, Eq.~\eqref{eq:IIaxion} is used in the third line, the definition of the effective action is used in the fourth line, and Eq.~\eqref{eq:Wgaxon} is used in the last line. 
       By taking to be $g=1$, the relative entopy between $P_{\rm NI}$ and $P_{\rm T}$ is given by
       \begin{align}
            S(P_{\rm NI}||P_{\rm T})&=W_{\rm NI}[A]-W_{\rm T}[A]=\frac{1}{2 f_a^2 m^2_a}\int (d^4x)_{\rm E}(F_{\mu\nu}\widetilde{F}^{\mu\nu})^2\geq 0\Rightarrow \frac{1}{2 f_a^2 m^2_a}\geq 0,
       \end{align}
       where $P_{\rm NI}=P_0$, $P_{\rm T}=P_{g=1}$, $W_{\rm NI}=W_0$, and $W_{\rm T}=W_{g=1}$ are used.
       The relative entropy denotes the dimension-eight term in Eq.~\eqref{eq:Wgaxon}, and the non-negativity of relative entropy yields the constraint on the Wilson coefficient of the dimension-eight operator.

\subsection{Massless scalar field with a shift symmetry}
\label{sec:Massle}
\begin{figure*}[t]
\centering
\includegraphics[width=0.65\textwidth]{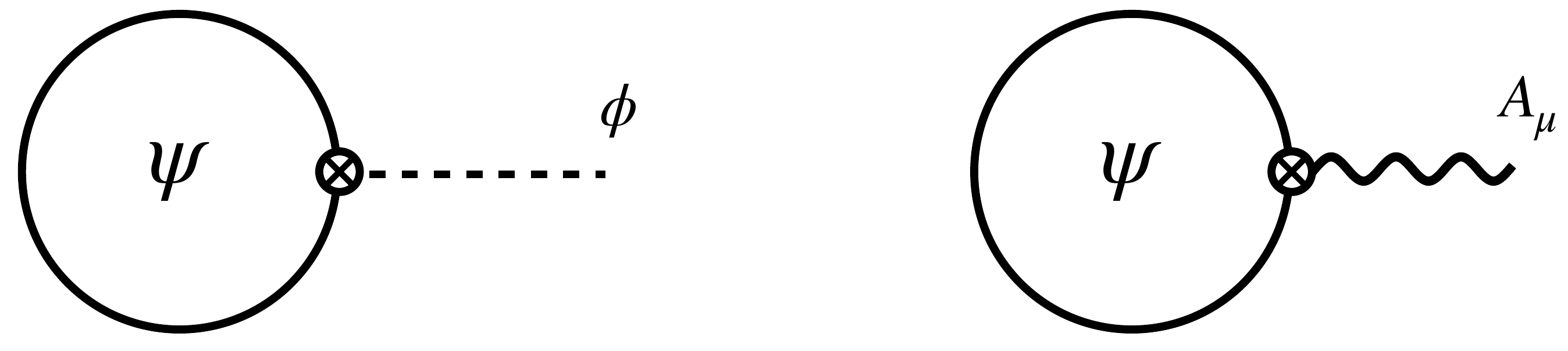}
\caption{
Left panel: Feynman diagrams of ${\langle I_{\rm I}\rangle}_{g=0}$ for the massless scalar field with a shift symmetry at the one-loop level.
Right panel: Feynman diagrams of ${\langle I_{\rm I}\rangle}_{g=0}$ for the Euler-Heisenberg theory with the charged fermion field at the one-loop level.
The solid lines denote the heavy fermion field. 
\label{fig:EH1}
}
\end{figure*}
Let us consider an action in the Minkowski space,
    \begin{align}
        I_{\rm T}[\phi, \psi,\bar{\psi}]\equiv \int d^4x \left(\frac{1}{2}\partial_\mu \phi \partial^\mu \phi +\bar{\psi}\left(i\slashed{\partial}-m\right)\psi-\frac{c}{\Lambda}\bar{\psi}\slashed{\partial}\phi \gamma_5 \psi\right),~~~~[{\rm \bf Target}]
    \end{align}
where $\psi$ is the heavy Dirac fermion, $\phi$ is the massless scalar field, $c$ is a dimensionless coupling constant, and $\Lambda$ is some energy scale.
The interacting and non-interacting terms of $I_{\rm T}$ is given by
\begin{align}
    &I_0[\phi,\psi,\bar{\psi}]\equiv I_{\rm T}[\phi,0,0]+I_{\rm T}[0,\psi,\bar{\psi}]=\int d^4x \left(\frac{1}{2}\partial_\mu \phi \partial^\mu \phi +\bar{\psi}\left(i\slashed{\partial}-m\right)\psi\right),
    \\
    &I_{\rm I}[\phi,\psi,\bar{\psi}]\equiv  I_{\rm T}[\phi,\psi,\bar{\psi}]-I_{0}[\phi,\psi,\bar{\psi}]=\int d^4x \left(-\frac{c}{\Lambda}\bar{\psi}\slashed{\partial}\phi \gamma_5 \psi\right).
\end{align}
By introducing the parameter $g$, define an action $I_g\equiv I_0+g\cdot I_{\rm I}$.
Then, the target theory is expressed as $I_{\rm T}=\lim_{g\to 1}I_g$.
To clarify the procedure of the wave function renormalization, we also perform the path integral over $\phi$. 
By calculating one-loop diagrams and performing the Wick rotation, the partition function of $I_g$ is obtained as
\begin{align}
Z_g[\widetilde{\phi}]&\equiv \int d[\phi]d[\psi]d[\bar{\psi}] e^{-I_g^{(\rm E)}[\phi,\psi,\bar{\psi}]}\notag
\\
&=\int d[\phi]{\rm exp}\bigg[
-\int (d^4x)_{\rm E}\left(-\frac{1}{2}\left(1+g^2\cdot\alpha_0\right)(\partial_{\mu}\phi\partial^{\mu}\phi)-g^4\cdot\frac{c^4}{12\pi^2 \Lambda^4}(\partial_{\mu}\phi\partial^{\mu}\phi)^2\right)
\bigg] \notag
\\
&={\rm exp}\bigg[
-\int (d^4x)_{\rm E}\left(-\frac{1}{2}\left(1+g^2\cdot\alpha_0\right)(\partial_{\mu}\widetilde{\phi'}\partial^{\mu}\widetilde{\phi'})-g^4\cdot\frac{c^4}{12\pi^2 \Lambda^4}(\partial_{\mu}\widetilde{\phi'}\partial^{\mu}\widetilde{\phi'})^2\right)
\bigg] \notag
\\
&={\rm exp}\bigg[
-\int (d^4x)_{\rm E}\left(-\frac{1}{2}(\partial_{\mu}\widetilde{\phi}\partial^{\mu}\widetilde{\phi})-g^4\cdot\frac{c^4}{12\pi^2 \Lambda^4}(\partial_{\mu}\widetilde{\phi}\partial^{\mu}\widetilde{\phi})^2\right)
\bigg],\label{eq:Zgscal}
\end{align}
where the first line is the definition of the partition function, $\widetilde{\phi'}$ is a classical solution satisfying $\partial\partial\widetilde{\phi'}=0$, and the quantum correction to the kinetic term of the scalar field is eliminated in the last line by the redefinition of the field, i.e., $\widetilde{\phi'}=(1-g^2\cdot \alpha_0/2)\widetilde{\phi}$.
In Eq.~\eqref{eq:Zgscal}, two-loop effects are neglected.
Also, the vacuum energy is omitted because it cancels in the relative entropy.
We also note that dimension-six terms and other dimension-eight terms, i.e., $(\partial\partial\widetilde{\phi})^2$ and $(\partial\partial\partial\widetilde{\phi})^2$, are generated, but they are now eliminated by a background field satisfying $\partial\widetilde{\phi}={\rm const.}$.
From Eq.~\eqref{eq:Zgscal}, the effective action of $I_g$ is obtained as follows:
\begin{align}
    W_g[\widetilde{\phi}]&\equiv-\ln Z_g[\widetilde{\phi}]=\int (d^4x)_{\rm E}\left(-\frac{1}{2}(\partial_{\mu}\widetilde{\phi}\partial^{\mu}\widetilde{\phi})-g^4\cdot\frac{c^4}{12\pi^2 \Lambda^4}(\partial_{\mu}\widetilde{\phi}\partial^{\mu}\widetilde{\phi})^2\right).\label{eq:Wgshift}
\end{align}
The Euclidean effective action of the target theory is given by
    \begin{align}
        W_{\rm T}[\widetilde{\phi}]=\lim_{g\to 1} W_{g}[\widetilde{\phi}].
    \label{eq:shift_symmetry_UV}
    \end{align}
By taking the limit of $g=0$, the action of the NIRT and MFFRT in the Minkowski space is defined as
\begin{align}
    I_{\rm NI}[\phi,\psi,\bar{\psi}]\equiv\lim_{g\to 0}I_{g}[\phi,\psi,\bar{\psi}]=\int d^4x \left(\frac{1}{2}\partial_\mu \phi \partial^\mu \phi +\bar{\psi}\left(i\slashed{\partial}-m\right)\psi\right).~~~~[{\rm \bf Reference}]
\end{align}
From Eqs.~\eqref{eq:Zgscal}, and \eqref{eq:Wgshift}, the partition function and effective action of the reference theory are obtained as follows:
\begin{align}
    &Z_{\rm NI}[\widetilde{\phi}]=\lim_{g\to 0}Z_g[\widetilde{\phi}], 
    \\
    &W_{\rm NI}[\widetilde{\phi}]=\lim_{g\to 0} W_{g}[\widetilde{\phi}]=\int (d^4x)_{\rm E}\frac{1}{2}(\partial_{\mu}\widetilde{\phi}\partial^{\mu}\widetilde{\phi}).
\end{align}
By defining a probability distribution function as $P_g[\phi,\psi,\bar{\psi}]\equiv e^{-I_g[\phi,\psi,\bar{\psi}]}/Z_g[\widetilde{\phi}]$, the derivative of $W_g$ with respect to $g$ is calculated as follows:
\begin{align}
    \left(\frac{dW_g}{dg}\right)_{g=0}&=\left(\frac{\partial W_g}{\partial g}\right)_{g=0}
    +
    \int (d^4 x)_{\rm E} \left( \frac{\delta W_g}{\delta \widetilde{\phi'}}\right)_{g=0}\left(\frac{d \widetilde{\phi'} }{dg}\right)_{g=0}\notag
    \\
    &=\left(\frac{\partial W_g}{\partial g}\right)_{g=0}\notag
    \\
    &=\int d[\phi]d[\psi]d[\bar{\psi}] P_0[\psi,\bar{\psi}] I_{\rm I}[\phi,\psi,\bar{\psi}]\notag
    \\
    &={\langle I_{\rm I}\rangle}_{g=0}=0,\label{eq:shiftsclII}
\end{align}
where the partial derivative means differentiating by $g$ while keeping $\widetilde{\phi'}$, $(d\widetilde{\phi'}/dg)_{g=0}=0$ is used in the second line, and $(dW_g/dg)_{g=0}$ is used in the last line.
The Feynman diagram for ${\langle I_{\rm I}\rangle}_{g=0}$ is shown in the left panel of Fig.~\ref{fig:EH1}.
From Eqs.~\eqref{eq:dyrel1}, \eqref{eq:Wgshift}, and \eqref{eq:shiftsclII}, the relative entropy between $P_0$ and $P_g$ is calculated as follows:
\begin{align}
    S(P_0||P_g)&\equiv \int d[\phi]d[\psi]d[\bar{\psi}] \left(
    P_0[\phi,\psi,\bar{\psi}]\ln P_0[\phi,\psi,\bar{\psi}]-P_0[\phi,\psi,\bar{\psi}]\ln P_g[\phi,\psi,\bar{\psi}]
    \right)\notag
    \\
    &=-\ln Z_0[\widetilde{\phi}]+\ln Z_g[\widetilde{\phi}]+g\cdot {\langle I_{\rm I}\rangle}_{g=0}\notag
    \\
    &=W_0[\widetilde{\phi}]-W_g[\widetilde{\phi}]\notag
    \\
    &=g^4\cdot\frac{c^4}{12\pi^2\Lambda^4}\int (d^4x)_{\rm E}\left(\partial_\mu\widetilde{\phi}\partial^\mu\widetilde{\phi}\right)^2\geq 0,
\end{align}
where the first line is the definition of the relative entropy, Eq.~\eqref{eq:shiftsclII} is used in the third line, and Eq.~\eqref{eq:Wgshift} is used in the last.
By taking $g=1$, the relative entropy between $P_{\rm NI}$ and $P_{\rm T}$ is given by
\begin{align}
    S(P_{\rm NI}||P_{\rm T})=W_{\rm NI}[\widetilde{\phi}]-W_{\rm T}[\widetilde{\phi}]=\frac{c^4}{12\pi^2\Lambda^4}\int (d^4x)_{\rm E}\left(\partial_\mu\widetilde{\phi}\partial^\mu\widetilde{\phi}\right)^2\geq 0 \Rightarrow \frac{c^4}{12\pi^2\Lambda^4}\geq 0,
\end{align}
where $P_{\rm NI}=P_0$, $P_{\rm T}=P_{g=1}$, $W_{\rm NI}=W_0$, and $W_{\rm T}=W_{g=1}$ are used.
Consequently, the relative entropy is non-negative in the UV action of \eqref{eq:shift_symmetry_UV}.
Although the above procedure is the top-down approach, we will consider a bound on the coefficient of the operator $(\partial\phi\partial\phi)^2$ in a bottom-up way in Sec. \ref{sec:bottomup_shift_symmetry}.

\subsection{Euler-Heisenberg theory}
\label{sec:EH}
We consider the Euler-Heisenberg theory, where the heavy particle is $U(1)$ charged field, and the light one is $U(1)$ gauge field.
For the charged scalar field, we show that the constraint on the Wilson coefficients arises by implementing a suitable gauge fixing procedure. 
\subsubsection{Heavy charged fermion field}
The action in the Mikowski space is described by
\begin{align}
I_{\rm T}[A,\psi,\bar{\psi}]\equiv\int d^4 x \left[-\frac{1}{4}F_{\mu\nu}F^{\mu\nu} +\bar{\psi} (i\slashed{D}-m)\psi\right],~~~~[{\rm \bf Target}]
\end{align}
where $\psi$ is the charged fermion field, $D_{\mu}=\partial_{\mu}+ie A_{\mu}$ is the covariant derivative, and $F_{\mu\nu}$ is the field strength of photon defined by $F_{\mu\nu}=\partial_{\mu}A_{\nu}-\partial_{\nu}A_{\mu}$.
We define $I_0$ and $I_{\rm I}$ as follows:
\begin{align}
    &I_0[A,\psi,\bar{\psi}]\equiv I_{\rm T}[A,0,0]+I_{\rm T}[0,\psi,\bar{\psi}]=\int d^4 x \left[-\frac{1}{4}F_{\mu\nu}F^{\mu\nu} +\bar{\psi} (i\slashed{\partial}-m)\psi\right],
    \\
    &I_{\rm I}[A,\psi,\bar{\psi}]\equiv I_{\rm T}[A,\psi,\bar{\psi}]-I_0[A,\psi,\bar{\psi}]=-e\cdot \int d^4x A_{\mu}\bar{\psi}\gamma^{\mu}\psi.
\end{align}
By introducing an auxiliary parameter $g$, define $I_{g}\equiv I_0 +g\cdot I_{\rm I}$.
Note here that the target theory is obtained as $I_{\rm T}=\lim_{g\to 1}I_g$.
To clarify the procedure of wave function renormalization, we also perform the path integral over the $U(1)$ gauge field.
By calculating one-loop diagrams and performing the Wick rotation, the partition function of $I_g$ is obtained in the following way~\cite{Quevillon:2018mfl}.
\begin{align}
Z_{g}[\overline{A}]&\equiv \int d[A]d[\psi]d[\bar{\psi}] e^{-I_{g}^{\rm (E)}[A,\psi,\bar{\psi}]},\notag
\\
&=\int d[A] {\rm exp}\bigg[
-\int (d^4 x)_{\rm E} \bigg(\frac{1}{4}\left(1+\alpha_0 \frac{g^2e^2}{4! \pi^2} \right)F_{\mu\nu}F^{\mu\nu}-\alpha_2 \frac{g^2{e}^2}{5!\pi^2 m^2}\partial^{\mu} F_{\mu\nu}\partial_{\rho}{F}^{\rho\nu}\notag
\\
&-\alpha_4\frac{g^2{e}^2}{6!\pi^2 m^4}\partial^{\mu}F_{\mu\nu}\Box \partial_{\rho}{F}^{\rho\nu}-\gamma_{4,1}\frac{g^4{e}^4}{6! \pi^2 m^4} (F_{\mu\nu}{F}^{\mu\nu})^2 -\gamma_{4,2} \frac{g^4{e}^4}{6! \pi^2 m^4} (F_{\mu\nu}\tilde{F}^{\mu\nu})^2 +\mathcal{O}(m^{-6})\bigg)
\bigg],\notag
\\
&={\rm exp}\bigg[
-\int (d^4 x)_{\rm E} \bigg(\frac{1}{4}\left(1+\alpha_0 \frac{g^2e^2}{4! \pi^2} \right)\overline{F'}_{\mu\nu}\overline{F'}^{\mu\nu}-\alpha_2 \frac{g^2{e}^2}{5!\pi^2 m^2}\partial^{\mu} \overline{F'}_{\mu\nu}\partial_{\rho}\overline{F'}^{\rho\nu}\notag
\\
&-\alpha_4\frac{g^2{e}^2}{6!\pi^2 m^4}\partial^{\mu}\overline{F'}_{\mu\nu}\Box \partial_{\rho}\overline{F'}^{\rho\nu}-\gamma_{4,1}\frac{g^4{e}^4}{6! \pi^2 m^4} (\overline{F'}_{\mu\nu}\overline{F'}^{\mu\nu})^2 -\gamma_{4,2} \frac{g^4{e}^4}{6! \pi^2 m^4} (\overline{F'}_{\mu\nu}\widetilde{\overline{F'}}^{\mu\nu})^2 +\mathcal{O}(m^{-6})\bigg)
\bigg],\notag
\\
&={\rm exp}\bigg[
-\int (d^4 x)_{\rm E} \bigg(\frac{1}{4}\overline{F}_{\mu\nu}\overline{F}^{\mu\nu}-\gamma_{4,1}\frac{g^4{e}^4}{6! \pi^2 m^4} (\overline{F}_{\mu\nu}\overline{F}^{\mu\nu})^2 -\gamma_{4,2} \frac{g^4{e}^4}{6! \pi^2 m^4} (\overline{F}_{\mu\nu}\widetilde{\overline{F}}^{\mu\nu})^2 +\mathcal{O}(m^{-6})\bigg)
\bigg],
\end{align}
where $\overline{F}'_{\mu\nu}\equiv \partial_{\mu}\overline{A}'_{\nu}-\partial_{\nu}\overline{A}'_{\mu}$ with the classical solution $\overline{A}'_{\mu}\equiv (1+\alpha_0 g^2e^2/4! \pi^2)\overline{A}_{\mu}$, and the last line is obtained by choosing a background field satisfying $\partial_\mu \overline{F}^{\mu\nu}=0$.
Here, we omit the vacuum energy in the Euclidean effective action because it cancels in the relative entropy. 
The effective action of $I_g$ is calculated as follows:
\begin{align}
W_{g}[\overline{A}]&\equiv-\ln Z_g[\overline{A}]\notag
\\
&=\int (d^4 x)_{\rm E}
\bigg(\frac{1}{4}\overline{F}_{\mu\nu}\overline{F}^{\mu\nu}-\gamma_{4,1}\frac{g^4{e}^4}{6! \pi^2 m^4} (\overline{F}_{\mu\nu}\overline{F}^{\mu\nu})^2 -\gamma_{4,2} \frac{g^4{e}^4}{6! \pi^2 m^4} (\overline{F}_{\mu\nu}\widetilde{\overline{F}}^{\mu\nu})^2 +\mathcal{O}(m^{-6})\bigg),\label{eq:lh1}
\end{align}
with~\cite{Quevillon:2018mfl}
\begin{align}
\alpha_2 = -1,~~\alpha_4=\frac{9}{14},~~\gamma_{4,1}=\frac{1}{2},~~\gamma_{4,2}=\frac{7}{8}. \label{eq:f1}
\end{align}
The Euclidean effective action of the target theory is calculated as
\begin{align}
    W_{\rm T}[\overline{A}]&= \lim_{g\to 1}W_{g}[\overline{A}].
\end{align}
By taking the limit of $g=0$, the action of the NIRT and MFFRT in the Minkowski space is defined as
\begin{align}
    I_{\rm NI}[A,\psi,\bar{\psi}]\equiv \lim_{g\to 0}I_{g}[A,\psi,\bar{\psi}]=\int d^4x \left[-\frac{1}{4}F_{\mu\nu}F^{\mu\nu}+\bar{\psi}(i\slashed{\partial}-m)\psi\right].~~~~[{\rm \bf Reference}]
\end{align}
Therefore, the partition function and effective action of the reference theory are respectively obtained as follows:
\begin{align}
    &Z_{\rm NI}[\overline{A}]=\lim_{g\to 0}Z_{g}[\overline{A}],
    \\
    &W_{\rm NI}[\overline{A}]=\lim_{g\to 0}W_{g}[\overline{A}]=\int (d^4x)_{\rm E} \frac{1}{4} \overline{F}_{\mu\nu}\overline{F}^{\mu\nu}.
\end{align}
By defining the probability distribution function $P_g[A,\psi,\bar{\psi}]\equiv e^{-I_{g}[A,\psi,\bar{\psi}]}/Z_{g}[\overline{A}]$, the derivative of $W_g$ with respect to $g$ is given by
\begin{align}
    \left(\frac{dW_g}{dg}\right)_{g=0}&=\left(\frac{\partial W_g}{\partial g}\right)_{g=0}+\int (d^4 x)_{\rm E} \left(\frac{\delta W_g}{\delta \overline{A}'}\right)_{g=0}\left(\frac{d\overline{A}'}{dg}\right)_{g=0}\notag
    \\
    &=\left(\frac{\partial W_g}{\partial g}\right)_{g=0}\notag
    \\
    &=\int d[A]d[\psi]d[\bar{\psi}]P_{0}[\psi,\bar{\psi}] I_{\rm I}[A,\psi,\bar{\psi}]\notag
    \\
    &={\langle I_{\rm I}\rangle}_{g=0}=0,\label{eq:IIEHfem1}
\end{align}
where the partial derivative means differentiating by $g$ while keeping $\overline{A}'$, $({d\overline{A}'}/{dg})_{g=0}$ is used in the second line, and $({dW_g}/{dg})_{g=0}=0$ is used in the last line.
The Feynman diagram for ${\langle I_{\rm I} \rangle}_{g=0}$ is shown in the right panel of Fig.~\ref{fig:EH1}.
From Eqs.~\eqref{eq:dyrel1}, \eqref{eq:lh1}, and \eqref{eq:IIEHfem1}, the relative entropy between $P_0$ and $P_g$ is calculated as follows:
\begin{align}
    S(P_0||P_g)&\equiv \int d[A]d[\psi]d[\bar{\psi}] \left(P_0[A,\psi,\bar{\psi}]\ln P_0[A,\psi,\bar{\psi}]-P_0[A,\psi,\bar{\psi}]\ln P_g[A,\psi,\bar{\psi}]\right)\notag
    \\
    &=-\ln Z_0[\overline{A}]+\ln Z_g[\overline{A}] +g\cdot \int d[A]d[\psi]d[\bar{\psi}]P_{0}[\psi,\bar{\psi}] I_{\rm I}[A,\psi,\bar{\psi}]\notag
    \\
    &=-\ln Z_0[\overline{A}]+\ln Z_g[\overline{A}]\notag
    \\
    &=W_0[\overline{A}]-W_g[\overline{A}]\notag
    \\
    &=\int (d^4 x)_{\rm E} \bigg[\gamma_{4,1}\frac{g^4{e}^4}{6! \pi^2 m^4} (\overline{F}_{\mu\nu}\overline{F}^{\mu\nu})^2 +\gamma_{4,2} \frac{g^4{e}^4}{6! \pi^2 m^4} (\overline{F}_{\mu\nu}\widetilde{\overline{F}}^{\mu\nu})^2 +\mathcal{O}(m^{-6})\bigg]\geq 0,\label{eq:relEHfer}
\end{align}
where the first line is the definition of the relative entropy, Eq.~\eqref{eq:IIEHfem1} is used in the third line, the fourth line is the definition of the effective action, and Eq.~\eqref{eq:lh1} is used in the last line. 
By taking $g=1$, the relative entropy between $P_{\rm NI}$ and $P_{\rm T}$ is given by
\begin{align}
    S(P_{\rm NI}||P_{\rm T})&=W_{\rm NI}[\overline{A}]-W_{\rm T}[\overline{A}]\notag
    \\
    &=\int (d^4 x)_{\rm E} \bigg[\gamma_{4,1}\frac{{e}^4}{6! \pi^2 m^4} (\overline{F}_{\mu\nu}\overline{F}^{\mu\nu})^2 +\gamma_{4,2} \frac{{e}^4}{6! \pi^2 m^4} (\overline{F}_{\mu\nu}\widetilde{\overline{F}}^{\mu\nu})^2 +\mathcal{O}(m^{-6})\bigg]\geq 0,\label{eq:relEH1}
\end{align}
where $P_{\rm NI}=P_0$, $P_{\rm T}=P_{g=1}$, $W_{\rm NI}=W_0$, and $W_{\rm T}=W_{g=1}$ are used.
According to Eq.~\eqref{eq:f1}, it is clear that the right-hand side of Eq.~\eqref{eq:relEH1} takes negative value up to dimension-eight terms. 
This result is consistent with the non-negativity of the relative entropy.

\subsubsection{Heavy charged scalar field}
\begin{figure*}[t]
\centering
\includegraphics[width=0.65\textwidth]{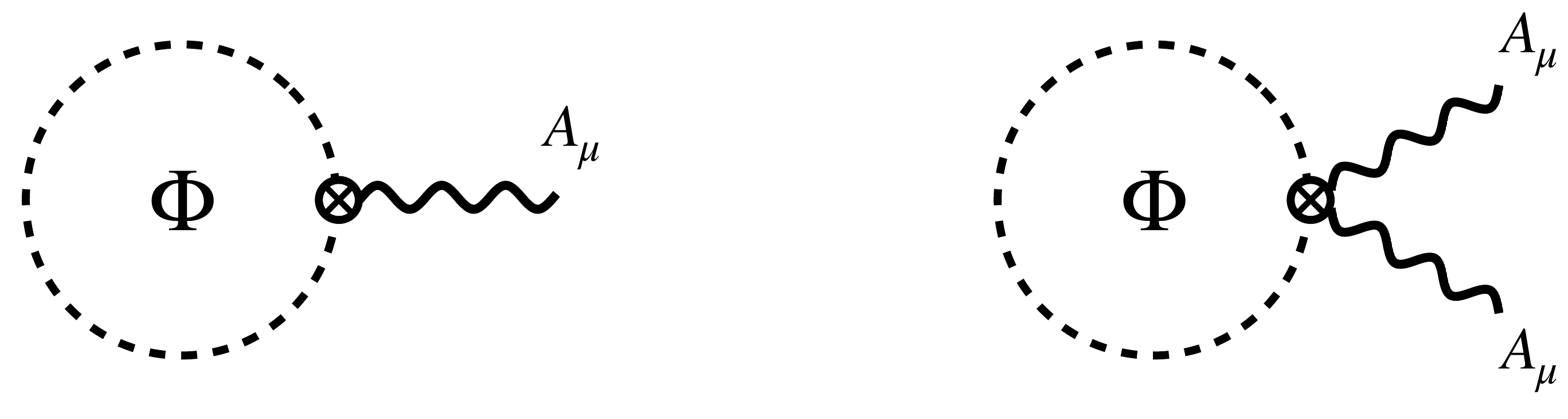}
\caption{
Feynman diagrams of ${\langle I_{\rm I}\rangle}_{g=0}$ for the Euler-Heisenberg theory with the charged scalar field at the one-loop level.
The dotted line denotes the heavy scalar field. 
The left and right diagrams represent $(\delta W_g/\delta A_{\mu})_{A=0}$ and $(\delta W_g/\delta (A_{\mu}A^{\mu}))_{A_{\mu}A^{\mu}=0}$, respectively.
The right one is a tadpole-like diagram for the composite field $A_{\mu}A^{\mu}$.
\label{fig:EHscl}
}
\end{figure*}

The action of the massive charged scalar field in the Minkowski space is described by
\begin{align}
I_{\rm T}[A,\Phi]\equiv \int d^4 x\left( -\frac{1}{4}F_{\mu\nu}F^{\mu\nu}+D_{\mu}\Phi D^{\mu}\Phi^{\ast}-m^2|\Phi|^2  \right),~~~~[{\rm \bf Target}]
\end{align}
where $\Phi$ is the charged massive scalar field.
We define $I_0$ and $I_{\rm I}$ as follows:
\begin{align}
    &I_0[A,\Phi]\equiv I_{\rm T}[A,0]+I_{\rm T}[0,\Phi]=\int d^4 x\left( -\frac{1}{4}F_{\mu\nu}F^{\mu\nu}+\partial_{\mu}\Phi \partial^{\mu}\Phi^{\ast}-m^2|\Phi|^2  \right),
    \\
    &I_{\rm I}[A,\Phi]\equiv I_{\rm T}[A,\Phi]-I_0[A,\Phi]= \int d^4x \left(-ie A^{\mu}(\partial_{\mu}\Phi)\Phi^{\ast}+i e A^{\mu}\Phi(\partial_{\mu}\Phi^{\ast})+e^2 A_{\mu}A^{\mu}|\Phi|^2 \right).
\end{align}
By introducing a parameter $g$, define an action as $I_{g}\equiv I_0 +g\cdot I_{\rm I}$.
Note here that the interaction $I_{\rm I}$ includes the first and second order of $e$, and the order of $g$ differs from that of $e$.
Similar to the massive fermion, the Euclidean effective action of the target theory is obtained as,
\begin{align}
W_{\rm T}[\overline{A}]&=\lim_{g\to 1} W_{g}[\overline{A}]\notag
\\
&= \int (d^4 x)_{\rm E} \bigg(\frac{1}{4}\overline{F}_{\mu\nu}\overline{F}^{\mu\nu}-\gamma_{4,1}\frac{{e}^4}{6! \pi^2 m^4} (\overline{F}_{\mu\nu}\overline{F}^{\mu\nu})^2 -\gamma_{4,2} \frac{{e}^4}{6! \pi^2 m^4} (\overline{F}_{\mu\nu}\widetilde{\overline{F}}^{\mu\nu})^2 +\mathcal{O}(m^{-6})\bigg),\label{eq:lh1ta}
\end{align}
with~\cite{Quevillon:2018mfl} 
\begin{align}
\alpha_2 &= \frac{37}{8},~~\alpha_4=\frac{159}{56},~~\gamma_{4,1}=\frac{261}{32},~~\gamma_{4,2}=\frac{243}{32}. \label{eq:scl1}
\end{align}
By taking the limit of $g=0$, the action of the NIRT and MFFRT in the Minkowski space is defined as
\begin{align}
    I_{\rm NI}[A,\Phi]=\lim_{g\to 0}I_{g}[A,\Phi]=\int d^4x \left( -\frac{1}{4}F_{\mu\nu}F^{\mu\nu}+\partial_{\mu}\Phi \partial^{\mu}\Phi^{\ast}-m^2|\Phi|^2  \right).~~~~[{\rm \bf Reference}]
\end{align}
Then, we obtain as follows:
\begin{align}
    &Z_{\rm NI}[\overline{A}]=\lim_{g\to 0} Z_g[\overline{A}]
    \\
    &W_{\rm NI}[\overline{A}]=\lim_{g\to 0}W_{g}[\overline{A}]=\int (d^4x)_E \frac{1}{4} \overline{F}_{\mu\nu}\overline{F}^{\mu\nu}.
\end{align}
By defining a probability distribution function as $P_g[A,\Phi]\equiv e^{-I_g^{\rm (E)}[A,\Phi]}/Z_g[\overline{A}]$, the expectation value of the interaction is given by
\begin{align}
    g\cdot {\langle I_{\rm I}\rangle}_{g=0}&=g\cdot\int d[A]d[\Phi] P_{0}[A,\Phi] I_{\rm I}^{\rm (E)}[A,\Phi]\notag
    \\
    &=\int (d^4 x)_{\rm E}\left[\left(\frac{\delta W_g}{\delta \overline{A'}_{\mu}}\right)_{g=0} \overline{A}_{\mu}+\left(\frac{\delta W_g}{\delta \overline{A'}_{\mu}\overline{A'}^{\mu}}\right)_{g=0} \overline{A}_{\mu}\overline{A}^{\mu}\right]\notag
    \\
    &=\int (d^4 x)_{\rm E}\left(\frac{\delta W_g}{\delta \overline{A'}_{\mu}\overline{A'}^{\mu}}\right)_{g=0} \overline{A}_{\mu}\overline{A}^{\mu},\label{eq:IIscl}
\end{align}
where the term proportional to $\overline{A}_{\mu}$ vanishes because of the Lorentz symmetry.
The Feynman diagram for ${\langle I_{\rm I} \rangle}_{g=0}$ is shown in Fig.~\ref{fig:EHscl}.
By taking the gauge-fixing condition $\overline{A}_{\mu}\overline{A}^{\mu}=0$, which is called the non-linear gauge~\cite{Nambu:1968qk}, ${\langle I_{\rm I} \rangle}_{g=0}$ can vanish.
From Eqs.~\eqref{eq:relEHfer} and \eqref{eq:IIscl}, the relative entropy between $P_{\rm NI}$ and $P_{\rm T}$ is given by
\begin{align}
    S(P_{\rm NI}||P_{\rm T})&=W_{\rm NI}[\overline{A}]-W_{\rm T}[\overline{A}]+{\langle I_{\rm I} \rangle}_{g=0}\notag
    \\
    &=W_{\rm NI}[\overline{A}]-W_{\rm T}[\overline{A}]\notag
    \\
    &=\int (d^4 x)_{\rm E} \bigg[\gamma_{4,1}\frac{{e}^4}{6! \pi^2 m^4} (\overline{F}_{\mu\nu}\overline{F}^{\mu\nu})^2 +\gamma_{4,2} \frac{{e}^4}{6! \pi^2 m^4} (\overline{F}_{\mu\nu}\widetilde{\overline{F}}^{\mu\nu})^2 +\mathcal{O}(m^{-6})\bigg]\geq 0,\label{eq:relEHscl2}
\end{align}
where the first line is the definition of the relative entropy, Eq.~\eqref{eq:IIscl} and the gauge fixing condition $A_{\mu}A^{\mu}=0$ are used in the second line, and Eq.~\eqref{eq:lh1} is used in the last line.
According to Eq.~\eqref{eq:scl1}, it is clear that the right-hand side of Eq.~\eqref{eq:relEHscl2} takes a negative value up to dimension-eight terms. 
This result is consistent with the non-negativity of the relative entropy.

The non-negativity of the relative entropy holds even for the massive charged vector field.
Since the interaction between the massive charged vector and $U(1)$ gauge fields arise from the covariant derivative of the kinetic term, Eq.~\eqref{eq:IIscl} holds even for the massive charged vector field.
By using the non-linear gauge $\overline{A}_{\mu}\overline{A}^{\mu}=0$, Eq.~\eqref{eq:relEHscl2} holds for the massive charged vector field with~\cite{Quevillon:2018mfl},
\begin{align}
\alpha_2 &= \frac{37}{8},~~\alpha_4=\frac{159}{56},~~\gamma_{4,1}=\frac{261}{32},~~\gamma_{4,2}=\frac{243}{32}.\label{eq:f2}
\end{align}
Consequently, the inequality of \eqref{eq:relEHscl2} holds even for the massive charged vector field.

\subsection{Gravitationally coupled massive scalar field at tree level}
\label{sec:Gratree}
Let us consider a theory~\cite{Cheung:2018cwt} in the Minkowski space:
\begin{align}
    I_{\rm T}[g_{\mu\nu};R_{\mu\nu\rho\sigma},A_{\mu},\Phi]&\equiv \int d^4 x \sqrt{-g}\bigg(\frac{M^2_{\rm Pl}}{2}R -\frac{1}{4}F_{\mu\nu}F^{\mu\nu}\notag
    \\
    &\quad\quad\quad\quad- (a_{\Phi}R +b_{\Phi}F_{\mu\nu}F^{\mu\nu})\Phi+
    \frac{1}{2}g^{\mu\nu}\partial_{\mu}\Phi \partial_{\nu}\Phi -\frac{1}{2}m^2_{\Phi}\Phi^2
    \bigg),~~~~[{\rm \bf Target}]
\end{align}
where $R_{\mu\nu\rho\sigma}$ is the Riemann tensor, $R$ is the scalar curvature of the metric $g_{\mu\nu}$, $F_{\mu\nu}$ is the field strength of $U(1)$ gauge field, $\Phi$ is a massive real scalar field, and $a_{\Phi}, b_{\Phi}$ are dimensionful coupling constants.
The action in the Euclidean space is expressed as
\begin{align}
    I^{\rm (E)}_{\rm T}[g_{\mu\nu};R_{\mu\nu\rho\sigma},A_{\mu},\Phi_{\rm E}]
    &=\int (d^4x)_{\rm E} \sqrt{g_{\rm E}}\bigg(-\frac{M^2_{\rm Pl}}{2}R +\frac{1}{4}F_{\mu\nu}F^{\mu\nu} \notag \\
    &\qquad+(a_{\Phi}R +b_{\Phi}F_{\mu\nu}F^{\mu\nu})\Phi_{\rm E}+\frac{1}{2}g_{{\rm E},IJ}\partial_I \Phi_{\rm E}\partial_J \Phi_{\rm E}+\frac{1}{2}m^2_{\Phi}(\Phi_{\rm E})^2  \bigg),
\end{align}
where the $R$ and $F_{\mu\nu}$ are defined from background fields in the Minkowski space; see Appendix~\ref{app:Wick_rotation}.
We define the actions $I_0^{\rm (E)}$ and $I_{\rm I}^{\rm (E)}$ in the Euclidean space as follows:
\begin{align}
    I_0^{\rm (E)}[g_{\mu\nu};R_{\mu\nu\rho\sigma},A_{\mu},\Phi_{\rm E}]&=I^{\rm (E)}_{\rm T}[g_{\mu\nu};R_{\mu\nu\rho\sigma},A_{\mu},0]+I^{\rm (E)}_{\rm T}[g_{\mu\nu};0,0,\Phi_{\rm E}]\notag
    \\
    &=\int (d^4x)_{\rm E} \sqrt{g_{\rm E}}\left(-\frac{M^2_{\rm Pl}}{2}R +\frac{1}{4}F_{\mu\nu}F^{\mu\nu} +\frac{1}{2}g_{{\rm E},IJ}\partial_I \Phi_{\rm E}\partial_I \Phi_{\rm E}+\frac{1}{2}m^2_{\Phi}(\Phi_{\rm E})^2\right), \\
    I_{\rm I}^{\rm (E)}[g_{\mu\nu};R_{\mu\nu\rho\sigma},A_{\mu},\Phi_{\rm E}]&=I^{\rm (E)}_{\rm T}[g_{\mu\nu};R_{\mu\nu\rho\sigma},A_{\mu},\Phi_{\rm E}]-I_0^{\rm (E)}[g_{\mu\nu};R_{\mu\nu\rho\sigma},A_{\mu},\Phi_{\rm E}]\notag
    \\
    &=\int (d^4x)_{\rm E} \sqrt{g_{\rm E}}(a_{\Phi}R +b_{\Phi}F_{\mu\nu}F^{\mu\nu})\Phi_{\rm E}.
\end{align}
The action of the NIRT is the same as that of the MFFRT and is defined as 
\begin{align}
    I_{\rm NI}^{\rm (E)}[g_{\mu\nu};R_{\mu\nu\rho\sigma},A_{\mu},\Phi_{\rm E}]&=\int (d^4x)_{\rm E} \sqrt{g_{\rm E}}\left(-\frac{M^2_{\rm Pl}}{2}R +\frac{1}{4}F_{\mu\nu}F^{\mu\nu} +\frac{1}{2}g_{{\rm E},IJ}\partial_I \Phi_{\rm E}\partial_I \Phi_{\rm E}+\frac{1}{2}m^2_{\Phi}(\Phi_{\rm E})^2\right).
\end{align}
Note here that the NIRT does not include the interaction between $\Phi$ and $A_{\mu}, R_{\mu\nu\rho\sigma}$, but the interaction between $g_{\mu\nu}$ and $\Phi$.
By introducing a parameter $g$, define an action as $I_g^{\rm (E)}\equiv I_0^{\rm (E)}+g\cdot I_{\rm I}^{\rm (E)}$.
Then, the target theory is reproduced as $I_{\rm T}^{\rm (E)}=\lim_{g\to 1} I_g^{\rm (E)}$.
The target and reference theories are expressed as follows:
\begin{align}
    &I^{\rm (E)}_{\rm T}[g_{\mu\nu};R_{\mu\nu\rho\sigma},A_{\mu},\Phi_{\rm E}]=\lim_{g\to 1} I_g^{\rm (E)}[g_{\mu\nu};R_{\mu\nu\rho\sigma},A_{\mu},\Phi_{\rm E}],
    \\
    &I^{\rm (E)}_{\rm NI}[g_{\mu\nu};R_{\mu\nu\rho\sigma},A_{\mu},\Phi_{\rm E}]=\lim_{g\to 0} I_g^{\rm (E)}[g_{\mu\nu};R_{\mu\nu\rho\sigma},A_{\mu},\Phi_{\rm E}].
\end{align}
The partition function and effective action of $I_g^{\rm (E)}$ are respectively calculated as follows:
\begin{align}
    Z_g[g_{\mu\nu};R_{\mu\nu\rho\sigma},A_{\mu}]&\equiv\int d[\Phi_{\rm E}] e^{-I_g^{\rm (E)}[g_{\mu\nu};R_{\mu\nu\rho\sigma},A_{\mu},\Phi_{\rm E}]}\notag
    \\
    &={\rm exp}\bigg(
    \int (d^4x)_{\rm E} \sqrt{g_{\rm E}}\left(
    -\frac{M^2_{\rm Pl}}{2}R +\frac{1}{4}F_{\mu\nu}F^{\mu\nu} -g^2\cdot \frac{1}{2 m_{\Phi}^2} \left(a_{\Phi} R  +b_{\Phi} F_{\mu\nu}F^{\mu\nu} \right)^2
    \right)
    \bigg),
    \\
    W_g[g_{\mu\nu};R_{\mu\nu\rho\sigma},A_{\mu}]&\equiv-\ln Z_g[g_{\mu\nu};R_{\mu\nu\rho\sigma},A_{\mu}]\notag
    \\
    &= \int (d^4x)_{\rm E} \sqrt{g_{\rm E}}\left(
    -\frac{M^2_{\rm Pl}}{2}R +\frac{1}{4}F_{\mu\nu}F^{\mu\nu} -g^2\cdot \frac{1}{2 m_{\Phi}^2} \left(a_{\Phi} R  +b_{\Phi} F_{\mu\nu}F^{\mu\nu} \right)^2
    \right).\label{eq:WgGR}
\end{align}
By defining a probability distribution function as
\begin{align}
    P_g[g_{\mu\nu};R_{\mu\nu\rho\sigma},A_{\mu},\Phi_{\rm E}]\equiv \frac{e^{-I_g^{\rm (E)}[g_{\mu\nu};R_{\mu\nu\rho\sigma},A_{\mu},\Phi_{\rm E}]}}{Z_g[g_{\mu\nu};R_{\mu\nu\rho\sigma},A_{\mu}]},
\end{align}
we obtain
\begin{align}
    {\langle I_{\rm I}\rangle}_{g=0}&=\int d[\Phi_{\rm E}] P_0[g_{\mu\nu};R_{\mu\nu\rho\sigma},A_{\mu},\Phi_{\rm E}] I_{\rm I}^{\rm (E)}[g_{\mu\nu};R_{\mu\nu\rho\sigma},A_{\mu},\Phi_{\rm E}]\notag
    \\
    &=I_{\rm I}^{\rm (E)}[g_{\mu\nu};R_{\mu\nu\rho\sigma},A_{\mu},0]\notag
    \\
    &=0.\label{eq:IIGR}
\end{align}
From Eqs.~\eqref{eq:sABF}, \eqref{eq:WgGR}, and \eqref{eq:IIGR}, the relative entropy between $P_0$ and $P_g$ is calculated as follows:
\begin{align}
    S(P_0||P_g)&\equiv \int d[\Phi_{\rm E}] \left(
    P_0 \ln P_0- P_0\ln P_g
    \right)\notag
    \\
    &=-\ln Z_0 [g_{\mu\nu};R_{\mu\nu\rho\sigma},A_{\mu}]+\ln Z_g[g_{\mu\nu};R_{\mu\nu\rho\sigma},A_{\mu}]+g\cdot{\langle I_{\rm I}\rangle}_{g=0}\notag
    \\
    &=-\ln Z_0 [g_{\mu\nu};R_{\mu\nu\rho\sigma},A_{\mu}]+\ln Z_g[g_{\mu\nu};R_{\mu\nu\rho\sigma},A_{\mu}]\notag
    \\
    &=W_0 [g_{\mu\nu};R_{\mu\nu\rho\sigma},A_{\mu}]-W_g[g_{\mu\nu};R_{\mu\nu\rho\sigma},A_{\mu}]\notag
    \\
    &= g^2\cdot \frac{1}{2 m_{\Phi}^2}\int (d^4x)_{\rm E} \sqrt{g_{\rm E}}
     \left(a_{\Phi} R  +b_{\Phi} F_{\mu\nu}F^{\mu\nu} \right)^2
    \geq 0,
\end{align}
where the first line is the definition of the relative entropy, Eq.~\eqref{eq:IIGR} is used in the third line, the fourth line is the definition of the effective actions, and Eq.~\eqref{eq:WgGR} is used in the last line. 
By taking $g=1$, the relative entropy between $P_{\rm NI}$ and $P_{\rm T}$ is given by
\begin{align}
    S(P_{\rm NI}||P_{\rm T})&=W_{\rm NI} [g_{\mu\nu};R_{\mu\nu\rho\sigma},A_{\mu}]-W_{\rm T}[g_{\mu\nu};R_{\mu\nu\rho\sigma},A_{\mu}]\notag
    \\
    &=\frac{1}{2 m_{\Phi}^2}\int (d^4x)_{\rm E} \sqrt{g_{\rm E}}
     \left(a_{\Phi} R  +b_{\Phi} F_{\mu\nu}F^{\mu\nu} \right)^2
    \geq 0,
\end{align}
where $P_{\rm NI}=P_0$, $P_{\rm T}=P_{g=1}$, $W_{\rm NI}=W_0$, and $W_{\rm T}=W_{g=1}$ are used.
This result represents the inequalities of \eqref{eq:uplowBG} and \eqref{eq:wt}.
Therefore, even in the gravitational theory, the relative entropy yields constraints on the EFT.

\subsection{Dimension-six four-fermion operators from tree-level UV completions}
\label{sec:ForFrm}
Let us consider a tree-level UV completion of dimension-six four-fermion operators in the Minkowski space:
\begin{align}
    I_{\rm T}[\psi,\bar{\psi},\Phi_i]\equiv\int d^4 x \left(\bar{\psi}(i\slashed{\partial}-m)\psi-(-1)^{s_i} M^2\Phi_i^2+ a_i\Phi_i(\bar{\psi}\Gamma_i \psi) \right),~~~[{\rm \bf Target}]\label{eq:fourf}
\end{align}
where the kinetic terms of the heavy degrees of freedom are omitted for simplicity, and the sum over the index $i$ is not performed.
Here, we defined as
\begin{gather}
    \Gamma_1=\hat{1},~~\Gamma_2=\gamma^{\mu},~~\Gamma_3=i\gamma_5,~~\Gamma_4=\gamma^{\mu}\gamma_5,~~\Gamma_5=\sigma^{\mu\nu}, \\
    s_1=0,~~ s_2=1,~~ s_3=0,~~ s_4=1,~~ s_5=0,
\end{gather}
where $\sigma^{\mu\nu}\equiv i[\gamma^{\mu},\gamma^{\nu}]/2$. 
The sign $s_i$ is determined so that the solution of the equation of motion for $\Phi_{\rm i}$ becomes a local minimum of the Euclidean action.
Otherwise, the validity of the Euclidean path integral method is violated.
We choose the time components of the background field $\bar{\psi}\Gamma_i \psi$ to be zero values since the non-negativity can be broken when the probability distribution functions of Eq.~\eqref{eq:non-neg} are not Hermitian.
Note here that the background field of the Euclidean space is defined by that of the Minkowski space.
The massive fields are defined as follows:
\begin{align}
    \Phi_1=\phi_{\rm S},~~\Phi_2=A^{\mu}_V,~~\Phi_3=\phi_P,~~\Phi_4=A^{\mu}_A,~~\Phi_5=T^{\mu\nu}.
\end{align}
Then, $I_0$ and $I_{\rm I}$ are given by,
\begin{align}
    &I_{0}[\psi,\bar{\psi},\Phi_i]\equiv\int d^4 x \left(\bar{\psi}(i\slashed{\partial}-m)\psi-(-1)^{s_i} M^2\Phi_i^2 \right),
    \\
    &I_{\rm I}[\psi,\bar{\psi},\Phi_i]\equiv \int d^4 x  a_i\Phi_i(\bar{\psi}\Gamma_i \psi).
\end{align}
For this target theory, the NIRT is the same as the MFFRT, and the action of the NIRT and MFFRT is defined as
\begin{align}
    I_{\rm NI}[\psi,\bar{\psi},\Phi_i]\equiv I_{0}[\psi,\bar{\psi},\Phi_i].~~~[{\rm \bf Reference}]
\end{align}
By using the auxiliary parameter $g$, define $I_g\equiv I_0+ g\cdot I_{\rm I}$.
Then, the target and reference theories are reproduced as follows:
\begin{align}
     I_{\rm T}[\psi,\bar{\psi},\Phi_i]=\lim_{g\to 1} I_{g}[\psi,\bar{\psi},\Phi_i],~~~I_{\rm NI}[\psi,\bar{\psi},\Phi_i]=\lim_{g\to 0} I_{g}[\psi,\bar{\psi},\Phi_i].
\end{align}
After the Wick rotation $I_{g}\to I^{\rm (E)}_{g}$, the partition function and effective action of $I^{\rm (E)}_{g}$ are respectively calculated as follows:
\begin{align}
    &Z_g[\psi,\bar{\psi}]\equiv\int d [\Phi_{{\rm E},i}] e^{-I_g^{(\rm E)}[\psi,\bar{\psi},\Phi_{{\rm E},i}]},
    \\
    &W_g[\psi,\bar{\psi}]\equiv -\ln Z_g[\psi,\bar{\psi}]=\int (d^4x)_{\rm E} \left(\bar{\psi}(i\slashed{\partial}+m)\psi-g^2\cdot(-1)^{s_i}\frac{a_i^2}{4M^2}(\bar{\psi}\Gamma_i \psi)^2 \right).\label{eq:Wgfermi}
\end{align}
By defining a probability distribution function as,
\begin{align}
    P_g[\Phi_i]\equiv \frac{e^{-I_g^{\rm (E)}[\psi,\bar{\psi},\Phi_{{\rm E},i}]}}{Z_g[\psi,\bar{\psi}]},
\end{align}
the expectation value of the interaction is calculated as
\begin{align}
    {\langle I_{\rm I}\rangle}_{g=0}&\equiv 
    \int d[\Phi_{{\rm E},i}] P_0[\Phi_i] I_{\rm I}^{\rm (E)}[\psi,\bar{\psi},\Phi_{{\rm E},i}]\notag
    \\
    &=I_{\rm I}^{\rm (E)}[\psi,\bar{\psi},0]\notag
    \\
    &=0.\label{eq:IIfermi}
\end{align}
From Eqs.~\eqref{eq:sABF}, \eqref{eq:Wgfermi}, and \eqref{eq:IIfermi}, the relative entropy between $P_0$ and $P_g$ is calculated as
\begin{align}
    S(P_0||P_g)&\equiv\int d[\Phi_{{\rm E},i}] \left(P_0[\Phi_{{\rm E},i}] \ln P_0[\Phi_{{\rm E},i}]-P_0[\Phi_{{\rm E},i}]\ln P_g[\Phi_{{\rm E},i}] \right)\notag
    \\
    &=-\ln Z_0[\psi,\bar{\psi}]+\ln Z_g[\psi,\bar{\psi}]+g\cdot {\langle I_{\rm I}\rangle}_{g=0}\notag
    \\
    &=W_0[\psi,\bar{\psi}] - W_g[\psi,\bar{\psi}]\notag
    \\
    &=g^2\cdot (-1)^{s_i}\frac{ a_i^2}{4 M^2}\int (d^4 x)_{\rm E} (\bar{\psi}\Gamma_i \psi)^2\geq 0
\end{align}
By taking to be $g=1$, we obtain the relative entropy between $P_{\rm NI}$ and $P_{\rm T}$ as follows.
\begin{align}
    S(P_{\rm NI}||P_{\rm T})=W_{\rm NI}[\psi,\bar{\psi}] - W_{\rm T}[\psi,\bar{\psi}]=(-1)^{s_i}\frac{ a_i^2}{4 M^2}\int (d^4 x)_{\rm E} (\bar{\psi}\Gamma_i \psi)^2\geq 0,
\end{align}
where $P_{\rm NI}=P_0$, $P_{\rm T}=P_{g=1}$, $W_{\rm NI}=W_0$, and $W_{\rm T}=W_{g=1}$ are used, and $(-1)^{s_i} (\bar{\psi}\Gamma_i \psi)^2$ takes positive values because the time components of $\bar{\psi}\Gamma_i \psi$ are assumed to be zero. 
This result is consistent with Ref.~\cite{Adams:2008hp} and the non-negativity of the relative entropy.

\subsection{Tree level UV completions with unstable field}
\label{sec:incon}
So far, we have studied target theories where the Euclidean path integral method is valid, i.e., the saddle point approximation works well.
In the following, we consider target theories with unstable auxiliary fields where the saddle point approximation is not valid in the Euclidean path integral method.
\begin{itemize}
    \item Unstable dilaton-like particle ---
    Consider a target theory defined by the action in the Minkowski space as follows:
    \begin{align}
        I_{\rm T}[A,\Phi]\equiv \int d^4x \left(-\frac{1}{4}F_{\mu\nu}F^{\mu\nu}+m^2_{\Phi} \Phi^2 +\frac{1}{f_{\Phi}}\Phi F_{\rho\sigma}F^{\rho\sigma}\right),~~~[{\rm\bf Target} ]\label{eq:un1}
    \end{align}
    where $m_{\Phi}$ and $f_{\Phi}$ are the mass and the decay constant of the heavy auxiliary field, respectively, and $F_{\mu\nu}$ is the field strength of photon field.
    The action in the Euclidean space is given by
    \begin{align}
        I_{\rm T}^{\rm (E)}[A,\Phi_{{\rm E}}]=\int (d^4x)_{\rm E} \left(\frac{1}{4}F_{\mu\nu}F^{\mu\nu}-m_{\Phi}^2 \Phi^2_{\rm E} -\frac{1}{f_{\Phi}}\Phi_{\rm E} F_{\rho\sigma}F^{\rho\sigma}\right),
    \end{align}
    where $F_{\mu\nu}$ is assumed to be a background field.
    Note here that the auxiliary field is unstable because of the negative mass term.
    The non-interacting and interacting terms in the Minkowski space are respectively expressed as follows:
    \begin{align}
      &I_0[A,\Phi]\equiv \int d^4x \left(-\frac{1}{4} F_{\mu\nu}F^{\mu\nu} +m^2_{\Phi}\Phi^2\right)
    \\
    &I_{\rm I}[A,\Phi]\equiv\int d^4x \left(\frac{1}{f_{\Phi}}  \Phi F_{\mu\nu}F^{\mu\nu} \right)
    \end{align}
    For this target theory, the NIRT is the same as the MFFRT, and the action of the NIRT and MFFRT is defined as
    \begin{align}
        I_{\rm NI}[A,\Phi]\equiv I_0[A,\Phi].~~~[{\rm \bf Reference}]
    \end{align}
    By using the auxiliary parameter $g$, define $I_g\equiv I_0 +g\cdot I_{\rm I}$.
    Then, the target and reference theories are reproduced as follows:
    \begin{align}
        I_{\rm T}[A,\Phi]=\lim_{g\to 1} I_g[A,\Phi],~~~I_{\rm NI}[A,\Phi]=\lim_{g\to 0} I_g[A,\Phi].
    \end{align}
    The solution of the equation of motion of $I_{g}^{\rm (E)}$ for $\Phi_{\rm E}$ is calculated as
    \begin{align}
     \widetilde{\Phi}_{g} =-g\cdot\frac{1}{2 m^2_{\Phi} f_{\Phi}}F_{\mu\nu}F^{\mu\nu},
    \end{align}
    which is not a local maximum of $I_{g}^{\rm (E)}$.
    Therefore, the saddle point approximation around this solution is not valid, and the Euclidean effective action cannot be calculated.
    In other words, the calculation procedures of the relative entropy of this work are not applicable.
    To check that the non-negativity of the relative entropy is violated, evaluate the effective action of the target theory in the Minkowski space.
    Consider a field redefinition of $\Phi$ as follows:
    \begin{align}
        \Phi\to \Phi-g\cdot\frac{1}{2m_{\Phi}^2 f_{\Phi}}F_{\mu\nu}F^{\mu\nu}. \notag
    \end{align}
    By performing the above field redefinition, $I_g$ yields the effective action in the Minkowski space as follows:
    \begin{align}
        W_{g}^{\rm (M)}[A]=\int d^4 x \left(-\frac{1}{4}F_{\mu\nu}F^{\mu\nu}-g^2\cdot\frac{1}{4 m^2_{\Phi} f_{\Phi}^2}(F_{\mu\nu}F^{\mu\nu})^2\right).
    \end{align}
    After the Wick rotation, we obtain the Euclidean effective action as follows:
    \begin{align}
        W_{g}[A]=\int (d^4 x)_{\rm E} \left(\frac{1}{4}F_{\mu\nu}F^{\mu\nu}+g^2\cdot\frac{1}{4 m^2_{\Phi} f_{\Phi}^2}(F_{\mu\nu}F^{\mu\nu})^2\right).\label{eq:Wguns1}
    \end{align}
    Note here that the second term of the right-hand side is positive.
    From Eqs.~\eqref{eq:sABF}, and \eqref{eq:Wguns1}, the relative entropy between $P_0$ and $P_g$ is calculated as follows:
    \begin{align}
      S(P_0||P_g)&\equiv W_0[A]-W_g[A] +g\cdot{\langle I_{\rm I}\rangle}_{g=0}\notag
      \\
      &=-g^2\cdot\frac{1}{4 m^2_{\Phi} f_{\Phi}^2}\int (d^4 x)_{\rm E} (F_{\mu\nu}F^{\mu\nu})^2,
    \end{align}
    where ${\langle I_{\rm I}\rangle}_{g=0}=(\partial W_g/\partial g)_{g=0}=0$ is used.
    By taking $g=1$, we obtain
    \begin{align}
      S(P_{\rm NI}||P_{\rm T})&=W_{\rm NI}[A]-W_{\rm T}[A]=-\frac{1}{4 m^2_{\Phi} f_{\Phi}^2}\int (d^4 x)_{\rm E} (F_{\mu\nu}F^{\mu\nu})^2,
    \end{align}
    where $P_{\rm NI}=P_0$, $P_{\rm T}=P_{g=1}$, $W_{\rm NI}=W_0$, and $W_{\rm T}=W_{g=1}$ are used.
    The right-hand side of the above equation takes a negative value, which is inconsistent with the non-negativity of the relative entropy.
    This is because the Euclidean path integral method does not work, and the relative entropy in the Euclidean space can not be defined in this theory.

    \item Doublet of real, shift-symmetric, massless scalar fields theory ---
    Consider a target theory~\cite{Arkani-Hamed:2021ajd} in the Minkowski space, 
    \begin{align}
        I_{\rm T}[\phi,X]\equiv \int d^4x \left(
        \frac{1}{2} (\partial_{\mu}\phi_{i}\partial^{\mu}\phi_{i})+m^2_X X_{\mu\nu}X^{\mu\nu}
        -\frac{\epsilon^{il}}{M}(\partial_{\mu}\phi_{i} \partial_{\nu}\phi_l)X^{\mu\nu}
        \right),~~~{[\rm \bf Target]}\label{eq:doblet}
    \end{align}
    where $\phi_i$, $i=1,2$ is a doublet of real, shift-symmetric, massless scalar fields, $X_{\mu\nu}$ is an auxiliary field, and $\epsilon^{12}=-\epsilon^{21}=1$.
    Similar to the previous example, the auxiliary field is unstable because of the mass term.
        The non-interacting and interacting terms in the Minkowski space are respectively expressed as follows:
    \begin{align}
      &I_0[\phi,X]\equiv\int d^4x \left(
        \frac{1}{2} (\partial_{\mu}\phi_{i}\partial^{\mu}\phi_{i})+m^2_X X_{\mu\nu}X^{\mu\nu}
        \right)
        \\
        &I_{\rm I}[\phi,X]\equiv
        \int d^4x \left(
      -\frac{\epsilon^{il}}{M}(\partial_{\mu}\phi_{i} \partial_{\nu}\phi_l)X^{\mu\nu}
        \right).
    \end{align}
        For this target theory, the NIRT is the same as the MFFRT, and the action of the NIRT and MFFRT is defined as
    \begin{align}
        I_{\rm NI}[\phi,X]\equiv I_0[\phi,X].~~~[{\rm \bf Reference}]
    \end{align}
    By the auxiliary parameter $g$, define $I_g\equiv I_0 +g\cdot I_{\rm I}$.
    The target and reference theories are reproduced as follows:
    \begin{align}
        I_{\rm T}[\phi,X]=\lim_{g\to 1} I_g[\phi,X],~~~I_{\rm NI}[\phi,X]=\lim_{g\to 0} I_g[\phi,X].
    \end{align}
    The solution of the equation of motion of $I_g^{\rm (E)}$ for $X_{\mu\nu}$ is calculated as
    \begin{align}
      \widetilde{X}_{g,\mu\nu}=-g\cdot\frac{\epsilon^{il}}{2m^2_X M}(\partial_{\mu}\phi_{ i}\partial_{\nu}\phi_l),
    \end{align}
    which is a classical solution not being a local minimum because of the sign of the mass term of Eq.~\eqref{eq:doblet}.
    Therefore, in the Euclidean path integral method, the saddle point approximation around this solution is not valid, and the relative entropy can not be calculated.
    To check the violation of the non-negativity of the relative entropy, evaluate the effective action of the target theory in the Minkowski space.
    By performing the following field redefinition,
    \begin{align}
        X_{\mu\nu}\to X_{\mu\nu} +g\cdot \frac{\epsilon^{il}}{2m^2_X M} (\partial_{\mu}\phi_i\partial_{\nu}\phi_l),
    \end{align}
    $I_g$ yields the effective action in the Minkowski space as follows:
    \begin{align}
        W^{\rm (M)}_g [\phi]=\int d^4x \left(
      \frac{1}{2}(\partial_{\mu}\phi_{i} \partial^{\mu}\phi_{i})
  +g^2\cdot\frac{1}{4m^2_X M^2}\epsilon^{il}\epsilon^{kj} (\partial_{\mu}\phi_{i} \partial^{\mu}\phi_j)(\partial_{\nu}\phi_k \partial^{\nu}\phi_l)\right).
    \end{align}
    After the Wick rotation, we obtain the Euclidean effective action as follows:
    \begin{align}
        W_g [\phi]=\int (d^4x)_{\rm E} \left(
      -\frac{1}{2}(\partial_{\mu}\phi_{i} \partial^{\mu}\phi_{i})
  -g^2\cdot\frac{1}{4m^2_X M^2}\epsilon^{il}\epsilon^{kj} (\partial_{\mu}\phi_{i} \partial^{\mu}\phi_j)(\partial_{\nu}\phi_k \partial^{\nu}\phi_l)\right).\label{eq:WGuns2}
    \end{align}
    From Eq.~\eqref{eq:sABF}, and \eqref{eq:WGuns2}, the relative entropy between $P_0$ and $P_g$ is expressed as follows:
    \begin{align}
      S(P_0||P_g)&\equiv W_0[\phi]-W_g[\phi]+g\cdot {\langle I_{\rm I}\rangle}_{g=0}\notag
      \\
      &=g^2\cdot\frac{1}{4m^2_X M^2}\int (d^4x)_{\rm E}\epsilon^{il}\epsilon^{kj} (\partial_{\mu}\phi_{i} \partial^{\mu}\phi_j)(\partial_{\nu}\phi_k \partial^{\nu}\phi_l),
    \end{align}
    where ${\langle I_{\rm I}\rangle}_{g=0}=(\partial W_g/\partial g)_{g=0}=0$ is used.
    By taking $g=1$, we obtain
    \begin{align}
        S(P_{\rm NI}||P_{\rm T})=W_{\rm NI}[\phi]-W_{\rm T}[\phi]=\frac{1}{4m^2_X M^2}\int (d^4x)_{\rm E}\epsilon^{il}\epsilon^{kj} (\partial_{\mu}\phi_{i} \partial^{\mu}\phi_j)(\partial_{\nu}\phi_k \partial^{\nu}\phi_l),\label{eq:doubrel}
    \end{align}
    where $P_{\rm NI}=P_0$, $P_{\rm T}=P_{g=1}$, $W_{\rm NI}=W_0$, and $W_{\rm T}=W_{g=1}$ are used.
    By choosing a background field, e.g., $\partial_{\mu}\widetilde{\phi}_1=(0,1,0,0)$, and $\partial_{\mu}\widetilde{\phi}_2=(0,0,1,0)$, we obtain
    \begin{align}
    \epsilon^{il}\epsilon^{kj} (\partial_{\mu}\widetilde{\phi}_i \partial^{\mu}\widetilde{\phi}_j)(\partial_{\nu}\widetilde{\phi}_k \partial^{\nu}\widetilde{\phi}_l)=2\left((\partial_{\mu}\widetilde{\phi}_1\partial^{\mu}\widetilde{\phi}_2)^2 -(\partial_{\mu}\widetilde{\phi}_1 \partial^{\mu}\widetilde{\phi}_1)(\partial_{\nu}\widetilde{\phi}_2 \partial^{\nu}\widetilde{\phi}_2)
        \right)=-2.
    \end{align}
    Then, the right-hand side of Eq.~\eqref{eq:doubrel} takes a negative value, which is inconsistent with the non-negativity of the relative entropy.
    This is because the Euclidean path integral method is not valid because of the sign of the mass term of Eq.~\eqref{eq:doblet}, and the relative entropy in the Euclidean space can not be defined in this theory.

\end{itemize}

\section{Bottom-up approach: bounds on EFTs}
\label{sec:bot}
We often face situations where the UV theory involving the heavy degrees of freedom is unknown in contrast to the top-down approach of the previous section.
In this section, we consider zero-temperature systems and take a bottom-up approach, where the UV theory involving the interactions between heavy and light degrees of freedom are unknown.
We focus on a class of EFTs, where the corrections to the non-higher derivative terms can be removed by a field redefinition of the background fields, and the entropy constraint on such EFTs is provided by focusing on the NIRT.
As explained later, the dimension-eight term of a single massless scalar field, the SMEFT dimension-eight $SU(N)$ gauge bosonic operators, and Einstein-Maxwell theory with higher-derivative operators belong to such a class of theories due to the existence of symmetries.

Before presenting the details of the calculations, we list the assumptions used to derive the results in this section in the following.
\begin{enumerate}[(I)]
    \item Hermiticity of probability distribution functions ---
    We assume the target and reference theory are represented by the Hermitian probability distribution functions.
    To derive the non-negativity of the relative entropy in Eq.~\eqref{eq:non-neg}, we used the Hermiticity of probability distribution functions, i.e., $\rho_{\rm R,T}=\rho_{\rm R,T}^{\dagger}$.
    The non-negativity of the relative entropy can be broken when this condition is not satisfied.
    This is the reason why the time components of the background field are chosen as zero in Sec.~\ref{sec:ForFrm}.

    \item Validity of Euclidean path integral method ---
    We assume the EFTs are generated from the solution of the local minimum.
    As shown in Sec.~\ref{sec:incon}, the non-negativity of the relative entropy can be broken when the Euclidean path integral method is not valid, i.e., the saddle point approximation does not work because of the solution not being the local minimum.
    
    \item Higher-derivative operators generated from the interaction between heavy and light fields ---
    We assume the higher-derivative operators of the EFTs are generated from the interactions between heavy and light fields.
    The interaction of the UV theory is generally expressed as,
    \begin{align}
    I_{\rm I}[\phi,\Phi]=\int (d^4x)_{\rm E} \mathcal{O}[\Phi] \otimes J[\phi],\label{eq:IIGEN}
    \end{align}
    where $\mathcal{O}[\Phi] \otimes J[\phi]$ generally involves summations over some indices in $\mathcal{O}[\Phi]$ and $J[\phi]$, e.g., Lorentz indices.

    \item Leading order in the interaction between heavy and light fields ---
    We assume $J[\phi]$ does not include the higher-derivative operators\footnote{For example, $(\partial\phi)^4$, $(F_{\mu\nu}F^{\mu\nu})^2$, etc. belong to the higher-derivative operators, which are the higher-dimensional operators. Note that the Einstein-Hilbert term is allowed to be $J$ by Assumption (IV).}.
    This assumption is quantitatively reasonable because the higher-dimensional operator in $J[\phi]$ is suppressed by a heavier mass than $\Phi$ as follows:
    \begin{align}
        J[\phi]=J_{\rm dim\text{-}4}[\phi]+\sum_{i=5,\cdots}\frac{1}{\Lambda^{i-4}}J_{{\rm dim\text{-}}i}[\phi],
    \end{align}
    where $J_{{\rm dim\text{-}}4}$ and  $J_{{\rm dim\text{-}}i}$ respectively denote operators up to dimension-four and dimension-$i$ operators, and $\Lambda$ is a mass scale satisfying $M \ll\Lambda$, where $M$ is the mass of $\Phi$.
    Note here that this assumption does not prohibit higher-dimensional interacting terms. 
    For example, in Secs.~\ref{sec:sfiftscal1}, \ref{sec:dilat}, and \ref{sec:axion}, we discussed $J[\phi]\propto (\partial_{\mu}\phi\partial^{\mu}\phi)$, $J[A]\propto F_{\mu\nu}F^{\mu\nu}$, and $J[A]\propto F_{\mu\nu}\widetilde{F}^{\mu\nu}$, respectively, and the interacting terms $I_{\rm I}$ were the dimension-five operator.
    Also, we assume the renormalizable terms in $I_{\rm I}$ are dominant effects on the Euclidean effective action, and the non-renormalizable terms are negligible when $I_{\rm I}$ includes both renormalizable and non-renormalizable terms.
    In other words, we consider the leading order of $1/\Lambda$ expansion for the interaction effects on the EFTs and assume as follows:
    \begin{align}
        J[\phi]\simeq J_{{\rm dim\text{-}}i}[\phi]/\Lambda^{i-4},
    \end{align}
     where $i$ denotes the leading order of $1/\Lambda$ expansion.

\end{enumerate}

    The first two assumptions are also imposed in Sec.~\ref{sec:entr}; see Sec.~\ref{sec:sum_en}.
    The main assumptions in this section are the third and fourth ones.
    
    In the following sections, we focus on two cases: tree-level UV completion and loop-level UV completion.
    For the tree-level UV completion, we assume the tree level effects dominate the perturbative corrections from the heavy degrees of freedom to the Euclidean effective action.
    On the other hand, for the loop-level UV completion, we assume the loop level effects dominate the perturbative corrections to the Euclidean effective action.
     For some examples, i.e., the single massless scalar field with the dimension-eight term, SMEFT dimension-eight gauge bosonic operators, and Einstein-Maxwell theory with higher-derivative terms, we provide constraints from the relative entropy in the following way.

\subsection{Single massless scalar field with dimension-eight operator}
\label{sec:bottomup_shift_symmetry}
Consider an EFT described by a single massless scalar field theory with a dimension-eight operator as follows:
\begin{align}
W_{\rm T}[\phi]=\int (d^4x)_{\rm E} \left[-\frac{1}{2}(\partial_\mu \phi \partial^\mu \phi)-\frac{c}{M^4}(\partial_\mu \phi \partial^\mu \phi)^2 \right].\label{eq:Wtsign}
\end{align}
Because of the shift symmetry: $\phi\to\phi+{\rm const.}$, Eq.~\eqref{eq:Wtsign} involves only the kinetic term, i.e., the non-higher derivative term, as the renormalizable term, and corrections to the kinetic term can be removed by redefining $\phi$.
Let us stand in the bottom-up approach and assume the second term of Eq.~\eqref{eq:Wtsign} is generated by integrating out the heavy fields of the theory of Eq.~\eqref{eq:Ignon}.
According to Assumption (IV), $J[\phi]$ can be $\partial_{\mu}\phi$ or $\partial_{\mu}\phi\partial^{\mu}\phi$, which preserve the shift symmetry, but $\partial_{\mu}\phi$ effects on ${\langle I_{\rm I}\rangle}_{g=0}$ vanish because ${\langle I_{\rm I}\rangle}_{g=0}$ preserves the Lorentz symmetry.
When we suppose that the EFT is generated by integrating out heavy degrees of freedom, the first order corrections for $g$ to the Euclidean effective action are expressed as
\begin{align}
    g\cdot {\langle I_{\rm I}\rangle}_{g=0}&=g\cdot \left(\frac{\partial W_g}{\partial g}\right)_{g=0}\notag
    \\
    &=\int (d^4 x)_{\rm E} \left(\frac{\delta W_g}{\delta J}\right)_{J=0} J[\phi]\notag
    \\
    &\propto \int (d^4x)_{\rm E} (\partial_{\mu}\phi\partial^{\mu}\phi),\label{eq:sigII}
\end{align}
where $({\delta W_g}/{\delta J})_{J=0}$ denotes a tadpole-like diagram for the composite field $J[\phi]$ and does not depend on space-time.
For both tree and loop-level UV completions, we consider the constraints on the Wilson coefficient of the dimension-eight operator in the following way.

\begin{itemize}
    \item Tree-level UV completion ---
    Consider the EFT generated by the tree-level UV completion.
    The partition function of $I_0+g\cdot I_{\rm I}$ is generally calculated as follows:
    \begin{align}
        Z_g[\widetilde{\phi}]&\equiv \int d[\phi]d[\Phi] e^{-I_g[\phi,\Phi]}\notag
        \\
        &=\int d[\phi]{\rm exp}\bigg[
        -\int (d^4 x)_{\rm E} \bigg(-\frac{1}{2}(1+\alpha_2^{\rm tree})(\partial_{\mu}\phi\partial^{\mu}{\phi})-\beta_2^{\rm tree}(\partial_{\mu}{\phi}\partial^{\mu}{\phi})^2 \bigg)
        \bigg]\notag
        \\
        &={\rm exp}\bigg[
        -\int (d^4 x)_{\rm E} \bigg(-\frac{1}{2}(1+\alpha_2^{\rm tree})(\partial_{\mu}\widetilde{\phi'}\partial^{\mu}\widetilde{\phi'})-\beta_2^{\rm tree}(\partial_{\mu}\widetilde{\phi'}\partial^{\mu}\widetilde{\phi'})^2 
        \bigg)
        \bigg]\notag
        \\
        &={\rm exp}\bigg[
        -\int (d^4 x)_{\rm E} \bigg(-\frac{1}{2}(\partial_{\mu}\widetilde{\phi}\partial^{\mu}\widetilde{\phi})-\beta_2^{\rm tree}\cdot(1+\alpha_2^{\rm tree})^{-2}(\partial_{\mu}\widetilde{\phi}\partial^{\mu}\widetilde{\phi})^2 
        \bigg)
        \bigg],\label{eq:Zgtreemass}
    \end{align}
    where $\alpha_2^{\rm tree}$ and $\beta_2^{\rm tree}$ denote the second or higher order corrections for $g$.
    Note here that $\beta_2^{\rm tree}$ does not include the first order correction for $g$ because of Eq.~\eqref{eq:sigII}.
    We assumed $\alpha_1^{\rm tree}$, $\alpha_2^{\rm tree}$, and $\beta_2^{\rm tree}$ are generated at the tree level. 
    Also, in the second line, according to the procedure in Eqs.~\eqref{eq:linrem1}, \eqref{eq:linrem2}, and \eqref{eq:linrem3}, the first order correction for $g$ is eliminated in $\alpha_2^{\rm tree}$ and absorbed into the definition of $\phi$. 
    As discussed in Sec.~\ref{sec:Massle}, the dimension-six operators and other dimension-eight operators, e.g., $(\partial\partial\phi)^2$ and $(\partial\partial\partial\phi)^2$, generally arise, but they are eliminated by the background field satisfying $\partial_{\mu}\widetilde{\phi'}={\rm const.}$, where $\widetilde{\phi'}$ denotes the classical solution of the effective action.
    To remove the dimension-six operators, we choose the background fields as follows:
    \begin{align}
        \widetilde{\phi'}=(1+\alpha^{\rm tree}_2)^{-1/2}\cdot\widetilde{\phi},\label{eq:treephipri}
    \end{align}
    with $\partial_{\mu}\widetilde{\phi}={\rm const.}$.
    From Eq.~\eqref{eq:Zgtreemass}, the Euclidean effective actions are calculated as follows:
    \begin{align}
        &W_g[\widetilde{\phi}]\equiv -\ln Z_g[\widetilde{\phi}]=\int (d^4 x)_{\rm E} \bigg(-\frac{1}{2}(\partial_{\mu}\widetilde{\phi}\partial^{\mu}\widetilde{\phi})-\beta_2^{\rm tree}\cdot\left(1+\alpha^{\rm tree}_2 \right)^{-2}(\partial_{\mu}\widetilde{\phi}\partial^{\mu}\widetilde{\phi})^2 
        \bigg),\label{eq:sigWg}
        \\
        &W_0[\widetilde{\phi}]=\lim_{g\to 0}W_g[\widetilde{\phi}]=\int (d^4 x)_{\rm E} \bigg(-\frac{1}{2}(\partial_{\mu}\widetilde{\phi}\partial^{\mu}\widetilde{\phi}) 
        \bigg).\label{eq:sigW0}
    \end{align}
    Combining Eq.~\eqref{eq:sigWg} and \eqref{eq:sigW0}, the shift of the Euclidean effective action by the interacting term is given by
    \begin{align}
         W_g[\widetilde{\phi}]-W_0[\widetilde{\phi}]=-\beta_2^{\rm tree}\cdot\left(1+\alpha^{\rm tree}_2 \right)^{-2} \int (d^4x)_{\rm E} (\partial_{\mu}\widetilde{\phi}\partial^{\mu}\widetilde{\phi})^2.\label{eq:DelWgsig}
    \end{align}
    Also, Eq.~\eqref{eq:sigWg} yields following relations:
    \begin{align}
        \left(\frac{dW_g}{dg}\right)_{g=0}&=\left(\frac{\partial W_g}{\partial g}\right)_{g=0}+\int (d^4 x)_{\rm E} \left(\frac{\delta W_g}{\delta \widetilde{\phi'}}\right)\cdot \left(\frac{d\widetilde{\phi'}}{dg}\right)_{g=0}\notag
        \\
        &=\left(\frac{\partial W_g}{\partial g}\right)_{g=0}\notag
        \\
        &={\langle I_{\rm I}\rangle}_{g=0}=0,\label{eq:sigdWgdg}
    \end{align}
    where the partial derivative means differentiating by $g$ while keeping $\widetilde{\phi'}$, and   $({d\widetilde{\phi'}}/{dg})_{g=0}=0$ is used because $\alpha_2^{\rm tree}$ in Eq.~\eqref{eq:treephipri} denotes the second or higher order corrections for $g$.
    Combining Eq.~\eqref{eq:upg}, \eqref{eq:DelWgsig}, and \eqref{eq:sigdWgdg}, we obtain
    \begin{align}
        S(P_0||P_g)&=W_0[\widetilde{\phi}]-W_g[\widetilde{\phi}]+ g{\langle I_{\rm I}\rangle}_{g=0}\notag
        \\
        &=\beta_2^{\rm tree}\cdot\left(1+\alpha^{\rm tree}_2 \right)^{-2} \int (d^4x)_{\rm E} (\partial_{\mu}\widetilde{\phi}\partial^{\mu}\widetilde{\phi})^2\geq 0.\label{eq:treesin1}
    \end{align}
    By taking $g=1$, Eq.~\eqref{eq:treesin1} represents the relative entropy between the reference and target theories and yields the following inequality.
    \begin{align}
        S(P_{\rm NI}||P_{\rm T})=\left(\beta_2^{\rm tree}\cdot\left(1+\alpha^{\rm tree}_2 \right)^{-2}\right)_{g=1}\cdot \int (d^4x)_{\rm E} (\partial_{\mu}\widetilde{\phi}\partial^{\mu}\widetilde{\phi})^2\geq 0&\Rightarrow \left(\beta_2^{\rm tree}\cdot\left(1+\alpha^{\rm tree}_2 \right)^{-2}\right)_{g=1}\geq 0,\label{eq:Tartreesin1}
    \end{align}
    where $P_{\rm NI}=P_0$ and $P_{\rm T}=P_{g=1}$ are used.
    From Eq.~\eqref{eq:sigWg}, this inequality represents the constraint on the coefficient of the dimension-eight operator of the effective action of the target theory.
    
    \item Loop-level UV completion ---
    Consider the EFT generated by the loop-level UV completion.
    The partition function of the theory $I_0+g\cdot I_{\rm I}$ is calculated as follows:
        \begin{align}
        Z_g[\widetilde{\phi}]&\equiv\int d[\phi]d[\Phi] e^{-I_g[\phi,\Phi]}\notag
        \\
        &=\int d[\phi] {\rm exp}\bigg[
        -\int (d^4 x)_{\rm E} \bigg(-\frac{1}{2}(1+ \alpha_1^{\rm loop}+\alpha_2^{\rm loop})(\partial_{\mu}\phi\partial^{\mu}\phi)-\beta_2^{\rm loop}(\partial_{\mu}\phi\partial^{\mu}\phi)^2 \bigg)+E_{\rm vac}^{\Phi}
        \bigg]\notag
        \\
        &={\rm exp}\bigg[
        -\int (d^4 x)_{\rm E} \bigg(-\frac{1}{2}(1+\alpha^{\rm loop}_1+\alpha_2^{\rm loop})(\partial_{\mu}\widetilde{\phi'}\partial^{\mu}\widetilde{\phi'})-\beta_2^{\rm loop}(\partial_{\mu}\widetilde{\phi'}\partial^{\mu}\widetilde{\phi'})^2+E_{\rm vac} \bigg)
        \bigg]\notag
        \\
        &={\rm exp}\bigg[
        -\int (d^4 x)_{\rm E} \bigg(-\frac{1}{2}(1+\alpha^{\rm loop}_1)(\partial_{\mu}\widetilde{\phi}\partial^{\mu}\widetilde{\phi})-\beta_2^{\rm loop}(\partial_{\mu}\widetilde{\phi}\partial^{\mu}\widetilde{\phi})^2+E_{\rm vac} 
        \bigg)
        \bigg],\label{eq:appZgSMEFT}
    \end{align}
    where $\alpha_1^{\rm loop}$ is the first order correction for $g$, $\alpha_2^{\rm loop}$ and $\beta_2^{\rm loop}$ are the second or higher order correction for $g$, 
    $E_{\rm vac}^{\Phi}$ is the vacuum energy coming from the one-loop level correction of $\Phi$, and $E_{\rm vac}$ is the vacuum energy of $\Phi$ and $\phi.$
    We neglect two-loop effects and assume $\alpha_1^{\rm loop}$, $\alpha_2^{\rm loop}$, and $\beta_2^{\rm loop}$ are generated from the one-loop corrections of $\Phi$.
    Note here that $\alpha_1^{\rm loop}$ cannot be removed by redefining $\Phi$ in contrast to the tree-level UV completion.
    Similar to the tree-level UV completion, dimension-six and other dimension-eight operators are eliminated by suitable $\widetilde{\phi'}$, which represents the classical solution of the effective action.
    We choose the background field as follows:
    \begin{align}
        \widetilde{\phi'}=\left(1-\frac{1}{2}\alpha_2^{\rm loop}\right)\cdot\widetilde{\phi},
    \end{align}
    with $\partial_{\mu}\widetilde{\phi}={\rm const.}$ to remove $(\partial\partial\phi)^2$ and $(\partial\partial\partial\phi)^2$.
    From Eq.~\eqref{eq:appZgloopSMEFT}, the Euclidean effective actions are given by
        \begin{align}
        W_g[\widetilde{\phi}]&\equiv -\ln Z_g[\tilde{\phi}]=\int (d^4 x)_{\rm E} \bigg(-\frac{1}{2}(1+\alpha^{\rm loop}_1)(\partial_{\mu}\widetilde{\phi}\partial^{\mu}\widetilde{\phi})-\beta_2^{\rm loop}(\partial_{\mu}\widetilde{\phi}\partial^{\mu}\widetilde{\phi})^2+E_{\rm vac} 
        \bigg),\label{eq:wgloop2}
        \\
        W_0[\widetilde{\phi}]&=\lim_{g\to 0}W_g[\widetilde{\phi}]=\int (d^4 x)_{\rm E} \bigg(-\frac{1}{2}(\partial_{\mu}\widetilde{\phi}\partial^{\mu}\widetilde{\phi})+E_{\rm vac} 
        \bigg).\label{eq:wgloop20}
    \end{align}
    From Eq.~\eqref{eq:wgloop2} and \eqref{eq:wgloop20}, the shift of the Euclidean effective action is calculated as
    \begin{align}
    W_g[\widetilde{\phi}]-W_0[\widetilde{\phi}]=-\frac{1}{2}\alpha^{\rm loop}_1\cdot\int (d^4 x)_{\rm E} (\partial_{\mu}\widetilde{\phi}\partial^{\mu}\widetilde{\phi})-\beta_2^{\rm loop}\int (d^4 x)_{\rm E} (\partial_{\mu}\widetilde{\phi}\partial^{\mu}\widetilde{\phi})^2.\label{eq:DelWgloop3}
    \end{align}
    Also, Eq.~\eqref{eq:wgloop2} yields the following relations.
    \begin{align}
        \left(\frac{dW_g}{dg}\right)_{g=0}&=\left(\frac{\partial W_g}{\partial g}\right)_{g=0} +\int (d^4x)_{\rm E} \left(\frac{\delta W_g}{\delta \widetilde{\phi'}}\right)\cdot \left(\frac{d \widetilde{\phi'}}{dg}\right)_{g=0}\notag
        \\
        &=\left(\frac{\partial W_g}{\partial g}\right)_{g=0}\notag
        \\
        &={\langle I_{\rm I}\rangle}_{g=0}\notag
        \\
        &=-\frac{1}{2}\frac{d\alpha^{\rm loop}_1}{dg}\cdot\int (d^4x)_{\rm E} (\partial_{\mu}\widetilde{\phi}\partial^{\mu}\widetilde{\phi})
        ,\label{eq:dWgdgloop2}
    \end{align}
    where $({d \widetilde{\phi'}}/{dg})_{g=0}=0$ is used in the second line, and the line is derived from Eq.~\eqref{eq:wgloop2}.
    Note here that $\alpha_1^{\rm loop}$ denotes the first order correction for $g$ and satisfies a relation of the form $g\cdot ({d\alpha^{\rm loop}_1}/{dg})=\alpha^{\rm loop}_1$.
    Combining Eqs.~\eqref{eq:upg}, \eqref{eq:DelWgloop3}, and \eqref{eq:dWgdgloop2}, we obtain
    \begin{align}
       S(P_0||P_g)&=W_0[\widetilde{\phi}]-W_g[\widetilde{\phi}]+ g{\langle I_{\rm I}\rangle}_{g=0}\notag
       \\
       &=W_0[\widetilde{\phi}]-W_g^{\rm non\text{-}lin}[\widetilde{\phi}]\notag
       \\
       &=\beta_2^{\rm loop}\int (d^4 x)_{\rm E} (\partial_{\mu}\widetilde{\phi}\partial^{\mu}\widetilde{\phi})^2\geq 0,
       \label{eq:loopboubd1}
    \end{align}
    where $g\cdot (d\alpha_1^{\rm loop}/dg)=\alpha_1^{\rm loop}$ was used, and we defined as follows:
    \begin{align}
        W_g^{\rm non\text{-}lin}[\widetilde{\phi}]&\equiv W_g[\widetilde{\phi}]- g{\langle I_{\rm I}\rangle}_{g=0}\notag
        \\
        &=\int (d^4 x)_{\rm E} \bigg(-\frac{1}{2}(\partial_{\mu}\widetilde{\phi}\partial^{\mu}\widetilde{\phi})-\beta_2^{\rm loop}(\partial_{\mu}\widetilde{\phi}\partial^{\mu}\widetilde{\phi})^2+E_{\rm vac} 
        \bigg).
    \end{align}
    By taking $g=1$, Eq.~\eqref{eq:loopboubd1} represents the relative entropy between the reference and target theories and yields the following inequality,
    \begin{align}
        S(P_{\rm NI}||P_{\rm T})=\beta_2^{\rm loop}|_{g=1}\cdot \int (d^4 x)_{\rm E} (\partial_{\mu}\widetilde{\phi}\partial^{\mu}\widetilde{\phi})^2\geq 0 \Rightarrow  \beta_2^{\rm loop}|_{g=1}\geq 0,\label{eq:Tarloopsin1}
    \end{align}
    where $P_{\rm NI}=P_0$ and $P_{\rm T}=P_{g=1}$ are used.
    This inequality represents the constraint on the dimension-eight operator generated at the loop-level in the target theory.

\end{itemize}

According to Eq.~\eqref{eq:Tartreesin1} and \eqref{eq:Tarloopsin1}, for both tree and loop level-UV completion, the relative entropy between the reference and target theories denotes the dimension-eight operator effects on the effective action.
By demanding $\partial_{\mu}\partial^{\mu}\widetilde{\phi}=0$ with constant $\partial^{\mu}\widetilde{\phi}$, after the Wick rotation, the non-negativity of the relative entropy gives rise to a constraint on Eq.~\eqref{eq:Wtsign} as follows: 
\begin{align}
S(P_{\rm NI}||P_{\rm T})=\frac{c}{M^4}\int (d^4x)_{\rm E} (\partial_{\mu} \widetilde{\phi} \partial^{\mu} \widetilde{\phi})^2\geq  0\Rightarrow \frac{c}{M^4}\geq 0.\label{eq:sclpos}
\end{align}
Consequently, the coefficient $c$ must be positive to respect the entropy constraint for both tree and loop-level UV completion.
This result is consistent with the positivity bound from the unitarity and causality~\cite{Adams:2006sv}.

\subsection{SMEFT dimension-eight gauge bosonic operators}
In this section, using the technique provided in Sec.~\ref{sec:entr}, we consider the entropy constraint on the SMEFT dimension-eight gauge bosonic operators.
We list the operator basis and evaluate its constraints in the following way.

\subsubsection{Operator Basis}\label{sec:Op}
In this section, we consider the EFT described by the following Lagrangian in the Minkowski space,
\begin{align}
    \mathcal{L}_{\rm SMEFT}=-\frac{1}{4}B_{\mu\nu}B^{\mu\nu} -\frac{1}{4}W^I_{\mu\nu}W^{I,\mu\nu}-\frac{1}{4}G^a_{\mu\nu}G^{a,\mu\nu}+\frac{1}{M^4}\sum_i c_i \mathcal{O}_i,\label{eq:SMEFTLa}
\end{align}
where $F^a_{\mu\nu}\equiv \partial_{\mu}A^a_{\nu}-\partial_{\nu}A^a_{\mu}+g f^{abc} A^b_{\mu}A^c_{\nu}$ is the field strength of the gauge field $A^a_{\mu}$, and $g$ denotes the gauge coupling of $SU(N)$.
The gauge fields for $U(1)_Y$ hypercharge, $SU(2)_L$ weak isospin, and $SU(3)_C$ color are expressed as $B_{\mu}$, $W^I_{\mu}$, and $G^a_{\mu}$, respectively.
Also, their gauge field strengths are defined as
\begin{align}
    B_{\mu\nu}&=\partial_{\mu}B_{\nu}-\partial_{\nu}B_{\mu},
    \\
    W^I_{\mu\nu}&=\partial_{\mu}W^I_{\nu}-\partial_{\nu}W^I_{\mu}+g_2\epsilon^{IJK} W^J_{\mu}W^K_{\nu},
    \\
    G^a_{\mu\nu}&=\partial_{\mu}G^a-\partial_{\nu} G^a_{\mu}+g_3 f^{abc} G^b_{\mu}G^c_{\nu},
\end{align}
where $g_2$, and $g_3$ denote the gauge couplings of $SU(2)_L$, and $SU(3)_C$, respectively.
The Greek letters stand for Lorentz indices, the Italic letters represent $SU(N)$ color indices, and totally antisymmetric and symmetric structure constants are defined by
\begin{align}
    [T^a,T^b]&=i f^{abc} T^c,\label{eq:deff}
    \\
    \{T^a,T^b\}&=\delta^{ab}\frac{\hat{1}}{N} +d^{abc} T^c,\label{eq:defd}
\end{align}
with the generator $T^a$ of $SU(N)$ Lie algebra.
In general, the $CP$ violating renormalizable term arises, but we assume such terms are removed by some mechanism, e.g.,  axion-like degrees of freedom in the UV theory.
The last term of Eq.~\eqref{eq:SMEFTLa} denotes the SMEFT dimension-eight gauge bosonic operators.
The basis of independent dimension-eight gauge bosonic operators for the gauge fields $B_{\mu}$, $W^I_{\mu}$, and $G^a_{\mu}$ are listed in the following~\cite{Morozov:1984goy,Remmen:2019cyz,Li:2020gnx,Murphy:2020rsh}.

\begin{itemize}
    \item $U(1)_Y$ ---
    The single field strength quartics with the gauge fields of the $U(1)_Y$ hypercharge are listed as follows:
    \begin{align}
        &\mathcal{O}^{B^4}_1 =(B_{\mu\nu}B^{\mu\nu})(B_{\rho\sigma}B^{\rho\sigma}),
        \\
        &\mathcal{O}^{B^4}_2 =(B_{\mu\nu} \widetilde{B}^{\mu\nu})(B_{\rho\sigma}\widetilde{B}^{\rho\sigma}),
        \\
        &\widetilde{\mathcal{O}}^{B^4}_1 =(B_{\mu\nu} {B}^{\mu\nu})(B_{\rho\sigma}\widetilde{B}^{\rho\sigma}).
    \end{align}

    \item $SU(2)_L$ ---
    The single field strength quartics with the gauge fields of the $SU(2)_L$ weak isospin are listed as follows:
    \begin{align}
        &\mathcal{O}^{W^4}_1 = (W^I_{\mu\nu}W^{I,\mu\nu})(W^J_{\rho\sigma}W^{I,\rho\sigma}),
        \\
        &\mathcal{O}^{W^4}_2 =(W^I_{\mu\nu}\widetilde{W}^{I,\mu\nu})(W^J_{\rho\sigma}\widetilde{W}^{J,\rho\sigma}),
        \\
        &\mathcal{O}^{W^4}_3 =(W^I_{\mu\nu}W^{J,\mu\nu})(W^I_{\rho\sigma}W^{J,\rho\sigma}),
        \\
        &\mathcal{O}^{W^4}_4 =(W^I_{\mu\nu}\widetilde{W}^{J,\mu\nu})(W^I_{\rho\sigma}\widetilde{W}^{J,\rho\sigma}),
        \\
        &\widetilde{\mathcal{O}}^{W^4}_1 =(W^I_{\mu\nu}{W}^{I,\mu\nu})(W^J_{\rho\sigma}\widetilde{W}^{J,\rho\sigma}),
        \\
        &\widetilde{\mathcal{O}}^{W^4}_2 =(W^I_{\mu\nu}{W}^{J,\mu\nu})(W^I_{\rho\sigma}\widetilde{W}^{J,\rho\sigma}).
    \end{align}
    
    \item $SU(3)_C$ ---
    The single field strength quartics with the gauge fields of the $SU(3)_C$ color are listed as follows:
    \begin{align}
        &\mathcal{O}^{G^4}_1=(G^a_{\mu\nu} G^{a,\mu\nu})(G^b_{\rho\sigma} G^{b,\rho\sigma}),
        \\
        &\mathcal{O}^{G^4}_2=(G^a_{\mu\nu} \widetilde{G}^{a,\mu\nu})(G^b_{\rho\sigma} \widetilde{G}^{b,\rho\sigma}),
        \\
        &\mathcal{O}^{G^4}_3=(G^a_{\mu\nu} G^{b,\mu\nu})(G^a_{\rho\sigma} G^{b,\rho\sigma}),
        \\
        &\mathcal{O}^{G^4}_4=(G^a_{\mu\nu} \widetilde{G}^{b,\mu\nu})(G^a_{\rho\sigma} \widetilde{G}^{b,\rho\sigma}),
        \\
        &\mathcal{O}^{G^4}_5=d^{abe} d^{cde} (G^a_{\mu\nu}G^{b,\mu\nu})(G^c_{\rho\sigma}G^{d,\rho\sigma}),
        \\
        &\mathcal{O}^{G^4}_6=d^{abe}d^{cde}(G^a_{\mu\nu}\widetilde{G}^{b,\mu\nu})(G^c_{\rho\sigma}\widetilde{G}^{d,\rho\sigma}),
        \\
        &\widetilde{\mathcal{O}}^{G^4}_1=(G^a_{\mu\nu}G^{a,\mu\nu})(G^b_{\rho\sigma}\widetilde{G}^{b,\rho\sigma}),
        \\
        &\widetilde{\mathcal{O}}^{G^4}_2=(G^a_{\mu\nu}G^{b,\mu\nu})(G^a_{\rho\sigma}\widetilde{G}^{b,\rho\sigma}),
        \\
        &\widetilde{\mathcal{O}}^{G^4}_3=d^{abe}d^{cde}(G^a_{\mu\nu}G^{b,\mu\nu})(G^c_{\rho\sigma}\widetilde{G}^{d,\rho\sigma}).
    \end{align}
    
\end{itemize}

In addition to the dimension-eight operators, the dimension-six gauge bosonic operators may arise as follows:
\begin{align}
    \mathcal{O}^{W^3}&=\epsilon^{IJK} W^{I,\nu}_{\mu} W^{J,\rho}_{\nu} W^{K,\mu}_{\rho},
    \\
    \widetilde{\mathcal{O}}^{W^3}&=\epsilon^{IJK} W^{I,\nu}_{\mu} W^{J,\rho}_{\nu} \widetilde{W}^{K,\mu}_{\rho},
    \\
    \mathcal{O}^{G^3}&=f^{abc}G^{a,\nu}_{\mu}G^{b,\rho}_{\nu} G^{c,\mu}_{\rho},
    \\
    \widetilde{\mathcal{O}}^{G^3}&=f^{abc} G^{a,\nu}_{\mu}G^{b,\rho}_{\nu}\widetilde{G}^{c,\mu}_{\rho}.
\end{align}
As discussed later, we can remove the above dimension-six operator corrections to the Euclidean effective action by choosing suitable background fields. 
Throughout this section, $CP$-violating operators are denoted with a tilde.

\subsubsection{Linear combination of single field strength quartics}
For later convenience, we summarize building blocks to calculate the relative entropy.
The dimension-eight gauge bosonic operators with a single type of the $SU(N)$ gauge field are expressed as~\cite{Morozov:1984goy,Remmen:2019cyz,Li:2020gnx,Murphy:2020rsh}
\begin{align}
    &\mathcal{O}_1^{F^4} =(F^a_{\mu\nu}F^{a,\mu\nu})(F^b_{\rho\sigma}F^{b,\rho\sigma}),\label{eq:O1F4}
\\
&\mathcal{O}_2^{F^4} =(F^a_{\mu\nu}\widetilde{F}^{a,\mu\nu})(F^b_{\rho\sigma}\widetilde{F}^{b,\rho\sigma}),
\\
&\mathcal{O}_3^{F^4} =(F^a_{\mu\nu}{F}^{b,\mu\nu})(F^a_{\rho\sigma}F^{b,\rho\sigma}),
\\
&{\mathcal{O}}_4^{F^4} =(F^a_{\mu\nu}\widetilde{F}^{b,\mu\nu})(F^a_{\rho\sigma}\widetilde{F}^{b,\rho\sigma}),
\\
&{\mathcal{O}}_5^{F^4} =d^{abe}d^{cde}(F^a_{\mu\nu}F^{b,\mu\nu})(F^c_{\rho\sigma}F^{d,\rho\sigma}),
\\
&{\mathcal{O}}_6^{F^4} =d^{abe}d^{cde}(F^a_{\mu\nu}\widetilde{F}^{b,\mu\nu})(F^c_{\rho\sigma}\widetilde{F}^{d,\rho\sigma}),
\\
&{\mathcal{O}}_7^{F^4} =d^{ace}d^{bde}(F^a_{\mu\nu}F^{b,\mu\nu})(F^c_{\rho\sigma}{F}^{d,\rho\sigma}),
\\
&{\mathcal{O}}_8^{F^4} =d^{ace}d^{bde}(F^a_{\mu\nu}\widetilde{F}^{b,\mu\nu})(F^c_{\rho\sigma}\widetilde{F}^{d,\rho\sigma}),
\\
&\widetilde{\mathcal{O}}_1^{F^4} =(F^a_{\mu\nu}F^{a,\mu\nu})(F^b_{\rho\sigma}\widetilde{F}^{b,\rho\sigma}),
\\
&\widetilde{\mathcal{O}}_2^{F^4} =(F^a_{\mu\nu}F^{b,\mu\nu})(F^a_{\rho\sigma}\widetilde{F}^{b,\rho\sigma}),
\\
&\widetilde{\mathcal{O}}_3^{F^4} =d^{abe}d^{cde}(F^a_{\mu\nu}F^{b,\mu\nu})(F^c_{\rho\sigma}\widetilde{F}^{d,\rho\sigma}),
\\
&\widetilde{\mathcal{O}}_4^{F^4} =d^{ace}d^{bde}(F^a_{\mu\nu}F^{b,\mu\nu})(F^c_{\rho\sigma}\widetilde{F}^{d,\rho\sigma}),\label{eq:Otil4F4}
\end{align}
where the totally antisymmetric and symmetric structure constants were defined in Eq.~\eqref{eq:deff} and \eqref{eq:defd}.
We assume the above higher dimensional operators are produced by integrating out the heavy degrees of freedom through the interaction of Eq.~\eqref{eq:IIGEN}.
We apply the entropy inequalities of Sec.~\ref{sec:entr} and provide constraints on the Wilson coefficients of these operators.

To derive the bounds on the Wilson coefficients, we need the classical solution of the leading-order equation of motion of Eq.~\eqref{eq:SMEFTLa},
\begin{align}
    \partial^{\mu}F^a_{\mu\nu} +g f^{abc} A^{\mu,b}F^c_{\mu\nu}=0.\label{eq:smeftEoM}
\end{align}
Although the result of the inequalities derived from the relative entropy would depend on the choice of classical solutions, for simplicity, we focus on a class of solutions defined as follows:
\begin{align}
    \overline{A}^a_{\mu}=u^a_1 \epsilon_{1,\mu}w_1+u^a_2 \epsilon_{2,\mu}w_2,\label{eq:sol1}
\end{align}
where $u_{1,2}$ is a constant real vector in $SU(N)$ color space, $\epsilon_{1,2}$ is a constant four-vector, and $w_{1,2}$ is an arbitrary Cartesian coordinate in spacetime satisfying $\partial_{\mu}w_1=l_{\mu}$ and $\partial_{\mu}w_2=k_{\mu}$ with $l_{\mu}$ and $k_{\mu}$ being constant four-vectors. 
%
%
%
In this work, we consider the classical solutions satisfying a condition $f^{abc}u^a_1 u^b_2=0$, and then Eq.~\eqref{eq:smeftEoM} can be expressed as
\begin{align}
    \partial^{\mu}\overline{F}^a_{\mu\nu}=0,
\end{align}
where $\overline{F}^a_{\mu\nu}=\partial_{\mu} \overline{A}^a_{\nu}-\partial_{\nu} \overline{A}^a_{\mu}$.
The effects from the dimension-six operators on the effective action also vanish by this condition.
Substituting Eq.~\eqref{eq:sol1} into the field and dual field strengths yields
\begin{align}
    &\overline{F}^a_{\mu\nu}\overline{F}^{b,\mu\nu}=2\left[u_1^a u_1^b\cdot \mathrm{A}_F+u_2^a u_2^b \cdot \mathrm{B}_F+\left(u_1^a u_2^b+u_2^a u_1^b\right)\cdot \Gamma_F \right],\label{eq:FFBG}
    \\
    &\overline{F}^a_{\mu\nu}\widetilde{\overline{F}}^{b,\mu\nu}=2\left(u_1^a u_2^b+u_2^a u_1^b\right)\cdot \Delta_F,\label{eq:FFtilBG}
\end{align}
with
\begin{align}
    &\mathrm{A}_F\equiv \epsilon_1^2l^2-(\epsilon_1\cdot l)^2,\label{eq:AF}
    \\
    &\mathrm{B}_F\equiv \epsilon_2^2k^2-(\epsilon_2\cdot k)^2,\label{eq:BF}
    \\
    &\Gamma_F\equiv (\epsilon_1\cdot \epsilon_2)(l\cdot k)-(\epsilon_1\cdot k)(\epsilon_2\cdot l),\label{eq:GaF}
    \\
    &\Delta_F \equiv \epsilon^{\mu\nu\rho\sigma}\epsilon_{1,\mu}l_{\nu}\epsilon_{2,\rho}k_{\sigma}.\label{eq:DeF}
\end{align}
From Eqs.~\eqref{eq:O1F4}-\eqref{eq:Otil4F4}, and \eqref{eq:FFBG}-\eqref{eq:FFtilBG}, the SMEFT operators involving the background fields are expressed as follows:
\begin{align}
    \mathcal{O}^{F^4}_1&=4\left[u_1^2 \mathrm{A}_F +u_2^2\mathrm{B}_F +2(u_1\cdot u_2)\Gamma_F \right]^2,\label{eq:O1}
    \\
    \mathcal{O}^{F^4}_2&=16(u_1\cdot u_2)^2 \Delta_F^2,\label{eq:O2}
    \\
    \mathcal{O}^{F^4}_3&=4\bigg[(u_1^2)^2 \mathrm{A}^2_F + (u_2^2)^2 \mathrm{B}_F^2 +2(u_1\cdot u_2)^2 \mathrm{A}_F \mathrm{B}_F \notag \\
    &\qquad \qquad +2 \left((u_1\cdot u_2)^2 +u_1^2 u_2^2\right)\Gamma_F^2+4 (u_1\cdot u_2)\left(u_1^2 \mathrm{A}_F +u_2^2 \mathrm{B}_F\right)\Gamma_F\bigg],\label{eq:O3}
    \\
    \mathcal{O}^{F^4}_4&=8 \left[(u_1\cdot u_2)^2 +u_1^2 u_2^2 \right]\Delta^2_F,\label{eq:O4}
    \\
    \mathcal{O}^{F^4}_5&=4\bigg[\mathrm{A}_F^2 V_F^2 +\mathrm{B}_F^2 W_F^2 +4\Gamma_F^2 U_F^2\notag
    \\
    &\qquad \qquad+2 \mathrm{A}_F \mathrm{B}_F V_F\cdot W_F +4\mathrm{A}_F\Gamma_F V_F\cdot U_F+4 \mathrm{B}_F\Gamma_F W_F\cdot U_F\bigg],\label{eq:O5}
    \\
    \mathcal{O}^{F^4}_6&=16 U_F^2\Delta_F^2,\label{eq:O6}
    \\
    \mathcal{O}^{F^4}_7&=4\bigg[\mathrm{A}_F^2 V_F^2 + \mathrm{B}_F^2 W_F^2 + 2\Gamma_F^2 \left(U_F^2+ V_F\cdot W_F\right) \notag
    \\
    &\qquad \qquad +2A_F B_F U_F^2 + 4 A_F\Gamma_F V_F\cdot U_F+ 4 B_F\Gamma_F W_F\cdot U_F \bigg],\label{eq:O7}
    \\
    \mathcal{O}^{F^4}_8&= 8 \left(V_F\cdot W_F +U_F^2\right)\Delta_F^2,\label{eq:O8}
    \\
    \tilde{\mathcal{O}}^{F^4}_1&=8 (u_1\cdot u_2)\left(\mathrm{A}_F u_1^2 +2 \Gamma_F(u_1\cdot u_2)+\mathrm{B}_F u_2^2\right)\Delta_F,\label{eq:tilO1}
    \\
    \tilde{\mathcal{O}}^{F^4}_2&=8 \left[(u_1\cdot u_2)\left(\Gamma_F (u_1\cdot u_2)+\mathrm{B}_F u_2^2\right)+u_1^2 \left(\mathrm{A}_F(u_1\cdot u_2)+\Gamma_F u_2^2\right)  \right]\Delta_F,\label{eq:tilO2}
    \\
    \tilde{\mathcal{O}}^{F^4}_3&=8 \left(\mathrm{A}_F V_F\cdot U_F +2\Gamma_F U_F^2 +\mathrm{B}_F U_F\cdot W_F \right)\Delta_F,\label{eq:tilO3}
    \\
    \tilde{\mathcal{O}}^{F^4}_4&=8 \left[ \mathrm{A}_F V_F\cdot U_F+\mathrm{B}_FU_F\cdot W_F+\Gamma_F\left(U_F^2+V_F\cdot W_F\right)\right]\Delta_F,\label{eq:tilO4}
\end{align}
where we defined as $U_F^a=d^{abc}u^b_1 u^c_2$, $V_F^a=d^{abc}u^b_1 u^c_1$, and $W_F^a=d^{abc}u^b_2 u^c_2$.
Then, the SMEFT operator effects on the effective action of the target theory in the Minkowski space are expressed as
\begin{align}
    \frac{1}{M^4}\sum_i\int d^4x c_i \mathcal{O}_i
    &=\frac{1}{M^4}\int d^4x \bigg[
    a_F \cdot \mathrm{A}_F^2 +b_F\cdot \mathrm{B}_F^2+c_F\cdot \Gamma_F^2 +d_F\cdot \Delta_F^2 + e_F\cdot  \mathrm{A}_F\mathrm{B}_F\notag
    \\
    &\quad \quad+f_F\cdot\mathrm{A}_F\Gamma_F+g_F\cdot\mathrm{B}_F\Gamma_F+h_F\cdot \mathrm{A}_F\Delta_F+i_F\cdot\mathrm{B}_F\Delta_F+ j_F\cdot\Gamma_F\Delta_F\bigg],\label{eq:Wsemeft}
\end{align}
where the second and last lines are obtained by substituting Eq.~\eqref{eq:O1}-\eqref{eq:tilO4} into the first line, and we defined linear combinations of the Wilson coefficients as follows:
\begin{align}
        &a_F=4 (u_1^2)^2 c_1^{F^4}+4 (u_1^2)^2 c_3^{F^4}+4 V_F^2 c_5^{F^4}+4 V_F^2 c_7^{F^4},\label{eq:aF}
        \\
        &b_F=4 (u_2^2)^2 c_1^{F^4}+4 (u_2^2)^2 c_3^{F^4}+4 W_F^2 c_5^{F^4}+4 W_F^2 c_7^{F^4},
        \\
        &c_F=8\left[
        2c_1^{F^4} (u_1\cdot u_2)^2+c_3^{F^4} \left((u_1\cdot u_2)^2+u_1^2 u_2^2\right)+2 c_5^{F^4}U_F^2
        +c_7^{F^4}\left(U_F^2+V_F\cdot W_F \right)
        \right],
        \\
        &d_F=8\left[
        2c_2^{F^4} (u_1\cdot u_2)^2+
        c_4^{F^4}\left((u_1\cdot u_2)^2+u_1^2 u_2^2\right)
        +2 c_6^{F^4} U^2_F
        +c_8^{F^4}\left(U_F^2 +V_F\cdot W_F\right)
        \right],
        \\
        &e_F=8 u_1^2 u_2^2 c_1^{F^4}+8 (u_1\cdot u_2)^2 c_3^{F^4} +8 V_F\cdot W_F c_5^{F^4}+8U_F^2 c_7^{F^4},
        \\
        &f_F=16 u_1^2 (u_1\cdot u_2)c_1^{F^4} +16 (u_1\cdot u_2)u_1^2 c_3^{F^4} +16 V_F\cdot U_F c_5^{F^4}+16V_F\cdot U_F c_7^{F^4},
        \\
        &g_F=16 u_2^2 (u_1\cdot u_2)c_1^{F^4} +16 u_2^2 (u_1\cdot u_2)c_3^{F^4} +16 W_F\cdot U_F c_5^{F^4}+16 U_F\cdot W_F c_7^{F^4},
        \\
        &h_F=8u_1^2(u_1\cdot u_2) \left(\tilde{c}_1^{F^4}+\tilde{c}_2^{F^4} \right)+8V_F\cdot U_F (\tilde{c}_3^{F^4}+\tilde{c}_4^{F^4}),
        \\
        &i_F=8 u_2^2(u_1\cdot u_2)\left(\tilde{c}_1^{F^4}+\tilde{c}_2^{F^4}\right)+8U_F\cdot W_F (\tilde{c}_3^{F^4}+\tilde{c}_4^{F^4}),
        \\
        &j_F=8\left[
        2 \tilde{c}_1^{F^4} (u_1\cdot u_2)^2
        +\tilde{c}_2^{F^4}\left((u_1\cdot u_2)^2+u^2_1 u^2_2 \right) +2\tilde{c}_3^{F^4} U^2_F +\tilde{c}_4^{F^4}\left(U^2_F+V_F\cdot W_F \right)
        \right].\label{eq:jF}
\end{align}
For $U(1)_Y$, $SU(2)_L$, and $SU(3)_C$, the above quantities are respectively listed as follows:
\begin{itemize}
    \item $U(1)_Y$ ---
    For the $U(1)_Y$ gauge field $B_{\mu}$, Eqs.~\eqref{eq:aF}-\eqref{eq:jF} are calculated as
    \begin{align}
        &a_B=b_B=c_B/4=e_B/2=f_B/4=g_B/4=4 c_1^{B^4},\label{eq:Bab}
        \\
        &d_B=16 c_2^{B^4},\label{eq:Bd}
        \\
        &h_B=i_B=j_B/2=8\tilde{c}_1^{B^4}.\label{eq:Bhi}
    \end{align}

     \item $SU(2)_L$ ---
     For the $SU(2)_L$ gauge field $W^I_{\mu}$, Eqs.~\eqref{eq:aF}-\eqref{eq:jF} are calculated as 
     \begin{align}
         &a_{W}=b_{W}=c_{W}/4=e_{W}/2=f_{W}/4=g_{W}/4=4(u_1^2)^2(c^{W^4}_1+c^{W^4}_3),\label{eq:Wab}
         \\
         &d_W=16(u_1^2)^2(c_2^{W^4}+c_4^{W^4}),\label{eq:Wd}
         \\
         &h_W=i_W=j_W/2=8(u_1^2)^2 (\tilde{c}_1^{W^4}+\tilde{c}_2^{W^4}),\label{eq:Whj}
     \end{align}

     \item $SU(3)_C$ ---
     Using an identity for $SU(N)$,
\begin{align}
    f^{abe}f^{cde}=\frac{2}{N}\left(\delta^{ac}\delta^{bd}-\delta^{ad}\delta^{bc}\right)+d^{ace}d^{bde}-d^{bce}d^{ade},
\end{align}
we obtain
\begin{align}
    V_F W_F=U^2_F +\frac{2}{N}\left[(u_1\cdot u_2)^2-u_1^2 u_2^2\right].
\end{align}
For $SU(3)_C$, additional identities hold as follows:
\begin{align}
    &3 d^{abe}d^{cde}-f^{ace}f^{bde}-f^{ade}f^{bce}=\delta^{ac}\delta^{bd}+\delta^{ad}\delta^{bc}-\delta^{ab}\delta^{cd},
    \\
    &3\left(d^{abe}d^{cde}+d^{ace}d^{bde}+d^{ade}d^{bce}\right)=
    \delta^{ab}\delta^{cd}+\delta^{ac}\delta^{bd}+\delta^{ad}\delta^{bc}.
\end{align}
From the above identities, we obtain
\begin{align}
    &V^2_G=\frac{1}{3}u_1^4,~~W^2_G=\frac{1}{3}u_2^4,~~U^2_G=\frac{1}{3}u_1^2 u_2^2,\label{eq:VWU}
    \\
    &V_G\cdot U_G=\frac{1}{3}u_1^2 (u_1\cdot u_2),~W_{G} \cdot U_G=\frac{1}{3}u_2^2 (u_1\cdot u_2),~V_G \cdot W_G=\frac{1}{3}\left(-u_1^2 u_2^2+2 (u_1\cdot u_2)^2\right),\label{eq:VW}
    \\
    &2 U^2_G+V_G W_G=\frac{1}{3}\left(u_1^2 u_2^2 +2(u_1\cdot u_2)^2\right).\label{eq:U2VW}
\end{align}
Also, from the above identities, $\mathcal{O}^{G^4}_7$, $\mathcal{O}^{G^4}_8$, and $\widetilde{\mathcal{O}}^{G^4}_4$ can be rewritten by the other operators, so we omit these operators. 
From Eqs.~\eqref{eq:aF}-\eqref{eq:jF}, and \eqref{eq:VWU}-\eqref{eq:U2VW}, we obtain
     \begin{align}
         &a_G=4(u_1^2)^2\left(c_1^{G^4}+c_3^{G^4}+\frac{1}{3}c_5^{G^4}+\frac{1}{3}c_7^{G^4}\right),\label{eq:Ga}
         \\
         &b_G=4(u_2^2)^2\left(c_1^{G^4}+c_3^{G^4}+\frac{1}{3}c_5^{G^4}+\frac{1}{3}c_7^{G^4}\right),\label{eq:Gb}
         \\
         &c_G=8\left[
         (2 c_1^{G^4} +c_3^{G^4}) (u_1\cdot u_2)^2+c_3^{G^4} u_1^2 u_2^2 +2c_5^{G^4} U^2_G
         \right],\label{eq:Gc}
         \\
         &d_G=8
         \left[
         (2 c_2^{G^4} +c_4^{G^4}) (u_1\cdot u_2)^2+c_4^{G^4} u_1^2 u_2^2 +2c_6^{G^4}U^2_G
         \right],\label{eq:Gd}
         \\
         &e_G=8 u_1^2 u_2^2 \left(c_1^{G^4}-\frac{1}{3}c_5^{G^4}\right)+8 (u_1\cdot u_2)^2 \left(c_3^{G^4}+\frac{2}{3}c_5^{G^4}\right),\label{eq:Ge}
         \\
         &f_G=16 u_1^2 (u_1\cdot u_2)\left(c_1^{G^4}+c_3^{G^4}+\frac{1}{3}c_5^{G^4}
         \right),\label{eq:Gf}
         \\
         &g_G=16 u_2^2 (u_1\cdot u_2)\left(c_1^{G^4}+c_3^{G^4}+\frac{1}{3}c_5^{G^4}
         \right),\label{eq:Gg}
         \\
         &h_G=8 u_1^2 (u_1\cdot u_2)\left(\tilde{c}_1^{G^4}+\tilde{c}_2^{G^4}+\frac{1}{3}\tilde{c}_3^{G^4}
         \right),\label{eq:Gh}
         \\
         &i_G =8 u_2^2 (u_1\cdot u_2)\left(\tilde{c}_1^{G^4}+\tilde{c}_2^{G^4}+\frac{1}{3}\tilde{c}_3^{G^4}
         \right),\label{eq:Gi}
         \\
         &j_G= 8
         \left[(2\tilde{c}_1^{G^4}+\tilde{c}_2^{G^4})(u_1\cdot u_2)^2+\tilde{c}_2^{G^4}u_1^2 u_2^2 +2\tilde{c}_3^{G^4}U^2_G\right].\label{eq:Gj}
     \end{align}

\end{itemize}

In the next section, we derive some constraints on the SMEFT Wilson coefficients from the above building blocks.

\subsubsection{Bounds from non-interacting reference theory}
We consider the NIRT as the reference theory, and evaluate the relative entropy.
We assume the dimension-eight operators listed in the previous sections are generated through the interaction between heavy and light fields, i.e., $I_{\rm I}[A^a_{\mu},\Phi]=\int d^4 x \mathcal{O}[\Phi]\otimes J[A^a_{\mu}]$.
Then, the first order corrections for $g$ to the Euclidean effective action are expressed as
\begin{align}
    g\cdot{\langle I_{\rm I}\rangle}_{g=0}&=g\cdot\left(\frac{\partial W_g}{\partial g}\right)_{g=0}\notag
    \\
    &=\int (d^4x)_{\rm E} \left(\frac{\delta W_g}{\delta J}\right)_{J=0}J[A^a_{\mu}]
    ,\label{eq:IISMEFTM}
\end{align}
where $(\delta W_g/\delta J)_{J=0}$ is a tadpole-like diagram for the composite field $J[A^a_{\mu}]$, e.g., Fig.~\ref{fig:EHscl}, and does not depend on space-time.
Because $J[A^a_{\mu}]$ does not include the higher-dimensional operators according to Assumption (IV), there are two cases: (i) $J[A^a_{\mu}]$ preserves the gauge symmetry or (ii) not.
For case (i), Eq.~\eqref{eq:IISMEFTM} is proportional to $\int (d^4x)_{\rm E} F^a_{\mu\nu}F^{a,\mu\nu}$ because ${\langle I_{\rm I}\rangle}_{g=0}$ preserves the Lorentz symmetry from the definition, and $(\delta W_g/\delta J)_{J=0}$ does not depend on space-time.
In general, the $CP$ violating term arises, but we omit such terms because they can be removed by  some mechanism, such as axion-like degrees of freedom in the UV theory.
For case (ii), $J[A^a_{\mu}]$ can be proportional to $A^a_{\mu}$, and $A^a_{\mu}A^{a}_{\nu}$ because of the covariant derivative of the kinetic term.
According to Assumption (IV), we focus on the leading order of the interacting term, which are corrections from the kinetic terms of heavy fields.   
However, $J[A^a_{\mu}]\propto A^a_{\mu}$ effects on Eq.~\eqref{eq:IISMEFTM} vanish because ${\langle I_{\rm I}\rangle}_{g=0}$ keeps the Lorentz symmetry.
Although terms proportional to $\int (d^4 x)_{\rm E} A^a_{\mu}A^{a,\mu}$ in Eq.~\eqref{eq:IISMEFTM} may remain, they can be eliminated by implementing the gauge fixing condition, the so-called non-linear gauge~\cite{Nambu:1968qk}.
Therefore, we focus on the case of ${\langle I_{\rm I}\rangle}_{g=0}\propto \int (d^4x)_{\rm E} F^a_{\mu\nu}F^{a,\mu\nu}$ below.
For each tree and loop-level UV completions, the constraints on the SMEFT Wilson coefficients from the relative entropy are evaluated as follows: 

\begin{itemize}
    \item Tree-level UV completion ---
    Consider the SMEFT operators generated by the tree-level UV completions.
    The partition function of the theory $I_0+g\cdot I_{\rm I}$ is generally calculated as follows:
    \begin{align}
        Z_g[\overline{A}]&\equiv \int d[A] d[\Phi] e^{-I_g[A,\Phi]}\notag
        \\
        &=\int d[A] {\rm exp}\left[
        -\int (d^4x)_{\rm E} \left(\frac{1}{2}(1+\alpha^{\rm tree}_2)F^a_{\mu\nu}F^{a,\mu\nu}-\sum_i\beta_{i,2}^{\rm tree}\mathcal{O}_i[A] \right)
        \right]\notag
        \\
        &= {\rm exp}\left[
        -\int (d^4x)_{\rm E} \left(\frac{1}{2}(1+\alpha^{\rm tree}_2)\overline{F'}^a_{\mu\nu}\overline{F'}^{a,\mu\nu}-\sum_i\beta_{i,2}^{\rm tree}\mathcal{O}_i[\overline{A'}] \right)
        \right]\notag
        \\
        &= {\rm exp}\left[
        -\int (d^4x)_{\rm E} \left(\frac{1}{2}\overline{F}^a_{\mu\nu}\overline{F}^{a,\mu\nu}-\sum_i\beta_{i,2}^{\rm tree}\cdot (1+\alpha^{\rm tree}_2)^{-2}\mathcal{O}_i[\overline{A}] \right)
        \right],\label{eq:ZgSMEFTtree}
    \end{align}
    where $\mathcal{O}_i[A]$ is the dimension-eight SMEFT operators,  $\alpha_2^{\rm tree}$ and $\beta_{i,2}^{\rm tree}$ denote the second or higher order corrections for $g$, and $\beta_{i,2}^{\rm tree}$ does not include the first order correction for $g$ because of Assumption (IV).
    The corrections $\alpha^{\rm tree}_{2}$ and $\beta^{\rm tree}_{i,2}$ are generated at the tree level. 
    Following the procedure in Eq.~\eqref{eq:linrem1}, \eqref{eq:linrem2}, and \eqref{eq:linrem3}, the first order correction for $g$ is eliminated in $\alpha_2^{\rm tree}$. 
    The background field $\overline{A'}_{\mu}^a$ denotes the classical solution of the effective action. 
    We choose the background fields as follows:
    \begin{align}
        \overline{A'}^a_{\mu}=(1+\alpha_2^{\rm tree})^{-1/2} \cdot \overline{A}^a_{\mu},\label{eq:Abartree}
    \end{align}
    with $ \overline{F}^a_{\mu\nu}={\rm const.}$
    In general, the dimension-six operators arise in Eq.~\eqref{eq:ZgSMEFTtree} but can be eliminated by choosing the suitable background fields of Eq.~\eqref{eq:sol1}.
    From Eq.~\eqref{eq:ZgSMEFTtree}, the Euclidean effective actions are calculated as follows: 
    \begin{align}
        &W_g[\overline{A}]\equiv-\ln Z_g[\overline{A}]=\int (d^4x)_{\rm E} \left(\frac{1}{2}\overline{F}^a_{\mu\nu}\overline{F}^{a,\mu\nu}-\sum_i\beta_{i,2}^{\rm tree}\cdot (1+a^{\rm tree}_2)^{-2}\mathcal{O}_i[\overline{A}] \right),\label{eq:Wgsmefttrree}
        \\
        &W_0[\overline{A}]=\lim_{g\to 0}W_g[\overline{A}]=\int (d^4x)_{\rm E} \left(\frac{1}{2}\overline{F}^a_{\mu\nu}\overline{F}^{a,\mu\nu} \right),\label{eq:W0smefttrree}
    \end{align}
    From Eqs.~\eqref{eq:Wgsmefttrree} and \eqref{eq:W0smefttrree}, the shift of the Euclidean effective action is calculated as follows:
    \begin{align}
     W_g[\overline{A}]-W_0[\overline{A}]=-\sum_i\beta_{i,2}^{\rm tree}\cdot (1+a^{\rm tree}_2)^{-2}\int (d^4x)_{\rm E} \mathcal{O}_i[\overline{A}]. \label{eq:DeltreeSMEFT}
    \end{align}
    From Eq.~\eqref{eq:Wgsmefttrree}, the first order corrections for $g$ is also calculated as
    \begin{align}
        \left(\frac{dW_g}{dg}\right)_{g=0}&=\left(\frac{\partial W_g}{\partial g}\right)_{g=0}+
        \int (d^4x)_{\rm E} \left(\frac{\delta W_g}{\delta \overline{A'}}\right)\cdot \left(\frac{d \overline{A'}}{dg}\right)_{g=0}\notag
        \\
        &=\left(\frac{\partial W_g}{\partial g}\right)_{g=0}\notag
        \\
        &={\langle I_{\rm I}\rangle}_{g=0}=0,\label{eq:sigdWgdgSMEFT}
    \end{align}
    where $({d \overline{A'}}/{dg})_{g=0}=0$ holds because of Eq.~\eqref{eq:Abartree}.
     From Eqs.~\eqref{eq:upg}, \eqref{eq:DeltreeSMEFT}, and \eqref{eq:sigdWgdgSMEFT}, we obtain
    \begin{align}
       S(P_0||P_g)&=W_0[\overline{A}]-W_g[\overline{A}]+ g{\langle I_{\rm I}\rangle}_{g=0}\notag
       \\
       &=W_0[\overline{A}]-W_g[\overline{A}]\notag
       \\
       &=\sum_i\beta_{i,2}^{\rm tree}\cdot (1+a^{\rm tree}_2)^{-2}\int (d^4x)_{\rm E} \mathcal{O}_i[\overline{A}]\geq 0.\label{eq:boundSMEFT1tree}
    \end{align}
    Note here that the relative entropy does not change even if we add the same term to both $W_0$ and $W_g$ simultaneously\footnote{We can also put the same boundary terms to both $W_0$ and $W_g$.}.
    By taking $g=1$, Eq.~\eqref{eq:boundSMEFT1tree} represents the relative entropy between the reference and target theories and yields the following inequality.
    \begin{align}
        S(P_{\rm NI}||P_{\rm T})=\sum_i\left(\beta_{i,2}^{\rm tree}\cdot (1+a^{\rm tree}_2)^{-2}\right)_{g=1}\int (d^4x)_{\rm E} \mathcal{O}_i[\overline{A}]\geq 0,\label{eq:SMEFTTree1}
    \end{align}
    where $P_{\rm NI}=P_0$ and $P_{\rm T}=P_{g=1}$ are used.
    The right-hand side of this inequality denotes the linear combination of the coefficients of the dimension-eight operators of the target theory.

    \item Loop-level UV completion ---
    Consider the SMEFT operators generated by the loop-level UV completion.
    The partition function of the theory $I_0+g\cdot I_{\rm I}$ is generally calculated as follows\footnote{Strictly speaking,  Eq.~\eqref{eq:appZgloopSMEFT} holds in the limit of $g=1$ or $g=0$.
    This is because the gauge symmetry may be broken by the auxiliary parameter $g$.
    However, this subtle point does not affect the following discussions because we focus only on the reference $(g=0)$ and target $(g=1)$ theories, where the gauge symmetry is restored.
    }:
        \begin{align}
        Z_g[\overline{A}]&\equiv \int d[A] d[\Phi] e^{-I_g[A,\Phi]}\notag
        \\
        &=\int d[A] {\rm exp}\left[
        -\int (d^4x)_{\rm E} \left(
        \frac{1}{2} \left(1+\alpha^{\rm loop}_1+\alpha^{\rm loop}_2\right)F^a_{\mu\nu}F^{a,\mu\nu}
        -\sum_i \beta^{\rm loop}_{2,i}\mathcal{O}_i[{A}]+E^{\Phi}_{\rm vac}\right)
        \right]\notag
        \\
        &={\rm exp}\bigg[
        -\int (d^4x)_{\rm E} \bigg(
        \frac{1}{2} \left(1+\alpha^{\rm loop}_1+\alpha^{\rm loop}_2\right)\overline{F'}^a_{\mu\nu}\overline{F'}^{a,\mu\nu}-\sum_i\beta^{\rm loop}_{2,i}\mathcal{O}_i[\overline{A}']+E_{\rm vac}\bigg)
        \bigg]\notag
        \\
        &={\rm exp}\bigg[
        -\int (d^4x)_{\rm E} \bigg(
        \frac{1}{2} \left(1+\alpha^{\rm loop}_1\right)\overline{F}^a_{\mu\nu}\overline{F}^{a,\mu\nu}-\sum_i\beta^{\rm loop}_{2,i} \mathcal{O}_i[\overline{A}]+E_{\rm vac}\bigg)
        \bigg].\label{eq:appZgloopSMEFT}
    \end{align}
    where $\mathcal{O}_i[A]$ is the dimension-eight SMEFT operators,  $\alpha_1^{\rm loop}$ is the first order correction for $g$, $\alpha_2^{\rm loop}$ and $\beta_{2,i}^{\rm loop}$ are the second or higher order correction for $g$,  
    $E_{\rm vac}^{\Phi}$ is the vacuum energy coming from the one-loop level correction of $\Phi$, and $E_{\rm vac}$ is the vacuum energy of $\Phi$ and $A^a_{\mu}$.
    $\alpha_1^{\rm loop}$, $\alpha_2^{\rm loop}$, and $\beta_{2,i}^{\rm loop}$ are generated from the one-loop corrections of $\Phi$.
    In Eq.~\eqref{eq:appZgloopSMEFT}, we neglected two-loop corrections.
    Note here that $\alpha_1^{\rm loop}$ cannot be remove by redefining $\Phi$ in contrast to the tree-level UV completion.
    The background field $\overline{A'}^a_{\mu}$ denotes Eq.~\eqref{eq:sol1}, which is the solution of the effective action.
    We choose the background field as follows:
    \begin{align}
        \overline{A'}^a_{\mu}=\left(1-\frac{1}{2}\alpha_2^{\rm loop} \right)\overline{A}^a_{\mu},\label{eq:Abarloop}
    \end{align}
    where $\overline{F}_{\mu\nu}^a={\rm const.}$ to remove the dimension-six operators; see Eq.~\eqref{eq:sol1}.
    From Eq.~\eqref{eq:appZgloopSMEFT}, the Euclidean effective actions are obtained as follows:
    \begin{align}
        W_g[\overline{A}]&\equiv -\ln Z_g[\overline{A}] =\int (d^4x)_{\rm E} \bigg(
        \frac{1}{2} \left(1+\alpha^{\rm loop}_1\right)\overline{F}^a_{\mu\nu}\overline{F}^{a,\mu\nu}-\sum_i\beta^{\rm loop}_{2,i} \mathcal{O}_i[\overline{A}]+E_{\rm vac}\bigg),\label{eq:appSMEFTWGloop}
        \\
        W_0[\overline{A}]&=\lim_{g\to 0}W_g[\overline{A}]=\int (d^4x)_{\rm E} \bigg(
        \frac{1}{2} \overline{F}^a_{\mu\nu}\overline{F}^{a,\mu\nu}+E_{\rm vac}\bigg).
    \end{align}
    The shift of the Euclidean effective action is calculated as follows:
    \begin{align}
         W_g[\overline{A}]-W_0[\overline{A}]=\int (d^4x)_{\rm E} \bigg(
        \frac{1}{2} \alpha^{\rm loop}_1\overline{F}^a_{\mu\nu}\overline{F}^{a,\mu\nu}-\sum_i\beta^{\rm loop}_{2,i} \mathcal{O}_i[\overline{A}]\bigg).\label{eq:loopDelWgSM}
    \end{align}
    Also, the first order corrections for $g$ is calculated as
    \begin{align}
        \left(\frac{dW_g}{dg}\right)_{g=0}&=\left(\frac{\partial W_g}{\partial g}\right)_{g=0}+\int (d^4x)_{\rm E} \left(\frac{\delta W_g}{\delta \overline{{A}'}}\right)\cdot \left(\frac{d \overline{{A}'}}{d g}\right)_{g=0}\notag
        \\
        &=\left(\frac{\partial W_g}{\partial g}\right)_{g=0}\notag
        \\
        &={\langle I_{\rm I}\rangle}_{g=0}\notag
        \\
        &=\frac{1}{2} \frac{d\alpha^{\rm loop}_1}{dg}\int (d^4x)_{\rm E} \overline{F}^a_{\mu\nu}\overline{F}^{a,\mu\nu},\label{eq:loopdelWdgSM}
    \end{align}
    where $({d \overline{{A}'}}/{d g})_{g=0}=0$ holds from Eq.~\eqref{eq:Abarloop}.
    This relation represents Eq.~\eqref{eq:IISMEFTM}.
    From Eqs.~\eqref{eq:upg}, \eqref{eq:loopDelWgSM}, and \eqref{eq:loopdelWdgSM}, we obtain
    \begin{align}
        S(P_0||P_g)&=W_0[\overline{A}]-W_g[\overline{A}]+ g{\langle I_{\rm I}\rangle}_{g=0}\notag
        \\
        &=W_0[\overline{A}]-W_g^{\rm non\text{-}lin}[\overline{A}]\notag
        \\
        &=\sum_i\beta^{\rm loop}_{2,i}\int (d^4x)_{\rm E} 
        \mathcal{O}_i[\overline{A}]\geq 0.\label{eq:SMEFTloopcons}
    \end{align}
    where $g\cdot (d\alpha_1^{\rm loop}/dg)=\alpha_1^{\rm loop}$ was used, and we defined as follows:
    \begin{align}
        W_g^{\rm non\text{-}lin}[\overline{A}]&\equiv W_g[\overline{A}]- g{\langle I_{\rm I}\rangle}_{g=0}\notag
        \\
        &=\int (d^4x)_{\rm E} \bigg(
        \frac{1}{2} \overline{F}^a_{\mu\nu}\overline{F}^{a,\mu\nu}-\sum_i\beta^{\rm loop}_{2,i} \mathcal{O}_i[\overline{A}]+E_{\rm vac}\bigg).
    \end{align}
    Similar to the tree-level UV completion, we can also add the same boundary terms to both $W_0$ and $W_g^{\rm non\text{-}lin}$ because it cancels in the relative entropy.
    By taking $g=1$, Eq.~\eqref{eq:SMEFTloopcons} represents the relative entropy between the reference and target theories and yields the following inequality.
    \begin{align}
        S(P_{\rm NI}||P_{\rm T})=\sum_i\beta^{\rm loop}_{2,i}|_{g=1}\int (d^4x)_{\rm E} 
        \mathcal{O}_i[\overline{A}]\geq 0,\label{eq:SMEFTLOOP}
    \end{align}
    where $P_{\rm NI}=P_0$ and $P_{\rm T}=P_{g=1}$ are used.
    This inequality yields the constraint on the dimension-eight operators generated at the one-loop level.

\end{itemize}

Consequently, for both tree and loop-level UV completions, it is found that the relative entropy denotes the linear combination of the dimension-eight operators generated from the interacting terms, i.e., Eqs.~\eqref{eq:SMEFTTree1} and \eqref{eq:SMEFTLOOP}.
Therefore, after Wick rotation, the inequalities~\eqref{eq:SMEFTTree1} and \eqref{eq:SMEFTLOOP} give rise to constraints on the dimension-eight operators of Eq.~\eqref{eq:SMEFTLa} as follows:
\begin{align}
    \frac{1}{M^4} \sum_i  \int (d^4 x)_{\rm E}c_i \mathcal{O}_i[\overline{A}]\geq 0.\label{eq:SMine}
\end{align}
Equation~\eqref{eq:SMine} and \eqref{eq:Wsemeft} yield bounds as follows:
\begin{align}
    &a_F \cdot \mathrm{A}_F^2 +b_F\cdot \mathrm{B}_F^2+c_F\cdot \Gamma_F^2 +d_F\cdot \Delta_F^2 \notag
    \\
    &+ e_F\cdot  \mathrm{A}_F\mathrm{B}_F+f_F\cdot\mathrm{A}_F\Gamma_F+g_F\cdot\mathrm{B}_F\Gamma_F+h_F\cdot \mathrm{A}_F\Delta_F+i_F\cdot\mathrm{B}_F\Delta_F+ j_F\cdot\Gamma_F\Delta_F\geq 0.\label{eq:SMBG}
\end{align}
The quantities $\mathrm{A}_F$, $\mathrm{B}_F$, $\Gamma_F$, and $\Delta_F$ are independent each other, and the inequality of \eqref{eq:SMBG} yields the following inequalities.
\begin{align}
    &a_F\geq 0,~~~~~~~~~~b_F\geq 0,~~~~~~~~~~c_F\geq 0,~~~~~~~~~~d_F\geq 0,\label{eq:puB1}
    \\
    &4a_F\cdot d_F-h_F^2\geq 0,~~~~~~~~4b_F\cdot d_F-i_F^2\geq 0,~~~~~~~4c_F\cdot d_F-j_F^2\geq 0,\label{eq:puB2}
    \\
    &a_F-\frac{h_F^2}{4d_F}-\frac{\left(f_F-\frac{h_F\cdot j_F}{2d_F}\right)^2}{4\left(c_F-\frac{j_F^2}{4d_F}\right)}\geq 0,~~~~~~~~~~b_F-\frac{i_F^2}{4d_F}-\frac{\left(g_F-\frac{i_F\cdot j_F}{2d_F}\right)^2}{4\left(c_F-\frac{j_F^2}{4d_F}\right)}\geq 0,\label{eq:puB3}
    \\
    &4\left[a_F-\frac{h_F^2}{4d_F}-\frac{\left(f_F-\frac{h_F\cdot j_F}{2d_F}\right)^2}{4\left(c_F-\frac{j_F^2}{4d_F}\right)}\right]\cdot
    \left[b_F-\frac{i_F^2}{4d_F}-\frac{\left(g_F-\frac{i_F\cdot j_F}{2d_F}\right)^2}{4\left(c_F-\frac{j_F^2}{4d_F}\right)}\right]\notag
    \\
    &\quad \quad\quad\geq 
    \left[e_F-\frac{h_F\cdot i_F}{2d_F}-\frac{\left(f_F-\frac{h_F\cdot j_F}{2d_F}\right)\left(g_F-\frac{i_F\cdot j_F}{2d_F}\right)}{2\left(c_F-\frac{j_F^2}{4d_F}\right)}\right]^2.\label{eq:puB4}
\end{align}
For $U(1)_Y$, $SU(2)_L$, and $SU(3)_C$ gauge fields, the above inequalities are listed as follows:

\begin{itemize}
    \item $U(1)_Y$ ---
    Substituting Eqs.~\eqref{eq:Bab}-\eqref{eq:Bhi} into the inequalities of \eqref{eq:puB1}-\eqref{eq:puB4}, for the $U(1)_Y$ gauge field $B_{\mu}$, we obtain the following constraints.
    \begin{align}
        c_1^{B^4}\geq 0,~~~c_2^{B^4}\geq 0,~~~4c_1^{B^4} c_2^{B^4}\geq (\tilde{c}_1^{B^4})^2.
    \end{align}
    Note here that the bounds of Eqs.~\eqref{eq:puB3} and \eqref{eq:puB4} vanish by substituting Eqs.~\eqref{eq:Bab}-\eqref{eq:Bhi}.

     \item $SU(2)_L$ ---
     From Eqs.~\eqref{eq:Wab}-\eqref{eq:Whj}, and the inequalities of \eqref{eq:puB1}-\eqref{eq:puB4}, for the $SU(2)_L$ gauge field $W^I_{\mu}$, we obtain the following constraints. 
     \begin{align}
         c_1^{W^4}+c_3^{W^4}\geq 0,~~c_2^{W^4}+c_4^{W^4}\geq 0,~~4(c_1^{W^4}+c_3^{W^4})(c_2^{W^4}+c_4^{W^4})\geq (\tilde{c}_1^{W^4}+\tilde{c}_2^{W^4})^2.
     \end{align}
     Similar to the case of the $U(1)_Y$ gauge field, the bounds of Eqs.~\eqref{eq:puB3} and \eqref{eq:puB4} vanish by substituting Eqs.~\eqref{eq:Wab}, \eqref{eq:Wd}, and \eqref{eq:Whj}.
     
     \item $SU(3)_C$ ---
     For simplicity, we assume $u_1^2=u_2^2=1$ and $u_1\cdot u_2=\cos\xi$.
     By considering the two cases of $\cos^2\xi=0$ and $1$, from Eqs.~\eqref{eq:Ga}-\eqref{eq:Gj}, and the inequalities of \eqref{eq:puB1}-\eqref{eq:puB4}, we obtain 
    \begin{align}
        &3c_1^{G^4}+3c_3^{G^4}+c_5^{G^4}\geq 0,\label{eq:boun1}
        \\
        &3c_3^{G^4}+2c_5^{G^4}\geq 0,\label{eq:boun2}
        \\
        &3 c_2^{G^4}+3 c_4^{G^4}+c_6^{G^4}\geq 0,
        \\
        &3c_4^{G^4}+2 c_6^{G^4}\geq 0,  
        \\
        &4 \left(3 c_1^{G^4}+3 c_3^{G^4}+ c_5^{G^4}\right)\left(3 c_2^{G^4}+3 c_4^{G^4}+c_6^{G^4}\right)\geq \left(3 \tilde{c}_1^{G^4}+3 \tilde{c}_2^{G^4}+\tilde{c}_3^{G^4}\right)^2,
        \\
        &4\left(3 c_3^{G^4}+2 c_5^{G^4}\right)\left(3 c_4^{G^4}+2 c_6^{G^4}\right)\geq \left(3 \tilde{c}^{G^4}_2+2 \tilde{c}^{G^4}_3\right)^2,
        \\
        &2 c_1^{G^4}+c_3^{G^4}\geq 0.\label{eq:bounadd}
    \end{align}
    The above first six bounds are the same as the positivity bounds from unitarity and causality considerations in Ref.~\cite{Remmen:2019cyz}, and the last inequality newly arises from the entropy constraint.
    Since, however, Eq.~\eqref{eq:boun1} is derived from Eqs.~\eqref{eq:boun2} and \eqref{eq:bounadd}, the results of the entropy constraints are consistent with the positivity bounds from unitarity and causality considerations in Ref.~\cite{Remmen:2019cyz}. 
     
\end{itemize}

\subsection{Einstein-Maxwell theory with higher-derivative operators}

Consider the Einstein-Maxwell theory with higher-derivative operators in the Minkowski space  as follows:
\begin{align}
    W_{\rm EM}&=\int d^4 x\sqrt{-g} \bigg(\frac{M_{\rm Pl}^2}{2} R -\frac{1}{4}F_{\mu\nu}F^{\mu\nu}+\frac{\alpha_1}{4 M_{\rm Pl}^4} (F_{\mu\nu}F^{\mu\nu})^2+\frac{\alpha_2}{4 M_{\rm Pl}^4} (F_{\mu\nu} \widetilde{F}^{\mu\nu})^2+\frac{\alpha_3}{2 M_{\rm Pl}^2}F_{\mu\nu}F_{\rho\sigma} R^{\mu\nu\rho\sigma} \bigg),\label{eq:EMeff}
\end{align}
where other operators up to four-derivative are eliminated by the field redefinition of $g_{\mu\nu}$; see Ref.~\cite{Cheung:2018cwt} and Appendix~\ref{app:EMred}. 
Also, the Gauss-Bonnet combination, i.e., $R_{\mu\nu\rho\sigma}R^{\mu\nu\rho\sigma}-4 R_{\mu\nu}R^{\mu\nu}+R^2$, is a topological term that does not contribute to the extremal black hole entropy in four dimensions, so we omit it throughout this section.
Similar to the previous subsections, we focus on the field theoretical description, which breaks down at some high energy scale $\Lambda_{\rm QFT}$.
Generically, $\Lambda_{\rm QFT}$ is smaller than the Planck scale $\Lambda_{\rm QFT}\ll M_{\rm Pl}$.
In this work, we do not consider the dynamics of stringy particles in the high energy regime beyond $\Lambda_{\rm QFT}$ because the ordinary field theoretical descriptions break down by infinitely many local fields.\footnote{The graviton accompanied by Regge states can break the positivity bounds~\cite{Hamada:2018dde} on the Wilson coefficients of Eq.~\eqref{eq:EMeff}, but such a scenario would be beyond the applicability of procedures of this section based on the field theory.}
According to Assumption (III), we consider the higher-derivative operators generated from the target UV theory defined by $I_{\rm T}[g_{\mu\nu};R_{\mu\nu\rho\sigma},A,\Phi]$, where $g_{\mu\nu}$ is the metric of space-time, $R_{\mu\nu\rho\sigma}$ is the Riemann tensor, $A_{\mu}$ is the $U(1)$ gauge boson, and $\Phi$ is the heavy degrees of freedom.
For this EFT, define the non-interacting and interacting terms as follows:
\begin{align}
    &I_0[g_{\mu\nu};R_{\mu\nu\rho\sigma},A,\Phi]\equiv I_{\rm T}[g_{\mu\nu};R_{\mu\nu\rho\sigma},A,0]+I_{\rm T}[g_{\mu\nu};0,0,\Phi],
    \\
    &I_{\rm I}[g_{\mu\nu};R_{\mu\nu\rho\sigma},A,\Phi]\equiv I_{\rm T}[g_{\mu\nu};R_{\mu\nu\rho\sigma},A,\Phi]-I_0[g_{\mu\nu};R_{\mu\nu\rho\sigma},A,\Phi],
\end{align}
where the cosmological constant is omitted because it cancels in the relative entropy.
These definitions are also adopted in Sec.~\ref{sec:Gratree}. 
It should be noted that the action $I_0$ does not include the interaction between $\Phi$ and $A_{\mu}, R_{\mu\nu\rho\sigma}$, but the interaction between $g_{\mu\nu}$ and $\Phi$.
Although gravitational operators such as $R_{\mu\nu}^2$ are generated from $I_0$, such operators up to four-derivative can be eliminated by the field redefinition of $g_{\mu\nu}$; see Appendix~\ref{app:EMred}.

We assume the dimension-eight operators of Eq.~\eqref{eq:EMeff} are generated through the interaction defined in Eq.~\eqref{eq:IIGEN}.
Then, the first order corrections for $g$ to the Euclidean effective action are expressed as follows:
\begin{align}
    g\cdot {\langle I_{\rm I}\rangle}_{g=0}&=g\cdot \left(\frac{\partial W_g }{\partial g}\right)_{g=0}\notag
    \\
    &=\int (d^4x)_{\rm E}\sqrt{g} \left(\frac{\delta W_g}{\delta J}\right)_{J=0} J[g_{\mu\nu};R_{\mu\nu\rho\sigma},A_{\mu}],\label{eq:IIEMaxwell}
\end{align}
where $({\delta W_g}/{\delta J})_{J=0}$ is a tadpole-like diagram for the composite field $J$. 
Similar to the SMEFT, $J[g_{\mu\nu};R_{\mu\nu\rho\sigma},A_{\mu}]$ does not include the higher-derivative operators according to Assumption (IV), so there are two cases: (i) $J[g_{\mu\nu};R_{\mu\nu\rho\sigma},A_{\mu}]$ preserves the gauge symmetry or (ii) not.
For case (i), Eq.~\eqref{eq:IIEMaxwell} is proportional to $\int (d^4x)_{\rm E}\sqrt{g} F_{\mu\nu}F^{\mu\nu}$ or $\int (d^4x)_{\rm E}\sqrt{g} R$ because ${\langle I_{\rm I}\rangle}$ is invariant under 
general coordinate transformations. 
We assume the interaction $I_{\rm I}$ does not involve the $CP$ violating terms.
For case (ii), $J[g_{\mu\nu};R_{\mu\nu\rho\sigma},A_{\mu}]\propto A_{\mu}$, or $A_{\mu}A_{\nu}$ because of the covariant derivative of the kinetic term.
According to Assumption (IV), we focus on the leading order of the interacting term, which arises from the kinetic terms of the heavy charged fields.
Then, $J\propto A_{\mu}$ effects on Eq.~\eqref{eq:IIEMaxwell} vanish from the invariance of ${\langle I_{\rm I}\rangle}_{g=0}$ under the 
general coordinate transformations.
Also, a term proportional to $\int (d^4x)_{\rm E} A_{\mu}A^{\mu}$ is generated in ${\langle I_{\rm I}\rangle}_{g=0}$ by $J\propto A_{\mu}A_{\nu}$ effects on Eq.~\eqref{eq:IIEMaxwell} because of the invariance of the  general coordinate transformations but can be eliminated by implementing the non-linear gauge fixing condition $A_{\mu}A^{\mu}=0$. 
Then, the first order corrections of the interaction to the effective action are expressed as follows:
\begin{align}
g\cdot{\langle I_{\rm I}\rangle}_{g=0}&=g\cdot\left(\frac{\partial W_g}{\partial g}\right)_{g=0}\notag
\\
&=\int (d^4x)_{\rm E}\sqrt{g}\left(\frac{\delta W_g}{\delta J}\right)_{J=0} J[g_{\mu\nu};R_{\mu\nu\rho\sigma},A_{\mu}]\notag
\\
&\propto \int (d^4 x)_{\rm E} \sqrt{g} F_{\mu\nu}F^{\mu\nu}~{\rm or}~\int (d^4 x)_{\rm E} \sqrt{g} R.\label{eq:IIEMM}
\end{align}
For each tree and loop-level UV completions, the constraints on the EFT from the relative entropy are evaluated as follows:

\begin{itemize}
    \item Tree-level UV completion ---
    Consider the EFT generated at the tree-level UV completion.
    The partition function of the theory $I_0+g\cdot I_{\rm I}$ is generally calculated as follows:
    \begin{align}
      Z_g[\overline{g}_{\mu\nu},\overline{A}]&\equiv\int d [g]d[A]d[\Phi]e^{-I_g[g_{\mu\nu};R_{\mu\nu\rho\sigma},A,\Phi]}\notag
        \\
        &=\int d [g]d[A] {\rm exp}\bigg[
        -\int (d^4x)_{\rm E} \sqrt{g}\bigg(-\frac{M^2_{\rm Pl}}{2}(1+\alpha_{2,R}^{\rm tree})R +\frac{1}{4}(1+\alpha_{2,F}^{\rm tree})F_{\mu\nu}F^{\mu\nu}\notag
        \\
      &-\beta_{2,1}^{\rm tree}(F_{\mu\nu}F^{\mu\nu})^2-\beta_{2,2}^{\rm tree}(F_{\mu\nu}\widetilde{F}^{\mu\nu})^2-\beta_{2,3}^{\rm tree} F_{\mu\nu}F_{\rho\sigma}R^{\mu\nu\rho\sigma}\bigg)
        \bigg]\notag
        \\
        &={\rm exp}\bigg[
        -\int (d^4x)_{\rm E} \sqrt{\overline{g'}}\bigg(-\frac{M^2_{\rm Pl}}{2}(1+\alpha_{2,R}^{\rm tree})\overline{R'} +\frac{1}{4}(1+\alpha_{2,F}^{\rm tree})\overline{F'}_{\mu\nu}\overline{F'}^{\mu\nu}\notag
        \\
        &-\beta_{2,1}^{\rm tree}(\overline{F'}_{\mu\nu}\overline{F'}^{\mu\nu})^2-\beta_{2,2}^{\rm tree}(\overline{F'}_{\mu\nu}\widetilde{\overline{F'}}^{\mu\nu})^2-\beta_{2,3}^{\rm tree} \overline{F'}_{\mu\nu}\overline{F'}_{\rho\sigma}\overline{R'}^{\mu\nu\rho\sigma}\bigg)
        \bigg]\notag
        \\
        &={\rm exp}\bigg[
        -\int (d^4x)_{\rm E} \sqrt{\overline{g}}\bigg(-\frac{M^2_{\rm Pl}}{2}\overline{R} +\frac{1}{4}\overline{F}_{\mu\nu}\overline{F}^{\mu\nu}\notag
        \\
        &-\beta_{2,1}^{\rm tree}\left(1+\frac{2}{3} \alpha^{\rm tree}_{2,R}-2 \alpha^{\rm tree}_{2,F}\right)(\overline{F}_{\mu\nu}\overline{F}^{\mu\nu})^2-\beta_{2,2}^{\rm tree}\left(1+2 \alpha^{\rm tree}_{2,R}-2\alpha^{\rm tree}_{2,F}\right)(\overline{F}_{\mu\nu}\widetilde{\overline{F}}^{\mu\nu})^2\notag
        \\
        &-\beta_{2,3}^{\rm tree}\left(1+\frac{1}{3}\alpha^{\rm tree}_{2,R}-\alpha^{\rm tree}_{2,F}\right) \overline{F}_{\mu\nu}\overline{F}_{\rho\sigma}\overline{R}^{\mu\nu\rho\sigma}\bigg)
    \bigg],\label{eq:ZgEMtree}
    \end{align}
    where $\alpha_{2,R}^{\rm tree}$, $\alpha_{2,F}^{\rm tree}$, $\beta_{2,1}^{\rm tree}$, $\beta_{2,2}^{\rm tree}$ and $\beta_{2,3}^{\rm tree}$\footnote{From causality consideration, in Ref.~\cite{Li:2017lmh,Afkhami-Jeddi:2018own}, it is argued that the tree level contribution to $\beta_{2,3}^{\rm tree}$ requires stringy particles in UV theories. If we focus on the field theoretical descriptions, there is no contribution to $\beta_{2,3}^{\rm tree}$ at the tree level.} denote the second or higher order corrections for $g$, and $\beta_{2,1}^{\rm tree}$, $\beta_{2,2}^{\rm tree}$ and $\beta_{2,3}^{\rm tree}$ do not include the first order correction for $g$ because of Eq.~\eqref{eq:IIEMM}.
    The corrections are assumed to be generated at the tree level. 
    According to the procedure in Eq.~\eqref{eq:linrem1}, \eqref{eq:linrem2}, and \eqref{eq:linrem3}, the first order correction for $g$ is eliminated in $\alpha_{2,R}^{\rm tree}$ and $\alpha_{2,F}^{\rm tree}$. 
    Since the gravitational operators only involving the Riemann tensors can be removed by the redefinition of $g_{\mu\nu}$; see Appendix~\ref{app:EMred}, and the Riemann-squared operator effects on the extremal black hole entropy can be dropped in four dimensions, we omit such irrelevant terms. %
    The background fields $\overline{A'}_{\mu}$ and $\overline{g'}_{\mu\nu}$ denote the classical solutions of the effective action. 
    We choose the background field as follows:
    \begin{align}
      &\overline{A'}_{\mu}=\left(1+\frac{1}{2}\left(\frac{4}{3}\alpha_{2,R}^{\rm tree}-\alpha_{2,F}^{\rm tree}\right)\right)\overline{A}_{\mu},\label{eq:Atree}
        \\
    &\overline{g'}_{\mu\nu}=\left(1-\frac{1}{3}\alpha_{2,R}^{\rm tree}\right)\overline{g}_{\mu\nu},~~~\overline{g'}^{\mu\nu}=\left(1+\frac{1}{3}\alpha_{2,R}^{\rm tree}\right)\overline{g}^{\mu\nu}.\label{eq:gtree}
    \end{align}
    From Eq.~\eqref{eq:ZgEMtree}, the effective actions are obtained as follows:
    \begin{align}
    W_g[\overline{g}_{\mu\nu},\overline{A}]&\equiv-\ln Z_g[\overline{g}_{\mu\nu},\overline{A}]\notag
    \\
    &=\int (d^4x)_{\rm E} \sqrt{\overline{g}}\bigg(-\frac{M^2_{\rm Pl}}{2}\overline{R} +\frac{1}{4}\overline{F}_{\mu\nu}\overline{F}^{\mu\nu}\notag
        \\
    &-\beta_{2,1}^{\rm tree}\left(1+\frac{2}{3}\alpha^{\rm tree}_{2,R}-2 \alpha^{\rm tree}_{2,F}\right)(\overline{F}_{\mu\nu}\overline{F}^{\mu\nu})^2-\beta_{2,2}^{\rm tree}\left(1+2\alpha^{\rm tree}_{2,R}-2\alpha^{\rm tree}_{2,F}\right)(\overline{F}_{\mu\nu}\widetilde{\overline{F}}^{\mu\nu})^2\notag
        \\
    &-\beta_{2,3}^{\rm tree}\left(1+\frac{1}{3}\alpha^{\rm tree}_{2,R}-\alpha^{\rm tree}_{2,F}\right) \overline{F}_{\mu\nu}\overline{F}_{\rho\sigma}\overline{R}^{\mu\nu\rho\sigma}\bigg),\label{eq:WgEMth}
        \\
    W_0[\overline{g}_{\mu\nu},\overline{A}]&=\lim_{g\to 0}W_g[\overline{g}_{\mu\nu},\overline{A}]=\int (d^4x)_{\rm E} \sqrt{\overline{g}}\bigg(-\frac{M^2_{\rm Pl}}{2}\overline{R} +\frac{1}{4}\overline{F}_{\mu\nu}\overline{F}^{\mu\nu}\bigg).\label{eq:W0EMth}
    \end{align}
    Note here that solutions $\overline{A}_{\mu}$ and $\overline{g}_{\mu\nu}$ include the effects of the higher-derivative terms but the first order correction for the higher-derivative terms vanishes in $W_0$ by using the equation of motion.  
    From Eq.~\eqref{eq:WgEMth} and \eqref{eq:W0EMth}, $W_g[\overline{g}_{\mu\nu},\overline{A}]-W_0[\overline{g}_{\mu\nu},\overline{A}]$ denotes the shift of the Euclidean effective action by the higher-derivative terms.
    Also, from Eq.~\eqref{eq:WgEMth}, the first order correction for $g$ is calculated as
    \begin{align}
        \left(\frac{dW_g}{dg}\right)_{g=0}&=\left(\frac{\partial W_g}{\partial g}\right)_{g=0}
        +\int (d^4x)_{\rm E}\sqrt{g} \bigg(\left(\frac{\delta W_g}{\delta \overline{A'}}\right)\cdot \left(\frac{d \overline{A'}}{dg}\right)_{g=0}+\left(\frac{\delta W_g}{\delta \overline{g'}_{\mu\nu}}\right)\cdot \left(\frac{d \overline{g'}_{\mu\nu}}{dg}\right)_{g=0}\bigg)\notag
        \\
        &=\left(\frac{\partial W_g}{\partial g}\right)_{g=0}\notag
        \\
        &={\langle I_ {\rm I}\rangle}_{g=0}=0,\label{eq:EMdWdgtree}
    \end{align}
    where $(d\overline{A'}/dg)_{g=0}=0$ and $(d\overline{g'}_{\mu\nu}/dg)_{g=0}=0$ are used from Eq.~\eqref{eq:Atree} and \eqref{eq:gtree}.
    From Eq.~\eqref{eq:upg} and \eqref{eq:EMdWdgtree}, we obtain the relative entropy,
    \begin{align}
        S(P_0||P_g)&=W_0[\overline{g}_{\mu\nu},\overline{A}]-W_g[\overline{g}_{\mu\nu},\overline{A}]+ g{\langle I_{\rm I}\rangle}_{g=0}\notag
        \\
        &=W_0[\overline{g}_{\mu\nu},\overline{A}]-W_g[\overline{g}_{\mu\nu},\overline{A}] \geq 0.
    \end{align}
    Note here that we can put the same boundary terms to both $W_0$ and $W_g$ because of its cancellation in the relative entropy.
    By taking $g=1$, it is found that the relative entropy yields the negative shift of the effective action by the higher derivative terms generated at the tree level.

    \item Loop-level UV completion ---
    Consider the EFT generated by the loop-level UV completion.
    The partition function of the theory $I_0+g\cdot I_{\rm I}$ is generally calculated as follows\footnote{Similar to the SMEFT, strictly speaking,  Eq.~\eqref{eq:senEMFT} holds in the limit of $g=$ $1$ or $0$.
    However, this subtle point does not affect the following discussions because we focus only on the reference $(g=0)$ and target $(g=1)$ theories.
    }:
    \begin{align}
    Z_g[\overline{g}_{\mu\nu},\overline{A}]&=\int d [g]d[A]d[\Phi]e^{-I_g[g_{\mu\nu};R_{\mu\nu\rho\sigma},A,\Phi]}\notag
        \\
        &\equiv\int d [g]d[A] {\rm exp}\bigg[
        -\int (d^4x)_{\rm E} \sqrt{g}\bigg(\Lambda_{0,\Phi}^{\rm loop}-\frac{M^2_{\rm Pl}}{2}(1+\alpha_{1,R}^{\rm loop}+\alpha_{2,R}^{\rm loop})R\notag
        \\
        &+\frac{1}{4}(1+\alpha_{1,F}^{\rm loop}+\alpha_{2,F}^{\rm loop})F_{\mu\nu}F^{\mu\nu}\notag
        \\
        &-\beta_{2,1}^{\rm loop}(F_{\mu\nu}F^{\mu\nu})^2-\beta_{2,2}^{\rm loop}(F_{\mu\nu}\widetilde{F}^{\mu\nu})^2-\beta_{2,3}^{\rm loop} F_{\mu\nu}F_{\rho\sigma}R^{\mu\nu\rho\sigma}\bigg)
        \bigg]\notag
        \\
        &={\rm exp}\bigg[
        -\int (d^4x)_{\rm E} \sqrt{\overline{g'}}\bigg(\Lambda_{0,\Phi}^{\rm loop}-\frac{M^2_{\rm Pl}}{2}(1+\alpha_{1,R}^{\rm loop}+\alpha_{2,R}^{\rm loop})\overline{R'} +\frac{1}{4}(1+\alpha_{1,F}^{\rm loop}+\alpha_{2,F}^{\rm loop})\overline{F'}_{\mu\nu}\overline{F'}^{\mu\nu}\notag
        \\
        &-\beta_{2,1}^{\rm loop}(\overline{F'}_{\mu\nu}\overline{F'}^{\mu\nu})^2-\beta_{2,2}^{\rm loop}(\overline{F'}_{\mu\nu}\widetilde{\overline{F'}}^{\mu\nu})^2-\beta_{2,3}^{\rm loop} \overline{F'}_{\mu\nu}\overline{F'}_{\rho\sigma}\overline{R'}^{\mu\nu\rho\sigma}\notag
        \\
        &+({\rm correction~from}~R~{\rm and}~F_{\mu\nu}F^{\mu\nu})\bigg)
        \bigg]\notag
        \\
        &={\rm exp}\bigg[
        -\int (d^4x)_{\rm E} \sqrt{\overline{g}}\bigg(\Lambda_{0,\Phi}^{\rm loop}-\frac{M^2_{\rm Pl}}{2}\left(1+\alpha^{\rm loop}_{1,R}\right)\overline{R} +\frac{1}{4}\left(1+\alpha^{\rm loop}_{1,F}\right)\overline{F}_{\mu\nu}\overline{F}^{\mu\nu}\notag
        \\
        &-\beta_{2,1}^{\rm loop}(\overline{F}_{\mu\nu}\overline{F}^{\mu\nu})^2-\beta_{2,2}^{\rm loop}(\overline{F}_{\mu\nu}\widetilde{\overline{F}}^{\mu\nu})^2-\beta_{2,3}^{\rm loop} \overline{F}_{\mu\nu}\overline{F}_{\rho\sigma}\overline{R}^{\mu\nu\rho\sigma}\notag
        \\
        &+({\rm correction~from}~R~{\rm and}~F_{\mu\nu}F^{\mu\nu})\bigg)
        \bigg],\label{eq:senEMFT}
    \end{align}
    where $\alpha_{2,R}^{\rm loop}$, $\alpha_{2,F}^{\rm loop}$, $\beta_{2,1}^{\rm loop}$, $\beta_{2,2}^{\rm loop}$ and $\beta_{2,3}^{\rm loop}$ are the second or higher order corrections for $g$, $\alpha_{1,R}^{\rm loop}$ and $\alpha_{1,F}^{\rm loop}$ are the first order corrections for $g$,
    and $\Lambda_{0,\Phi}^{\rm loop}$ is the vacuum energy coming from $\Phi$.
    The last term of Eq.~\eqref{eq:senEMFT} denotes corrections from light fields in $M^2_{\rm Pl}R/2$ and $F_{\mu\nu}F^{\mu\nu}/4$ at the one loop level. 
    Since these corrections do not depend on $g$, they cancel in the relative entropy.
    We neglect two-loop effects in Eq.~\eqref{eq:senEMFT}.
    The background fields $\overline{A'}_{\mu}$ and $\overline{g'}_{\mu\nu}$ denote the classical solution of the effective action. 
    We choose the background field as follows:
    \begin{align}
        &\overline{A'}_{\mu}=\left(1+\frac{1}{2}\left(\frac{4}{3}\alpha_{2,R}^{\rm loop}-\alpha_{2,F}^{\rm loop}\right)\right)\overline{A}_{\mu},\label{eq:Aloop}
        \\
        &\overline{g'}_{\mu\nu}=\left(1-\frac{1}{3}\alpha_{2,R}^{\rm loop}\right)\overline{g}_{\mu\nu},~~~\overline{g'}^{\mu\nu}=\left(1+\frac{1}{3}\alpha_{2,R}^{\rm loop}\right)\overline{g}^{\mu\nu}.\label{eq:gloop}
    \end{align}
    The effective actions are obtained as follows:
    \begin{align}
        W_g[\overline{g}_{\mu\nu},\overline{A}]&=\int (d^4x)_{\rm E} \sqrt{\overline{g}}\bigg(\Lambda_{0,\Phi}^{\rm loop}-\frac{M^2_{\rm Pl}}{2}(1+\alpha^{\rm loop}_{1,R})\overline{R} +\frac{1}{4}(1+\alpha^{\rm loop}_{1,F})\overline{F}_{\mu\nu}\overline{F}^{\mu\nu}\notag
        \\
        &-\beta_{2,1}^{\rm loop}(\overline{F}_{\mu\nu}\overline{F}^{\mu\nu})^2-\beta_{2,2}^{\rm loop}(\overline{F}_{\mu\nu}\widetilde{\overline{F}}^{\mu\nu})^2-\beta_{2,3}^{\rm loop} \overline{F}_{\mu\nu}\overline{F}_{\rho\sigma}\overline{R}^{\mu\nu\rho\sigma}\notag
        \\
        &+({\rm correction~from}~R~{\rm and}~F_{\mu\nu}F^{\mu\nu})\bigg),\label{eq:WgEM}
        \\
        W_0[\overline{g}_{\mu\nu},\overline{A}]&=\lim_{g\to 0}W_g[\overline{g}_{\mu\nu},\overline{A}]\notag
        \\
        &=\int (d^4x)_{\rm E} \sqrt{\overline{g}}\bigg(\Lambda_{0,\Phi}^{\rm loop}-\frac{M^2_{\rm Pl}}{2}\overline{R} +\frac{1}{4}\overline{F}_{\mu\nu}\overline{F}^{\mu\nu}+({\rm correction~from}~R~{\rm and}~F_{\mu\nu}F^{\mu\nu})\bigg).\label{eq:W0EM}
    \end{align}
    Similar to the tree-level UV completion, the first order correction for the higher-derivative terms vanishes in $W_0$ by the equation of motion.
    Also, from Eq.~\eqref{eq:WgEM},  the first order correction for $g$ is calculated as
    \begin{align}
        \left(\frac{dW_g}{dg}\right)_{g=0}&=\left(\frac{\partial W_g}{\partial g}\right)_{g=0}
        +\int (d^4x)_{\rm E}\sqrt{\overline{g'}} \bigg(\left(\frac{\delta W_g}{\delta \overline{A'}}\right)\cdot \left(\frac{d \overline{A'}}{dg}\right)_{g=0}+\left(\frac{\delta W_g}{\delta \overline{g'}_{\mu\nu}}\right)\cdot \left(\frac{d \overline{g'}_{\mu\nu}}{dg}\right)_{g=0}\bigg)\notag
        \\
        &=\left(\frac{\partial W_g}{\partial g}\right)_{g=0}\notag
        \\
        &={\langle I_{\rm I}\rangle}_{g=0}\notag
        \\
        &=\int (d^4x)_{\rm E} \sqrt{\overline{g}}\bigg(-\frac{M^2_{\rm Pl}}{2} \frac{d \alpha^{\rm loop}_{1,R}}{dg}\overline{R}+\frac{1}{4} \frac{d \alpha^{\rm loop}_{1,F}}{dg}\overline{F}_{\mu\nu}\overline{F}^{\mu\nu}
        \bigg),\label{eq:EMFIRg}
    \end{align}
    where $(d\overline{A'}_{\mu}/dg)_{g=0}=0$ and $(d\overline{g'}_{\mu\nu}/dg)_{g=0}=0$ hold from Eq.~\eqref{eq:Aloop} and \eqref{eq:gloop}.
    Note here that the last term of Eq.~\eqref{eq:WgEM} does not depend on $g$.
    From Eq.~\eqref{eq:upg}, \eqref{eq:WgEM}, \eqref{eq:W0EM} and \eqref{eq:EMFIRg}, we obtain
     \begin{align}
        S(P_0||P_g)&=W_0[\overline{g}_{\mu\nu},\overline{A}]-W_g[\overline{g}_{\mu\nu},\overline{A}]+ g{\langle I_{\rm I}\rangle}_{g=0} \notag
        \\
        &=W_0[\overline{g}_{\mu\nu},\overline{A}]-W_g^{\rm non\text{-}lin}[\overline{g}_{\mu\nu},\overline{A}]\geq 0,\label{eq:EMbound}
    \end{align}
    where we used $g\cdot (d\alpha^{\rm loop}_{1,R}/dg)=\alpha^{\rm loop}_{1,R}$ and $g\cdot (d\alpha^{\rm loop}_{1,F}/dg)=\alpha^{\rm loop}_{1,F}$ and defined the effective action without the first order corrections for $g$ as follows:
        \begin{align}
        W_g^{\rm non\text{-}lin}[\overline{g}_{\mu\nu},\overline{A}]&=\int (d^4x)_{\rm E} \sqrt{\overline{g}}\bigg(\Lambda_{0,\Phi}^{\rm loop}-\frac{M^2_{\rm Pl}}{2}\overline{R} +\frac{1}{4}\overline{F}_{\mu\nu}\overline{F}^{\mu\nu}\notag
        \\
        &-\beta_{2,1}^{\rm loop}(\overline{F}_{\mu\nu}\overline{F}^{\mu\nu})^2-\beta_{2,2}^{\rm loop}(\overline{F}_{\mu\nu}\widetilde{\overline{F}}^{\mu\nu})^2-\beta_{2,3}^{\rm loop} \overline{F}_{\mu\nu}\overline{F}_{\rho\sigma}\overline{R}^{\mu\nu\rho\sigma}\notag
        \\
        &+({\rm correction~from}~R~{\rm and}~F_{\mu\nu}F^{\mu\nu})\bigg).
    \end{align}
        $W_0[\overline{g}_{\mu\nu},\overline{A}]-W_g^{\rm non\text{-}lin}[\overline{g}_{\mu\nu},\overline{A}]$ denotes the corrections from the higher-derivative terms, and the inequality of \eqref{eq:EMbound} means that the Euclidean effective action decreases by the higher-derivative operators.
        It should be noted that the one-loop correction from $R$ and $F_{\mu\nu}F^{\mu\nu}$ cancels in Eq.~\eqref{eq:EMbound}.
        By taking $g=1$ in Eq.~\eqref{eq:EMbound}, we found that the relative entropy yields the negative shift of the effective action by the higher derivative terms generated at the one-loop level.
        Especially for loop-level UV completions involving massive charged particles with large charge-to-mass ratios, this result is consistent with Ref.~\cite{Hamada:2018dde}.
    
\end{itemize}

It is found that, for both tree and loop-level UV completion, the non-negativity of the relative entropy yields the negative shift of the Euclidean effective action by the higher-derivative terms.
This argument holds when the assumptions at the beginning of this section are valid.
In the context of the WGC, the negative shift of the Euclidean effective action by the higher-derivative terms is demonstrated in a wide range of theories from unitarity and causality considerations~\cite{Hamada:2018dde}. 
Therefore, the results of the relative entropy considerations are consistent with that of the unitarity and causality, especially for loop-level UV completions involving massive charged particles with large charge-to-mass ratios and tree-level UV completions.
As discussed in the next chapter, this result is closely related to the WGC.

\subsection{Summary of bottom-up approach}
In this section, we focused on a class of EFTs where the corrections to the leading terms can be removed by redefining light fields.
For example, the single massless scalar field with the dimension-eight operator, the SMEFT dimension-eight gauge bosonic operators, and the Einstein-Maxwell theory with higher-derivative operators belong to such a class of EFTs.
Our arguments in this section are based on assumptions summarized at the beginning of this section.
In particular, Assumptions (III) and (IV) are the main assumptions in this section.
Under the assumptions, we found that the relative entropy is the linear combination of the higher-derivative operators generated from the interactions $I_{\rm I}$.
Strictly speaking, for the tree and loop level UV completions, we derive the following relations,
\begin{align}
    S(P_{\rm NI}||P_{\rm T})&=W_{\rm NI}[\widetilde{\phi}]-W_{\rm T}[\widetilde{\phi}]+{\langle I_{\rm I}\rangle}_{\rm NI}\notag
    \\
    &={(\rm linear~combination~of~higher\text{-}derivative~operators)}\geq 0.
\end{align}
For each EFTs, this inequality yields the constraints on the Wilson coefficients of the higher-derivative operators, which are consistent with the positivity bounds from unitarity and causality.

\section{Weak gravity conjecture and entropy constraint}
\label{sec:WGC}
We discuss a connection between the entropy constraints and the WGC.
We focus on the NIRT and consider the perturbative corrections to the Euclidean effective action from the interactions between heavy and light degrees of freedom.
In Sec.~\ref{sec:corrth}, we investigate a consequence of the non-negativity of the relative entropy to the corrections to thermodynamic entropy.
In Sec.~\ref{sec:wgcentr}, we explain a relation between the shift of mass of the extremal black hole by the perturbative corrections of interaction and the non-negativity of the relative entropy.
We will also comment on a connection between this work and Ref.~\cite{Cheung:2018cwt}.

\subsection{Corrections to thermodynamic entropy}
\label{sec:corrth}
The perturbative corrections to the thermodynamic entropy have been actively studied in the context of the WGC~\cite{Goon:2019faz}.
We summarize the standard thermodynamic relations with the notation in Ref.~\cite{Goon:2019faz} and investigate a connection between the non-negativity of the relative entropy and the corrections to thermodynamic entropy.
The free energy of the thermodynamic system is defined as
\begin{align}
    \beta\cdot G\equiv \beta\cdot \left(M-\beta^{-1}\cdot S -Q\cdot \mu\right),\label{eq:freeG}
\end{align}
where $G$ is the free energy, $\beta$ is the inverse temperature, $S$ is the thermodynamic entropy, $Q$ is the charge such as $U(1)$ charge, and $\mu$ is the chemical potential.
The first law of thermodynamics is expressed as follows:
\begin{align}
    dG=-S\cdot dT -Q\cdot d\mu.\label{eq:fist}
\end{align}
From Eq.~\eqref{eq:freeG}, we obtain
\begin{align}
    d G=d M- T\cdot dS-S\cdot  dT-Q\cdot d\mu-\mu \cdot dQ.\label{eq:dG}
\end{align}
Combining Eq.~\eqref{eq:fist} and \eqref{eq:dG}, we obtain
\begin{align}
    dM=T\cdot dS +\mu\cdot dQ\Rightarrow \beta =\left(\frac{\partial S}{\partial M}\right)_{Q},~~\mu=\left(\frac{\partial M}{\partial Q}\right)_S.\label{eq:combeta}
\end{align}
Now, let us assume the free energy of the thermodynamic system is shifted by perturbative effects, e.g., corrections from heavy degrees of freedom, as follows:
\begin{align}
    G(T,\mu,0)\to G(T,\mu,\epsilon)\equiv G(T,\mu,0)+\epsilon\cdot \Delta G(T,\mu),  
\end{align}
where $\epsilon$ is an auxiliary parameter to characterize the perturbative corrections.
Note here that $\epsilon$ is not the same as $g$ in Sec.~\ref{sec:entr} because $\Delta G$ involve the first or higher order corrections of $g$. 
Then, the thermodynamic entropy shift by the perturbative effects is defined as follows:
\begin{align}
    (\Delta S)_{\beta,\mu}\equiv S(T,\mu,\epsilon)-S(T,\mu,0)=\epsilon\cdot\left(\frac{\partial S}{\partial \epsilon}\right)_{\beta,\mu}+\mathcal{O}(\epsilon^2).\label{eq:delSep}
\end{align}
The leading correction of $\epsilon$ to the thermodynamic entropy is expressed as
\begin{align}
    \left(\frac{\partial S}{\partial \epsilon}\right)_{\beta,\mu}=\left(\frac{\partial S}{\partial M}\right)_{Q,\epsilon}\cdot \left(\frac{\partial M}{\partial \epsilon}\right)_{\beta,\mu}+\left(\frac{\partial S}{\partial Q}\right)_{M,\epsilon}\cdot \left(\frac{\partial Q}{\partial \epsilon}\right)_{\beta,\mu}+\left(\frac{\partial S}{\partial \epsilon}\right)_{M,Q}.\label{eq:parS}
\end{align}
Also, using the triple product rule, we obtain
\begin{align}
    \left(\frac{\partial S}{\partial Q}\right)_{M,\epsilon}=-\left(\frac{\partial S}{\partial M}\right)_{Q,\epsilon}\cdot \left(\frac{\partial M}{\partial Q}\right)_{S,\epsilon}.\label{eq:tri}
\end{align}
Combining Eqs.~\eqref{eq:combeta}, \eqref{eq:parS}, and \eqref{eq:tri}, we obtain
\begin{align}
    \left(\frac{\partial S}{\partial \epsilon}\right)_{\beta,\mu}=\beta\cdot \left(\frac{\partial M}{\partial \epsilon}\right)_{\beta,\mu}-\beta\cdot \mu\cdot \left(\frac{\partial Q}{\partial \epsilon}\right)_{\beta,\mu}+\left(\frac{\partial S}{\partial \epsilon}\right)_{M,Q}.\label{eq:Sep}
\end{align}
On the other hand, Eq.~\eqref{eq:freeG} yields
\begin{align}
    \left(\frac{\partial (\beta\cdot G)}{\partial \epsilon}\right)_{\beta,\mu}=\beta\cdot \left(\frac{\partial M}{\partial\epsilon}\right)_{\beta,\mu}-\left(\frac{\partial S}{\partial \epsilon}\right)_{\beta,\mu}-\beta\cdot \mu\cdot \left(\frac{\partial Q}{\partial \epsilon}\right)_{\beta,\mu}.\label{eq:beG}
\end{align}
From Eqs.~\eqref{eq:Sep} and \eqref{eq:beG}, we obtain
\begin{align}
    \left(\frac{\partial S}{\partial \epsilon}\right)_{M,Q}=-\left(\frac{\partial (\beta\cdot G)}{\partial \epsilon}\right)_{\beta,\mu}.
\end{align}
Consequently, the leading perturbative correction of $\epsilon$ to thermodynamic entropy at fixed energy and charge is expressed as follows:
\begin{align}
    \epsilon\cdot \left(\frac{\partial S}{\partial \epsilon}\right)_{M,Q}=-\epsilon\cdot \left(\frac{\partial (\beta\cdot G)}{\partial \epsilon}\right)_{\beta,\mu}=-\beta\cdot(\epsilon\cdot \Delta  G(T,\mu)),\label{eq:delGS}
\end{align}
where the most right-hand side of Eq.~\eqref{eq:delGS} denotes the leading perturbative correction of $\epsilon$ to the Euclidean effective action, which is relevant to the non-negativity of the relative entropy.
The point is that Eq.~\eqref{eq:delGS} takes a positive value for the negative free energy shift.

    In Eq.~\eqref{eq:uplowBG} of Sec.~\ref{sec:shif}, we have provided the lower and upper bounds on the perturbative corrections to the Euclidean effective action from the interaction between the heavy and light degrees of freedom.
Note here that the derivation of Eq.~\eqref{eq:uplowBG} does not depend on whether the temperature of the system is zero or not.
The perturbative corrections to the free energy of the thermodynamic system are expressed as follows:
\begin{align}
   \beta\cdot( \epsilon\cdot \Delta  G)\equiv W_{\rm T}[\beta,\phi]-W_{\rm NI}[\beta,\phi],\label{eq:betadelG}
\end{align}
    where the left-hand side denotes the most right-hand side of Eq.~\eqref{eq:delGS}.
    Combing Eqs.~\eqref{eq:uplowBG} and \eqref{eq:betadelG}, we obtain the following inequality.
    \begin{align}
        {\langle I_{\rm I}\rangle}_{\rm NI}\geq\beta\cdot ( \epsilon\cdot \Delta  G)\geq {\langle I_{\rm I}\rangle}_{\rm T}.\label{eq:delGuplow}
    \end{align}
    From Eqs.~\eqref{eq:delGS} and \eqref{eq:delGuplow}, we obtain
    \begin{align}
        -{\langle I_{\rm I}\rangle}_{\rm T}\geq \epsilon\cdot \left(\frac{\partial S}{\partial \epsilon}\right)_{M,Q}\geq -{\langle I_{\rm I}\rangle}_{\rm NI}.\label{eq:lowdelS}
    \end{align}
    Consequently, the relative entropy yields the upper and lower bounds on the thermodynamic entropy shift at fixed energy and charge by the interaction between heavy and light degrees of freedom.

    For the class of EFTs in Sec.~\ref{sec:bot}, under the assumptions summarized at the beginning of Sec.~\ref{sec:bot}, the relative entropy denotes the shift of the Euclidean effective action, i.e., the free energy at zero temperature, consisting of the linear combination of higher-derivative operators as follows:
    \begin{align}
         S(P_{\rm NI}||P_{\rm T})=-\beta\cdot( \epsilon\cdot \Delta  G)={(\rm linear~combination~of~higher\text{-}derivative~operators)}\geq 0.\label{eq:delSg0}
    \end{align}
    Combining Eq.~\eqref{eq:delGS} and \eqref{eq:delSg0}, we obtain
    \begin{align}
        \epsilon\cdot \left(\frac{\partial S}{\partial \epsilon}\right)_{M,Q}=S(P_{\rm NI}||P_{\rm T})\geq 0,\label{eq:enlast}
    \end{align}
    which is evaluated at zero temperature.
    Consequently, for a class of EFTs in Sec.~\ref{sec:bot}, it is found that the non-negativity of relative entropy yields the positive shift of the thermodynamic entropy of zero temperature system at fixed energy and charge by the higher-derivative operators. 
    It should be noted that, in contrast to the inequalities of \eqref{eq:lowdelS}, this argument is derived from the assumptions summarized at the beginning of Sec.~\ref{sec:bot}.
    Therefore, Eq.~\eqref{eq:lowdelS} is applicable in more theories than Eq.~\eqref{eq:enlast}.

\subsection{Entropy constraints and weak-gravity-conjecture}
\label{sec:wgcentr}
In Ref.~\cite{Goon:2019faz}, by using general thermodynamic considerations, it has demonstrated that the perturbations decrease the minimal energy of thermodynamic systems at a fixed charge when the perturbative correction to the thermodynamic entropy of the left-hand side of Eq.~\eqref{eq:delGS} is positive. 
For the sake of self-contained, we briefly review Ref.~\cite{Goon:2019faz} and then explain the connection between the WGC-like behavior in a shift of the extremal black hole mass and the relative entropy.
From Eq.~\eqref{eq:delSep}, and the third law of thermodynamics, i.e., $\lim_{T\to 0} S(T,Q,\epsilon)=0$, we obtain
\begin{align}
    \lim_{T\to 0} T\left(\frac{\partial S(T,Q,\epsilon)}{\partial \epsilon}\right)_{\beta,Q}=0.\label{eq:third}
\end{align}
Equations~\eqref{eq:freeG} and \eqref{eq:fist} yield the following thermodynamic relations:
\begin{align}
    M(T,\mu,\epsilon)&=G(T,\mu,\epsilon)+ T\cdot S +\mu\cdot Q,
    \\
    S(T,\mu)&=-\left(\frac{\partial G}{\partial T}\right)_{\mu,\epsilon},
    \\
    Q(T,\mu,\epsilon)&=-\left(\frac{\partial G}{\partial \mu}\right)_{\beta,\epsilon}.
\end{align}
By using the above relations, the energy shift at fixed temperature and charge is calculated as follows:
\begin{align}
    \left(\frac{\partial M}{\partial \epsilon}\right)_{\beta,Q}&=\left(\frac{\partial}{\partial \epsilon}\left(G+T\cdot S+\mu\cdot Q\right)\right)_{\beta,Q}\notag
    \\
    &=\left(\frac{\partial G}{\partial \mu}\right)_{\beta,\epsilon}\cdot\left(\frac{\partial \mu}{\partial \epsilon}\right)_{\beta,Q}+\left(\frac{\partial G}{\partial \epsilon}\right)_{\beta,\mu} +T\cdot \left(\frac{\partial S}{\partial \epsilon}\right)_{\beta,Q}+Q\cdot \left(\frac{\partial \mu}{\partial \epsilon}\right)_{\beta,Q}\notag
    \\
    &=\left(\frac{\partial G}{\partial \epsilon}\right)_{\beta,\mu} +T\cdot \left(\frac{\partial S}{\partial \epsilon}\right)_{\beta,Q}.\label{eq:delMep}
\end{align}
Combining Eqs.~\eqref{eq:third} and \eqref{eq:delMep}, we obtain
\begin{align}
    \lim_{T\to 0} \left(\frac{\partial M}{\partial \epsilon}\right)_{\beta,Q}=\lim_{T\to 0} \left(\frac{\partial G}{\partial \epsilon}\right)_{\beta,\mu}.\label{eq:delMdelG}
\end{align}
From Eqs.~\eqref{eq:delGS} and \eqref{eq:delMdelG}, we obtain
\begin{align}
    \lim_{T\to 0} \left(\frac{\partial M}{\partial \epsilon}\right)_{\beta,Q}= \lim_{T\to 0} - T\cdot \left(\frac{\partial S}{\partial \epsilon}\right)_{M,Q}.\label{eq:delMdelS}
\end{align}
Since the minimum energy of the system is defined as
\begin{align}
    M_{\rm ext}(Q,\epsilon)\equiv \lim_{T\to 0} M(T,Q,\epsilon),
\end{align}
Eq.~\eqref{eq:delMdelS} is expressed as follows:
\begin{align}
    \lim_{T\to 0} \left(\frac{\partial M}{\partial \epsilon}\right)_{\beta,Q}= \lim_{M\to M_{\rm ext}(Q,\epsilon)} - T(M,Q,\epsilon)\cdot \left(\frac{\partial S}{\partial \epsilon}\right)_{M,Q}.\label{eq:uni}
\end{align}
Therefore, we see that the perturbations decrease the minimal energy of thermodynamic systems at a fixed charge when the perturbative correction to the thermodynamic entropy is positive~\cite{Goon:2019faz}.
Particularly for the charged-BH described by the Einstein-Maxwell theory, $Q$ is the $U(1)$ charge, and the minimum mass of the BH is $M_{\rm ext}=\sqrt{2}M_{\rm Pl}Q$.
As studied in a large amount of literature~\cite{Arkani-Hamed:2006emk,Hamada:2018dde,Cheung:2018cwt,Banks:2006mm,Montero:2018fns,Arkani-Hamed:2021ajd}, a state with a charge-to-mass ratio larger than unity is motivated by the thought experiment of decay of the extremal black hole.
When the minimum mass of BH at fixed charge decrease by the perturbative correction from the higher dimensional operators, asymptotically large extremal black holes described by the Einstein-Maxwell theory with higher derivative operators are allowed to decay to a state whose charge exceeds its mass, e.g., extremal black hole.
In this work, we refer to the behavior that the minimum energy of a thermodynamic system at fixed charge decreases by the perturbative correction as {\it WGC-like behavior}.
Note here that, in this work, we use this terminology for general thermodynamic systems with and without gravity.

For the EFTs in Sec.~\ref{sec:bot}, the non-negativity of the relative entropy yields the negative free energy shift at zero temperature within the validity of the field theoretical descriptions. 
From Eq.~\eqref{eq:delGS}, this result means that the thermodynamic entropy shift at fixed energy and charge by the higher-derivative operators is positive in the extremal limit of $M=M_{\rm ext}$, i.e., the zero temperature limit.
Then, the right-hand side of Eq.~\eqref{eq:uni} is negative, and the WGC-like behavior arises.
For the Einstein-Maxwell theory, as studied in Ref.~\cite{Hamada:2018dde}, the unitarity and causality considerations also yield the positive BH entropy shift at fixed energy and charge by the higher-derivative operators generated in a wide range of theories in the extremal limit.
Therefore, the entropy constraints yield consistent results with the unitarity and causality considerations~\cite{Hamada:2018dde}, especially for loop-level UV completions involving massive charged particles with large charge-to-mass ratios $z=\sqrt{2}M_{\rm Pl}|q|/m$ and tree-level UV completions.
The entropy constraint implies that the extremal black hole would behave as a state with $z\geq 1$ even if massive charged particles with $z\geq 1$ do not exist.
Our argument is applicable when the thermodynamic relations and the assumptions at the beginning of Sec.~\ref{sec:bot} are held.
The amplitude considerations in Ref.~\cite{Hamada:2018dde} are also applicable beyond the field theoretical descriptions under the assumption that the higher spin states Reggeizing graviton exchange are subdominant.
However, our discussions in Sec.~\ref{sec:bot} rely on the field theory, so it is not clear if our results are applicable beyond the field theoretical descriptions. 
Detailed studies of the applicability of this study are expected in the future. 

Here, we comment on a relation between Ref.~\cite{Cheung:2018cwt} and this work.
In Ref.~\cite{Cheung:2018cwt}, it has been demonstrated that the Euclidean effective action decreases by higher-derivative operators generated at the tree level.
For convenience, we briefly review it by using the notation of this work.
Using the saddle point approximation, we obtain
\begin{align}
    I_0[\widetilde{\phi}_{0},0]=I_g[\widetilde{\phi}_{0},0]\geq I_g[\widetilde{\phi}_{g},\widetilde{\Phi}_{g}],\label{eq:prewo1}
\end{align}
where $\widetilde{\phi}_{0}$ is the classical solution of $I_0$, $\widetilde{\phi}_{g}$ and $\widetilde{\Phi}_{g}$ are that of $I_g$, and $I_g[\widetilde{\phi}_{0},0]=I_0[\widetilde{\phi}_{0},0]$ holds because the interaction $I_{\rm I}$ vanishes for $\Phi=0$.
The inequality of \eqref{eq:prewo1} arises because $\widetilde{\Phi}_g$ denotes the local minimum of $I_g$ and would take a small value by heavy field mass suppression. 
By taking $g=1$ in Eq.~\eqref{eq:prewo1}, we obtain
\begin{align}
    W_{\rm NI}[\widetilde{\phi}_{\rm NI}]=I_{\rm NI}[\widetilde{\phi}_{\rm NI},0]\geq I_{\rm T}[\widetilde{\phi}_{\rm T},\widetilde{\Phi}_{\rm T}]=W_{\rm T}[\widetilde{\phi}_{\rm T}],
\end{align}
where $I_{\rm NI}=I_0$, $I_{\rm T}=I_{g=1}$, $\widetilde{\phi}_{\rm NI}=\widetilde{\phi}_{0}$, $\widetilde{\phi}_{\rm T}=\widetilde{\phi}_{g=1}$, and $\widetilde{\Phi}_{\rm T}=\widetilde{\Phi}_{g=1}$ are used, and $\widetilde{\Phi}_{\rm T}$ is expressed as a linear combination of $\widetilde{\phi}_{\rm T}$.
The effective action $W_{\rm NI}$ does not generate the higher-dimensional operators, but the $W_{\rm T}$ yields them through the interacting term between $\phi$ and $\Phi$. 
Therefore, the inequality~\eqref{eq:prewo1} means that the Euclidean effective action decreases by higher-dimensional operators generated at the tree level.
In other words, at fixed temperature $\beta$, the free energy decreases by higher-dimensional operators generated at the tree level.
It should be noted that in Eq.~\eqref{eq:prewo1}, we used the following relation.
\begin{align}
\lim_{\widetilde{\phi}_{g}\to\widetilde{\phi}_{0},\widetilde{\Phi}_{g}\to 0}I_g[\widetilde{\phi}_{g},\widetilde{\Phi}_{g}]=I_0[\widetilde{\phi}_{0},0]=\lim_{g\to 0}I_g[\widetilde{\phi}_{g},\widetilde{\Phi}_{g}],
\end{align}
where procedures of the leftmost and rightmost sides are respectively performed by Ref.~\cite{Cheung:2018cwt}, and this work and Ref.~\cite{Cao:2022iqh}.
The point is that Ref.~\cite{Cheung:2018cwt} also derives the WGC-like behavior by comparing the NIRT and target theories.  
It should be emphasized that this work is essentially the same as Ref.~\cite{Cheung:2018cwt} at the tree level.

\section{Implication of entropy constraints}
\label{sec:imp}
%
We found that the non-negativity of relative entropy yields constraints on the class of EFTs in Sec.~\ref{sec:bot}, which are consistent with the conventional positivity bounds~\cite{Adams:2006sv,Remmen:2019cyz,Hamada:2018dde}. 
In this section, we investigate relations of the non-negativity of the relative entropy with the unitary time evolution of the system, causality, and the second law of thermodynamics.

\subsection{Unitary time-evolution}
The relative entropy, i.e., $S(\rho_{\rm T}||\rho_{\rm R})={\rm Tr}[\rho_{\rm T}\ln \rho_{\rm T}-\rho_{\rm T}\ln \rho_{\rm R}]$, is a non-negative quantity, which is a consequence of the Hermiticity of the probability distribution functions $\rho_{\rm T}$ and $\rho_{\rm R}$.  
In the quantum mechanical approach of the entropy constraints in Sec.~\ref{sec:relative}, to derive the bounds on EFTs, we have focused on the probability distribution functions described as follows:
\begin{align}
    \rho_{\rm T}\equiv\frac{e^{-\beta  H_{\rm T}}}{Z_{\rm T}},~~~\rho_{\rm R}\equiv\frac{e^{-\beta  H_{\rm R}}}{Z_{\rm R}},
\end{align}
with the partition functions $Z_{\rm T}\equiv {\rm Tr}[e^{-\beta  H_{\rm T}}]$ and $Z_{\rm R}\equiv {\rm Tr}[e^{-\beta  H_{\rm R}}]$.
In Sec.~\ref{sec:top} and \ref{sec:bot}, we have demonstrated that the positivity bounds on some EFTs are derived by the above type of probability distribution functions.
Since the non-negativity of the relative entropy is based on the Hermiticity of $\rho_{\rm T}$ and $\rho_{\rm R}$, the Hermiticity of the Hamiltonians $H_{\rm T}$ and $H_{\rm R}$ is also assumed to derive the entropy constraints on EFTs.
Therefore, the time evolution of the target theory defined by $H_{\rm T}$ is unitary, and the unitary S matrix is ensured because of the Hermiticity of $H_{\rm T}$.
The point is that the entropy constraint is based on the unitary time evolution of the target theory, i.e., the optical theorem in the target theory. %
The conventional positivity bounds from amplitude considerations~\cite{Adams:2006sv,Remmen:2019cyz} rely on the optical theorem.

\subsection{Causality}
We discuss the connection between the non-negativity of the relative entropy and the causality~\cite{Adams:2006sv} by using a simple example. 
Consider the EFT defined by the Euclidean Lagrangian as,
\begin{align}
    \mathcal{L}^{\rm (E)}[\phi]= \frac{1}{2}(\partial_{I} \phi\partial_{I} \phi)-\frac{c}{M^4}(\partial_{I} \phi\partial_{I} \phi)^2.
\end{align}
By setting $\phi=\widetilde{\phi}+\varphi$, with $\varphi$ the small dynamical perturbation, expand the Lagrangian by the perturbation as follows:
\begin{align}
    \mathcal{L}^{\rm (E)}&=\frac{1}{2}\left[\delta_{IJ}-\frac{c}{6M^4}\left(4 (\partial_{I}\widetilde{\phi})^2\delta_{IJ} +8 (\partial_{I}\widetilde{\phi})(\partial_{J}\widetilde{\phi}) \right)\right](\partial_{I}\widetilde{\phi}\partial_{J}\widetilde{\phi})\notag
    \\
    &+\left[1 -\frac{4 c}{M^4} (\partial_{J}\widetilde{\phi})^2 \right]\partial_{I}\widetilde{\phi} \partial_{I}\varphi\notag
    \\
    &+\frac{1}{2}\left[\delta_{IJ}-\frac{c}{M^4}\left(4 (\partial_{I}\widetilde{\phi})^2\delta_{IJ} +8 (\partial_{I}\widetilde{\phi})(\partial_{J}\widetilde{\phi}) \right) \right](\partial_{I}\varphi\partial_{J}\varphi)+\mathcal{O}(\varphi^3).\label{eq:varp}
\end{align}
To ensure the validity of the Euclidean path integral around $\widetilde{\phi}$, the last line of Eq.~\eqref{eq:varp} yields the following relation.
\begin{align}
    k^2-\frac{c}{M^4}\left(4 q^2 k^2 +8 (q\cdot k)^2 \right)\geq 0\Rightarrow v_{\rm E}^2 \leq 1-\frac{8c (q\cdot k)^2}{|\vec{k}|^2(M^4-4c q^2)},
\end{align}
where $\varphi$ is expanded in plane waves as $\varphi\propto e^{ik\cdot x}$, $v_{\rm E}^2\equiv-k_4^2/|\vec{k}|^2$ denotes the speed of propagation of $\varphi$ in the Euclidean space, and $\partial_I\widetilde{\phi}=q_{I}$ is a constant vector.
As discussed in Sec.~\ref{sec:bottomup_shift_symmetry}, the non-negativity of relative entropy yields $c\geq 0$.
Therefore, combining the above validity of the Euclidean path integral, and the non-negativity of relative entropy, we obtain the upper bound on $v_{\rm E}$ as follows,
\begin{align}
    v_{\rm E}^2\leq 1,
\end{align}
which ensures causality in the Minkowski space.
Consequently, the non-negativity of relative entropy yields causality in the EFT.
It should be noted that the entropy constraints are based on the Euclidean path integral method, and the validity of the Euclidean path integral around the classical solution is assumed.
As explained in Sec.~\ref{sec:incon}, the non-negativity of relative entropy may be broken when the Euclidean path integral is performed around a point not being a local minimum.

\subsection{Second law of thermodynamics}
\label{sec:second}
The non-negativity of relative entropy is closely related to the second law of thermodynamics~\cite{2000cond.mat..9244T,2012}.
We demonstrate that a simple derivation of the second law of thermodynamics is contained in the entropy constraints of this study.
Adopt the thermal reference theory of Sec.~\ref{sec:defTH} as the reference theory, and consider the system consisting of a thermodynamic system S and a heat bath system B.
We suppose that the Hamiltonian of the whole system is defined as
\begin{align}
    H_{\rm T}=H_{\rm S}+H_{\rm B}+H_{\rm SB},
\end{align}
 where $H_{\rm S}$ is the Hamiltonian of S, and $H_{\rm B}$ is that of B.
The interacting term $H_{\rm SB}$ denotes the interaction between S and B, which is generally a time-dependent operator.
Note here that the system S can involve both light and heavy degrees of freedom.

Now, let us assume that the initial quantum state of the whole system is defined as
\begin{align}
    \rho_{\rm ini}\equiv\rho_{\rm ini,S}\otimes e^{-\beta H_B}/Z_{\rm B}(\beta),
\end{align}
where $\rho_{\rm ini,S}$ is the initial state of S, and $e^{-\beta H_B}/Z_{\rm B}(\beta)$ is that of B at an inverse temperature $\beta$.
By tracing over the heat bath degrees of freedom, the partition function of B is defined as
\begin{align}
    Z_{\rm B}(\beta)\equiv {\rm Tr}_{\rm B}[e^{-\beta H_{\rm B}}].
\end{align}
After a time-evolution of the whole system described by a unitary operator $U$, the final state of the whole system is expressed as
\begin{align}
    \rho_{\rm fin}= U \rho_{\rm ini} U^{\dagger}. 
\end{align}
By tracing out the heat bath degrees of freedom, the final sate of S is expressed as
\begin{align}
    \rho_{\rm fin,S}\equiv {\rm Tr}_{\rm B}[\rho_{\rm fin}].
\end{align}
Note here that the time evolution of S is not unitary because the heat bath system is traced out.

Following Sec.~\ref{sec:defTH}, define the probability distribution functions of the target theory and reference theory as follows.
\begin{align}
    &\rho_{\rm T}\equiv \rho_{\rm ini},\label{eq:secterg}
    \\
    &\rho_{\rm R}\equiv U^{\dagger}\rho_{\rm fin, S}\otimes e^{-\beta H_{\rm B}}/Z_{\rm B}(\beta) U. \label{eq:secref}
\end{align}
The relative entropy between $\rho_{\rm T}$ and $\rho_{\rm R}$ is calculated as
\begin{align}
    S(\rho_{\rm T}||\rho_{\rm R})=S(\rho_{\rm fin,S})-S(\rho_{\rm ini,S})-\beta\cdot Q\geq 0,\label{eq:clau}
\end{align}
where $S(\rho)\equiv -{\rm Tr}_{\rm S}[\rho\ln \rho]$ denotes the von Neumann 
entropy of S, and $Q\equiv  {\rm Tr}\left[H_{\rm B} \rho_{\rm ini} \right]-{\rm Tr}\left[H_{\rm B} \rho_{\rm fin} \right]$ is the heat exchange between S and B.
The details of the derivations of Eq.~\eqref{eq:clau} are summarized in Appendix~\ref{app:seclaw}.
Equation~\eqref{eq:clau} is the Clausius inequality and denotes the second law of thermodynamics.
Consequently, we obtain the second law of thermodynamics in the above simple setup from the non-negativity of relative entropy.

Lastly, for convenience, we would like to discuss the connection between the target system of this section and that of Sec.~\ref{sec:shif}.
Although the heat bath system is not included in Sec.~\ref{sec:shif}, we can freely add it to the systems.
Let us take the case of NIRT as an example and consider the probability distribution functions of the target and reference theories $\rho'_{\rm T}$ and $\rho'_{\rm NI}$, where the heat bath system is not included.
We assume $\rho'_{\rm T}$ and $\rho'_{\rm NI}$ denote the probability distribution functions of the target and reference theories in Sec.~\ref{sec:shif}.
By including the heat bath system $e^{-\beta H_{\rm B}}/Z_{\rm B}(\beta)$, one can define as follows.
\begin{align}
    &\rho_{\rm T}\equiv \rho'_{\rm T}\otimes e^{-\beta H_{\rm B}}/Z_{\rm B}(\beta),
    \\
    &\rho_{\rm NI}\equiv \rho'_{\rm NI}\otimes e^{-\beta H_{\rm B}}/Z_{\rm B}(\beta),
\end{align}
where $\rho_{\rm T}$ is the same as the probability distribution of Eq.~\eqref{eq:secterg}, and $\rho'_{\rm T}$ and $\rho'_{\rm NI}$ represent the probability distribution in Sec.~\ref{sec:shif}.
Then, the relative entropy between $\rho_{\rm T}$ and $\rho_{\rm NI}$ is expressed as follows,
\begin{align}
    S(\rho_{\rm T}||\rho_{\rm NI})&={\rm Tr}[\rho_{\rm T} \ln \rho_{\rm T}-\rho_{\rm T}\ln \rho_{\rm NI}]=S(\rho'_{\rm T}||\rho'_{\rm NI})\label{eq:relS1},
    \\
    S(\rho_{\rm NI}||\rho_{\rm T})&={\rm Tr}[\rho_{\rm NI} \ln \rho_{\rm NI}-\rho_{\rm NI}\ln \rho_{\rm T}]=S(\rho'_{\rm NI}||\rho'_{\rm T})\label{eq:relS2}.
\end{align}
The right-hand sides of Eq.~\eqref{eq:relS1} and \eqref{eq:relS2} are the same quantity that was discussed in Sec.~\ref{sec:shif}.
The point is that the above derivation of the second law of thermodynamics is included as a special case of the entropy constraints of this work.
Here, it should be noted that the corrections to thermodynamic entropy, i.e., the left-hand side of Eq.~\eqref{eq:delGS} is a different quantity from the entropy shift due to the time-evolution discussed in this section.
It is remarkable that the non-negativity of both entropy shifts is derived from the relative entropy depending on the choice of the reference theory.

\section{Summary}
\label{sec:sum}
We studied constraints on the EFTs by evaluating the relative entropy between the target and reference theories.
In addition to providing the details of Ref.~\cite{Cao:2022iqh}, we updated the results in Ref.~\cite{Cao:2022iqh} by considering more theories and new reference theories.
Firstly, in Sec.~\ref{sec:entr}, we reviewed the details of the main idea of the entropy constraint of Ref.~\cite{Cao:2022iqh} and provided the procedures to calculate the relative entropy by introducing some new reference theories, i.e., massive free reference theory, and infinite mass reference theory.  
Following the procedures of Sec.~\ref{sec:entr}, in Secs.~\ref{sec:top}-\ref{sec:imp}, we evaluated the relative entropy and investigated the physical consequence of the non-negativity of the relative entropy in various EFTs.

In Sec.~\ref{sec:top}, we adopted the top-down approach and evaluated the relative entropy in various theories involving field theories, quantum mechanical models, and Gaussian distribution functions. 
The various examples satisfy the non-negativity of the relative entropy, which yields non-trivial constraints on the EFTs when the heavy degrees of freedom are integrated out. 
In Sec.~\ref{sec:incon}, we also discussed some examples where the non-negativity of the relative entropy is violated.
In contrast to Sec.~\ref{sec:top}, in Sec.~\ref{sec:bot}, we adopted the bottom-up approach, i.e., the EFTs are provided while the UV theories are unknown, and investigated the consequence of the non-negativity of the relative entropy.
We focused on a class of EFTs where the perturbative corrections to the non-higher derivative operators can be removed by redefinitions of the light fields, e.g., the single massless scalar field with the dimension-eight operator, SMEFT dimension-eight $SU(N)$ gauge bosonic operators and Einstein-Maxwell theory with higher-derivative operators.
Under the four assumptions, i.e., Hermiticity of probability distribution functions, the validity of Euclidean path integral method, higher-derivative operators generated from the interaction between heavy and light fields, and leading order of the interaction between heavy and light fields; see the beginning of Sec.~\ref{sec:bot}, we found that the non-negativity of the relative entropy yields the constraints on such EFTs, which are consistent with the positivity bounds from the unitarity and causality.
These constraints are derived for each of the tree-level and loop-level UV completions; see Sec.~\ref{sec:bot}.

In Secs.~\ref{sec:WGC} and \ref{sec:imp}, we discussed connections between the non-negativity of the relative entropy and various inequalities in physics, i.e., the WGC-like behavior, unitary time evolution, causality, and second law of thermodynamics.
In particular, the entropy constraints on the Einstein-Maxwell theory with higher-derivative operators imply that the minimum mass of the black hole at fixed charge decrease by the higher-derivative operators generated from the interactions between heavy and light fields.
This argument is applicable when the thermodynamic relations and the assumptions at the beginning of Sec.~\ref{sec:bot} are held.
It should be noted that the assumptions include that corrections from the interaction involving higher-derivative operators of light fields are not dominant in the EFTs. 
These results about Einstein-Maxwell theory rely on the field theoretical descriptions in Sec.~\ref{sec:bot}, so it is not clear if our results are applicable to UV theories involving stringy particles.
Detailed studies of the applicability of this study are expected in the future.
Also, our relative entropy consideration yields the second law of thermodynamics in a simple setup depending on the reference theory; see Sec.~\ref{sec:second}.

Our entropy consideration is applied to various theories, but further applications to more theories are required to understand its validity range.  
In particular, a derivation of the WGC-like behavior is subject to the assumptions at the beginning in Sec.~\ref{sec:bot}, and further studies based on various UV theories are essential to confirm the consistency of our results under the assumptions.
We also note that our entropy constraint is a different approach from the conventional unitarity and causality considerations, and therefore applicable theories would be different from each other.

In summary, we conclude that the relative entropy consideration provides a unified understanding of various inequalities in physics in addition to yielding a new approach to constraints on EFTs, and further studies are expected in various theories.

\appendix

\section*{Acknowledgement}\label{sec:ackn}
The work is supported in part by the National Science Foundation of China under Grant Nos. 11675002, 11635001, 11725520, and 12235001.

\section{Wick rotation}
\label{app:Wick_rotation}
We provide how to obtain the Euclidean effective action from the Minkowski one.
We denote the Euclidean indices by $I,J,\ldots$ and the Minkowski indices by $\mu, \nu,\ldots$.
The Euclidean effective action $I_{\rm E}[\phi_{\rm E},g_{{\rm E},IJ}]$ is obtained by the Wick rotation as follows:
\begin{align}
   I_{\rm E}[\phi_{\rm E},g_{{\rm E},IJ}]\equiv \frac{1}{i}I[\phi,g_{\mu\nu}]_{t=-i \tau},
\end{align}
where $I[\phi, g_{\mu\nu}]$ is the effective action in the Minkowski space defined as
\begin{align}
    I[\phi,g_{\mu\nu}]=\int \sqrt{-g} d^4 x \mathcal{L}[\phi,g_{\mu\nu}].
\end{align}
Here, $\mathcal{L}$ is the Lagrangian, $\phi$ is the background light fields in the Minkowski space, $g_{\mu\nu}$ is a metric tensor in the Minkowski signature, and $g={\rm det}g^{\mu\nu}$ is a determinant of the metric tensor. 

Let us consider a transformation of coordinates: $x^{\mu}=(x^0=t,\vec{x})\to {x'}^{\mu}=({x'}^0,\vec{x}')=(\lambda t,\vec{x})$, where $\lambda$ is a constant.
For convenience, we summarize the transformed scalar, vector, and metric tensor fields as follows:
\begin{align}
&\phi'(t',\vec{x})=\phi( t,\vec{x}),
\\
&{A'}^{\mu}(t',\vec{x})=\frac{\partial {x'}^{\mu}}{\partial {x}^{\alpha}}{A}^{\alpha}(t,\vec{x})=(\lambda\cdot A^0(x),A^i(x)),
\\
&{A'}_{\mu}(t',\vec{x})=\frac{\partial {x}^{\alpha}}{\partial {x'}^{\mu}}{A}_{\alpha}(t,\vec{x})=(\lambda^{-1}\cdot A_0(x),A_i(x)),
\\
&{g'}^{\mu\nu}(t',\vec{x})=\frac{\partial {x'}^{\mu}}{\partial {x}^{\alpha}}\frac{\partial {x'}^{\nu}}{\partial {x}^{\beta}}{g}^{\alpha\beta}(t,\vec{x})=\begin{pmatrix}
\lambda^{2} g^{00}(x) & \lambda g^{01}(x) & \lambda g^{02}(x) & \lambda g^{03}(x) 
\\
\lambda g^{10}(x) & g^{11}(x) & g^{12}(x) & g^{13}(x)
\\
\lambda g^{20}(x) & g^{21}(x) & g^{22}(x) & g^{23}(x)
\\
\lambda g^{30}(x) & g^{31}(x) & g^{32}(x) & g^{33}(x)
\end{pmatrix},
\\
&{g'}_{\mu\nu}(t',\vec{x})=\frac{\partial {x}^{\alpha}}{\partial {x'}^{\mu}}\frac{\partial {x}^{\beta}}{\partial {x'}^{\nu}}{g}_{\alpha\beta}(t,\vec{x})=\begin{pmatrix}
\lambda^{-2} g_{00}(x) & \lambda^{-1} g_{01}(x) & \lambda^{-1} g_{02}(x) & \lambda^{-1} g_{03}(x) 
\\
\lambda^{-1} g_{10}(x) & g_{11}(x) & g_{12}(x) & g_{13}(x)
\\
\lambda^{-1} g_{20}(x) & g_{21}(x) & g_{22}(x) & g_{23}(x)
\\
\lambda^{-1} g_{30}(x) & g_{31}(x) & g_{32}(x) & g_{33}(x)
\end{pmatrix}.
\end{align}
Also, the completely 
antisymmetric tensor is transformed as follows:
\begin{align}
    &\frac{1}{\sqrt{-g'}}{\epsilon'}^{\mu\nu\rho\sigma}= J\cdot\sqrt{\lambda^2}\frac{\partial {x'}^{\mu}}{\partial {x}^{\alpha}}\frac{\partial {x'}^{\nu}}{\partial {x}^{\beta}}\frac{\partial {x'}^{\rho}}{\partial {x}^{\gamma}}\frac{\partial {x'}^{\sigma}}{\partial {x}^{\delta}}\frac{1}{\sqrt{-g}}{\epsilon}^{\alpha\beta\gamma\delta},
\end{align}
where $\epsilon_{\mu\nu\rho\sigma}=\epsilon^{\mu\nu\rho\sigma}$, $\epsilon_{0123}=+1$, a determinant of the metric tensor $g={\rm det}g_{\mu\nu}$ is calculated as $g=\lambda^{2} g'$, and $J=\lambda^{-1}$ is defined as
\begin{align}
    J\equiv\epsilon_{\mu\nu\rho\sigma}\frac{\partial x^{\mu}}{\partial {x'}^{0}}\frac{\partial x^{\nu}}{\partial {x'}^1}\frac{\partial x^{\rho}}{\partial {x'}^2}\frac{\partial x^{\sigma}}{\partial {x'}^3}.
\end{align}
Also ${\epsilon'}^{\mu\nu\rho\sigma}$ satisfies the following relation.
\begin{align}
    J= {\epsilon'}^{\mu\nu\rho\sigma}\frac{\partial x^0}{\partial {x'}^{\mu}}\frac{\partial x^1}{\partial {x'}^{\nu}}\frac{\partial x^2}{\partial {x'}^{\rho}}\frac{\partial x^3}{\partial {x'}^{\sigma}}.
\end{align}
By the above transformation, the covariant volume element is not changed as follows:
\begin{align}
    \sqrt{-g}d^4 x = \sqrt{-\lambda^2 g'}\cdot\lambda^{-1} d^4 x' =\sqrt{-g'} d^4 x'. 
\end{align}
The action in the coordinate $x'$ is obtained as
\begin{align}
    I[\phi,{g}_{\mu\nu}]&=\int \sqrt{-g'} d^4 x' \mathcal{L}[\phi',{g'}_{\mu\nu}]=\lambda^{-1}\int \sqrt{-\lambda^2 g'} d^4 x' \mathcal{L}[\phi',{g'}_{\mu\nu}].
\end{align}
The Wick rotation is performed after taking $\lambda=i$.
Then, we obtain the analytically continued Euclidean action as
\begin{align}
    I^{\rm  (E)}[\phi_{\rm E},g_{{\rm E},IJ}]=\frac{1}{i}I[\phi,g_{\mu\nu}]_{t=-i \tau}=-\int \sqrt{g_{\rm E}} (d^4 x)_{\rm E} \mathcal{L}[\phi_{\rm E},{g}_{{\rm E},IJ}],
\end{align}
where we defined the Euclidean quantities as 
    \begin{gather}
        x^{ I}_{\rm E}=(\tau, \vec{x}_{\rm E})\equiv {x'}^\mu=(t', \vec{x}')=(it, \vec{x}), \\
        \partial_{{\rm E},I}=(\partial_\tau, \vec{\partial}_{\rm E})\equiv \partial'_\mu=(\partial'_t,\vec{\partial}')=(-i\partial_t,\vec{\partial}), \\
        (d^4x)_{\rm E}\equiv d^4x'=i\, d^4x, \\
        \sqrt{g_{\rm E}}\equiv\sqrt{g'}|_{\lambda=i} =\sqrt{-g}.
    \end{gather}
We also define the Euclidean scalar, vector and metric fields as follows:
    \begin{gather}
        \phi_{\rm E}(x_{\rm E})\equiv\phi'(t',\vec{x}) = \phi(t,\vec{x})=\phi(x), \\
        A^{ I}_{\rm E}=(A_{{\rm E},\tau}, \vec{A}_{\rm E})\equiv {A'}^\mu=(i A^0, \vec{A}), \\
       g_{{\rm E},IJ}=-{g'}_{\mu\nu}=
        \begin{pmatrix}
            g_{00} & ig_{0j} \\
            ig_{i0} & -g_{ij},
       \end{pmatrix},\quad g^{ IJ}_{{\rm E}}=-{g'}^{\mu\nu}=
        \begin{pmatrix}
            g^{00} & -ig^{0j} \\
            -ig^{i0} & -g^{ij}
        \end{pmatrix},
        \\
        \epsilon_{{\rm E},IJKL}=\epsilon_{{\rm E}}^{IJKL}=\epsilon_{\mu\nu\rho\sigma}=\epsilon^{\mu\nu\rho\sigma}
    \end{gather}
where the metrics in the flat Euclidean space become the Kronecker delta $\delta_{IJ}$ and $\delta^{IJ}$.

Here, we consider two examples: a single massless scalar field theory with a dimension-eight term and the Euler-Heisenberg theory in the following.

\begin{itemize}
    \item Massless scalar field theory with a dimension-eight term ---
First, let us consider an effective action in the Minkowski space:
    \begin{align}
        I_c[\phi]=\int d^4x  \left(\frac{1}{2}\partial_\mu \phi \partial^\mu \phi + \frac{c}{M^4}\left(\partial_\mu \phi \partial^\mu \phi\right)^2\right),
    \end{align}
where $\phi$ denotes a background field, and a metric tensor $g_{\mu\nu}=g^{\mu\nu}={\rm diag.}(+1,-1,-1,-1)$.
By the analytic continuation, the Euclidean effective action is obtained as
    \begin{align}
        I_c[\phi]&=\int d^4x' \sqrt{-g'} \left(\frac{1}{2}g'^{\mu\nu}{\partial'}_\mu \phi' {\partial'}_\nu \phi'+\frac{c}{M^4}\left(g'^{\mu\nu}{\partial'}_\mu \phi' {\partial'}_\nu \phi' \right)^2\right)\notag
        \\
        &=\int \lambda^{-1} d^4x' \sqrt{-\lambda^2 g'}\left(\frac{1}{2}g'^{\mu\nu}{\partial'}_\mu \phi' {\partial'}_\nu \phi'+\frac{c}{M^4}\left(g'^{\mu\nu}{\partial'}_\mu \phi' {\partial'}_\nu \phi' \right)^2\right)\notag \\
        &\to -i\int (d^4x)_{\rm E} \sqrt{g_{\rm E}}\left(-\frac{1}{2}g^{IJ}_{\rm E}\partial_{{\rm E},I} \phi_{\rm E}\partial_{{\rm E},J} \phi_{\rm E}+\frac{c}{M^4}\left(g^{IJ}_{\rm E}\partial_{{\rm E},I} \phi_{\rm E}\partial_{{\rm E},J} \phi_{\rm E}\right)^2\right)\notag \\
        &=i\int (d^4x)_{\rm E} \sqrt{g_{\rm E}}\left(\frac{1}{2} \partial_{{\rm E},I}\phi_{\rm E}\partial_{{\rm E},I}\phi_{\rm E}-\frac{c}{M^4}\left(\partial_{{\rm E},I}\phi_{\rm E}\partial_{{\rm E},I}\phi_{\rm E}\right)^2\right)\notag
        \\
        &=i\int (d^4x)_{\rm E}\left(-\frac{1}{2}(\partial_{\mu}\phi\partial^{\mu}\phi)-\frac{c}{M^4}\left(\partial_{\mu}\phi\partial^{\mu}\phi\right)^2 \right)\notag
        \\
        &=i I_c^{\rm (E)}[\phi_{\rm E}],
    \end{align}
where $g_{\rm E}=+1$, $\partial_{{\rm E},I}\phi_{\rm E}=(-i\cdot\partial_{t}\phi,\partial_{i}\phi )$, and the metric tensor with the Euclidean signature is the Kronecker delta, so we omitted them in the last line for short.

\item Dimension-eight $U(1)$ gauge bosonic operator ---
Consider a effective action in Minkowski space:
\begin{align}
    I_e[A]=\int d^4x \left(-\frac{1}{4}{F}_{\mu\nu}{F}^{\mu\nu} +\frac{c_1}{M^4} ({F}_{\mu\nu}{F}^{\mu\nu})^2+\frac{c_2}{M^4} ({F}_{\mu\nu}\tilde{{F}}^{\mu\nu})^2+\frac{c_3}{M^4}({F}_{\mu\nu}{{F}}^{\mu\nu})({F}_{\mu\nu}\widetilde{{F}}^{\mu\nu})\right),
\end{align}
where $\widetilde{{F}}^{\mu\nu}=\epsilon^{\mu\nu\rho\sigma}{F}_{\rho\sigma}/2$.
Then, the Euclidean effective action is obtained as
\begin{align}
I_e[{A}]&=\int \sqrt{-g}d^4x \bigg(-\frac{1}{4}g^{\mu\rho}g^{\nu\sigma}{F}_{\mu\nu}{F}_{\rho\sigma} +\frac{c_1}{M^4} (g^{\mu\rho}g^{\nu\sigma}{F}_{\mu\nu}{F}_{\rho\sigma} )^2+\frac{c_2}{M^4} \left(\frac{1}{\sqrt{-g}}\epsilon^{\mu\nu\rho\sigma}{F}_{\mu\nu}{F}_{\rho\sigma}\right)^2\notag
\\
&+\frac{c_3}{M^4}(g^{\mu\rho}g^{\nu\sigma}{F}_{\mu\nu}{{F}}_{\rho\sigma})\left(\frac{1}{\sqrt{-g}}\epsilon^{\mu\nu\rho\sigma}{F}_{\mu\nu}{F}_{\rho\sigma}\right)\bigg)\notag
\\
&=\int \sqrt{-g'}d^4x' \bigg(-\frac{1}{4}{g'}^{\mu\rho}{g'}^{\nu\sigma}{F}'_{\mu\nu}{F}'_{\rho\sigma} +\frac{c_1}{M^4} ({g'}^{\mu\rho}{g'}^{\nu\sigma}{F}'_{\mu\nu}{F}'_{\rho\sigma} )^2+\frac{c_2}{M^4} \left(\frac{1}{\sqrt{-g'}}{\epsilon'}^{\mu\nu\rho\sigma}{F}'_{\mu\nu}{F}'_{\rho\sigma}\right)^2\notag
\\
&+\frac{c_3}{M^4}({g'}^{\mu\rho}{g'}^{\nu\sigma}{F}'_{\mu\nu}{F}'_{\rho\sigma})\left(\frac{1}{\sqrt{-g'}}{\epsilon'}^{\mu\nu\rho\sigma}{F}'_{\mu\nu}{F}'_{\rho\sigma}\right)\bigg)\notag
\\
&\to -i\int \sqrt{g_{\rm E}}(d^4x)_{\rm E} \bigg(-\frac{1}{4}{g}_{{\rm E},{IK}}{g}_{{\rm E},{JL}}{F}_{{\rm E},IJ}{F}_{{\rm E},KL} +\frac{c_1}{M^4} ({g}_{{\rm E},{IK}}{g}_{{\rm E},{JL}}{F}_{{\rm E},IJ}{F}_{{\rm E},KL} )^2\notag
\\
&-\frac{c_2}{M^4} \left(\frac{1}{\sqrt{g_{\rm E}}}{\epsilon}_{{\rm E},{IJKL}}{F}_{{\rm E},IJ}{F}_{{\rm E},KL}\right)^2-i\frac{c_3}{M^4}({g}_{{\rm E},{IK}}{g}_{{\rm E},{JL}}{F}_{{\rm E},IJ}{F}_{{\rm E},KL})\left(\frac{1}{\sqrt{g_{\rm E}}}{\epsilon}_{{\rm E},{IJKL}}{F}_{{\rm E},IJ}{F}_{{\rm E},KL}\right) \bigg)\notag
\\
&= -i\int (d^4x)_{\rm E} \bigg(-\frac{1}{4}{g}_{{\rm E},{IK}}{g}_{{\rm E},{JL}}{F}_{{\rm E},IJ}{F}_{{\rm E},KL} +\frac{c_1}{M^4} ({g}_{{\rm E},{IK}}{g}_{{\rm E},{JL}}{F}_{{\rm E},IJ}{F}_{{\rm E},KL} )^2\notag
\\
&-\frac{c_2}{M^4} \left({\epsilon}_{{\rm E},{IJKL}}{F}_{{\rm E},IJ}{F}_{{\rm E},KL}\right)^2-i\frac{c_3}{M^4}({g}_{{\rm E},{IK}}{g}_{{\rm E},{JL}}{F}_{{\rm E},IJ}{F}_{{\rm E},KL})({\epsilon}_{{\rm E},{IJKL}}{F}_{{\rm E},IJ}{F}_{{\rm E},KL})\bigg)\notag
\\
&=i\int (d^4 x)_{\rm E} \bigg(\frac{1}{4}{F}^{\mu\nu}{F}_{\mu\nu} -\frac{c_1}{M^4} ({F}^{\mu\nu}{F}_{\mu\nu})^2-\frac{c_2}{M^4} (\epsilon^{\mu\nu\rho\sigma}{F}_{\mu\nu}{{F}}_{\rho\sigma})^2+\frac{c_3}{M^4}({F}^{\mu\nu}{F}_{\mu\nu})(\epsilon^{\mu\nu\rho\sigma}{F}_{\mu\nu}{{F}}_{\rho\sigma})\bigg)\notag
\\
&=i I_{e}^{(\rm E)}[{A}],
\end{align}
where we used following relation:
\begin{align}
    {F}^{E}_{\mu\nu}=\frac{\partial x^{\alpha}}{\partial {x'}^{\mu}}\frac{\partial x^{\beta}}{\partial {x'}^{\nu}}{F}_{\alpha\beta}\Bigg|_{\lambda=i}=\begin{pmatrix}
    0 & -i {F}_{01} & -i {F}_{02} & -i {F}_{03}
    \\
    -i {F}_{10} & 0 & {F}_{12} & {F}_{13}
    \\
     -i {F}_{20} & {F}_{21} & 0 & {F}_{23}
     \\
      -i {F}_{30} & {F}_{31} & {F}_{32} & 0
     \end{pmatrix}.
\end{align}

\end{itemize}

\section{Relative entropy under field redefinition}
\label{sec:redef}
To check the invariant formulation under the field redefinition, consider a target theory described by the following action in Euclidean space.
\begin{align}
    I_{\rm T}\equiv\int (d^4 x)_{\rm E}\left(\frac{1}{4}F_{\mu\nu}F^{\mu\nu}+m_A^2\phi_A^2 -\frac{1}{M} \phi_A F_{\rho\sigma}F^{\rho\sigma} \right),
\end{align}
where $\phi_A$ is an auxiliary field.
We define the non-interacting and interacting terms as follows:
\begin{align}
    &I_0\equiv\int (d^4 x)_{\rm E} \left(\frac{1}{4}F_{\mu\nu}F^{\mu\nu}+m_A^2\phi_A^2 \right),
    \\
    &I_{\rm I}\equiv-\int (d^4 x)_{\rm E}\left(\frac{1}{M} \phi_A F_{\rho\sigma}F^{\rho\sigma} \right).
\end{align}
By defining an action as $I_g\equiv I_0+g\cdot I_{\rm I}$ with the parameter $g$, we obtain as follows:
\begin{align}
    &Z_g[A]\equiv \int d[\phi_A] e^{-I_g},
    \\
    &W_g[A]\equiv -\ln Z_g[A]=\int (d^4 x)_{\rm E} \left(\frac{1}{4}F_{\mu\nu} F^{\mu\nu}-g^2\cdot \frac{1}{4 m_A^2 M^2}(F_{\rho\sigma}F^{\rho\sigma})^2 \right).
\end{align}
The expectation value of the interaction $I_{\rm I}$ is calculated as
\begin{align}
    {\langle I_{\rm I}\rangle}_{g=0}=\left(\frac{\partial W_g}{\partial g}\right)_{g=0}=\int d[\phi_A] P_0 I_{\rm I}=0.
\end{align}
Therefore, the relative entropy is calculated as
\begin{align}
    S(P_0||P_g) &=W_0[A]-W_g[A]+g {\langle I_{\rm I}\rangle}_{g=0}\notag
    \\
    &=W_0[A]-W_g[A]\notag
    \\
    &=g^2\cdot\int (d^4 x)_{\rm E} \left( \frac{1}{4 m_A^2 M^2}(F_{\rho\sigma}F^{\rho\sigma})^2 \right)\geq 0,\label{eq:relex1}
\end{align}
where $P_0\equiv e^{-I_0}/Z_0[A]$ and $P_g\equiv e^{-I_g}/Z_g[A]$ are used.
Here, consider the following field redefinition:
\begin{align}
    \phi_A\to \phi_A+g\cdot\frac{1}{2 m_A^2 M}F_{\rho\sigma}F^{\rho\sigma}.
\end{align}
Under this field redefinition, the actions are rewritten as
\begin{align}
    I_0&\to I'_0=\int (d^4x)_{\rm E} \left(\frac{1}{4}F_{\mu\nu}F^{\mu\nu}+m_A^2 \phi_A^2+g\cdot  \frac{1}{M}\phi_A F_{\rho\sigma}F^{\rho\sigma} +g^2 \cdot \frac{1}{4 m_A^2 M^2}( F_{\rho\sigma}F^{\rho\sigma})^2 \right), \\
    g\cdot I_{\rm I} &\to g\cdot I'_I= g\cdot I_{\rm I} -g^2\cdot \int (d^4 x)_E \frac{1}{2m_A^2 M^2}(F_{\rho\sigma}F^{\rho\sigma})^2,\label{eq:fireIN}
    \\
    I_g &\to I'_g=\int (d^4 x)_E \left(\frac{1}{4}F_{\mu\nu}F^{\mu\nu}+m_A^2 \phi_A^2 -g^2 \cdot \frac{1}{4 m_A^2 M^2}( F_{\rho\sigma}F^{\rho\sigma})^2 \right).
\end{align}
Then, the relative entropy is also rewritten as
\begin{align}
    S(P_0||P_g)\to S(P'_0||P'_g) 
\end{align}
where $P'_0\equiv e^{-I'_0}/Z'_0[A]$ and $P'_g\equiv e^{-I'_g}/Z'_g[A]$ with $Z'_0[A]\equiv\int d[\phi_A] e^{-I'_0}$ and $Z'_g[A]\equiv\int d[\phi_A] e^{-I'_g}$.
Then, the relative entropy $ S(P'_0||P'_g)$ is calculated as
\begin{align}
    S(P'_0||P'_g)&=\int d[\phi_A] \left(P'_0\ln P'_0-P'_0\ln P'_g \right)
    \\
    &=-\ln Z'_0[A]+\ln Z'_g[A]+ \int d[\phi_A] P'_0 \left(I'_g -I'_0 \right)\notag
    \\
    &=W'_0[A]- W'_g[A]-g^2 \cdot \frac{1}{2 m_A^2 M^2}(F_{\rho\sigma}F^{\rho\sigma})^2-g\cdot \int d[\phi_A] P'_0 \int (d^4 x)_E \frac{1}{M} \phi_A (F_{\rho\sigma}F^{\rho\sigma})\notag
    \\
    &=W'_0[A]- W'_g[A]\notag
    \\
    &=g^2\cdot\int (d^4 x)_E \left( \frac{1}{4 m_A^2 M^2}(F_{\rho\sigma}F^{\rho\sigma})^2 \right)\geq 0,\label{eq:relex2}
\end{align}
where we used following relations.
\begin{align}
    &\int d[\phi_A] P'_0 \int (d^4 x)_E \frac{1}{M} \phi_A (F_{\rho\sigma}F^{\rho\sigma})=-g \cdot \frac{1}{2 m_A^2 M^2}(F_{\rho\sigma}F^{\rho\sigma})^2,
    \\
    &W'_0[A]=W_0[A],
    \\
    &W'_g[A]=W_g[A].
\end{align}
From Eq.~\eqref{eq:relex1} and \eqref{eq:relex2}, we found that the relative entropy is invariant under the field redefinition.
Therefore, the relative entropy is invariant under the field redefinition once $I_0$ and $I_{\rm I}$ are defined.

\section{Calculation of the second law of thermodynamics}
\label{app:seclaw}
We provide details of the calculation of the Clausius inequality in Eq.~\eqref{eq:clau}.
The target and reference systems given in Eq.~\eqref{eq:secterg} and \eqref{eq:secref} are defined as
\begin{align}
    &\rho_{\rm T}\equiv \rho_{\rm ini},
    \\
    &\rho_{\rm R}\equiv U^{\dagger}\rho_{\rm fin, S}\otimes e^{-\beta H_{\rm B}}/Z_{\rm B}(\beta) U,
\end{align}
respectively.
The relative entropy between $\rho_{\rm T}$ and $\rho_{\rm R}$ is given by
\begin{align}
    S(\rho_{\rm T}||\rho_{\rm R})&={\rm Tr}[\rho_{\rm T}\ln\rho_{\rm T}-\rho_{\rm T}\ln \rho_{\rm R}]
    \\
    &={\rm Tr}[\rho_{\rm T}\ln \rho_{\rm T}]-{\rm Tr}[\rho_{\rm R}\ln \rho_{\rm R}]+{\rm Tr}[\rho_{\rm R}\ln \rho_{\rm R}]-{\rm Tr}[\rho_{\rm T}\ln \rho_{\rm R}].
\end{align}
We calculate calculate each term on the right-hand side as follows:
\begin{align}
    {\rm Tr}[\rho_{\rm T}\ln \rho_{\rm T}]&={\rm Tr}\left[
    \rho_{\rm ini,S}\otimes e^{-\beta H_{\rm B}}/Z_{\rm B}(\beta)\ln\rho_{\rm ini,S}\otimes e^{-\beta H_{\rm B}}/Z_{\rm B}(\beta)
    \right]\notag
    \\
    &={\rm Tr}_{\rm S}\left[
    \rho_{\rm ini,S}\ln\rho_{\rm ini,S}
    \right]+
    {\rm Tr}_{\rm B}\left[
     e^{-\beta H_{\rm B}}/Z_{\rm B}(\beta)\ln e^{-\beta H_{\rm B}}/Z_{\rm B}(\beta)
    \right],
\\
    {\rm Tr}[\rho_{\rm R}\ln \rho_{\rm R}]&={\rm Tr}\left[
    U^{\dagger}\rho_{\rm fin, S}\otimes e^{-\beta H_{\rm B}}/Z_{\rm B}(\beta) U\ln U^{\dagger}\rho_{\rm fin, S}\otimes e^{-\beta H_{\rm B}}/Z_{\rm B}(\beta) U
    \right]\notag
    \\
    &={\rm Tr}\left[
    \rho_{\rm fin, S}\otimes e^{-\beta H_{\rm B}}/Z_{\rm B}(\beta) \ln \rho_{\rm fin, S}\otimes e^{-\beta H_{\rm B}}/Z_{\rm B}(\beta) 
    \right]\notag
    \\
    &={\rm Tr}_{\rm S}\left[
    \rho_{\rm fin, S}\ln\rho_{\rm fin,S}
    \right]+
    {\rm Tr}_{\rm B}\left[
     e^{-\beta H_{\rm B}}/Z_{\rm B}(\beta)\ln e^{-\beta H_{\rm B}}/Z_{\rm B}(\beta)
    \right]\notag
    \\
    &={\rm Tr}_{\rm S}\left[
    \rho_{\rm fin, S}\ln\rho_{\rm fin,S}
    \right]-\ln Z_{\rm B}(\beta)-\beta\cdot {\rm Tr}\left[ H_{\rm B}\rho_{\rm ini,S}\otimes e^{-\beta H_{\rm B}}/Z_{\rm B}(\beta)\right],
\\
    {\rm Tr}[\rho_{\rm T}\ln \rho_{\rm R}]&={\rm Tr}[ \rho_{\rm ini,S}\otimes e^{-\beta H_{\rm B}}/Z_{\rm B}(\beta)\ln U^{\dagger} \rho_{\rm fin,S}\otimes e^{-\beta H_{\rm B}}/Z_{\rm B}(\beta)U]\notag
    \\
    &={\rm Tr}[ U\rho_{\rm ini,S}\otimes e^{-\beta H_{\rm B}}/Z_{\rm B}(\beta)U^{\dagger}\ln  \rho_{\rm fin,S}\otimes e^{-\beta H_{\rm B}}/Z_{\rm B}(\beta)]\notag
    \\
    &={\rm Tr}[ U\rho_{\rm ini,S}\otimes e^{-\beta H_{\rm B}}/Z_{\rm B}(\beta)U^{\dagger}\ln  \rho_{\rm fin,S}]+{\rm Tr}[ U\rho_{\rm S}\otimes e^{-\beta H_{\rm B}}/Z_{\rm B}(\beta)U^{\dagger}\ln  e^{-\beta H_{\rm B}}/Z_{\rm B}(\beta)]\notag
    \\
    &={\rm Tr}_{\rm S}[ \rho_{\rm fin,S}\ln  \rho_{\rm fin,S}]+{\rm Tr}[ U\rho_{\rm ini,S}\otimes e^{-\beta H_{\rm B}}/Z_{\rm B}(\beta)U^{\dagger}\ln  e^{-\beta H_{\rm B}}/Z_{\rm B}(\beta)]\notag
    \\
    &={\rm Tr}_{\rm S}[ \rho_{\rm fin,S}\ln  \rho_{\rm fin,S}]-\ln Z_{\rm B}(\beta)-\beta\cdot {\rm Tr}\left[U^{\dagger}H_{\rm B} U \rho_{\rm ini,S}\otimes e^{-\beta H_{\rm B}}/Z_{\rm B}(\beta) \right].
\end{align}
Then we obtain the Clausius inequality from the non-negativity of the relative entropy:
\begin{align}
    S(\rho_{\rm T}||\rho_{\rm R})&={\rm Tr}_{\rm S}[\rho_{\rm ini,S}\ln\rho_{\rm ini,S}]-{\rm Tr}_{\rm S}[\rho_{\rm fin, S}\ln\rho_{\rm fin,S}]-\beta\cdot \left({\rm Tr}\left[H_{\rm B} \rho_{\rm ini} \right]-{\rm Tr}\left[H_{\rm B}\rho_{\rm fin} \right]\right)\label{eq:secthir}
    \\
    &=S(\rho_{\rm fin,S})-S(\rho_{\rm ini,S})-\beta\cdot Q\geq 0,
\end{align}
where $S(\rho)\equiv -{\rm Tr}_{\rm S}[\rho\ln \rho]$ denotes the von Neumann entropy of S, and $Q\equiv  {\rm Tr}\left[H_{\rm B} \rho_{\rm ini} \right]-{\rm Tr}\left[H_{\rm B} \rho_{\rm fin} \right]$ is the heat exchange between S and B.

\section{Einstein-Maxwell theory with higher-derivative operators under field redefinition}
\label{app:EMred}
We explain the field redefinitions to derive Eq.~\eqref{eq:EMeff} following the procedures and notations of Ref.~\cite{Cheung:2018cwt}.
The higher-derivative operators up to the four derivative terms of the Einstein-Maxwell theory are expressed as
\begin{align}
    \mathcal{L}&=\frac{M^2_{\rm Pl}}{2}R -\frac{1}{4}F_{\mu\nu}F^{\mu\nu}\notag
    \\
    &+c_1 R^2 +c_2 R_{\mu\nu}R^{\mu\nu}+c_3 R_{\mu\nu\rho\sigma}R^{\mu\nu\rho\sigma}\notag
    \\
    &+c_4 R F_{\mu\nu}F^{\mu\nu} +c_5 R_{\mu\nu} F^{\mu\rho}F^{\nu}~_{\rho} +c_6 R_{\mu\nu\rho\sigma} F^{\mu\nu}F^{\rho\sigma},\notag
    \\
    &+c_7 F_{\mu\nu}F^{\mu\nu}F_{\rho\sigma}F^{\rho\sigma} +c_8 F_{\mu\nu}F^{\nu\rho} F_{\rho\sigma}F^{\sigma\mu},\label{eqEMapp1}
\end{align}
where terms involving $\nabla_{\rho}F_{\mu\nu}$ or $\nabla_{\mu}F^{\mu\nu}$ vanish. 
Also, the Gauss-Bonnet combination, i.e., $R_{\mu\nu\rho\sigma}R^{\mu\nu\rho\sigma}-4 R_{\mu\nu}R^{\mu\nu}+R^2$, is a total derivative and vanishes for the extremal black hole in four dimensions.
Thus, in four dimensions, Eq.~\eqref{eqEMapp1} is expressed as
\begin{align}
    \mathcal{L}&=\frac{M^2_{\rm Pl}}{2}R -\frac{1}{4}F_{\mu\nu}F^{\mu\nu}\notag
    \\
    &+(c_1-c_3) R^2 +(c_2+4c_3) R_{\mu\nu}R^{\mu\nu}\notag
    \\
    &+c_4 R F_{\mu\nu}F^{\mu\nu} +c_5 R_{\mu\nu} F^{\mu\rho}F^{\nu}~_{\rho} +c_6 R_{\mu\nu\rho\sigma} F^{\mu\nu}F^{\rho\sigma}\notag
    \\
    &+c_7 F_{\mu\nu}F^{\mu\nu}F_{\rho\sigma}F^{\rho\sigma} +c_8 F_{\mu\nu}F^{\nu\rho} F_{\rho\sigma}F^{\sigma\mu},
    \\
    &=\frac{M^2_{\rm Pl}}{2}R -\frac{1}{4}F_{\mu\nu}F^{\mu\nu}\notag
    \\
    &+c_{13} R^2 +c_{23} R_{\mu\nu}R^{\mu\nu}\notag
    \\
    &+c_4 R F_{\mu\nu}F^{\mu\nu} +c_5 R_{\mu\nu} F^{\mu\rho}F^{\nu}~_{\rho} +c_6 R_{\mu\nu\rho\sigma} F^{\mu\nu}F^{\rho\sigma}\notag
    \\
    &+\left(c_7+\frac{c_8}{2}\right) (F_{\mu\nu}F^{\mu\nu})^2 +\frac{c_8}{4} (F_{\mu\nu}\widetilde{F}^{\mu\nu})^2,\label{eq:EMapp2}
\end{align}
where $c_{13}\equiv c_1-c_3$, $c_{23}\equiv c_2+4c_3$, and we used
\begin{align}
    F_{\mu\nu}F^{\nu\rho} F_{\rho\sigma}F^{\sigma\mu}=\frac{1}{2}(F_{\mu\nu}F^{\mu\nu})^2+\frac{1}{4}(F_{\mu\nu}\widetilde{F}^{\mu\nu})^2.
\end{align}
Consider a field redefinition of $g_{\mu\nu}$~\cite{Cheung:2018cwt},
\begin{align}
    g_{\mu\nu}\to g_{\mu\nu} +\delta g_{\mu\nu}, 
\end{align}
where
\begin{align}
    \delta g_{\mu\nu} =r_1 R_{\mu\nu} +r_2 g_{\mu\nu} R +r_3 M_{\rm Pl}^{-2} F_{\mu\rho}F_{\nu}~^{\rho} +r_4 M_{\rm Pl}^{-2} g_{\mu\nu} F_{\rho\sigma}F^{\rho\sigma},
\end{align}
with a set of four constants $r_i$.
Under this field redefinition, the coefficients of the higher-derivative operator in four dimensions are shifted as follows:
\begin{align}
    &c_{13}\to c_{13} -\frac{M_{\rm Pl}^2}{4}r_1 -\frac{M_{\rm Pl}^2}{2} r_2,
    \\
    &c_{23}\to c_{23} +\frac{M_{\rm Pl}^2}{2}r_1,
    \\
    &c_4\to c_4 +\frac{1}{8}r_1 -\frac{1}{4}r_3-\frac{1}{2}r_4,
    \\
    &c_5\to c_5-\frac{1}{2}r_1 +\frac{1}{2}r_3,
    \\
    &c_6 \to c_6,
    \\
    &c_7\to c_7 +\frac{M_{\rm Pl}^{-2}}{8}r_3,
    \\
    &c_8\to c_8 -\frac{M_{\rm Pl}^{-2}}{2}r_3.
\end{align}
Equation~\eqref{eq:EMeff} is derived by choosing the set of four constants as follows:
\begin{align}
    &r_1=-\frac{2}{M^2_{\rm Pl}}c_{23},
    \\
    &r_2=\frac{2}{M^2_{\rm Pl}}\left(c_{13}+\frac{1}{2}c_{23} \right),
    \\
    &r_3=-2 \left(c_5 +\frac{1}{M^2_{\rm Pl}}c_{23}\right),
    \\
    &r_4=2 \left(\frac{1}{4M^2_{\rm Pl}}c_{23}+c_4 +\frac{1}{2}c_5 \right).
\end{align}
After the above field redefinitions, Eq.~\eqref{eq:EMapp2} is rewritten as follows:
\begin{align}
    \mathcal{L}&=\frac{M^2_{\rm Pl}}{2}R -\frac{1}{4}F_{\mu\nu}F^{\mu\nu}+c_6 R_{\mu\nu\rho\sigma} F^{\mu\nu}F^{\rho\sigma}\notag
    \\
    &+\left(\frac{1}{4M^4_{\rm Pl}} c_{23}+\frac{1}{4 M^2_{\rm Pl}}c_5+\frac{1}{2}(2c_7 +c_8)\right) (F_{\mu\nu}F^{\mu\nu})^2 +\left(\frac{1}{4M_{\rm Pl}^4} c_{23}+\frac{1}{4M^2_{\rm Pl}}c_5+\frac{1}{4}c_8\right) (F_{\mu\nu}\widetilde{F}^{\mu\nu})^2,\notag
    \\
    &=\int d^4 x\sqrt{-g} \bigg(\frac{M_{\rm Pl}^2}{2} R -\frac{1}{4}F_{\mu\nu}F^{\mu\nu}+\frac{\alpha_1}{4 M_{\rm Pl}^4} (F_{\mu\nu}F^{\mu\nu})^2+\frac{\alpha_2}{4 M_{\rm Pl}^4} (F_{\mu\nu} \tilde{F}^{\mu\nu})^2+\frac{\alpha_3}{2 M_{\rm Pl}^2}F_{\mu\nu}F_{\rho\sigma} R^{\mu\nu\rho\sigma} \bigg),
\end{align}
with
\begin{align}
    &\alpha_1\equiv c_{23} +M^2_{\rm Pl}\cdot c_5 +2 M^4_{\rm Pl}\cdot (2c_7 +c_8),
    \\
    &\alpha_2\equiv  c_{23}+ M^2_{\rm Pl}\cdot c_5 +M^4_{\rm Pl}\cdot c_8,
    \\
    &\alpha_3\equiv 2 M^2_{\rm Pl} \cdot c_6.
\end{align}

\clearpage
\bibliographystyle{JHEP}
\bibliography{entropy.bib}

\end{document}